\documentclass[twocolumn,trackchanges]{aastex62}
\usepackage{amsmath}
\usepackage{xspace}
\usepackage{graphics,graphicx}
\usepackage{natbib}
\usepackage{multirow}
\usepackage{footnote}
\usepackage{float}
\usepackage{longtable}
\usepackage[percent]{overpic}
\usepackage{color}
\usepackage{rotating}

\usepackage{alphalph,etoolbox}
\appto\appendix{
  \renewcommand\thesection{\AlphAlph{\value{section}}}
}

\newcommand{\Msun}{$M_{\odot}$\xspace}
\newcommand{\Lsun}{$L_{\odot}$\xspace}
\newcommand{\msx}{{\it MSX}\xspace}
\newcommand{\spitzer}{{\it Spitzer}\xspace}
\newcommand{\iras}{{\it IRAS}\xspace}
\newcommand{\herschel}{{\it Herschel}\xspace}
\newcommand{\planck}{{\it Planck}\xspace}
\newcommand{\chisqr}{$\chi^2_r$\xspace}
\newcommand\HII{H$\;${\scshape{ii}}\xspace}

\received{XX YY, 2018}
\revised{XX YY, 2018}
\accepted{XX YY, 2018}
\submitjournal{AAS Journals}

\shorttitle{Galactic Massive Star Forming Regions}
\shortauthors{Binder \& Povich}

\begin{document}

\title{A Multiwavelength Look at Galactic Massive Star Forming Regions}


\author{Breanna A. Binder}
\affiliation{Department of Physics \& Astronomy, California State Polytechnic University, 3801 W. Temple Ave., Pomona, CA 91768, USA}

\author{Matthew S. Povich}
\affiliation{Department of Physics \& Astronomy, California State Polytechnic University, 3801 W. Temple Ave., Pomona, CA 91768, USA}

\begin{abstract}
We present a multiwavelength study of 28 Galactic massive star-forming \HII regions. For 17 of these regions, we present new distance measurements based on Gaia DR2 parallaxes. By fitting a multicomponent dust, blackbody, and power-law continuum model to the 3.6~\micron\ through 10~mm spectral energy distributions, we find that ${\sim}34\%$ of Lyman continuum photons emitted by massive stars are absorbed by dust before contributing to the ionization of \HII regions, while ${\sim}68\%$ of the stellar bolometric luminosity is absorbed and reprocessed by dust in the \HII regions and surrounding photodissociation regions. The most luminous, infrared-bright regions that fully sample the upper stellar initial mass function (ionizing photon rates $N_C \ge 10^{50}~{\rm s}^{-1}$ and dust-processed $L_{\rm TIR}\ge 10^{6.8}$~\Lsun) have on average higher percentages of absorbed Lyman continuum photons ($\sim$51\%) and reprocessed starlight ($\sim$82\%) compared to less luminous regions. Luminous \HII regions show lower average PAH fractions than less luminous regions, implying that the strong radiation fields from early-type massive stars are efficient at destroying PAH molecules. On average, the monochromatic luminosities at 8, 24, and 70~\micron\ combined carry 94\% of the dust-reprocessed $L_{\rm TIR}$. $L_{70}$ captures ${\sim}52\%$ of $L_{\rm TIR}$, and is therefore the preferred choice to infer the bolometric luminosity of dusty star-forming regions. We calibrate SFRs based on $L_{24}$ and $L_{70}$ against the Lyman continuum photon rates of the massive stars in each region. Standard extragalactic calibrations of monochromatic SFRs based on population synthesis models are generally consistent with our values.
\end{abstract}


\section{Introduction}
Dust plays several prominent roles in the physics of the interstellar medium (ISM) and star formation. Dust absorbs ultraviolet (UV) radiation emitted by young stars. This absorbed UV radiation is re-emitted at infrared (IR) wavelengths, cooling the dust and also cooling the gas via collisions with dust grains \citep{Draine78, Dwek86, Hollenbach+97}. Studies of nearby star-forming galaxies suggest that, on average, nearly half of emitting starlight is reprocessed by dust \citep{Draine03, Tielens+05}, and the thermal infrared (IR) emission from dust grains dominates the 10--100 \micron\xspace spectral energy distributions (SEDs) of galaxies.

Dust is an important tracer of star formation activity and provides an indirect measure of the star formation rate (SFR) in external galaxies. The bolometric, thermal IR luminosity ($L_{\rm TIR}$) is one of the most reliable tracers of dust-obscured star formation \citep{Kennicutt98, PerezGonzalez+06, Kennicutt+09, Kennicutt+12}. \HII regions are locations of recent, active star formation, where massive OB stars emit ionizing UV photons that can interact with the surrounding dust or escape into ISM. \HII regions are generally composed of multiple different components, including the central ionizing cluster(s) of OB stars, a surrounding photodissociation region (PDR), and the remnants of the giant molecular cloud from which the star cluster(s) formed. \HII regions are frequently seen in close proximity to one another, sometimes so much so that their components overlap. It is therefore common to regard \HII regions more generally as star-forming regions within a galaxy.

In general, star formation in the Milky Way cannot be studied using the same observational techniques as external galaxies \citep{Chomiuk+Povich11}. Sight-lines through the Galactic disk suffer very high extinction, so SFR diagnostics that depend upon optical/UV observational tracers (in particular, H$\alpha$) cannot be applied. Distances to Galactic \HII regions are often highly uncertain, and confusion arises from multiple star forming regions overlapping along a given line of sight. However, mid- and far-IR SFR tracers can be applied to Galactic and extragalactic regions \citep{Calzetti+07,Calzetti+10,Li+10,Li+13,Stephens+14,V+E13,VEH16}, and thermal radio continuum observations can easily resolve individual Galactic regions and be used as a substitute for recombination-line diagnostics to count ionizing photons \citep{Paladini+03}.

The Milky Way offers the unique opportunity to study individual massive star forming regions (MSFRs) resolved over sub-parsec distance scales, where the associated young stellar populations can be directly observed. The Massive Young Star-Forming Complex Study in Infrared and X-ray \citep[MYStIX;][]{Feigelson+13, Broos+13} has characterized hundreds of OB stars in $\sim$20 young Galactic star-forming regions within 4 kpc, and $\sim$100 more obscured OB stars in the MYStIX point-source catalog have recently been found by \citet{Povich+17}. The MSFRs Omnibus X-ray Catalog \citep[MOXC;][]{Townsley+14} produced X-ray point-source catalogs and diffuse emission maps from archival {\it Chandra X-ray Observatory} data on seven MYStIX MSFRs and four additional Galactic MSFRs out to 7 kpc (plus 30 Doradus in the Large Magellanic Cloud). 

To better understand the interplay between massive stars and IR/radio nebular tracers of star formation, we have conducted a study of 29 Galactic MSFRs, with 21 drawn from the MYStIX and MOXC surveys, and nine additional, prominent regions that have similar high-resolution X-ray through mid-IR archival data available. We construct SEDs by performing aperture photometry on data from the {\it Spitzer Space Telescope}, the {\it Midcourse Space Experiment (MSX)}, the {\it Infrared Astronomical Satellite (IRAS)}, the {\it Herschel Space Observatory}, and the \planck satellite. We then fit a multi-component \citet{Draine+Li07} dust, blackbody, and power-law continuum model to the mid-IR through radio SEDs for each region to measure $L_{\rm TIR}$, constrain dust properties, and search for evidence of supernova contamination in the radio continuum. We use MYStIX point-source database of X-ray and IR-detected OB stars, along with supplementary lists of massive stars from the literature for non-MYStIX targets, to predict the ionizing photon rate injected into each region and to calculate the fraction of emitted luminosity that is reprocessed by dust. 

This paper is organized as follows: Section~\ref{section:observations} describes the data sources used in this paper. Section~\ref{section:SED_model_description} describes our SED modeling procedure, while Section~\ref{section:SED_fit_results} summarizes the trends observed in the resulting fits. In Section~\ref{section:reprocessing} we discuss the relationship between the MSFRs and their ionizing stellar clusters. In Section~\ref{section:lum_and_SFRs} we discuss commonly used SFR indicators that rely on monochromatic luminosities, and investigate the differences in predicted SFRs these indicators yield when applied to our sample of MSFRs.

\section{Targets and Observations}\label{section:observations}
We targeted MSFRs that could plausibly appear as compact IR sources to an extragalactic observer studying the Milky Way with a spatial resolution of ${\sim}100$~pc. The essential criteria for selecting MSFRs for inclusion in this study were (1) bright, localized mid-IR nebular emission and (2) availability of high-resolution X-ray through IR imaging data that provide spatially-resolved information about of the nebular morphology and associated stellar populations. We included as many high-luminosity regions hosting rich massive clusters as possible. Some prominent regions, notably the very massive Arches and Quintuplet clusters near the Galactic center, were omitted because any localized nebular emission is indistinguishable from the extremely bright diffuse IR--radio background. Table~\ref{table:targets} lists the basic properties of our MSFR sample, including Galactic coordinates, distance from the Sun, and spectral type(s) of the dominant ionizing star(s). See Appendix~\ref{appendix:region_discussion} for details of the OB stellar population in each region.  While these MSFRs represent a wide range of masses, luminosities, heliocentric distances, and spatial morphologies, we caution that our sample cannot be considered an unbiased sample of Galactic \HII regions or young massive clusters. Our selection criteria favor younger, more nearby, and more massive regions.

Distances to Galactic MSFRs and their associated young clusters have historically been difficult to measure. A handful of regions (e.g., the Orion Nebula, M17, W3, W51A, and NGC~6334) have had distances measured from multi-epoch very long baseline interferometry (VLBI) parallax measurements of maser spots associated with high-mass protostars. Other techniques to estimate distances include fitting of the high-mass main sequence on the HR diagram, utilizing extinction maps of background stars, or deriving distance constraints from the X-ray luminosity function or molecular cloud radial velocities. All of these techniques are subject to considerable uncertainties (e.g., incorrect accounting for binarity, differential absorption for individual stars, or peculiar velocities deviating from Galactic rotation). For example, over a dozen estimates for the distance to the Lagoon Nebula are presented in \citet{Tothill+08}, most of which fall in the range of 1.3--1.8 kpc.  

We searched the Gaia DR2 database \citep{Gaia2018} and found reliable parallax measurements for 193 cataloged OB stars associated with 19 of our MSFRs. Reliable parallaxes had Gaia $g$-band average magnitudes brighter than 15 and typical relative parallax uncertainties ${<}10\%$, with a few exceptions showing larger uncertainties. We computed the uncertainty-weighted mean parallax distance among the OB stars within each MSFR, rejecting ${>}3\sigma$ outliers.

New, reliable parallax distances are available for 17 of the 28 MSFRs. In all cases these distances fall within the (sometimes very wide) range of previously-published distance estimates and provide a significant improvement in precision. In four cases (the Flame Nebula, W40, the Trifid Nebula, and Berkeley~87) these distances are based on a single star, but we nevertheless judge them to be more reliable than previous distance estimates. Cases where the distance appears to have been obtained to greater accuracy for regions without Gaia parallaxes may only represent fewer distance estimates available in the literature. We adopt the distances listed in Table~\ref{table:targets}.

\begin{table}[ht]\setlength{\tabcolsep}{2.5pt}
\centering
\scriptsize
\caption{Massive Star-Forming Region Sample}
\begin{tabular}{ccccc}
\hline \hline
			& ($l, b$)		& \multicolumn{2}{c}{Distance}	& Earliest		\\ \cline{3-4}
Name		& (J2000)		& (kpc)	& Reference		& Sp. Type	\\
 (1)			& (2)			& (3)	 	& (4)		 		& (5)		\\
\hline
Flame Nebula			& 206.5--16.3	& 0.33$\pm$0.01			& 1	& O9V 	\\ 
Orion Nebula			& 209.0--19.4	& 0.41$\pm$0.01			& 1	& O7V 	\\ 
W40					& 028.8+03.5	& 0.49$\pm$0.05 			& 1	& O9.5V	\\ 
RCW~36				& 265.1+01.4	& 1.09$\pm$0.09		 	& 1	& O9V	\\
Lagoon Nebula			& 006.0--01.2	& 1.17$\pm$0.10			& 1	& O4V	\\
Trifid Nebula			& 007.1--00.3	& 1.57$\pm$0.21 			& 1	& O7V	\\
NGC~6334			& 351.1+00.5	& 1.63$\pm$0.16	 		& 1	& O7V	\\
RCW~38				& 268.0--01.1	& 1.7$\pm$0.9				& 2	& O5.5V	\\ 
Eagle Nebula			& 017.0+00.8	& 1.71$\pm$0.18 			& 1	& O5V	\\
Berkeley~87			& 075.7+00.3	& 1.74$\pm$0.09			& 1 	& WC5	\\
NGC~6357			& 353.2+00.9	& 1.78$\pm$0.18 			& 1	 & O3.5III	\\
M17					& 015.1--00.7	& 1.82$\pm$0.16		 	& 1	& O4V 	\\
W3					& 133.9+01.1	& 2.18$\pm$0.12			& 1	& O6V	\\ 
W42					& 025.4--00.2	& 2.2						& 3	 & O5V	\\
W4					& 134.7+00.9	& 2.24$\pm$0.17 			& 1	& O4I 	\\ 
W33					& 012.8--00.2	& 2.40$^{+0.17}_{-0.15}$ 		& 4 & O5I	\\
G333				& 333.6--00.2	& 2.6 					& 5	& O5V	\\
NGC~7538			& 111.5+00.8	& 2.65$^{+0.12}_{-0.11}$ 		& 6 & O5V	\\
Carina Nebula			& 287.7--00.8	& 2.69$\pm$0.40 			& 1	& LBV	\\
NGC~3576			& 291.3--00.8	& 2.77$\pm$0.31 			& 1	& O7.5V	\\ 
G305				& 305.3+00.1	& 3.59$\pm$0.85			& 1	& O5.5I	\\
Westerlund~1 (Wd~1)	& 339.5--00.4	& 3.9$\pm$0.6				& 7	& O9.5I	\\
RCW~49				& 284.3--00.3	& 4.4$\pm$1.0				& 1	& O3V	\\
W51A				& 049.5--00.3	& 5.1$^{+2.9}_{-1.4}$ 		& 8	& O4V	\\
W43					& 030.8--00.0	& 5.5$^{+0.4}_{-0.3}$ 		& 9	& O3.5III	\\
G29.96--0.02			& 030.0--00.0	& 6.2 					& 10	& O5III	\\
NGC~3603			& 291.6--00.5	& 7.0 					& 11	& O3V	\\ 
W49A				& 043.2+00.0	& 11.4$\pm$1.2				& 12	& O3I	\\
\hline \hline
\end{tabular}\label{table:targets}
\tablecomments{MSFRs are listed in order of increasing heliocentric distance. Distance references are: 1: \citet{Gaia2018}, 2: \citet{Schneider+10}, 3: \citet{Blum+00}, 4: \citet{Immer+13}, 5: \citet{Figueredo+05}, 6: \citet{Moscadelli+09}, 7: \citet{Koumpia+Bonanos12}, 8: \citet{Xu+09}, 9: \citet{Zhang+14}, 10: \citet{Russeil+11}, 11: \citet{Harayama+08}, and 12: \citet{Gwinn+92}. For a discussion of the spectral types found in each stellar population (including references), see Appendix~\ref{appendix:region_discussion}.}
\end{table}

Although the MSFRs in our sample are very young ($<$5 Myr), even during this short timescale dramatic, evolutionary changes to the density, temperature, and morphology of the gas and dust can and do occur. The MSFRs in our sample range from highly-embedded \HII regions where the bulk of the stellar luminosity is reprocessed by dust (e.g., the Flame Nebula and W51A) to relatively unobscured \HII regions that have been largely evacuated of dust (e.g., W4, Wd~1). However, age-dating methods for \HII regions and their associated stellar populations are heterogeneous and suffer from large uncertainties, so we will not attempt to place the MSFRs in our sample into an evolutionary sequence. A detailed analysis of the age and SFRs in these regions will be the subject of a forthcoming paper.

\subsection{Observations} \label{subsec:observations}
IR data from \spitzer, \msx, and \iras and radio data from \planck were retrieved using the NASA/IPAC Infrared Science Archive (IRSA)\footnote{See \url{http://irsa.ipac.caltech.edu}}. 

                \subsubsection{\spitzer}
		The majority (23) of our target MSFRs were included in the Galactic Legacy Infrared Mid-Plane Survey Extraordinaire \citep[GLIMPSE;][]{Benjamin+03} or follow-up surveys \citep[GLIMPSE II, GLIMPSE 3D, GLIMPSE 360, or Vela-Carina surveys;][]{Churchwell+09,Zasowski+09,Povich+11a} using the four \spitzer IRAC bands, centered at 3.6, 4.5, 5.8, and 8.0 \micron\xspace\citep{Fazio+04}. High-resolution (1\farcs2 pixels) mosaics were created by the GLIMPSE pipeline\footnote{See \url{http://www.astro.wisc.edu/glimpse/}.} from Basic Calibrated Data (BCD) image frames processed by the \spitzer Science Center (SSC). The GLIMPSE pipeline removes artifacts such as stray light (from all bands), muxbleed (3.6 and 4.5 \micron~bands), and banding (5.8 and 8.0 \micron~bands). The SSC Mopex package \citep{Makovoz+06} is used to mask image artifacts (primarily cosmic rays), and the IPAC Montage packages were used to mosaic the images \citep{Berriman+02}. 

The remaining 7 MSFRs were included in the MYStIX survey, and for these we use mosaic images produced by \citet{Kuhn+13} from publicly-available \spitzer/IRAC archival observations. The majority of our targets were also observed at 24 and 70~\micron\ using the Multiband Infrared Photometer for \spitzer\ \citep[many as part of the MIPSGAL survey;][]{Carey+09}. Because many of our MSFRs are extremely mid-IR bright, the MIPS 24~\micron\ images frequently become saturated, and \herschel offers superior sensitivity and photometric calibration at 70~\micron. For these reasons we do not use MIPS data for this study.
		
		\subsubsection{MSX}
		The Spirit III instrument on board the \msx satellite surveyed the Galactic plane in four IR bands \citep{Price+01}: A (8.28 \micron), C (12.13 \micron), D (14.65 \micron), and E (21.3 \micron). The spatial resolution of Spirit~III was $\sim$18\farcs3. Although its resolution and sensitivity are inferior to that of \spitzer, the absolute flux calibration of \msx, determined in-flight by measuring the fluxes from projectiles fired away from the spacecraft, is reliable to $\sim$1\% (Price et al. 2004). Hence \msx mid-IR fluxes are the most accurate currently available. \msx A images provide the benchmark against which IRAC [8.0] fluxes can be compared \citep{Cohen+07}, and \msx E provides a substitute for saturated \spitzer/MIPS 24~\micron\ images.
		
		\subsubsection{IRAS}
		From January to November 1983 the {\it Infrared Astronomy Satellite} (\iras) mapped 98\% of the sky in four IR bands. These bands have effective wavelengths of 12, 25, 60, and 100 \micron\xspace\citep{Beichman+98}. Although the sensitivity of \iras is comparable to that of \msx, its resolution was much lower, with 1\farcm5 pixels. We use the Improved Reprocessing of the \iras Survey (IRIS) products available through IRSA, which benefits from improved zodiacal light subtraction, calibration and zero level compatible with {\it DIRBE}, and better destriping, particularly in the 12 and 25 \micron\xspace bands \citep{Miville+Lagache05}. All IR images were downsampled to the 1\farcm5 pixel scale of \iras before aperture photometry and analysis was performed.
		
		\subsubsection{Herschel}
		All target MSFRs were included in one or more of the \herschel Hi-Gal, HOBYS, or Gould Belt surveys \citep{Molinari+10,Motte+10,Andre+10}. Far-infrared and submillimeter images were downloaded from the \herschel Space Observatory Science Archive\footnote{See \url{http://archives.esac.esa.int/hsa/whsa/}} using the \herschel Interactive Processing Environment \citep[HIPE;][]{Ott10}. Level 2.5 images were retrieved for both PACS and SPIRE observations. The default JScanam map-maker was selected for all PACS observations. The beam sizes of the 70 \micron, 160 \micron, 250 \micron, 350 \micron, and 500 \micron\xspace images are 5\farcs8, 12\farcs0, 18\farcs0, 25\farcs0, and 37\farcs0, respectively \citep{Griffin+10,Poglitsch+10}.
		
		\subsubsection{Planck}
		We retrieved \planck cut-out images\footnote{See \url{https://irsa.ipac.caltech.edu/applications/planck/}} at 30 GHz, 44 GHz, 70 GHz, and 100 GHz centered on each MSFR. The images were scaled to be four or eight times the FWHM of the effective beam at each frequency to provide an adequate estimate of the background. The effective FWHM in the 30, 44, 70, and 100 GHz channels are 32\farcm41, 27\farcm10, 13\farcm32, and 9\farcm69, respectively. The flux density is estimated by integrating the data in a circular aperture centered at the position of the source (see next section for details).
		
		Many of the MSFRs in our sample have been studied with a variety of radio facilities at different frequencies. We do not use these historical measurements, which frequently give disparate results even for a single region \citep[e.g., as in M17;][]{Povich+07}, in our analysis. Instead, our analysis utilizes the homogeneous \planck data, which had excellent absolute surface brightness calibration and covered the entire sky with sufficient angular resolution to measure the radio continuum for individual star forming regions. Cases where previous measurements of the radio continuum are available for MSFRs are discussed in Appendix~\ref{appendix:region_discussion}.

	\subsection{Aperture Photometry}\label{subsection:aperture_photometry}
To construct multiwavelength SEDs, we performed aperture photometry on the IR and radio images of all MSFRs in our sample. Circular apertures sizes were determined by first extracting the surface brightness profile of each MSFR in the IRAC 8.0 \micron\xspace band, centered on the cluster location given in Table~\ref{table:targets}. The surface brightness as a function of distance was then fit with a decaying exponential function plus a constant background. The ``global'' MSFR aperture was defined by the radius within which 99\% of the source surface brightness was enclosed. Circular background apertures were selected by visual inspection near the outer edge of the source apertures, and were required to possess an average surface brightness consistent with that of the constant background level found in the full surface brightness profile. In a few cases, the spatial extent of the MSFR made extracting IRAC 8 \micron\xspace surface brightness profiles out to the background level impossible, as large segments of the nebula run off the edge of the \spitzer field of view before the background is reached. In other cases, \spitzer 8 \micron\xspace images were missing completely. For these regions, we use the \msx A channel (8.28 \micron) to extract the surface brightness profile and define source apertures.

In Figure~\ref{figure:surface_brightness}, we present an RGB-rendered finding chart of the Orion Nebula, with the extraction aperture superimposed; we additionally show the surface brightness profiles that were extracted using the IRAC 8 \micron\, \msx E channel (21.3 \micron), and PACS 70 \micron\ images. The \msx E and PACS 70 \micron profiles have been renormalized to equal the 8 \micron\ surface brightness at the radius of the extraction aperture. The remaining finding charts and surface brightness profiles are shown in Figures~\ref{figure:surface_brightness}.1 through \ref{figure:surface_brightness}.28 in the online figure set. A detailed analysis of the spatially-resolved SEDs will be presented in a forthcoming paper.

\figsetstart
\figsetnum{1}
\figsettitle{The RGB rendered finding charts and surface brightness profiles for the 28 MSFRs in our sample.}

\figsetgrpstart
\figsetgrpnum{1.1}
\figsetgrptitle{Flame Nebula Finding Chart and Surface Brightness Profile}
\figsetplot{\includegraphics[width=0.31\linewidth,clip,trim=1.2cm 12.4cm 3.5cm 3.7cm]{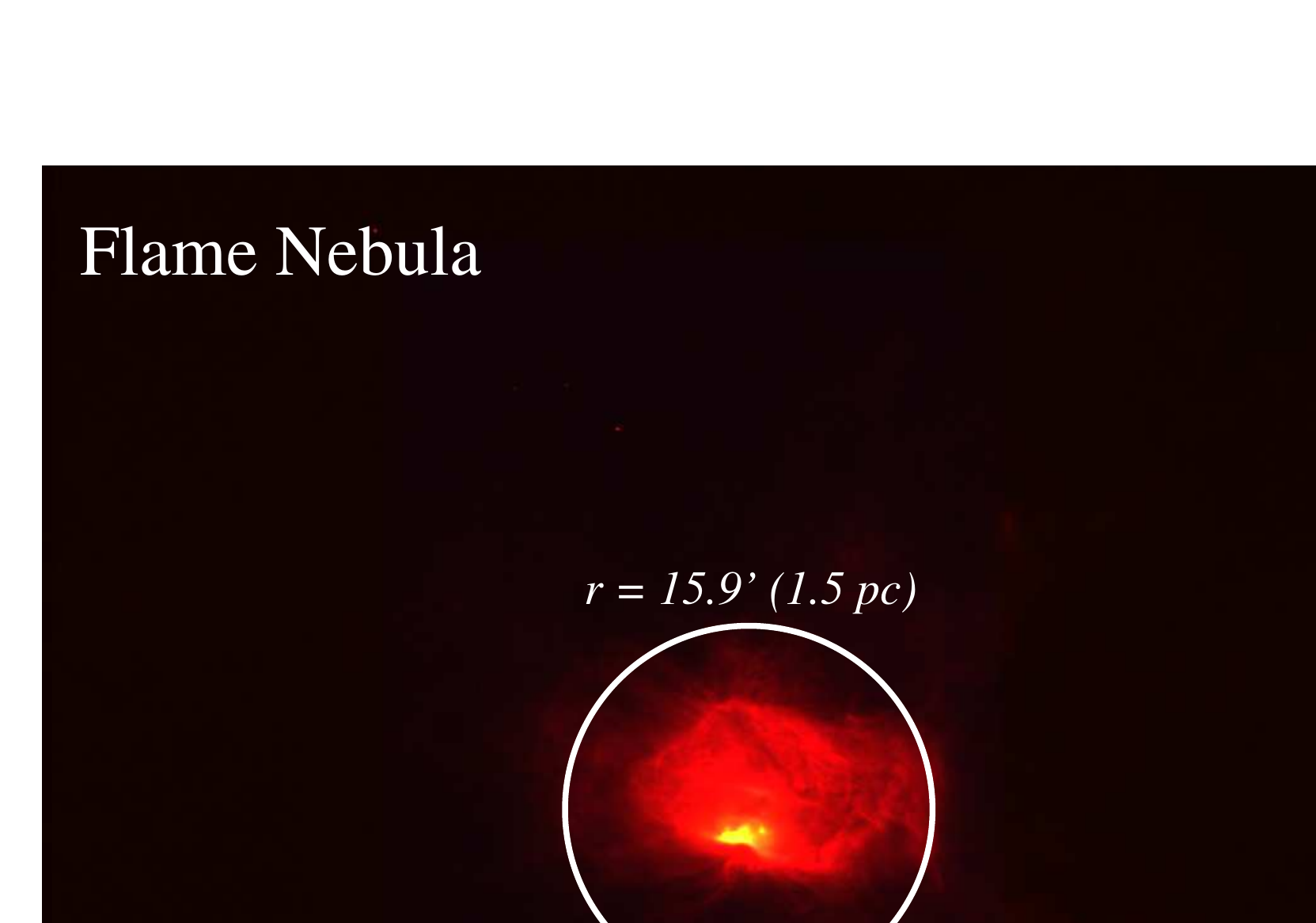}}
\figsetplot{\includegraphics[width=0.31\linewidth,clip,trim=1.2cm 12.4cm 3.5cm 3.7cm]{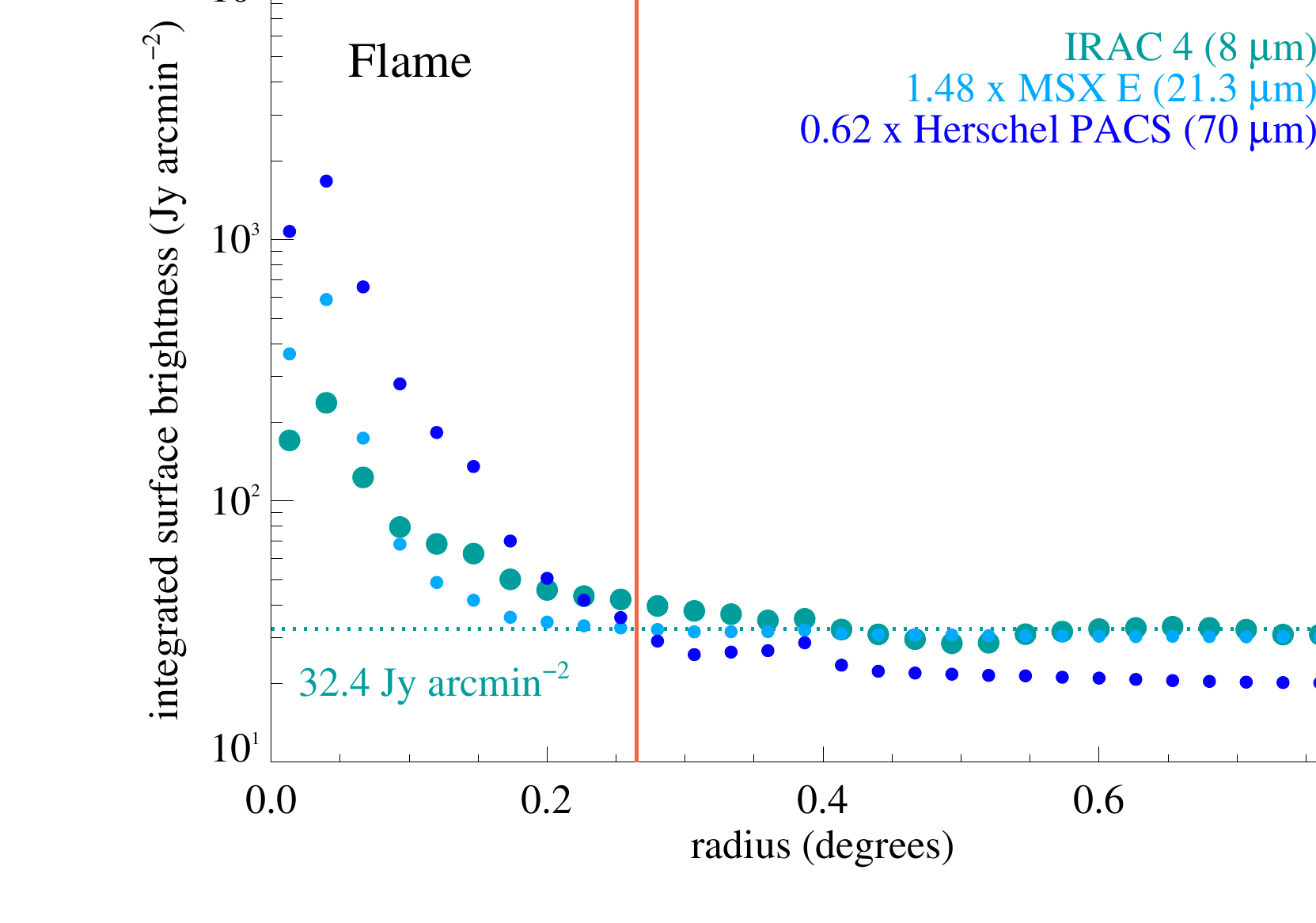}}
\figsetgrpnote{The RGB-rendered finding chart and surface brightness profile for the Flame Nebula. Blue is \spitzer IRAC 4 (8 \micron), green is \msx E (21.3 \micron), and red is \herschel PACS 70 \micron.}
\figsetgrpend

\figsetgrpstart
\figsetgrpnum{1.2}
\figsetgrptitle{W40 Finding Chart and Surface Brightness Profile}
\figsetplot{\includegraphics[width=0.31\linewidth,clip,trim=1.2cm 12.4cm 3.5cm 3.7cm]{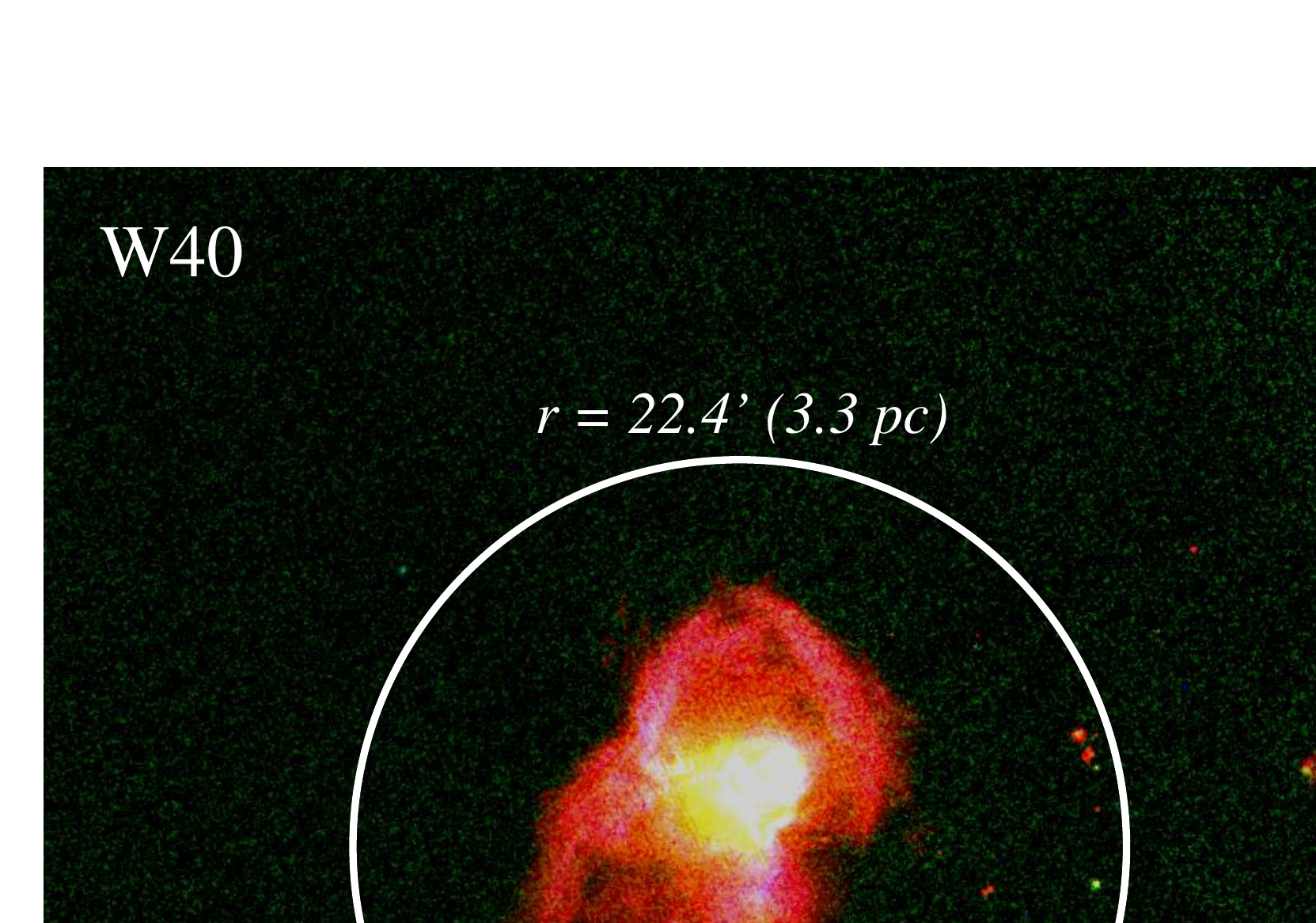}}
\figsetplot{\includegraphics[width=0.31\linewidth,clip,trim=1.2cm 12.4cm 3.5cm 3.7cm]{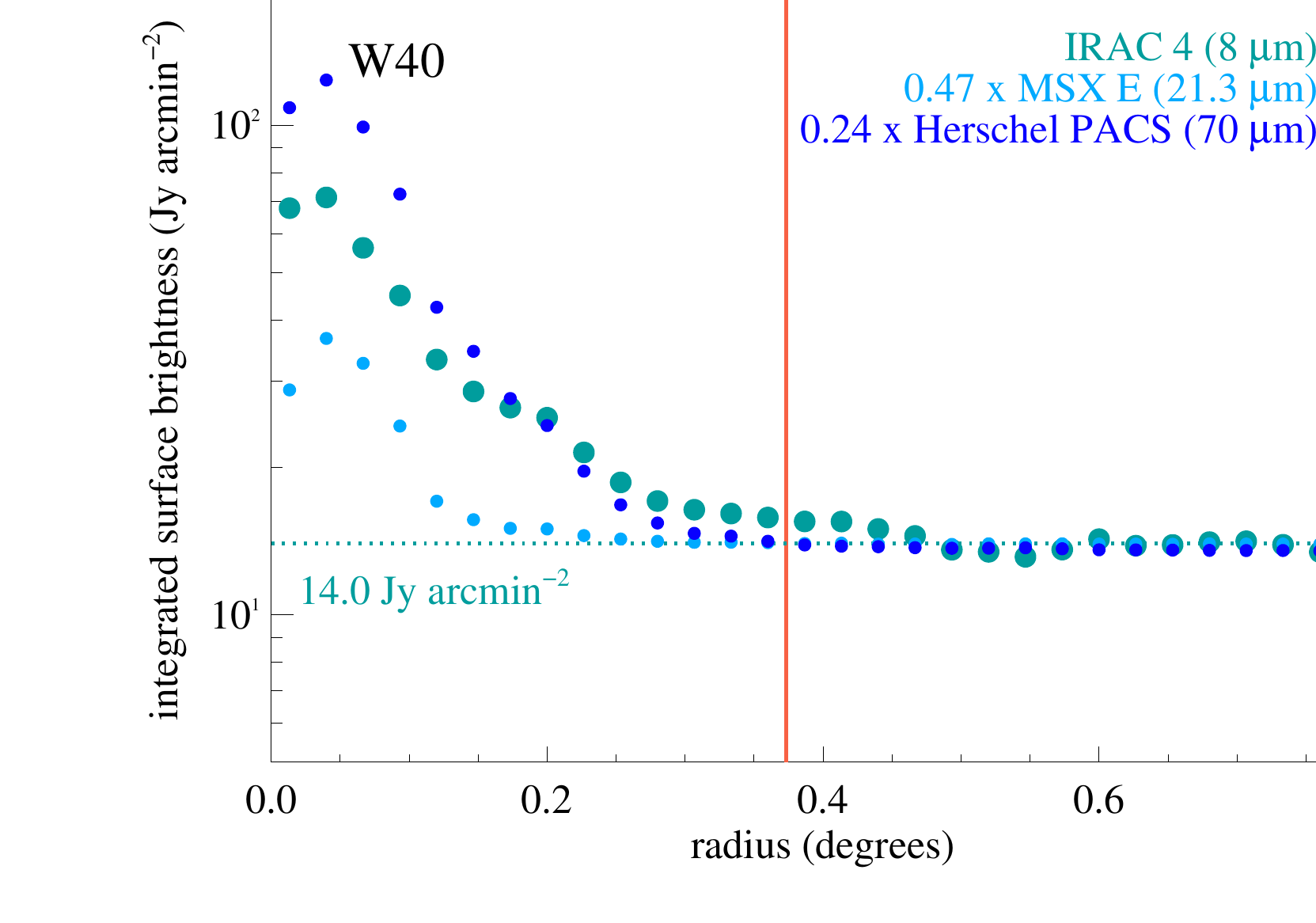}}
\figsetgrpnote{The RGB-rendered finding chart and surface brightness profile for W40. Blue is \spitzer IRAC 4 (8 \micron), green is \msx E (21.3 \micron), and red is \herschel PACS 70 \micron.}
\figsetgrpend

\figsetgrpstart
\figsetgrpnum{1.3}
\figsetgrptitle{Westerlund 1 Finding Chart and Surface Brightness Profile}
\figsetplot{\includegraphics[width=0.31\linewidth,clip,trim=1.2cm 12.4cm 3.5cm 3.7cm]{fc_Westerlund 1.pdf}}
\figsetplot{\includegraphics[width=0.31\linewidth,clip,trim=1.2cm 12.4cm 3.5cm 3.7cm]{sb_Westerlund 1.pdf}}
\figsetgrpnote{The RGB-rendered finding chart and surface brightness profile for Wd 1. Blue is \spitzer IRAC 4 (8 \micron), green is \msx E (21.3 \micron), and red is \herschel PACS 70 \micron.}
\figsetgrpend

\figsetgrpstart
\figsetgrpnum{1.4}
\figsetgrptitle{RCW36 Finding Chart and Surface Brightness Profile}
\figsetplot{\includegraphics[width=0.31\linewidth,clip,trim=1.2cm 12.4cm 3.5cm 3.7cm]{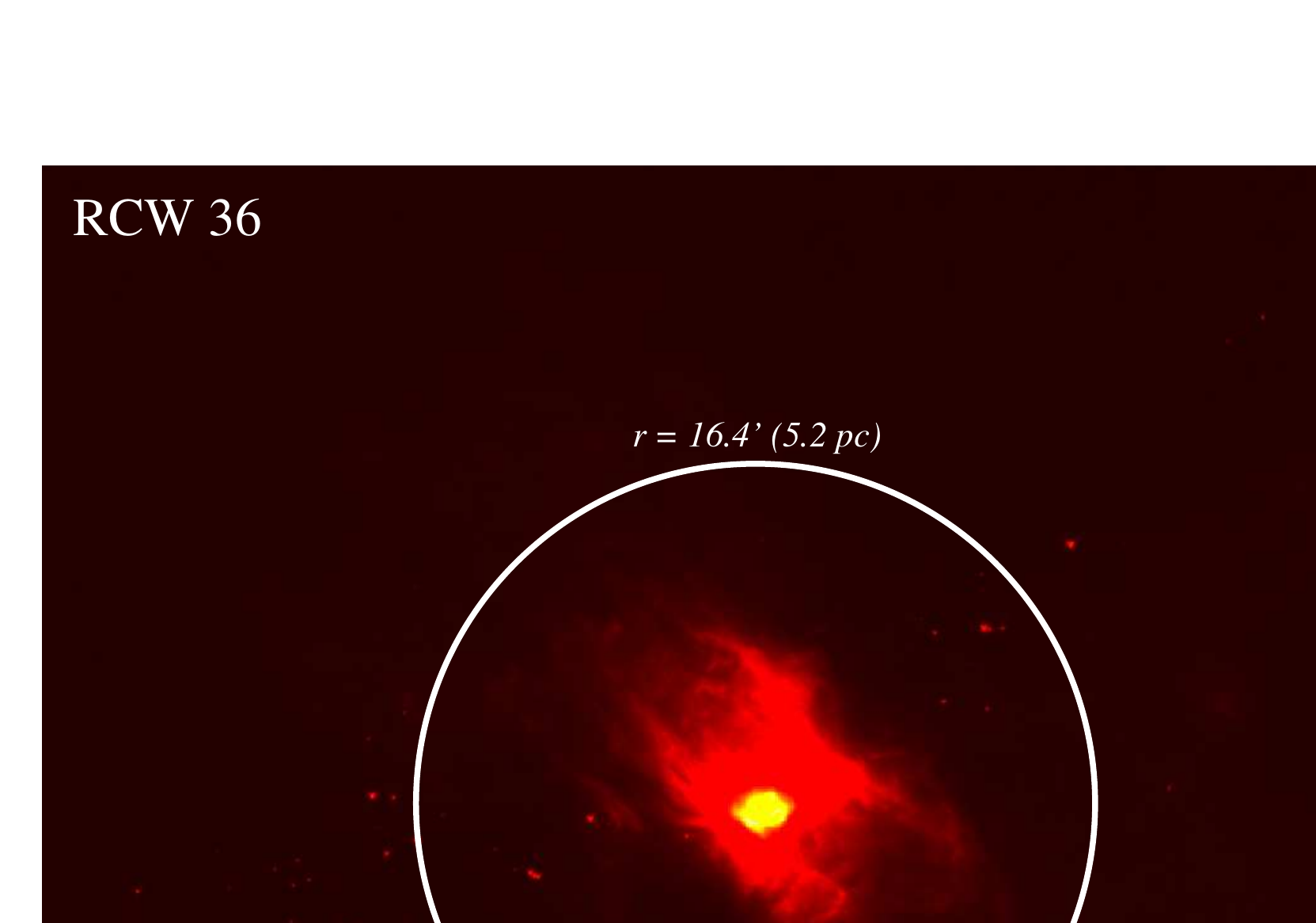}}
\figsetplot{\includegraphics[width=0.31\linewidth,clip,trim=1.2cm 12.4cm 3.5cm 3.7cm]{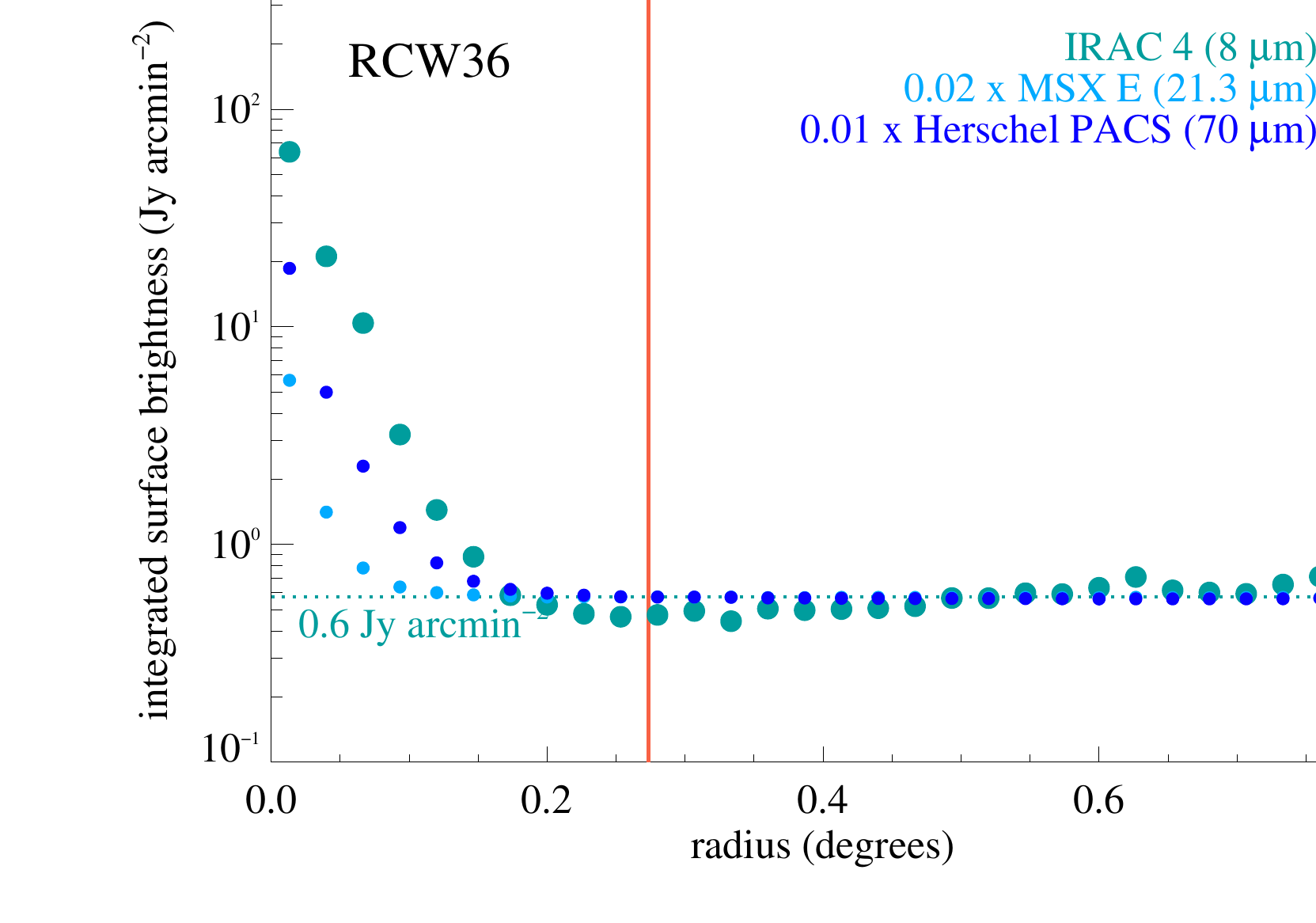}}
\figsetgrpnote{The RGB-rendered finding chart and surface brightness profile for RCW36. Blue is \spitzer IRAC 4 (8 \micron), green is \msx E (21.3 \micron), and red is \herschel PACS 70 \micron.}
\figsetgrpend

\figsetgrpstart
\figsetgrpnum{1.5}
\figsetgrptitle{Berkeley 87 Finding Chart and Surface Brightness Profile}
\figsetplot{\includegraphics[width=0.31\linewidth,clip,trim=1.2cm 12.4cm 3.5cm 3.7cm]{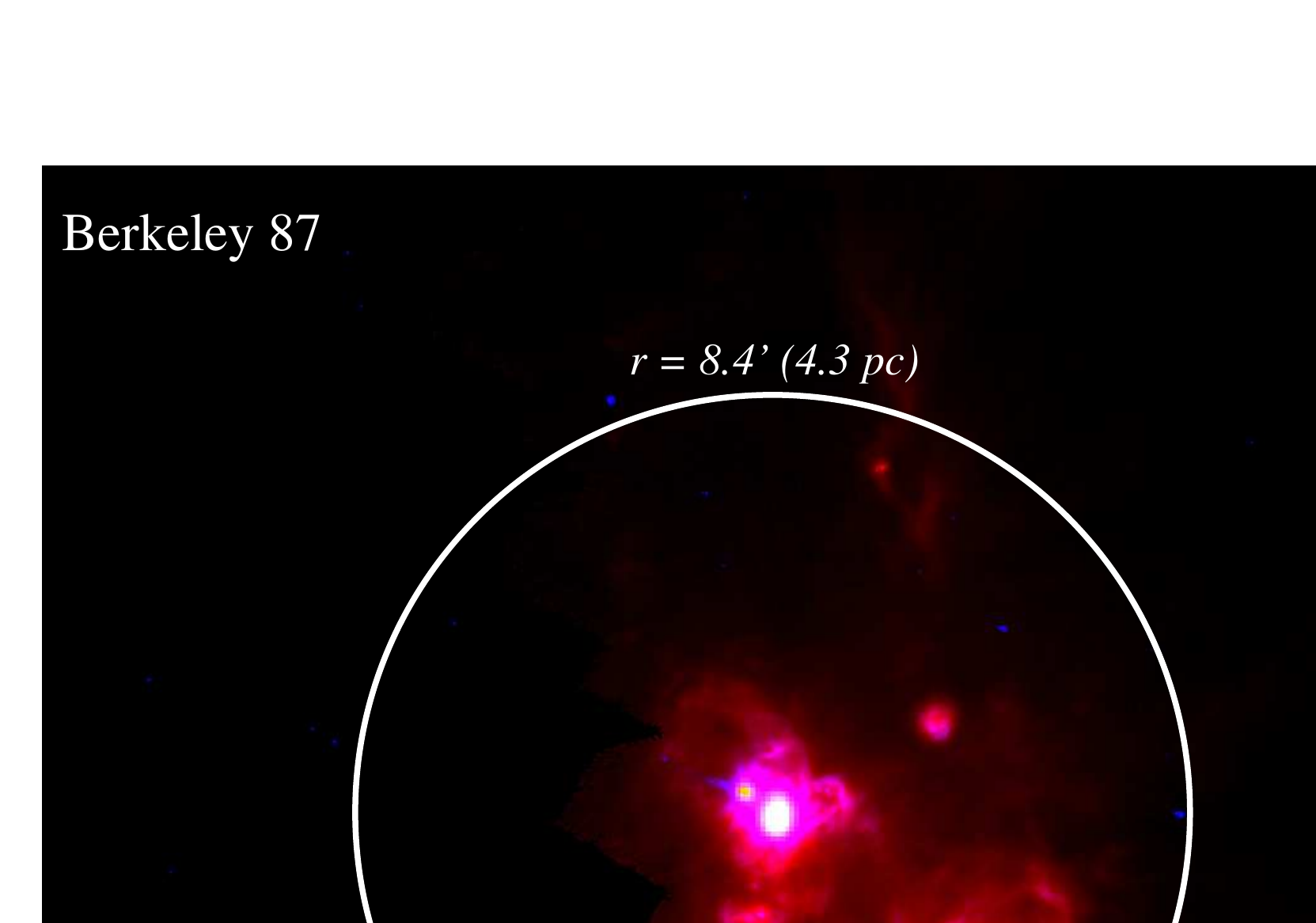}}
\figsetplot{\includegraphics[width=0.31\linewidth,clip,trim=1.2cm 12.4cm 3.5cm 3.7cm]{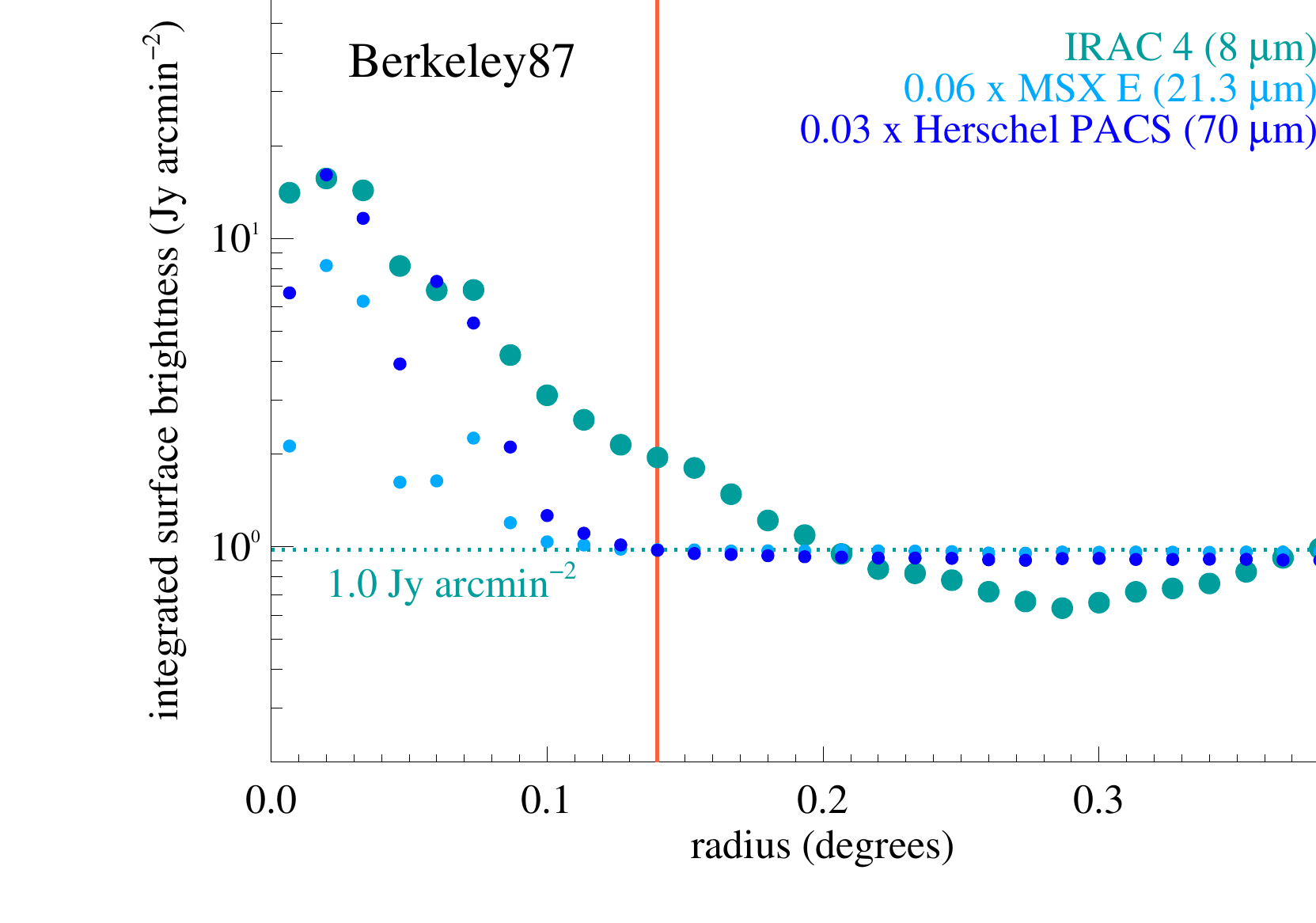}}
\figsetgrpnote{The RGB-rendered finding chart and surface brightness profile for Berkeley 87. Blue is \spitzer IRAC 4 (8 \micron), green is \msx E (21.3 \micron), and red is \herschel PACS 70 \micron.}
\figsetgrpend

\figsetgrpstart
\figsetgrpnum{1.6}
\figsetgrptitle{Orion Nebula Finding Chart and Surface Brightness Profile}
\figsetplot{\includegraphics[width=0.31\linewidth,clip,trim=1.2cm 12.4cm 3.5cm 3.7cm]{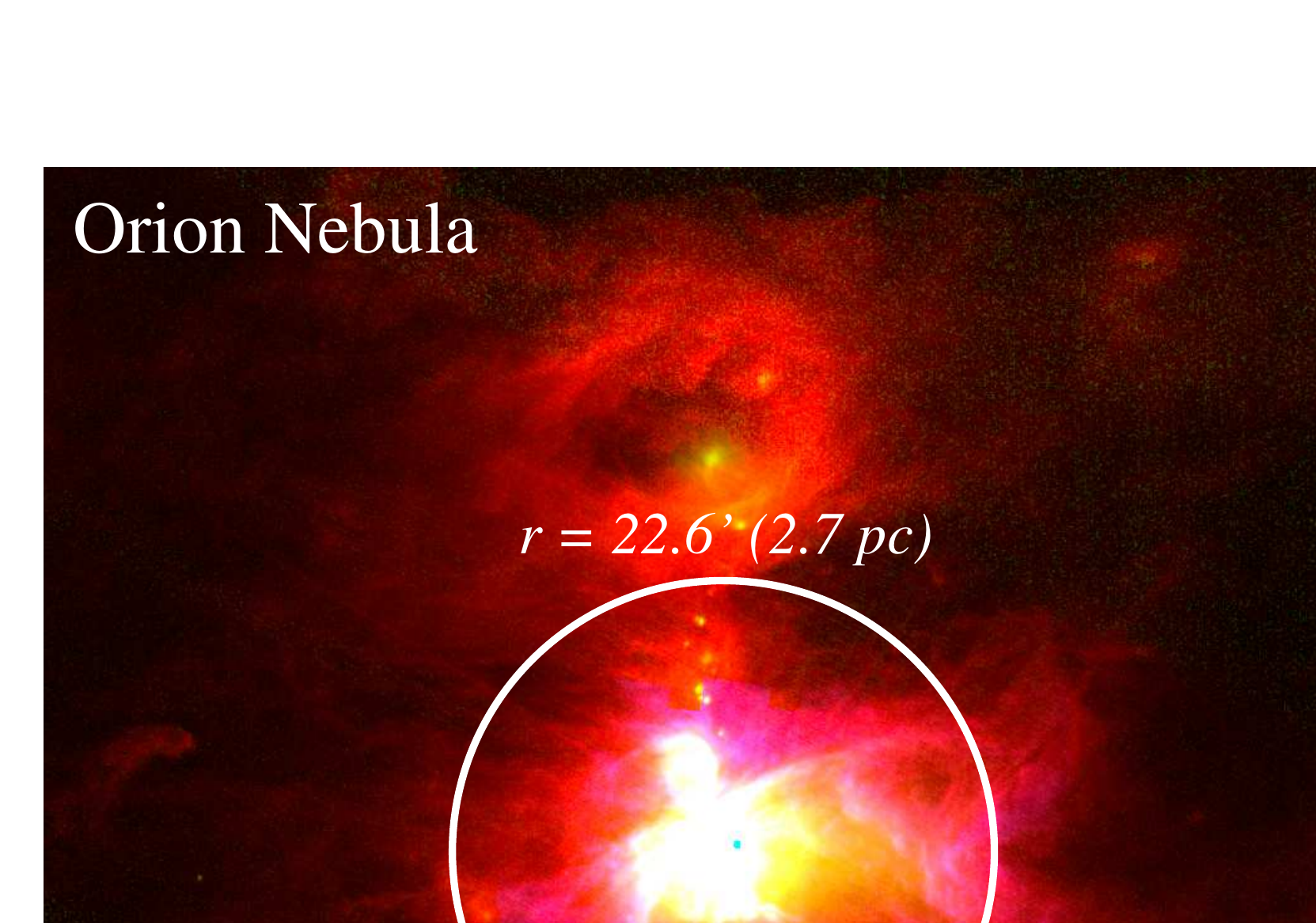}}
\figsetplot{\includegraphics[width=0.31\linewidth,clip,trim=1.2cm 12.4cm 3.5cm 3.7cm]{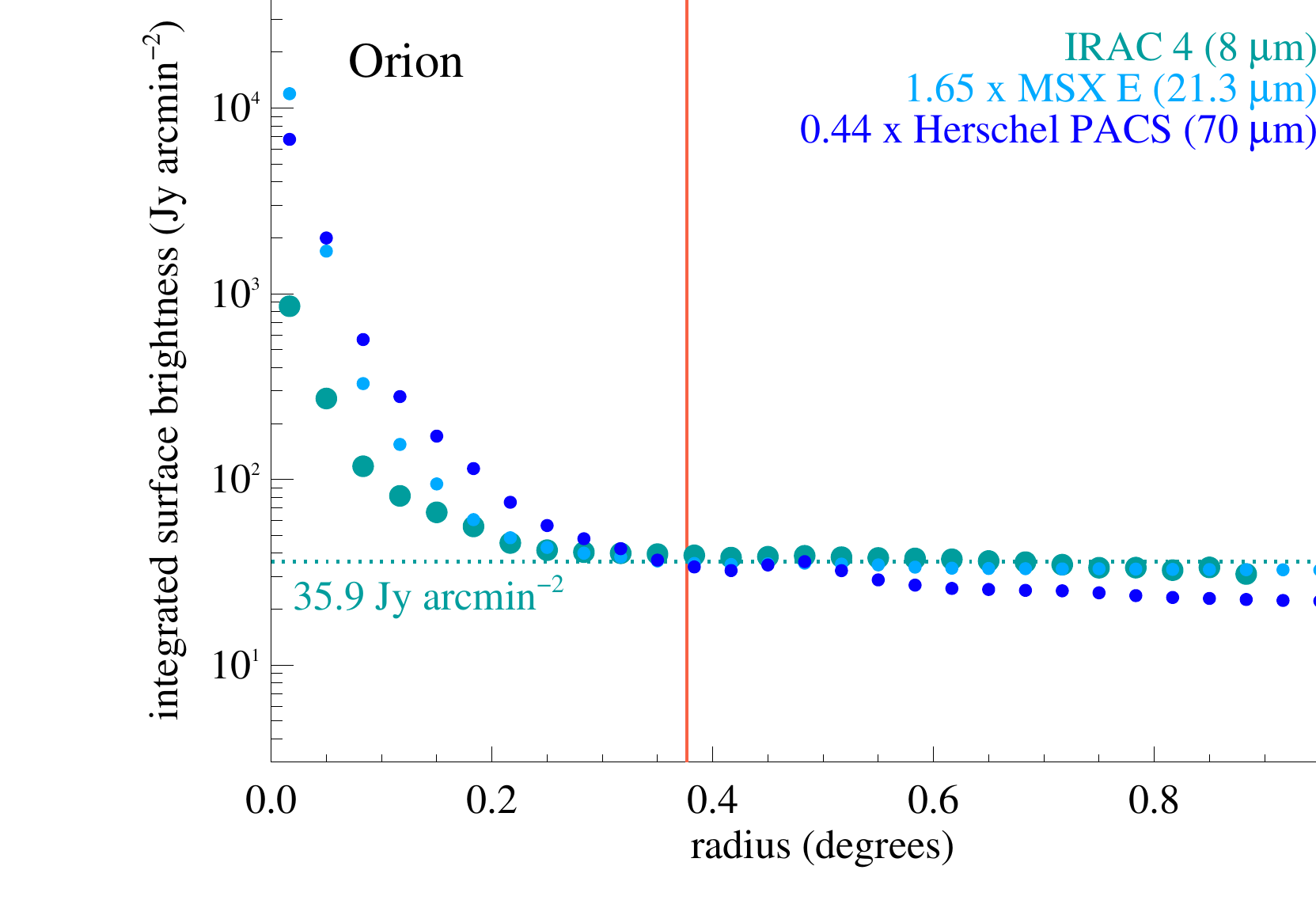}}
\figsetgrpnote{The RGB-rendered finding chart and surface brightness profile for the Orion Nebula. Blue is \spitzer IRAC 4 (8 \micron), green is \msx E (21.3 \micron), and red is \herschel PACS 70 \micron.}
\figsetgrpend

\figsetgrpstart
\figsetgrpnum{1.7}
\figsetgrptitle{Lagoon Nebula Finding Chart and Surface Brightness Profile}
\figsetplot{\includegraphics[width=0.31\linewidth,clip,trim=1.2cm 12.4cm 3.5cm 3.7cm]{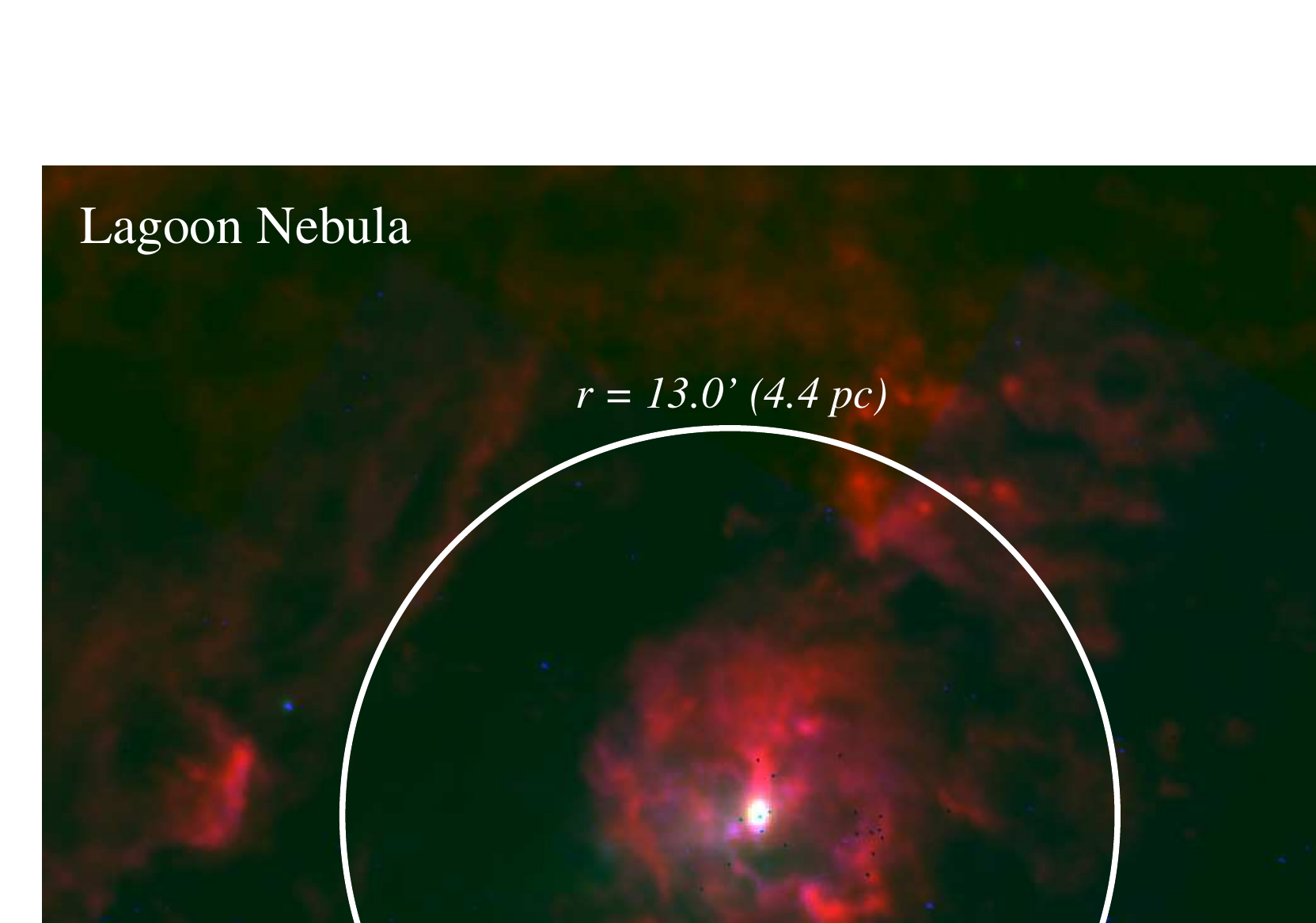}}
\figsetplot{\includegraphics[width=0.31\linewidth,clip,trim=1.2cm 12.4cm 3.5cm 3.7cm]{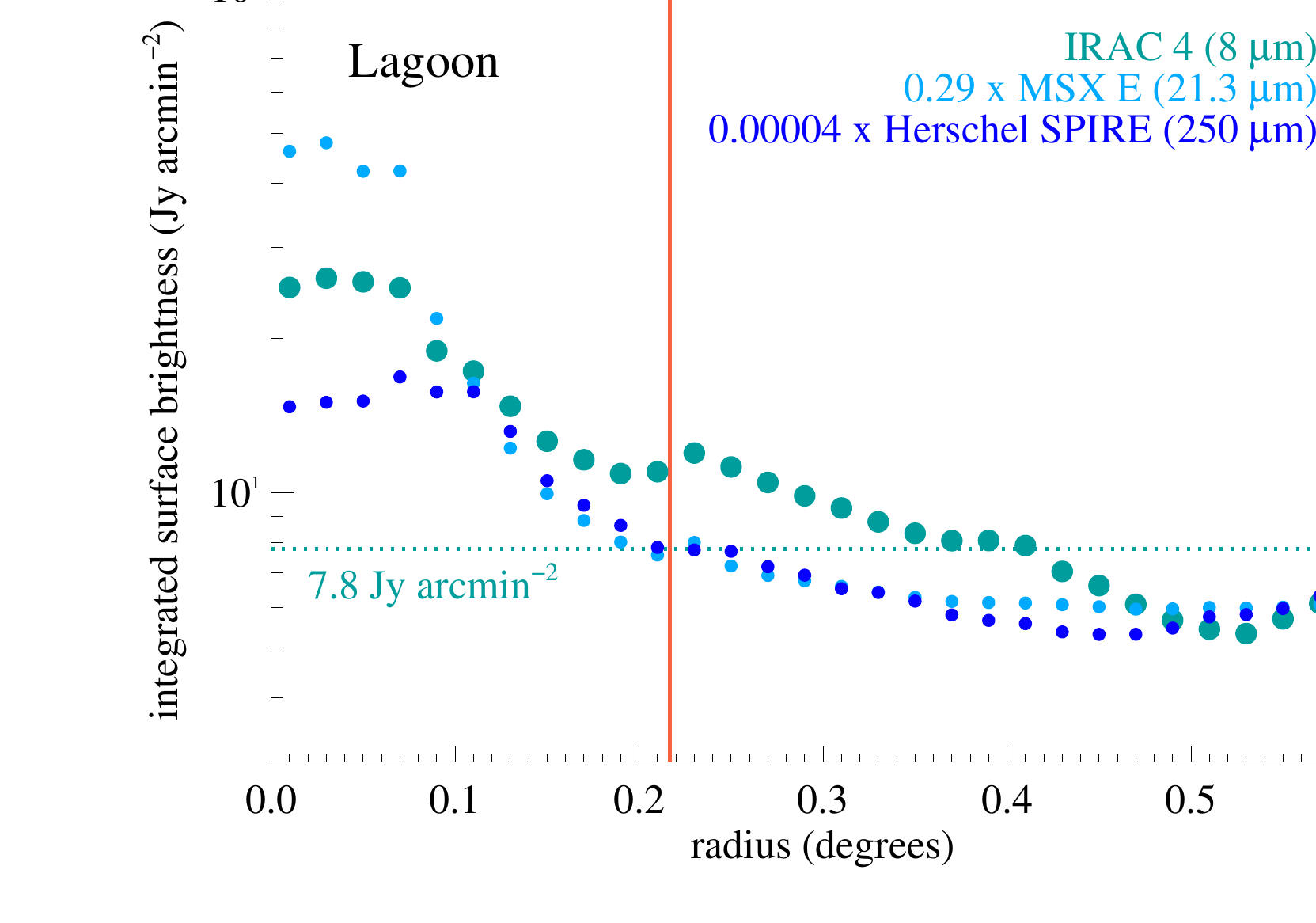}}
\figsetgrpnote{The RGB-rendered finding chart and surface brightness profile for the Lagoon Nebula. Blue is \spitzer IRAC 4 (8 \micron), green is \msx E (21.3 \micron), and red is \herschel SPIRE 250 \micron.}
\figsetgrpend

\figsetgrpstart
\figsetgrpnum{1.8}
\figsetgrptitle{Trifid Nebula Finding Chart and Surface Brightness Profile}
\figsetplot{\includegraphics[width=0.31\linewidth,clip,trim=1.2cm 12.4cm 3.5cm 3.7cm]{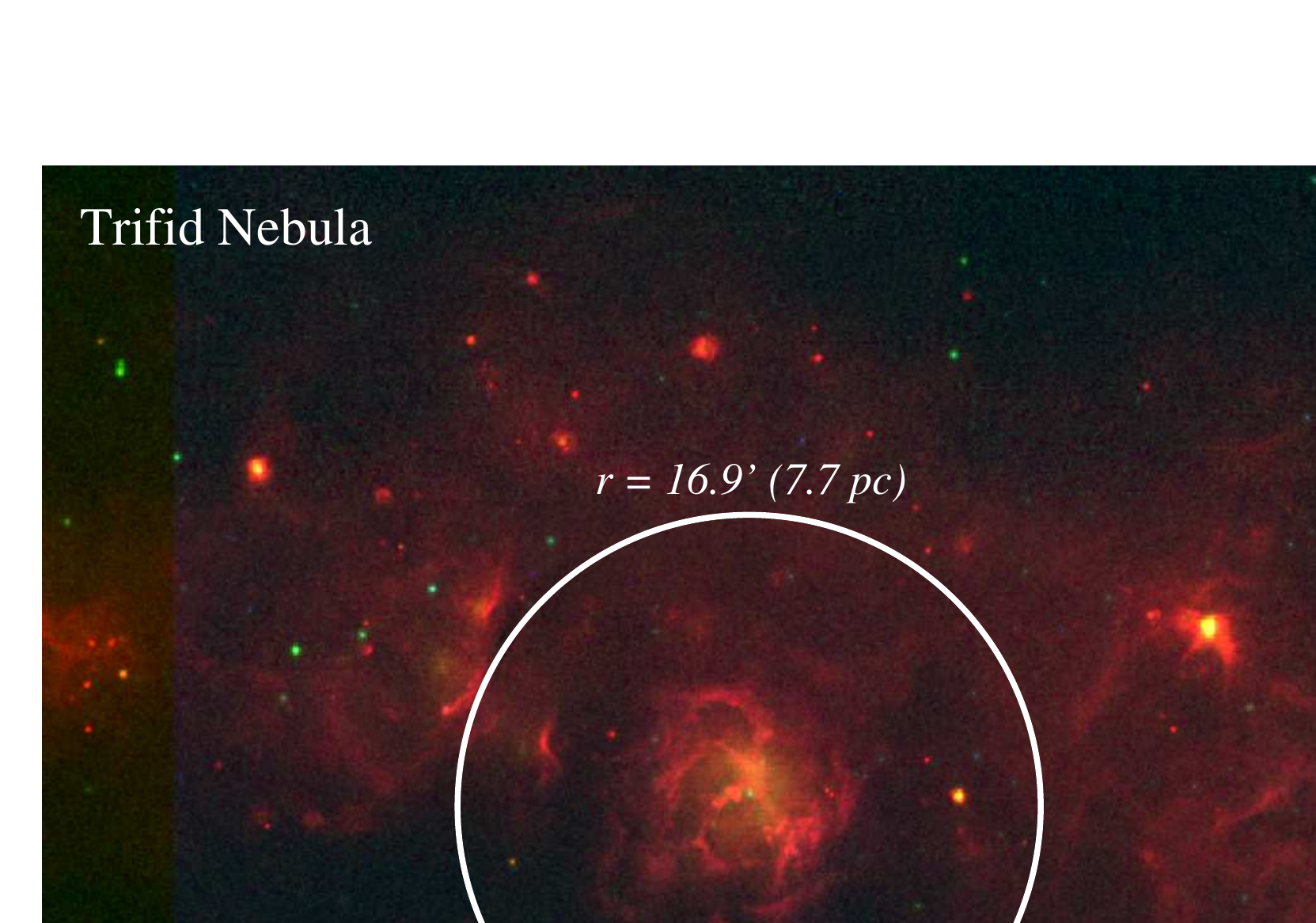}}
\figsetplot{\includegraphics[width=0.31\linewidth,clip,trim=1.2cm 12.4cm 3.5cm 3.7cm]{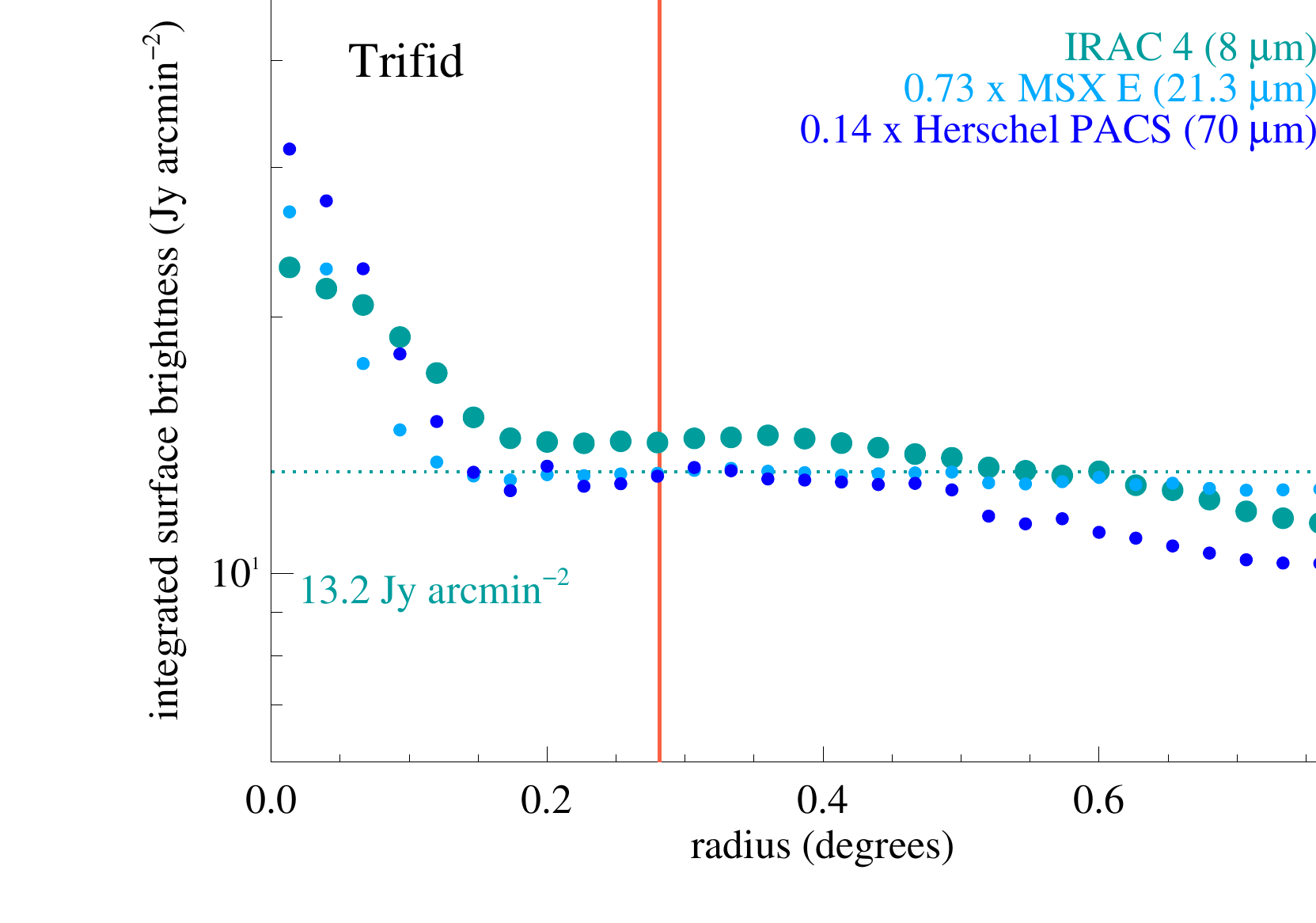}}
\figsetgrpnote{The RGB-rendered finding chart and surface brightness profile for the Trifid Nebula. Blue is \spitzer IRAC 4 (8 \micron), green is \msx E (21.3 \micron), and red is \herschel PACS 70 \micron.}
\figsetgrpend

\figsetgrpstart
\figsetgrpnum{1.9}
\figsetgrptitle{W42 Finding Chart and Surface Brightness Profile}
\figsetplot{\includegraphics[width=0.31\linewidth,clip,trim=1.2cm 12.4cm 3.5cm 3.7cm]{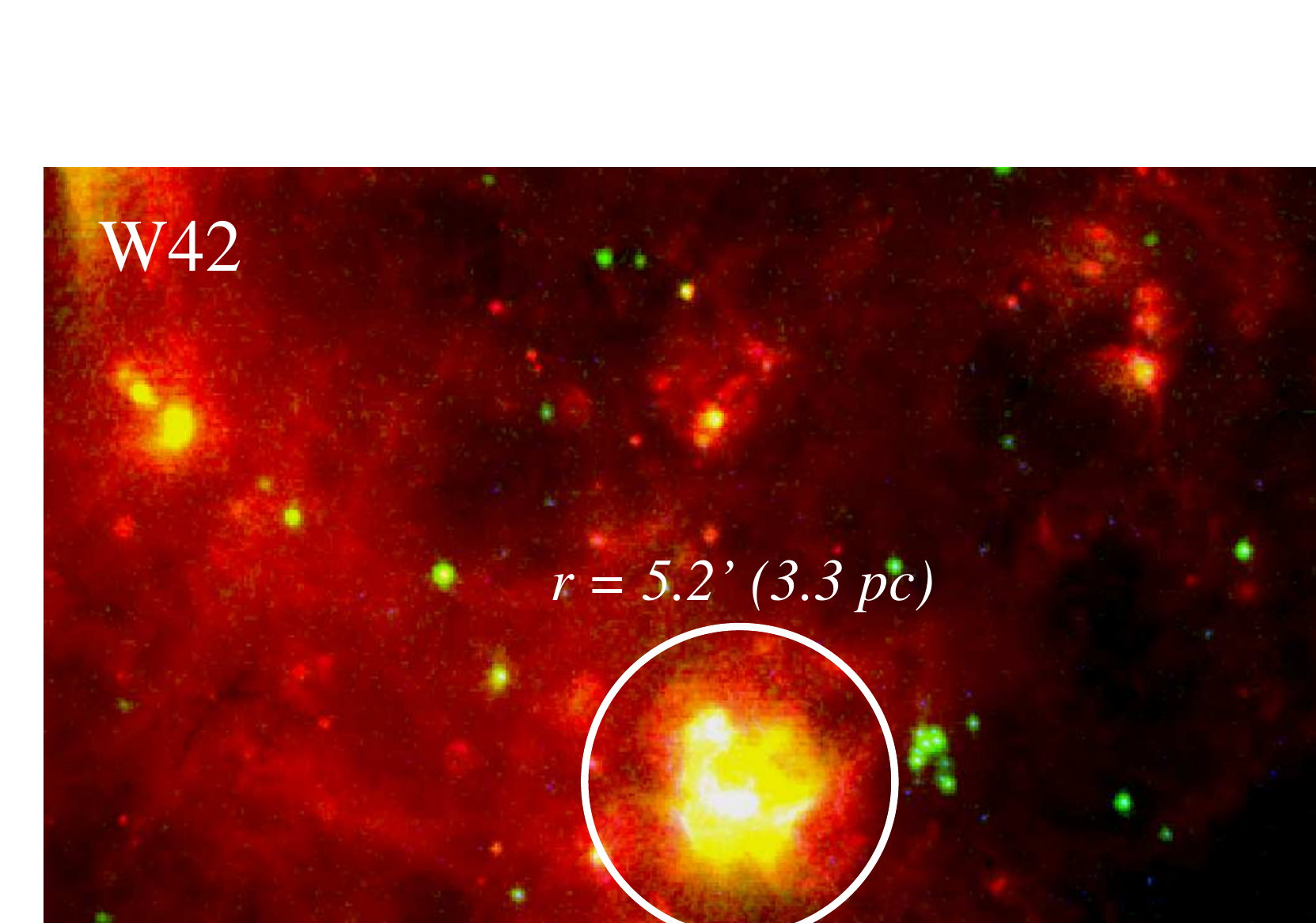}}
\figsetplot{\includegraphics[width=0.31\linewidth,clip,trim=1.2cm 12.4cm 3.5cm 3.7cm]{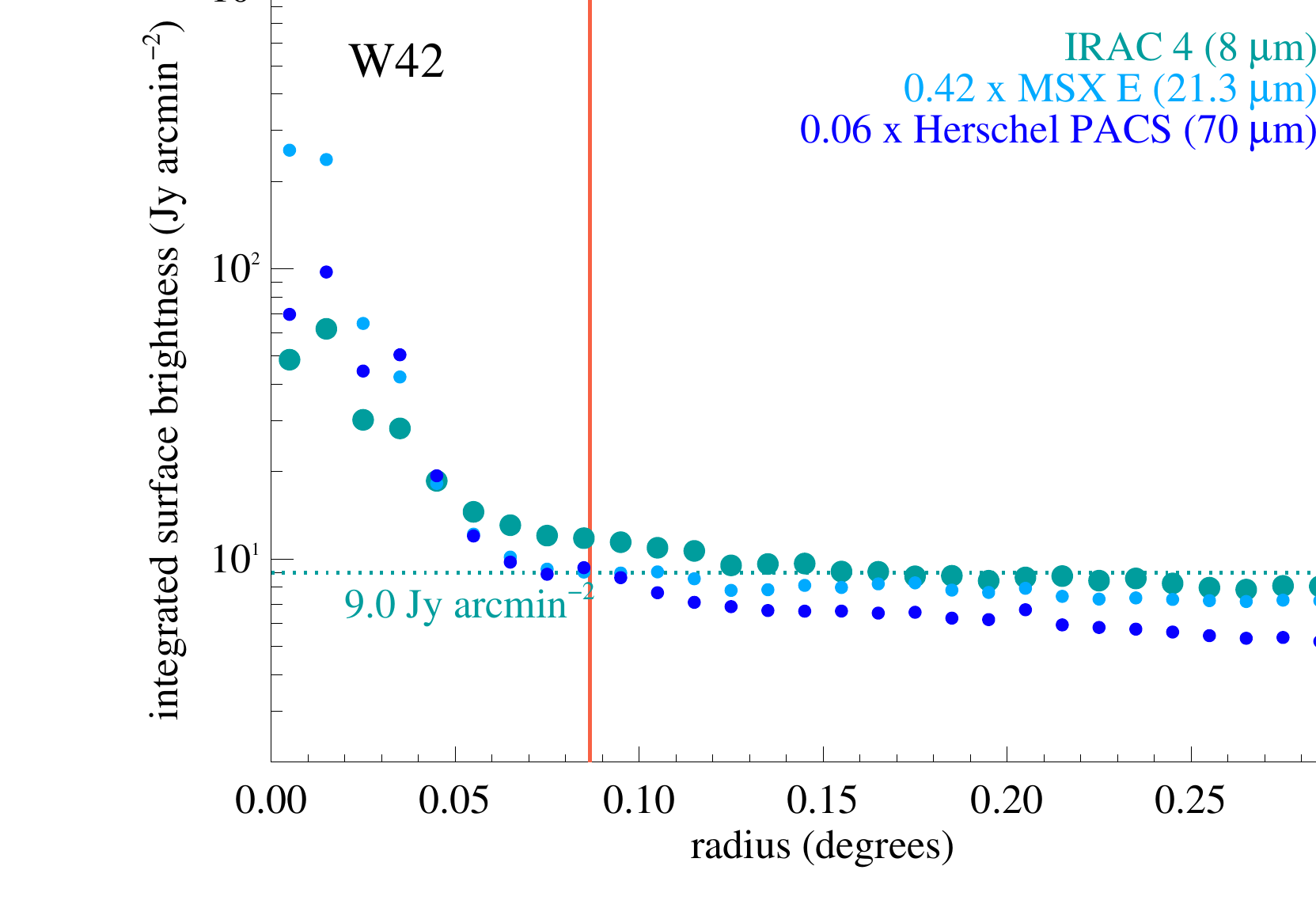}}
\figsetgrpnote{The RGB-rendered finding chart and surface brightness profile for W42. Blue is \spitzer IRAC 4 (8 \micron), green is \msx E (21.3 \micron), and red is \herschel PACS 70 \micron.}
\figsetgrpend

\figsetgrpstart
\figsetgrpnum{1.10}
\figsetgrptitle{NGC 7538 Finding Chart and Surface Brightness Profile}
\figsetplot{\includegraphics[width=0.31\linewidth,clip,trim=1.2cm 12.4cm 3.5cm 3.7cm]{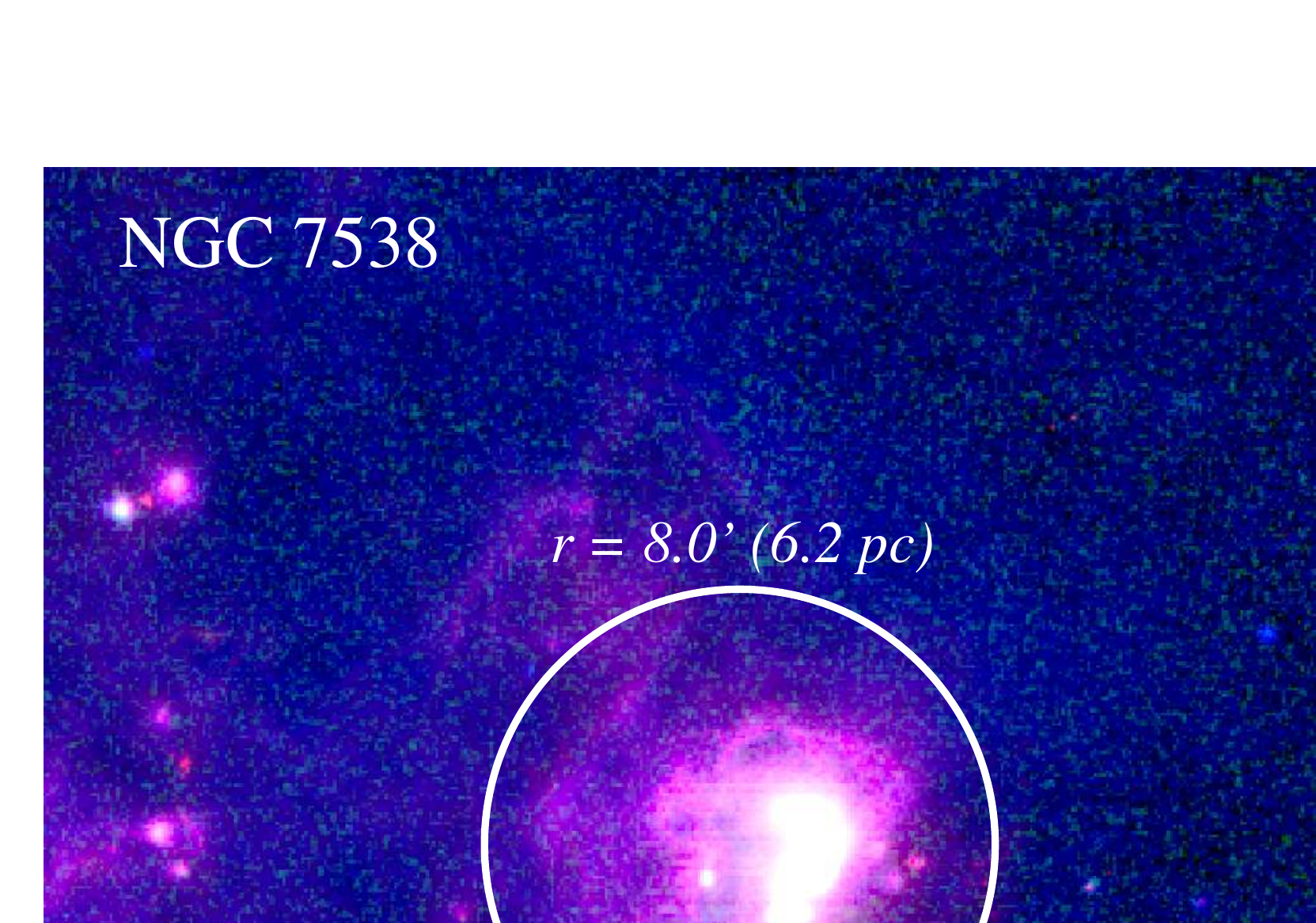}}
\figsetplot{\includegraphics[width=0.31\linewidth,clip,trim=1.2cm 12.4cm 3.5cm 3.7cm]{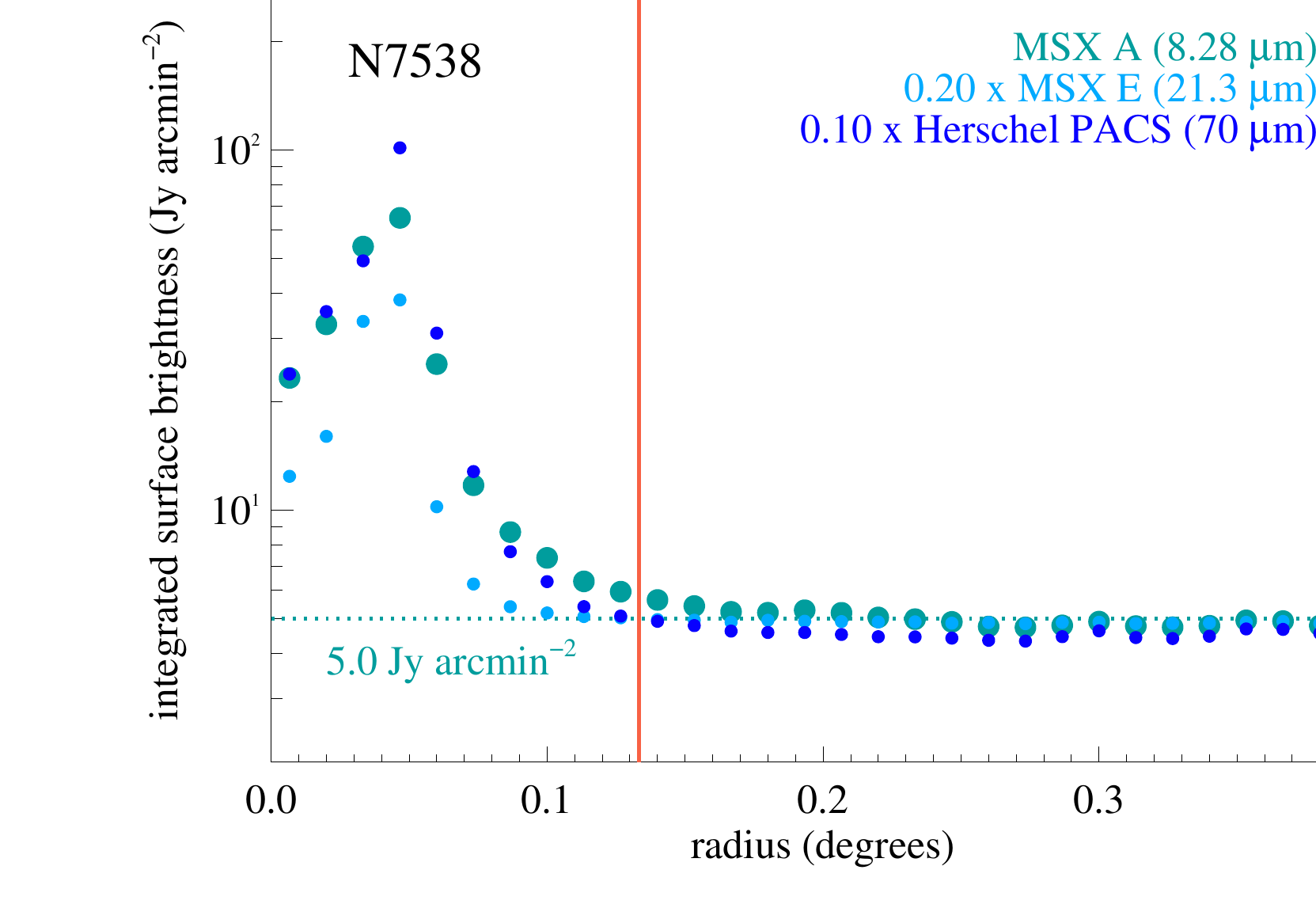}}
\figsetgrpnote{The RGB-rendered finding chart and surface brightness profile for NGC 7538. Blue is \msx A (8.28 \micron), green is \msx E (21.3 \micron), and red is \herschel PACS 70 \micron.}
\figsetgrpend

\figsetgrpstart
\figsetgrpnum{1.11}
\figsetgrptitle{W4 Finding Chart and Surface Brightness Profile}
\figsetplot{\includegraphics[width=0.31\linewidth,clip,trim=1.2cm 12.4cm 3.5cm 3.7cm]{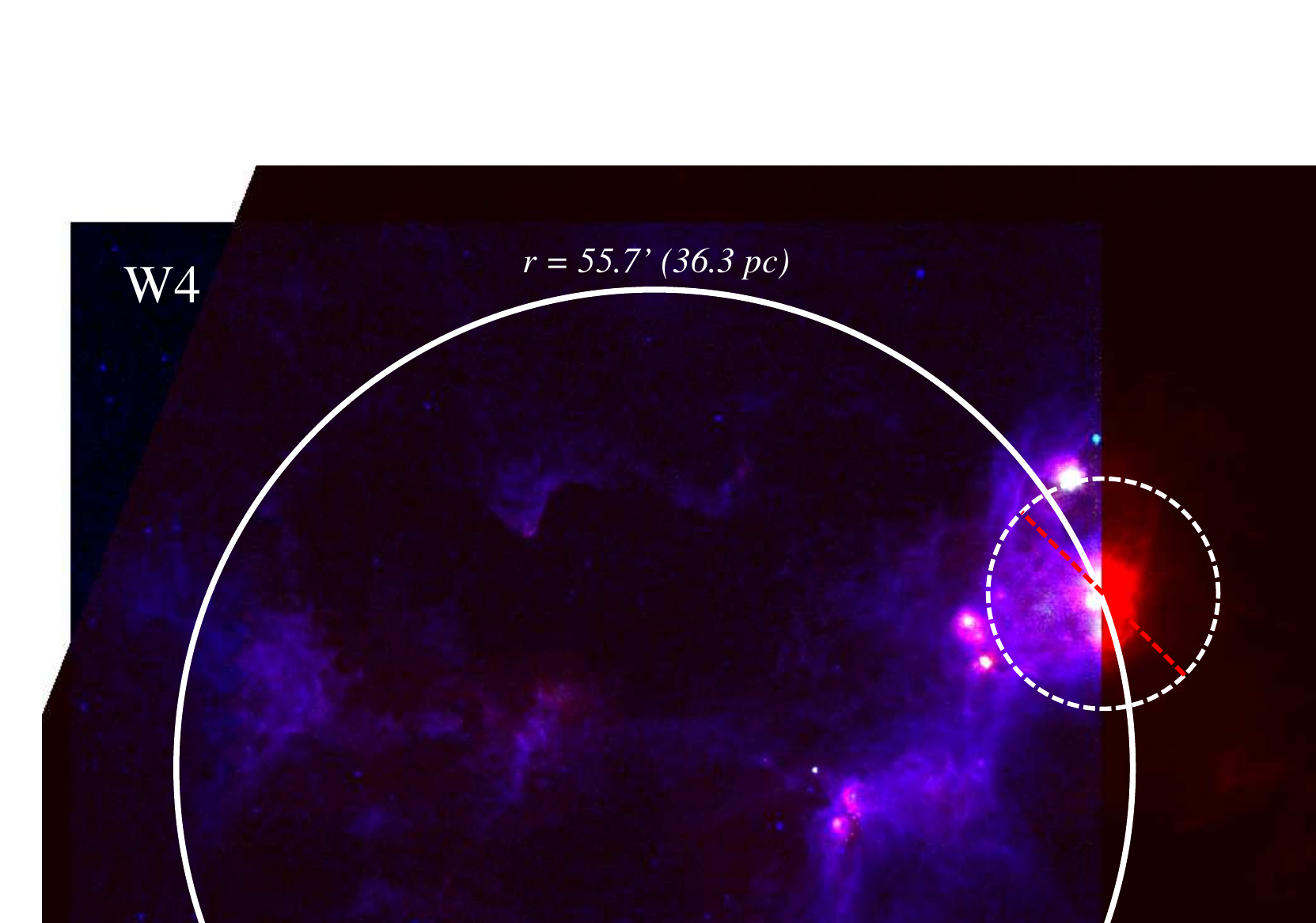}}
\figsetplot{\includegraphics[width=0.31\linewidth,clip,trim=1.2cm 12.4cm 3.5cm 3.7cm]{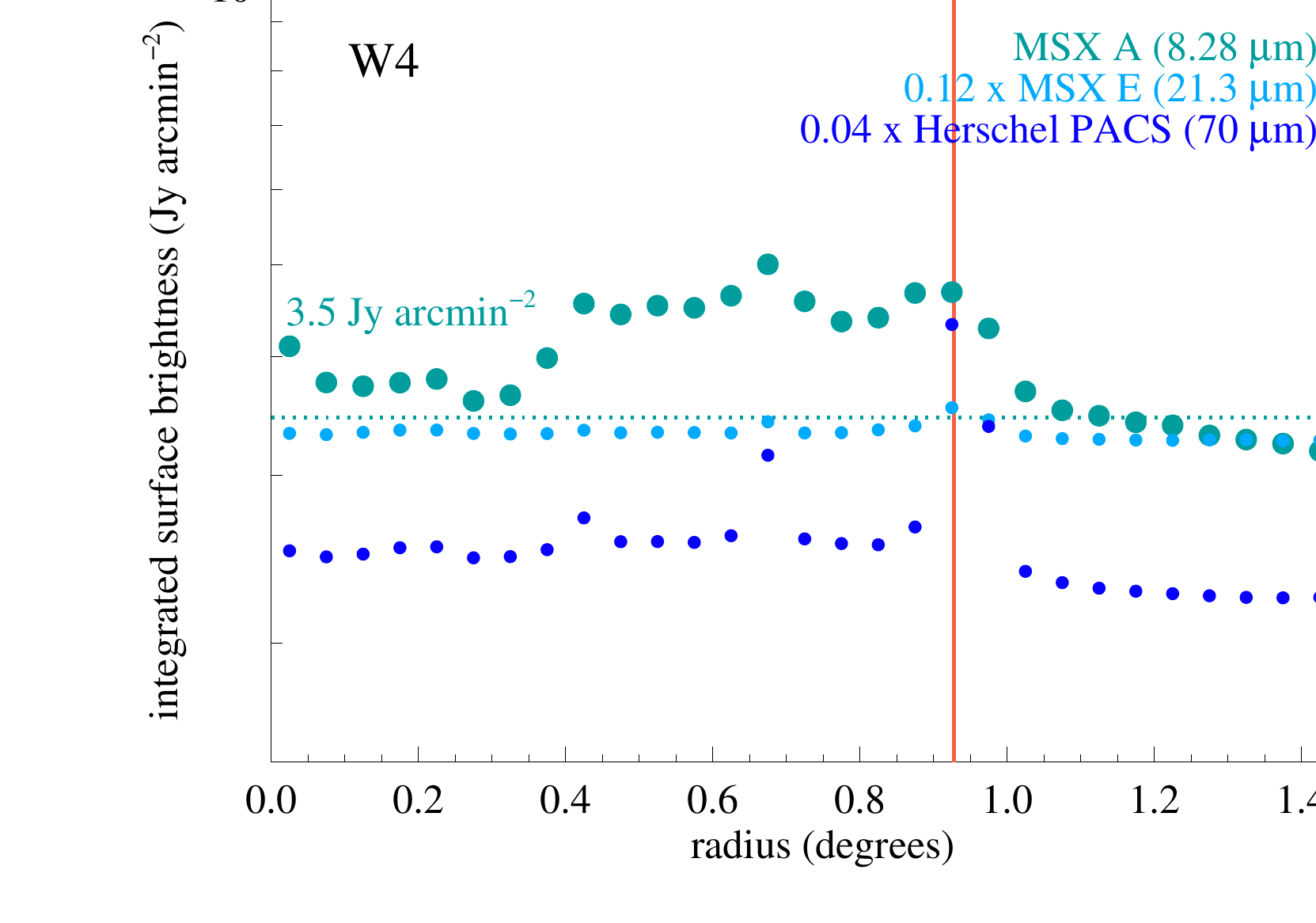}}
\figsetgrpnote{The RGB-rendered finding chart and surface brightness profile for W4. Blue is \msx A (8.28 \micron), green is \msx E (21.3 \micron), and red is \herschel PACS 70 \micron.}
\figsetgrpend

\figsetgrpstart
\figsetgrpnum{1.12}
\figsetgrptitle{Eagle Nebula Finding Chart and Surface Brightness Profile}
\figsetplot{\includegraphics[width=0.31\linewidth,clip,trim=1.2cm 12.4cm 3.5cm 3.7cm]{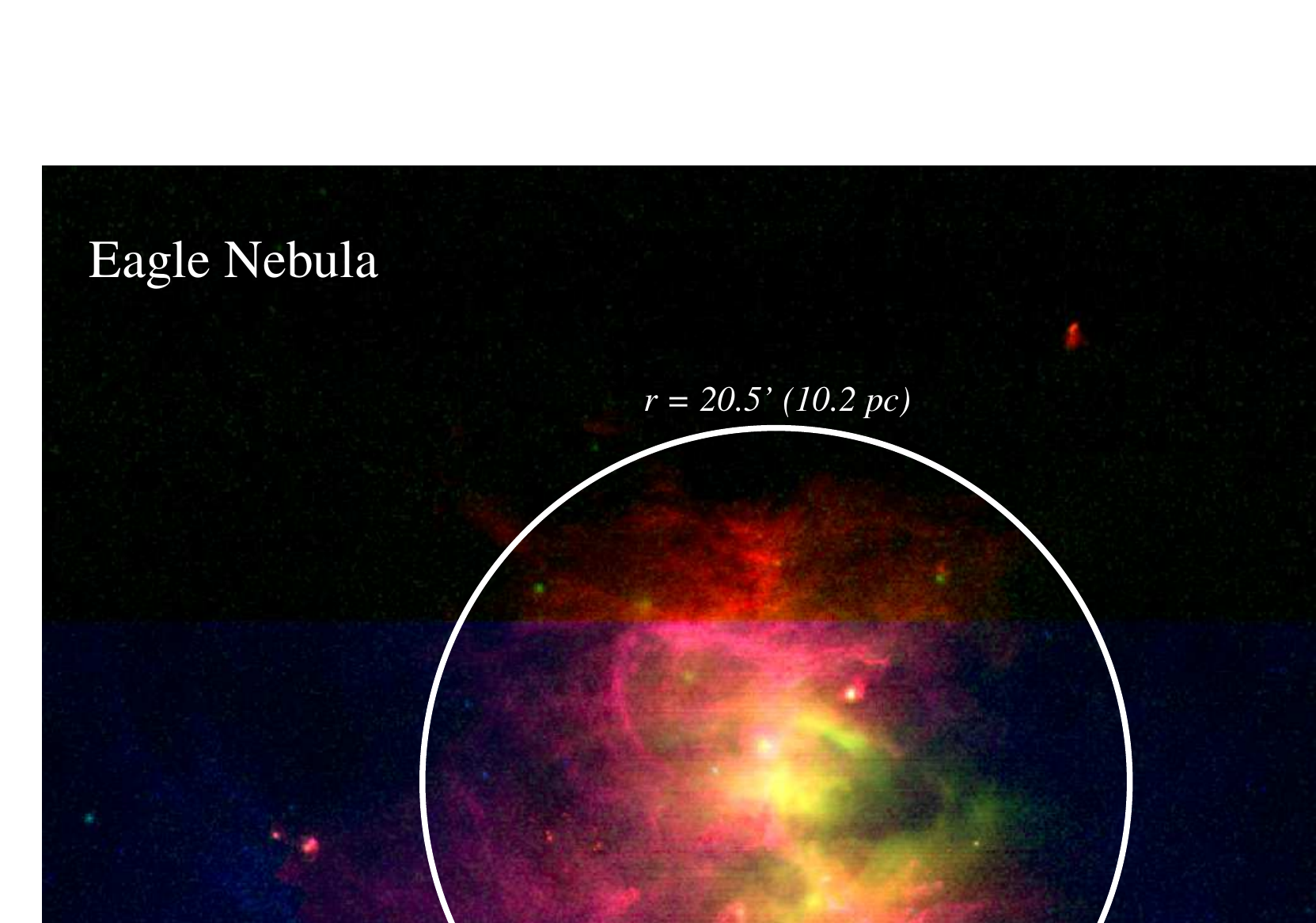}}
\figsetplot{\includegraphics[width=0.31\linewidth,clip,trim=1.2cm 12.4cm 3.5cm 3.7cm]{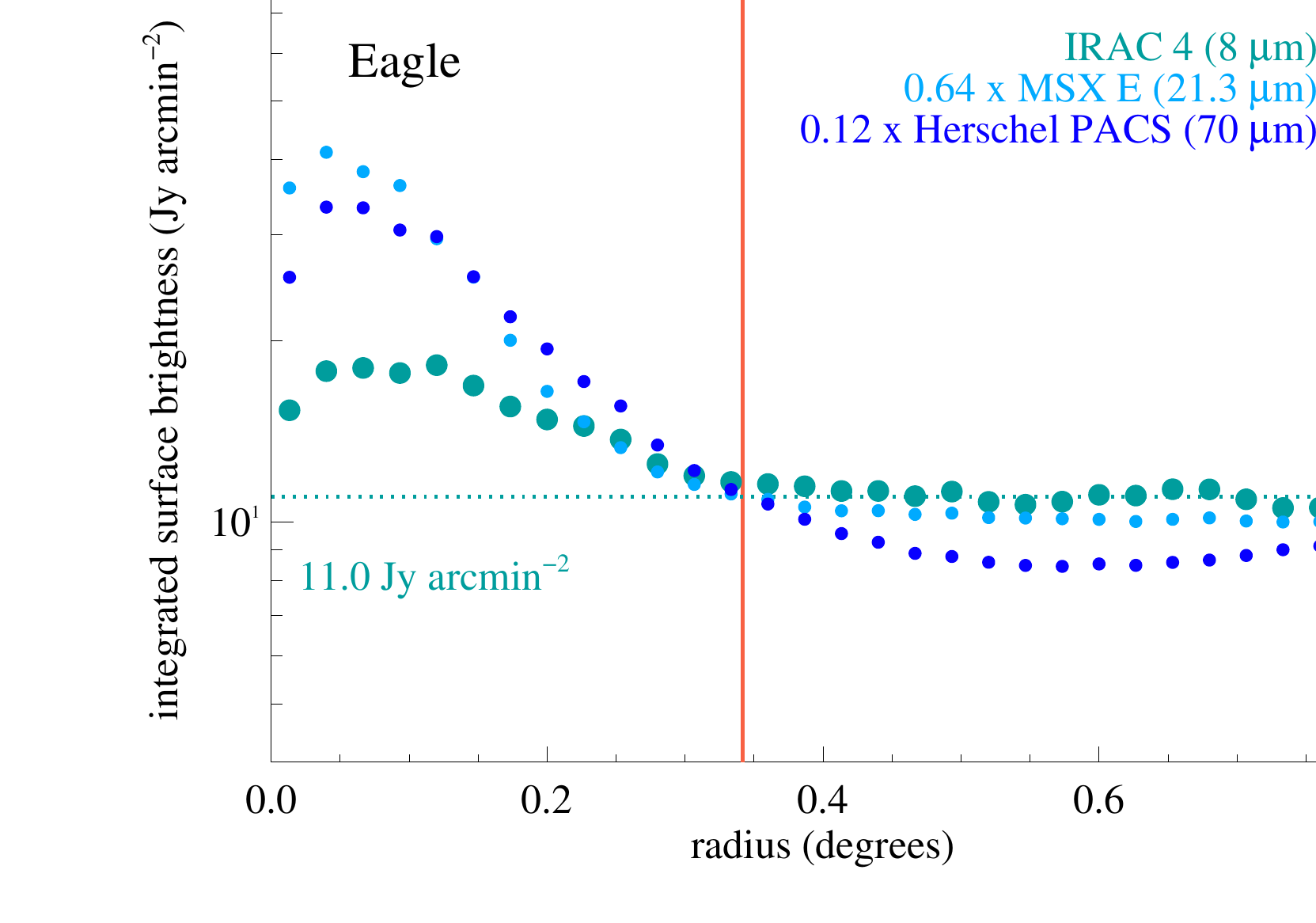}}
\figsetgrpnote{The RGB-rendered finding chart and surface brightness profile for the Eagle Nebula. Blue is \spitzer IRAC 4 (8 \micron), green is \msx E (21.3 \micron), and red is \herschel PACS 70 \micron.}
\figsetgrpend

\figsetgrpstart
\figsetgrpnum{1.13}
\figsetgrptitle{W33 Finding Chart and Surface Brightness Profile}
\figsetplot{\includegraphics[width=0.31\linewidth,clip,trim=1.2cm 12.4cm 3.5cm 3.7cm]{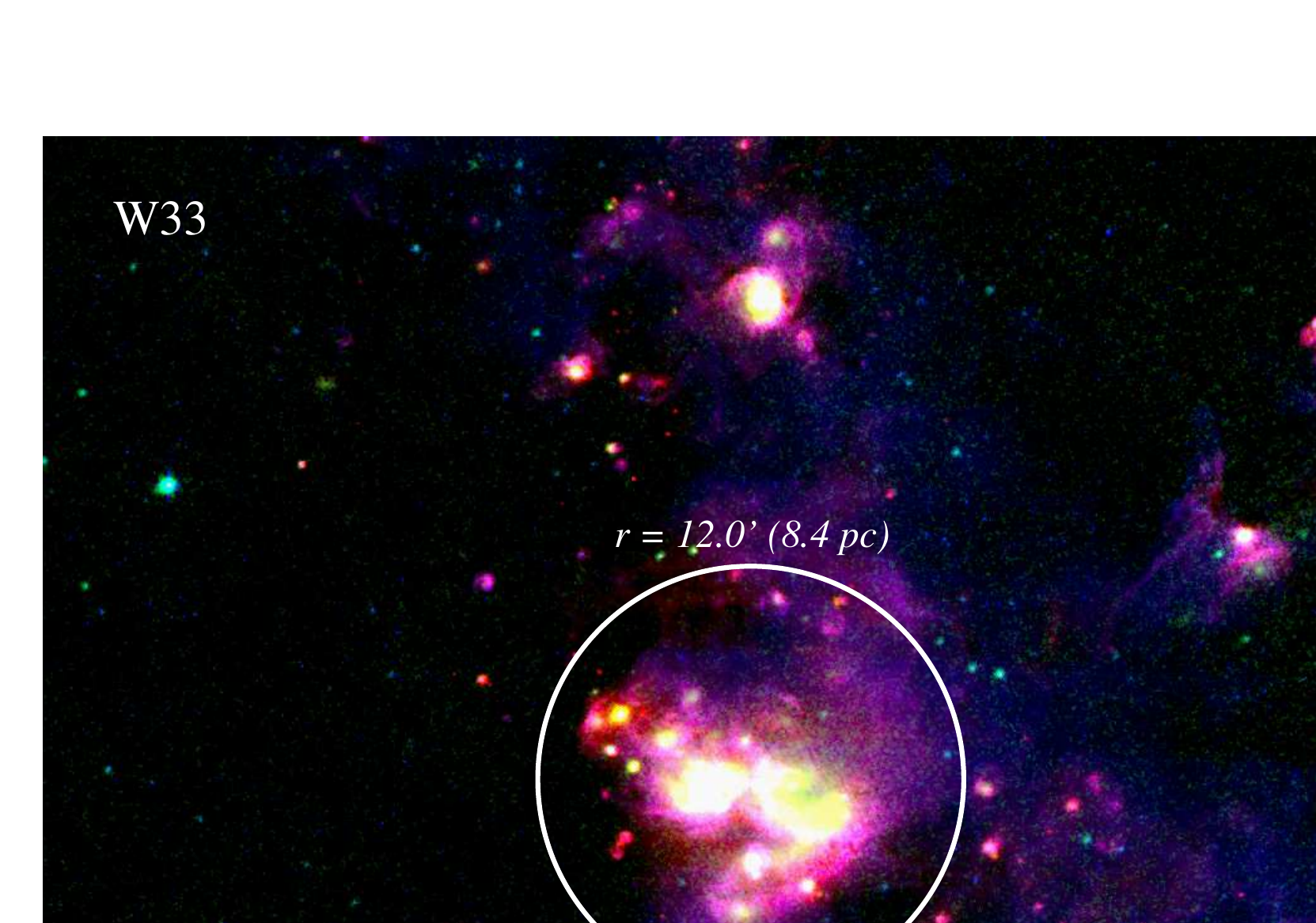}}
\figsetplot{\includegraphics[width=0.31\linewidth,clip,trim=1.2cm 12.4cm 3.5cm 3.7cm]{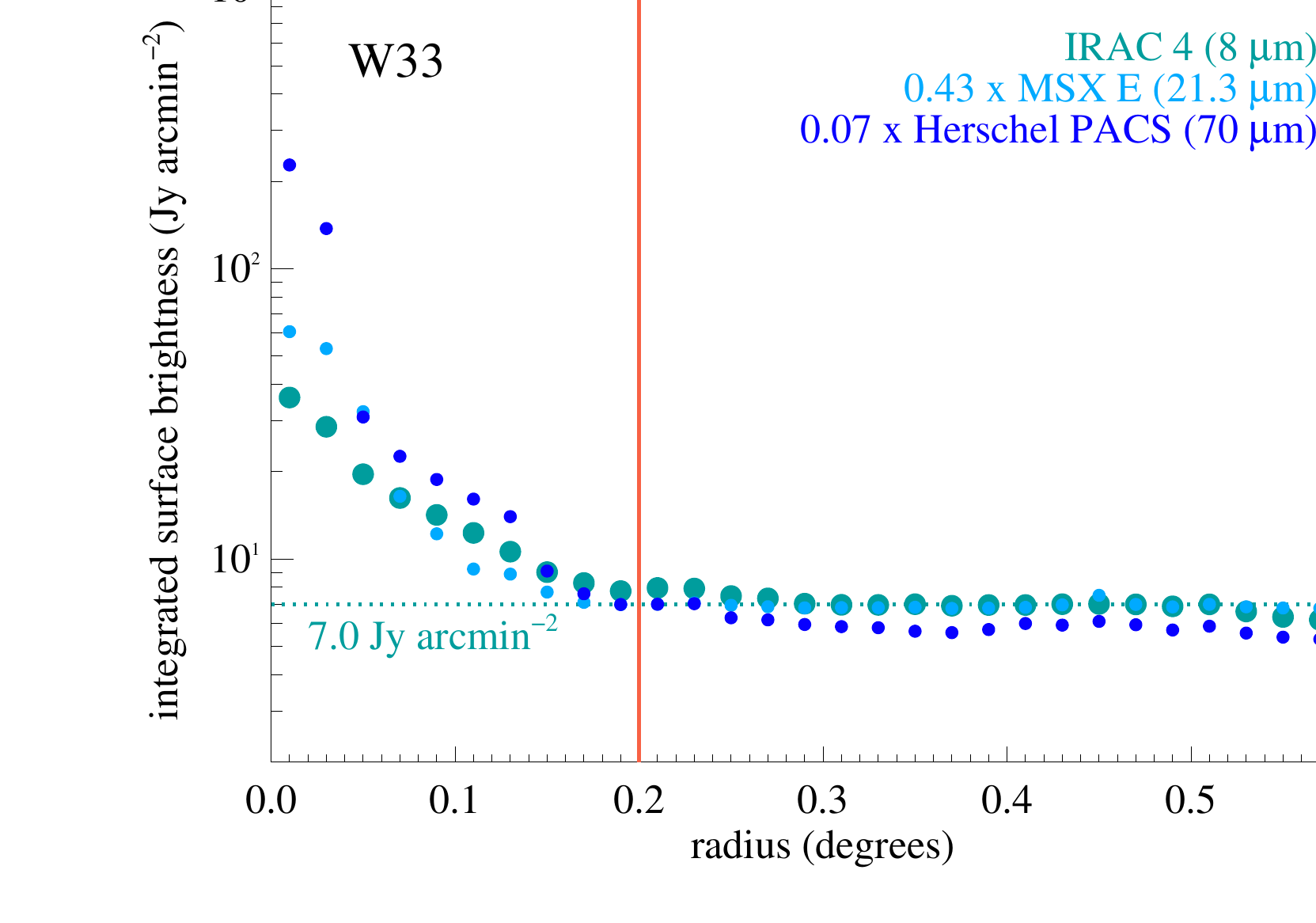}}
\figsetgrpnote{The RGB-rendered finding chart and surface brightness profile for W33. Blue is \spitzer IRAC 4 (8 \micron), green is \msx E (21.3 \micron), and red is \herschel PACS 70 \micron.}
\figsetgrpend

\figsetgrpstart
\figsetgrpnum{1.14}
\figsetgrptitle{RCW38 Finding Chart and Surface Brightness Profile}
\figsetplot{\includegraphics[width=0.31\linewidth,clip,trim=1.2cm 12.4cm 3.5cm 3.7cm]{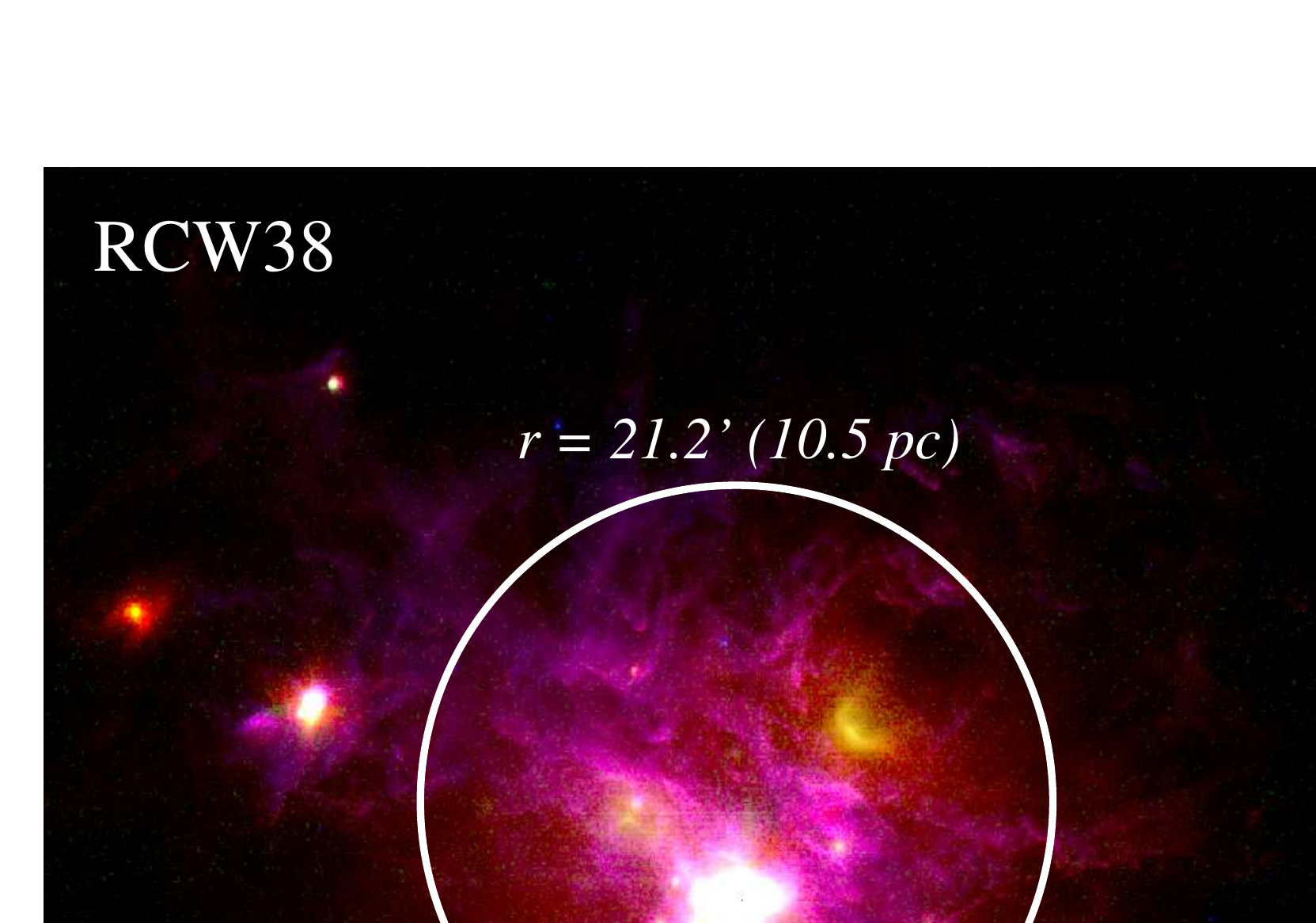}}
\figsetplot{\includegraphics[width=0.31\linewidth,clip,trim=1.2cm 12.4cm 3.5cm 3.7cm]{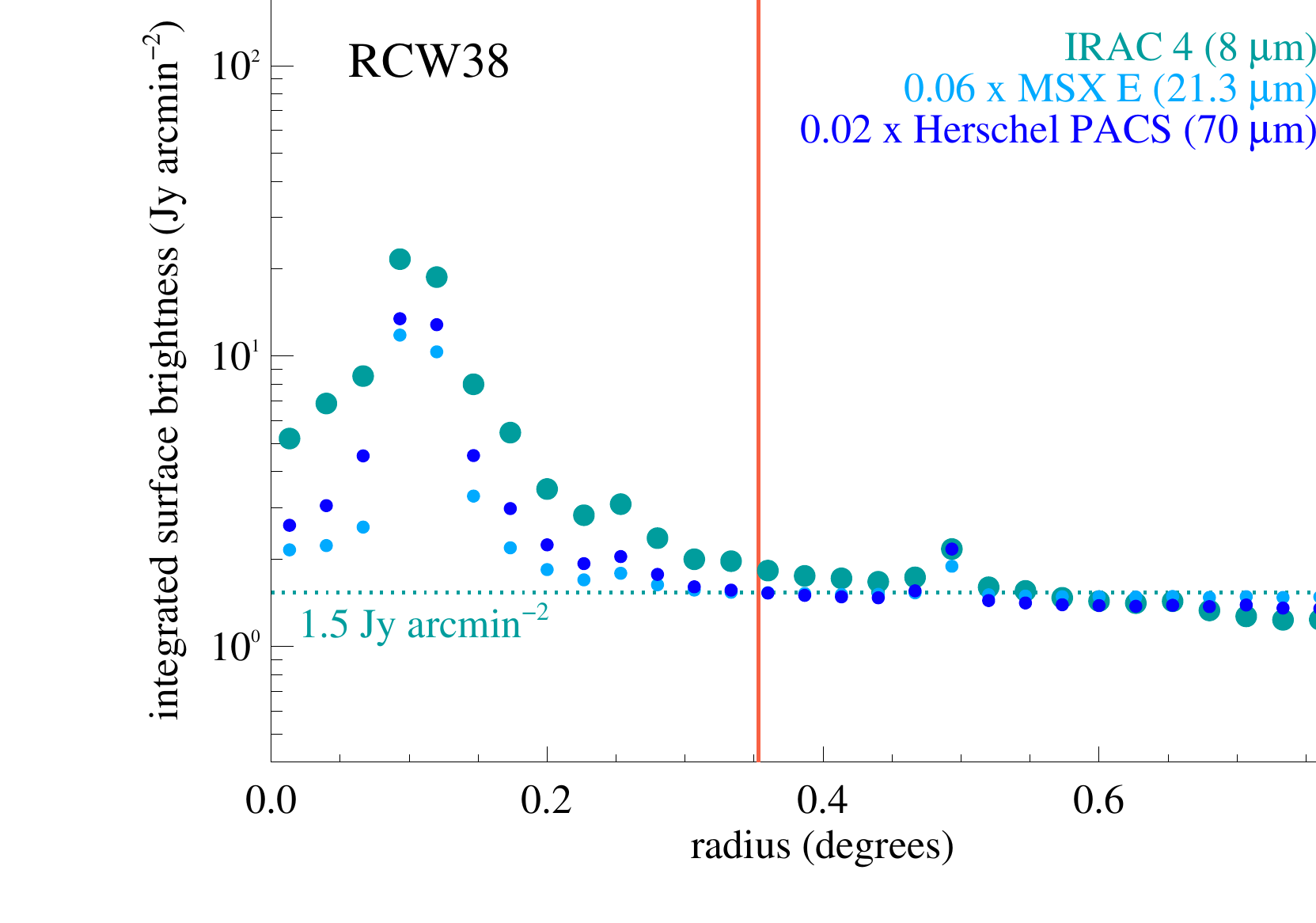}}
\figsetgrpnote{The RGB-rendered finding chart and surface brightness profile for RCW38. Blue is \spitzer IRAC 4 (8 \micron), green is \msx E (21.3 \micron), and red is \herschel PACS 70 \micron.}
\figsetgrpend

\figsetgrpstart
\figsetgrpnum{1.15}
\figsetgrptitle{W3 Finding Chart and Surface Brightness Profile}
\figsetplot{\includegraphics[width=0.31\linewidth,clip,trim=1.2cm 12.4cm 3.5cm 3.7cm]{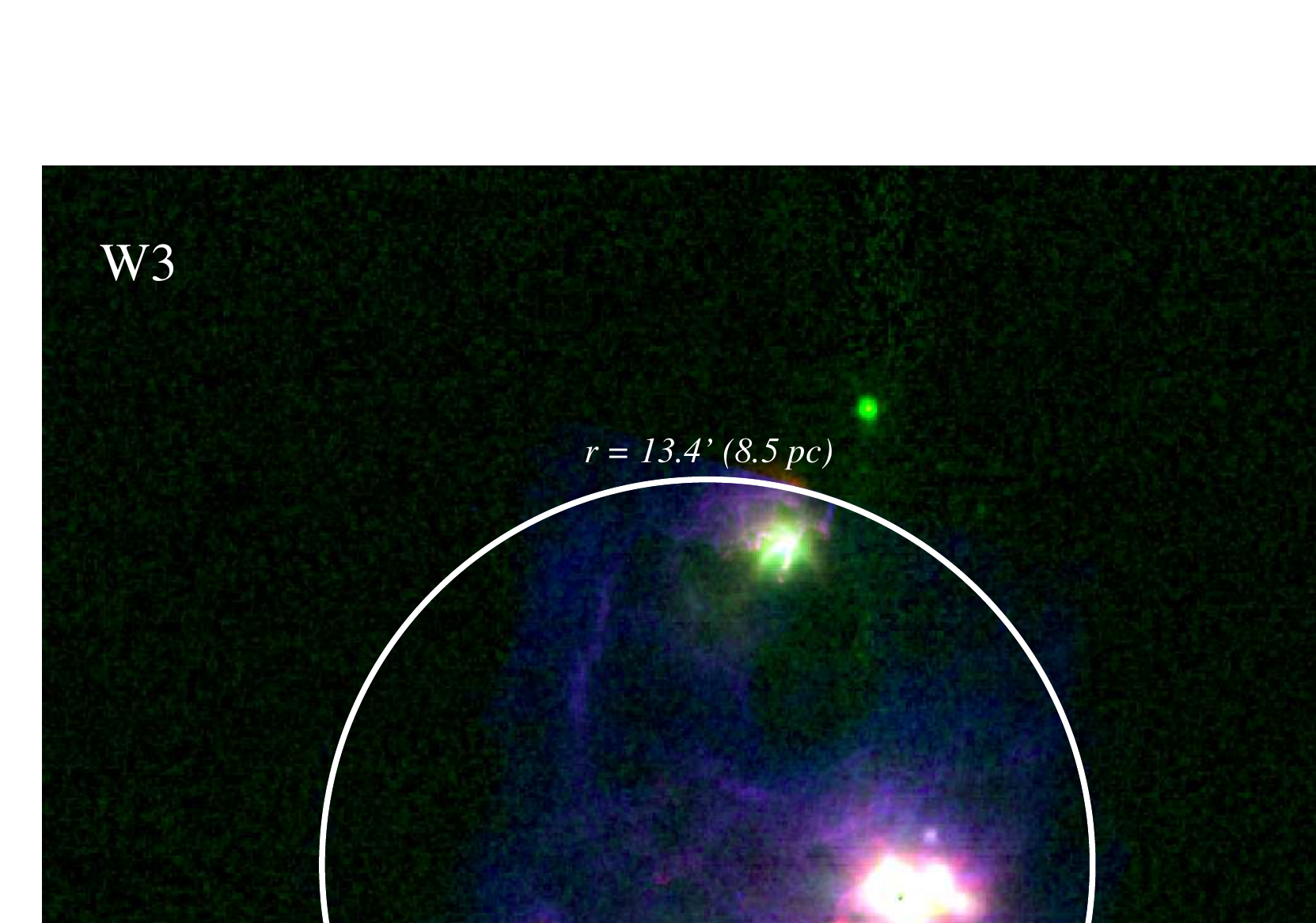}}
\figsetplot{\includegraphics[width=0.31\linewidth,clip,trim=1.2cm 12.4cm 3.5cm 3.7cm]{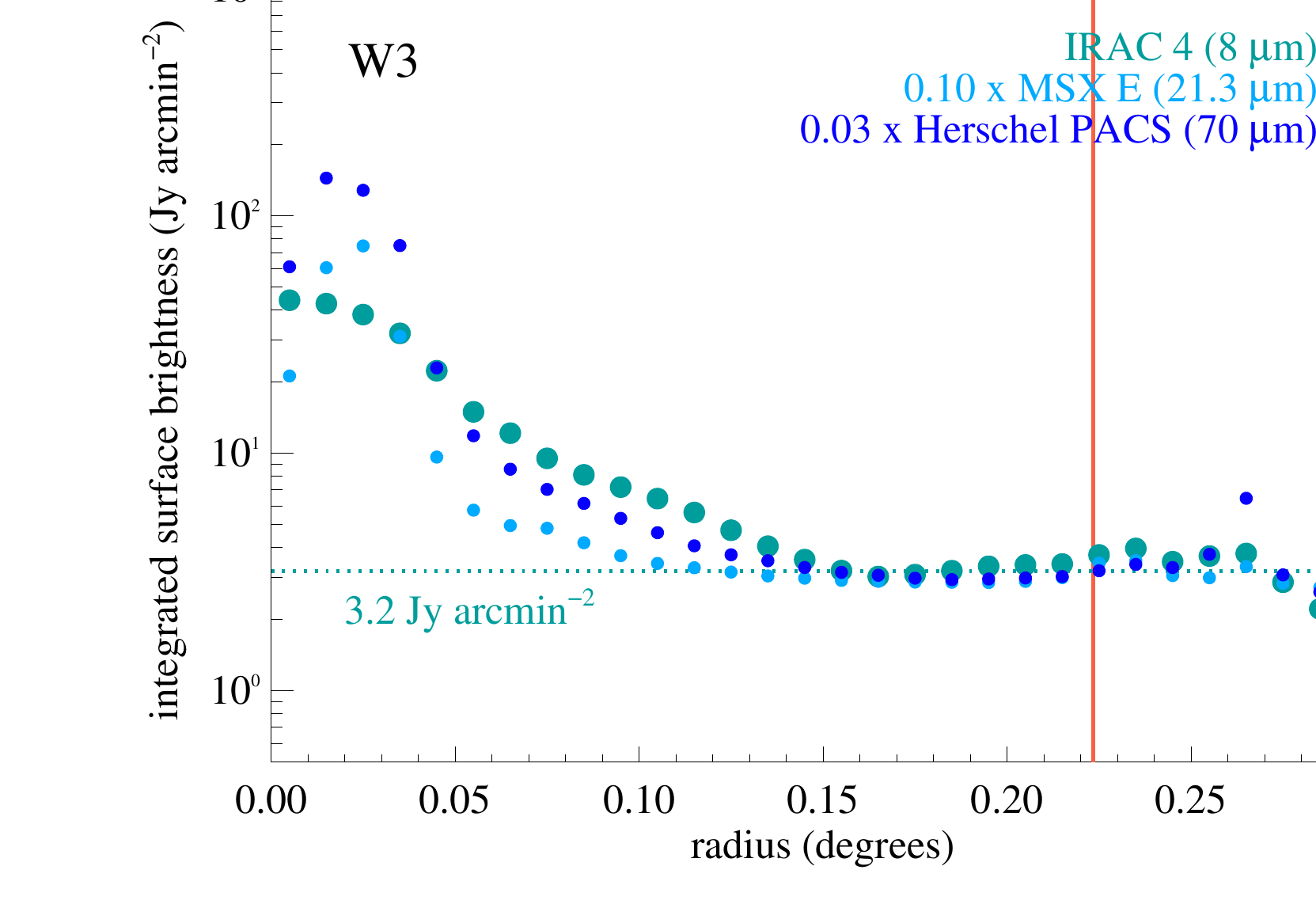}}
\figsetgrpnote{The RGB-rendered finding chart and surface brightness profile for W3. Blue is \spitzer IRAC 4 (8 \micron), green is \msx E (21.3 \micron), and red is \herschel PACS 70 \micron.}
\figsetgrpend

\figsetgrpstart
\figsetgrpnum{1.16}
\figsetgrptitle{NGC 3576 Finding Chart and Surface Brightness Profile}
\figsetplot{\includegraphics[width=0.31\linewidth,clip,trim=1.2cm 12.4cm 3.5cm 3.7cm]{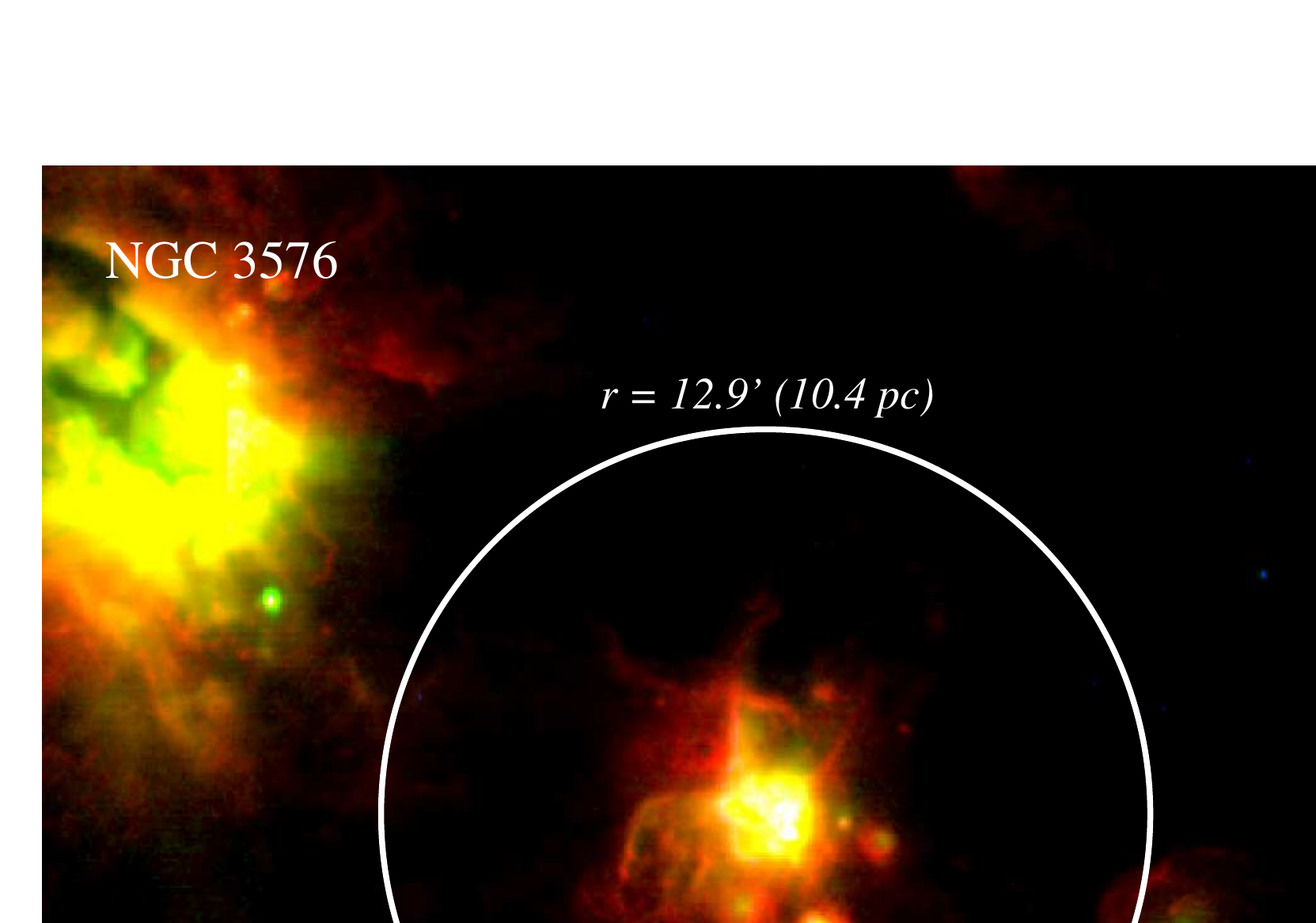}}
\figsetplot{\includegraphics[width=0.31\linewidth,clip,trim=1.2cm 12.4cm 3.5cm 3.7cm]{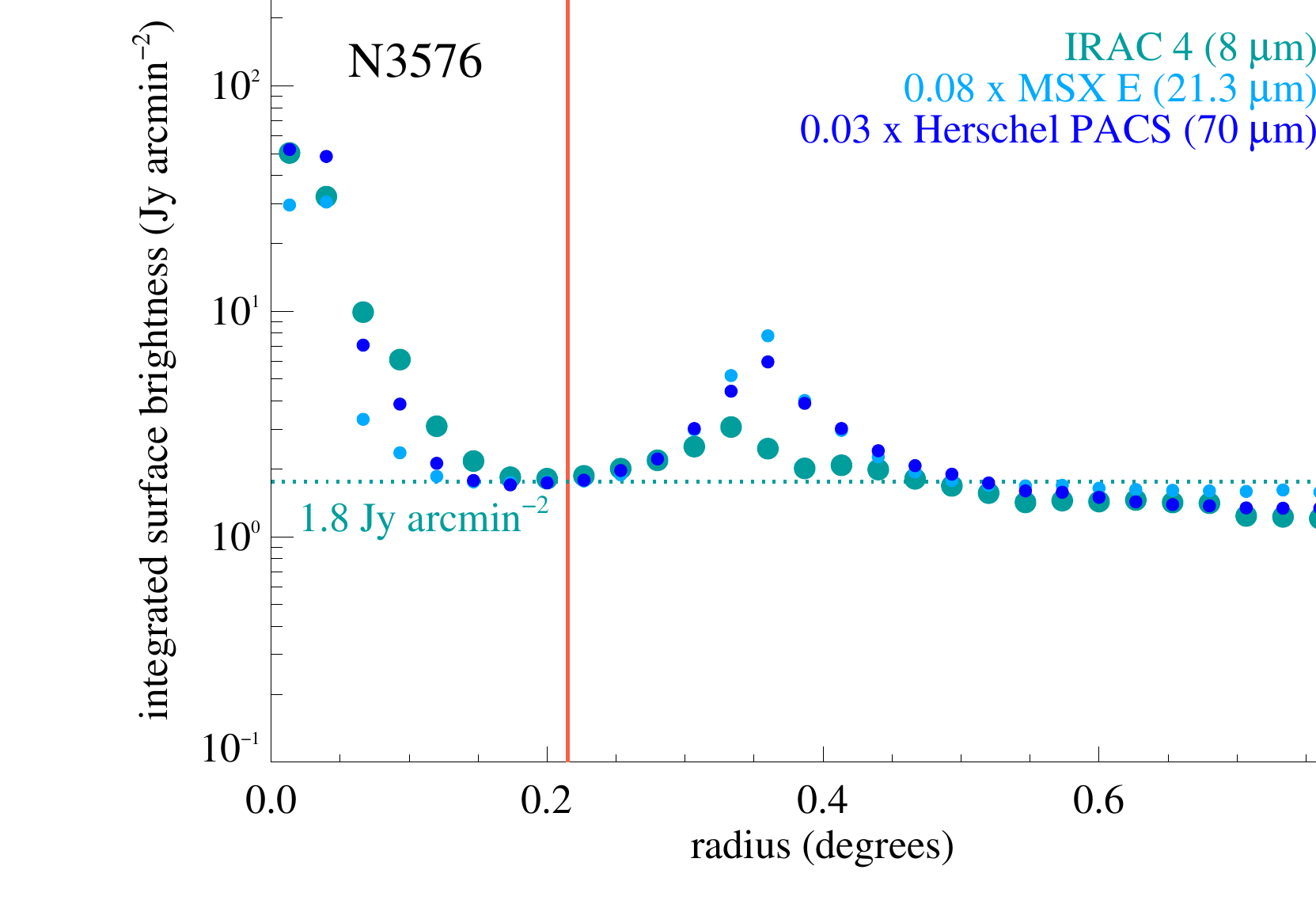}}
\figsetgrpnote{The RGB-rendered finding chart and surface brightness profile for NGC 3576. Blue is \spitzer IRAC 4 (8 \micron), green is \msx E (21.3 \micron), and red is \herschel PACS 70 \micron.}
\figsetgrpend

\figsetgrpstart
\figsetgrpnum{1.17}
\figsetgrptitle{NGC 6334 Finding Chart and Surface Brightness Profile}
\figsetplot{\includegraphics[width=0.31\linewidth,clip,trim=1.2cm 12.4cm 3.5cm 3.7cm]{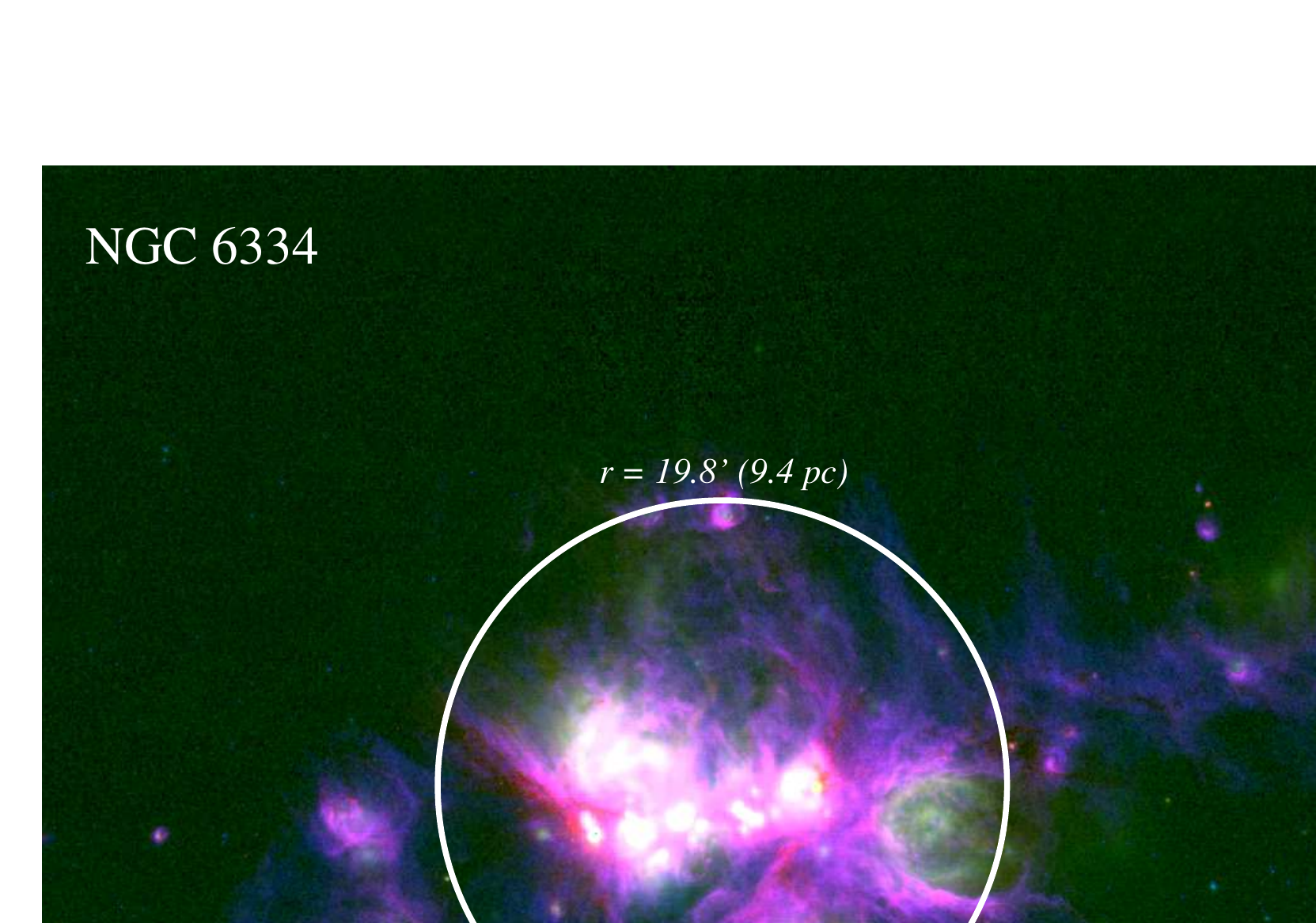}}
\figsetplot{\includegraphics[width=0.31\linewidth,clip,trim=1.2cm 12.4cm 3.5cm 3.7cm]{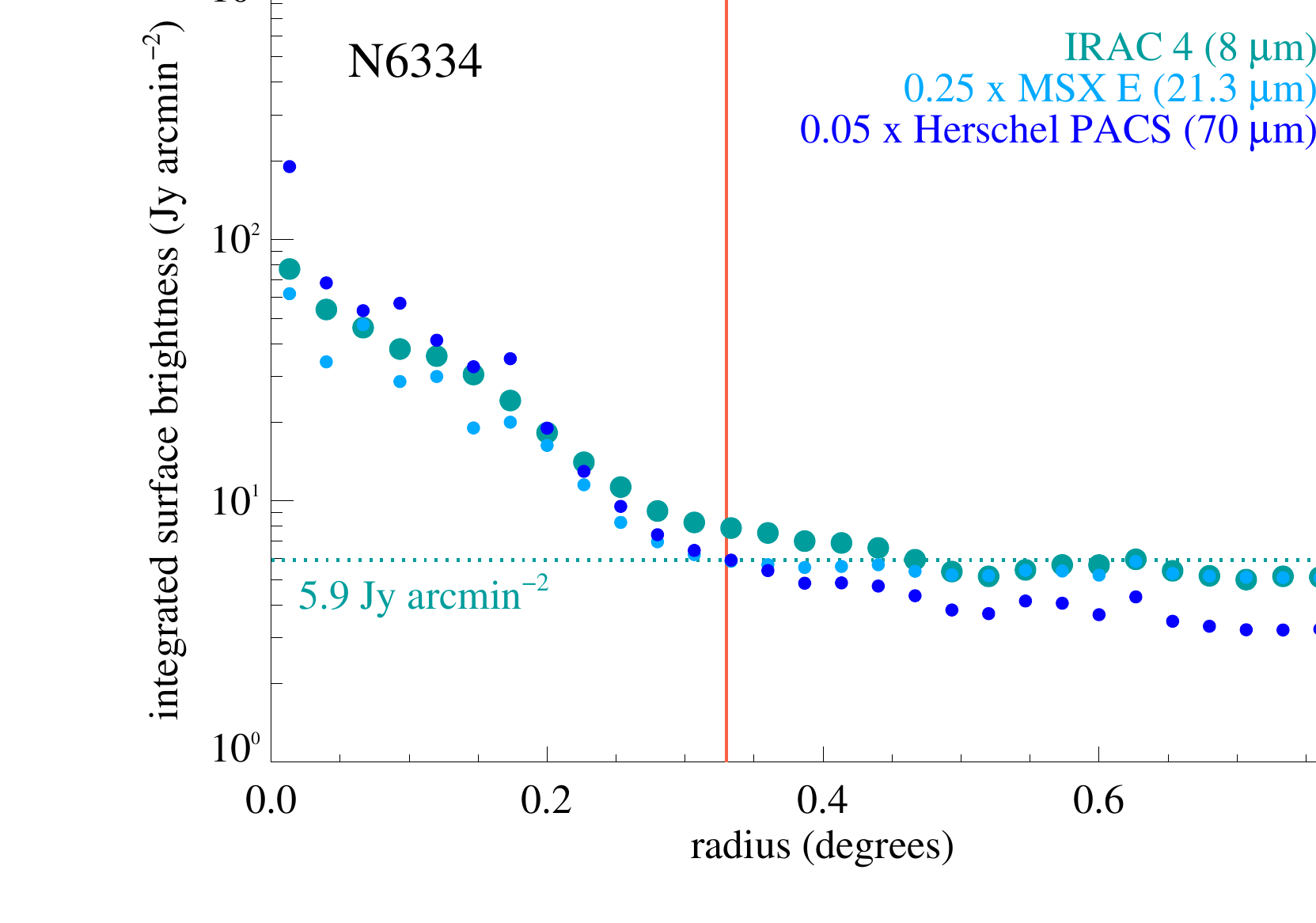}}
\figsetgrpnote{The RGB-rendered finding chart and surface brightness profile for NGC 6334. Blue is \spitzer IRAC 4 (8 \micron), green is \msx E (21.3 \micron), and red is \herschel PACS 70 \micron.}
\figsetgrpend

\figsetgrpstart
\figsetgrpnum{1.18}
\figsetgrptitle{G29.96-0.02 Finding Chart and Surface Brightness Profile}
\figsetplot{\includegraphics[width=0.31\linewidth,clip,trim=1.2cm 12.4cm 3.5cm 3.7cm]{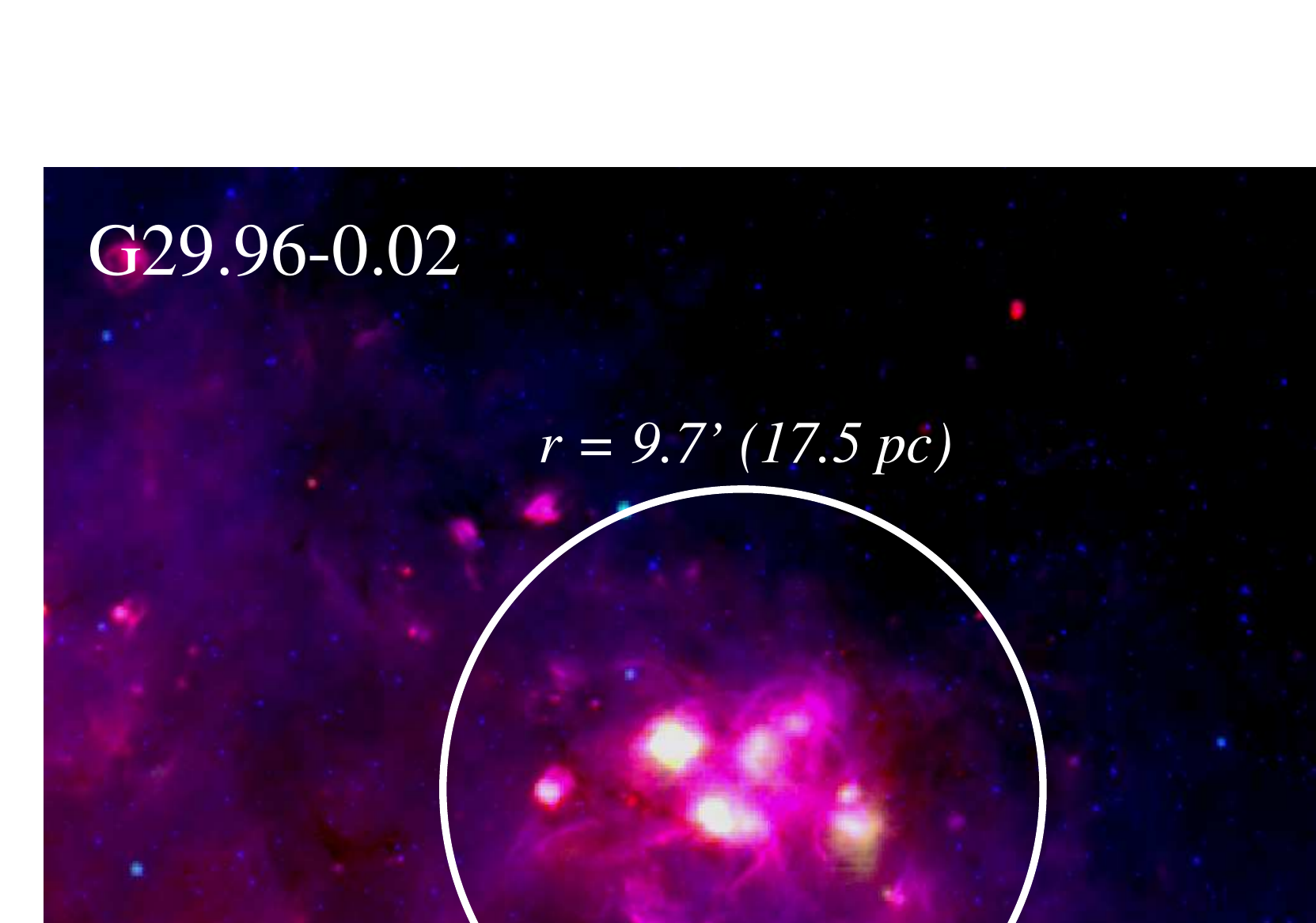}}
\figsetplot{\includegraphics[width=0.31\linewidth,clip,trim=1.2cm 12.4cm 3.5cm 3.7cm]{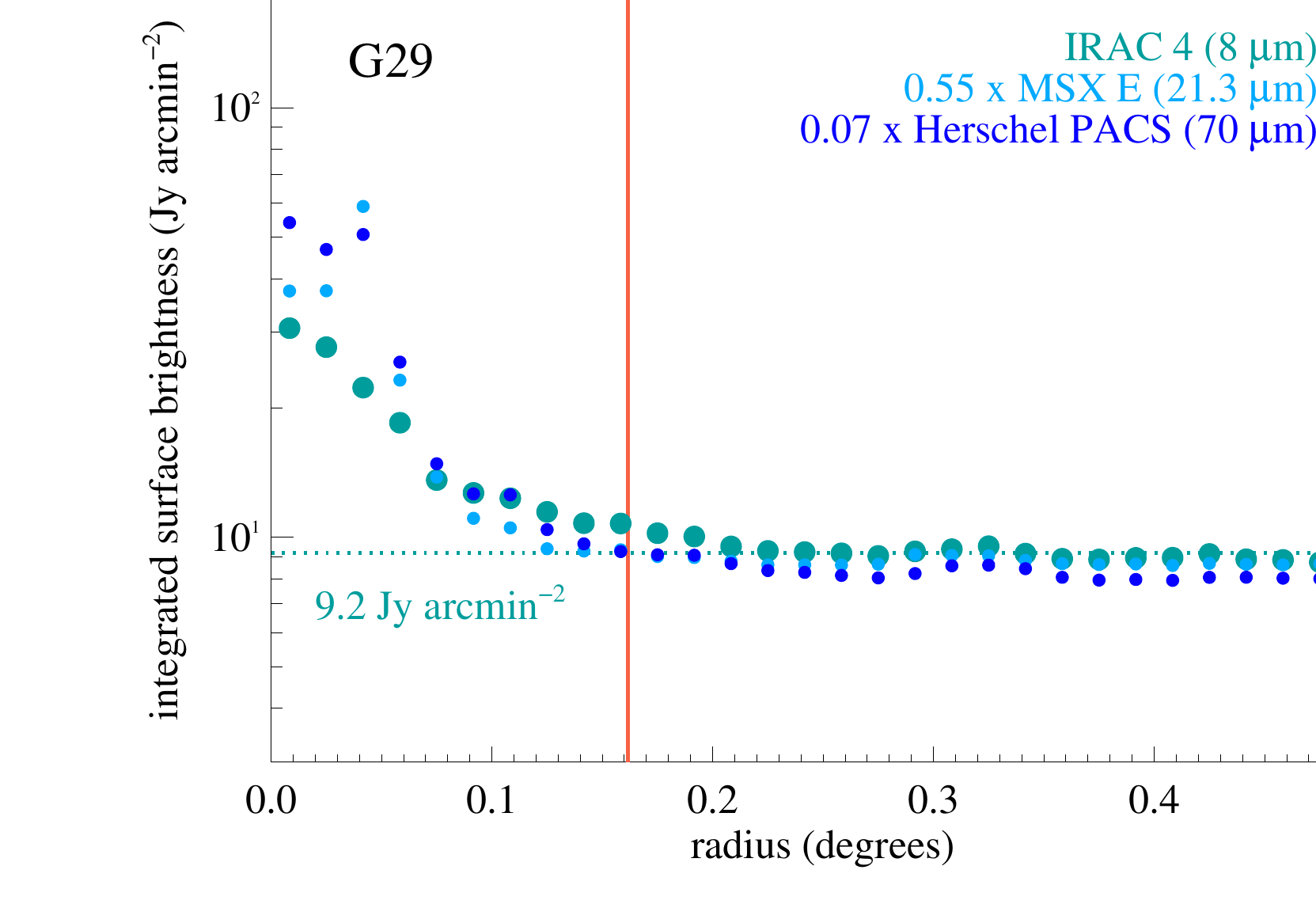}}
\figsetgrpnote{The RGB-rendered finding chart and surface brightness profile for G29.96-0.02. Blue is \spitzer IRAC 4 (8 \micron), green is \msx E (21.3 \micron), and red is \herschel PACS 70 \micron.}
\figsetgrpend

\figsetgrpstart
\figsetgrpnum{1.19}
\figsetgrptitle{NGC 6357 Finding Chart and Surface Brightness Profile}
\figsetplot{\includegraphics[width=0.31\linewidth,clip,trim=1.2cm 12.4cm 3.5cm 3.7cm]{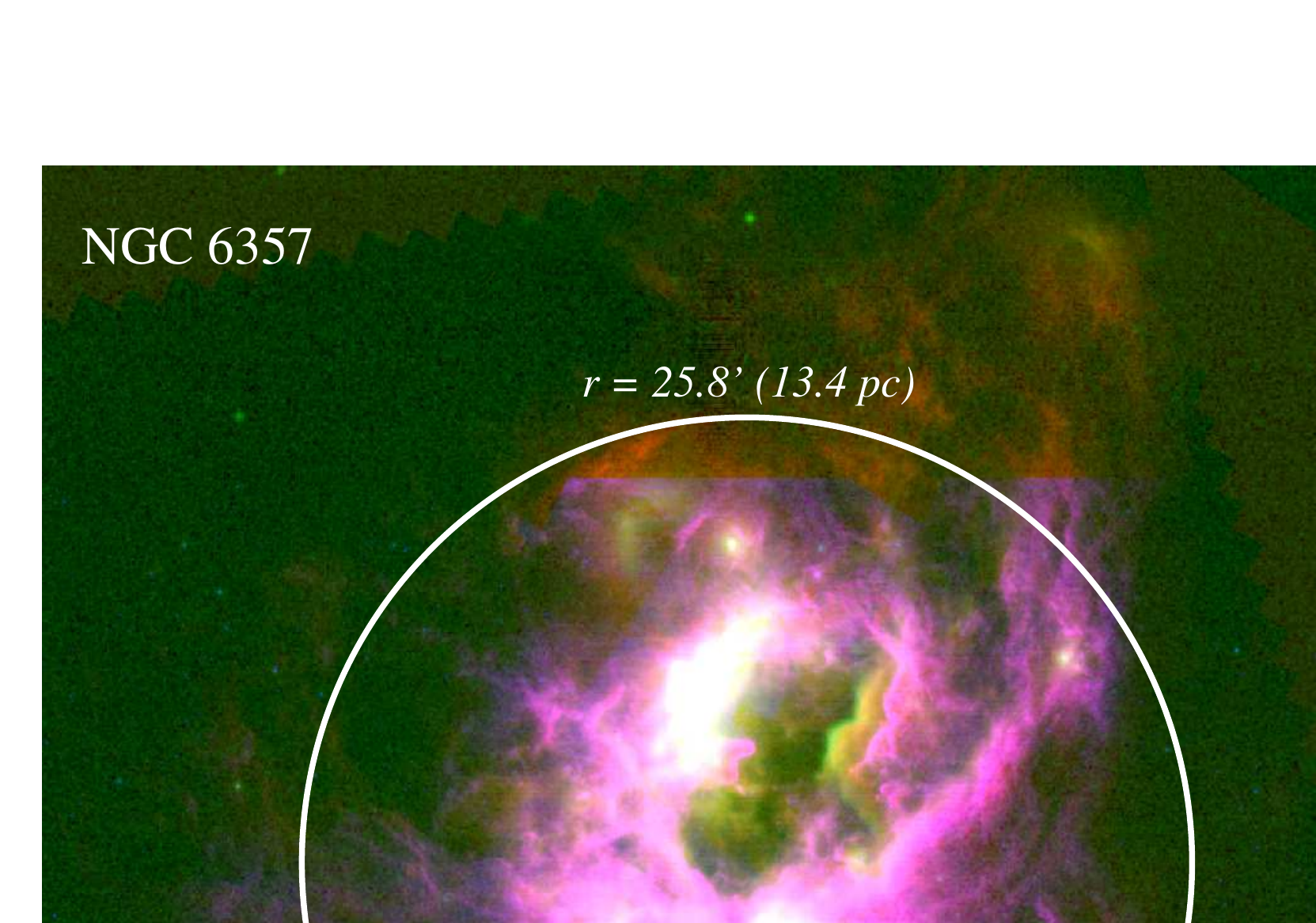}}
\figsetplot{\includegraphics[width=0.31\linewidth,clip,trim=1.2cm 12.4cm 3.5cm 3.7cm]{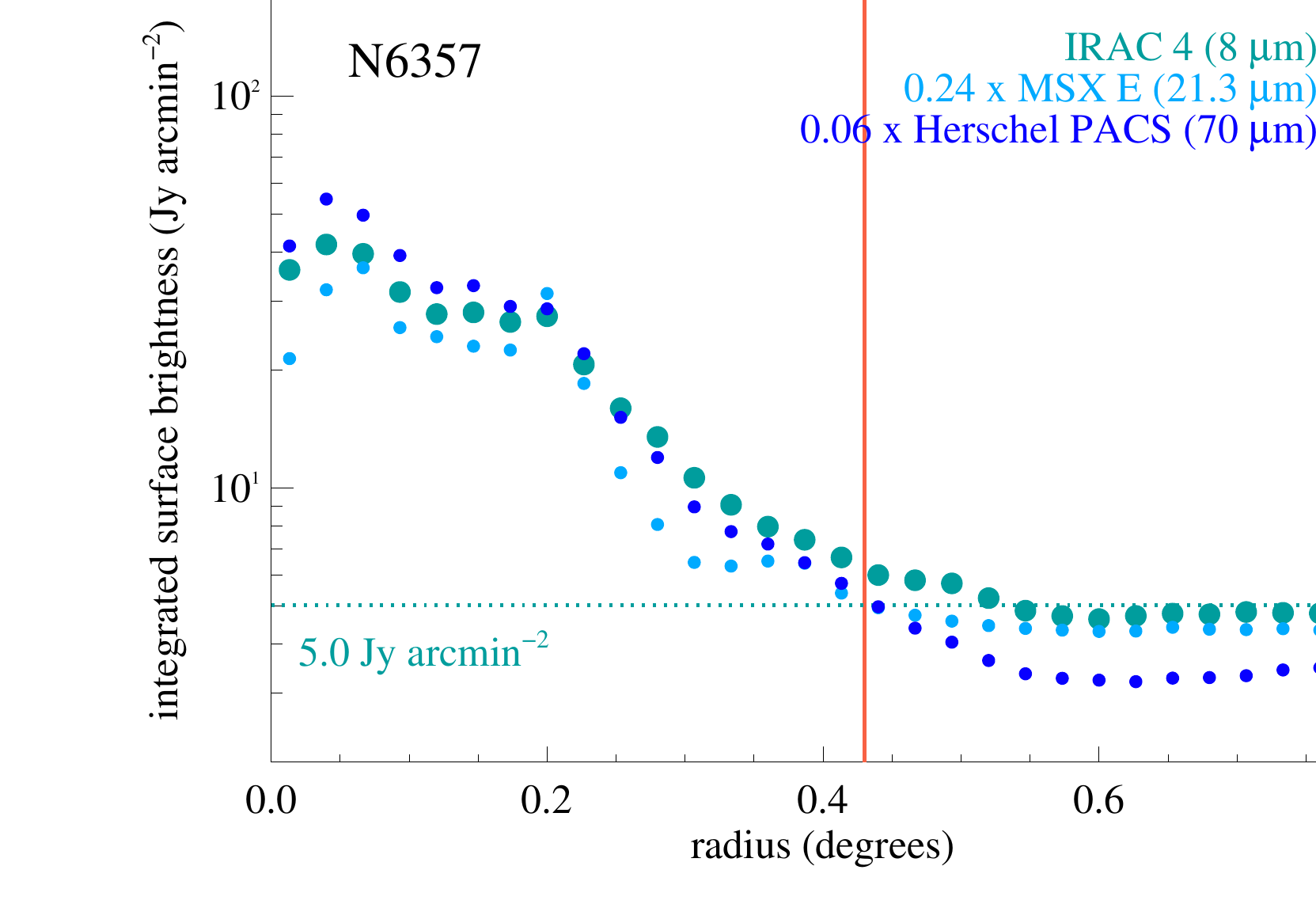}}
\figsetgrpnote{The RGB-rendered finding chart and surface brightness profile for NGC 6357. Blue is \spitzer IRAC 4 (8 \micron), green is \msx E (21.3 \micron), and red is \herschel PACS 70 \micron.}
\figsetgrpend

\figsetgrpstart
\figsetgrpnum{1.20}
\figsetgrptitle{M17 Finding Chart and Surface Brightness Profile}
\figsetplot{\includegraphics[width=0.31\linewidth,clip,trim=1.2cm 12.4cm 3.5cm 3.7cm]{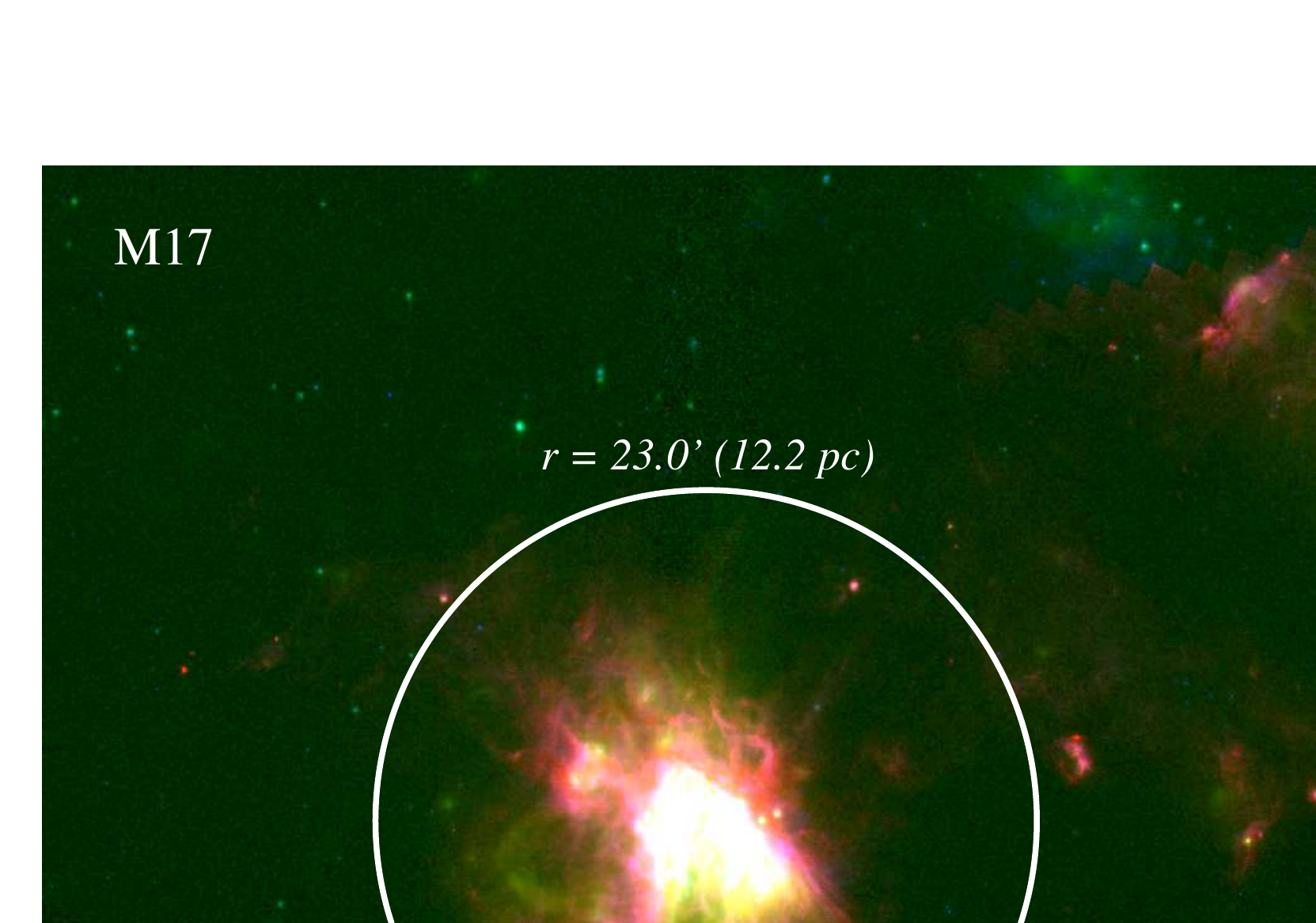}}
\figsetplot{\includegraphics[width=0.31\linewidth,clip,trim=1.2cm 12.4cm 3.5cm 3.7cm]{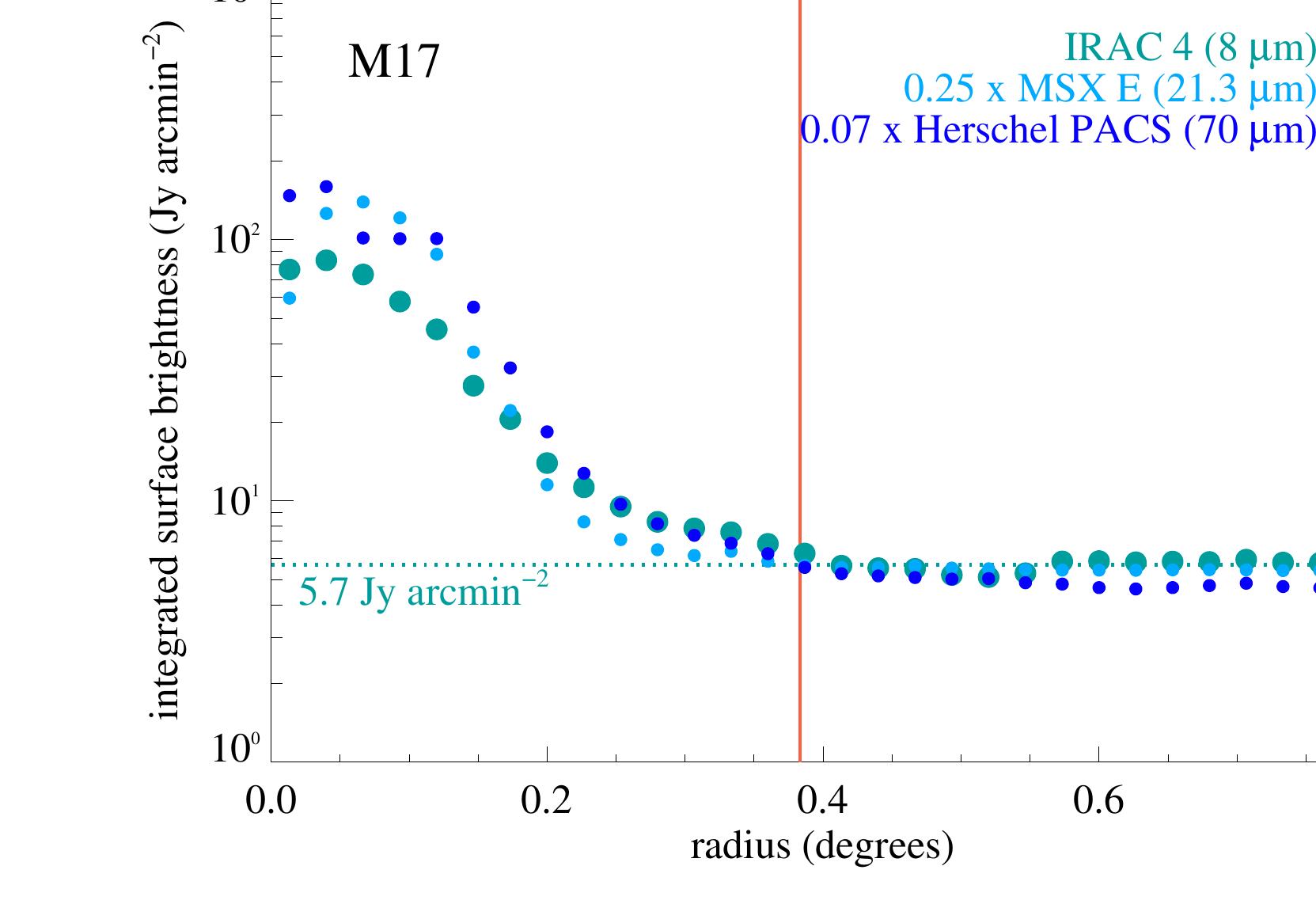}}
\figsetgrpnote{The RGB-rendered finding chart and surface brightness profile for M17. Blue is \spitzer IRAC 4 (8 \micron), green is \msx E (21.3 \micron), and red is \herschel PACS 70 \micron.}
\figsetgrpend

\figsetgrpstart
\figsetgrpnum{1.21}
\figsetgrptitle{G333 Finding Chart and Surface Brightness Profile}
\figsetplot{\includegraphics[width=0.31\linewidth,clip,trim=1.2cm 12.4cm 3.5cm 3.7cm]{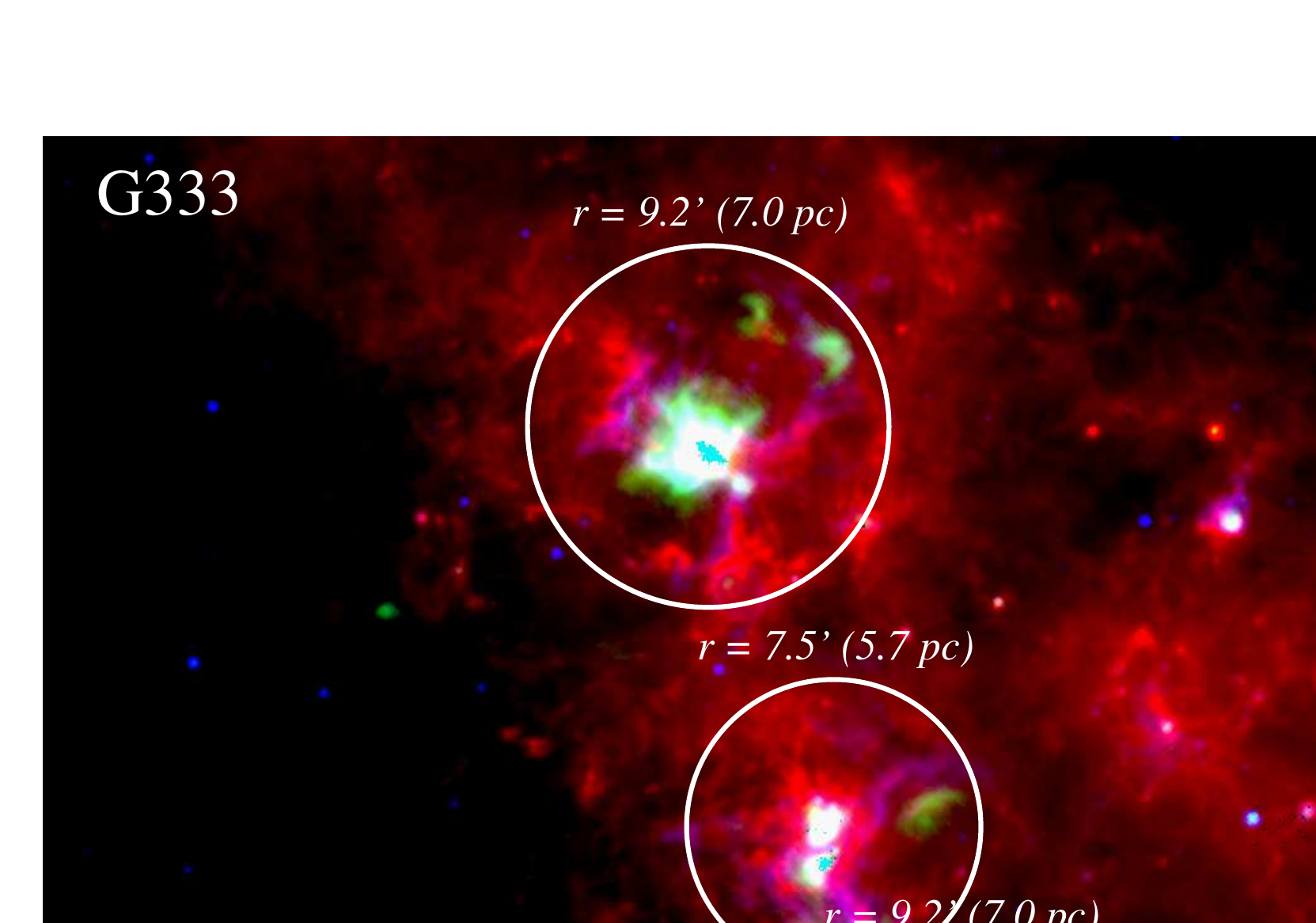}}
\figsetplot{\includegraphics[width=0.31\linewidth,clip,trim=1.2cm 12.4cm 3.5cm 3.7cm]{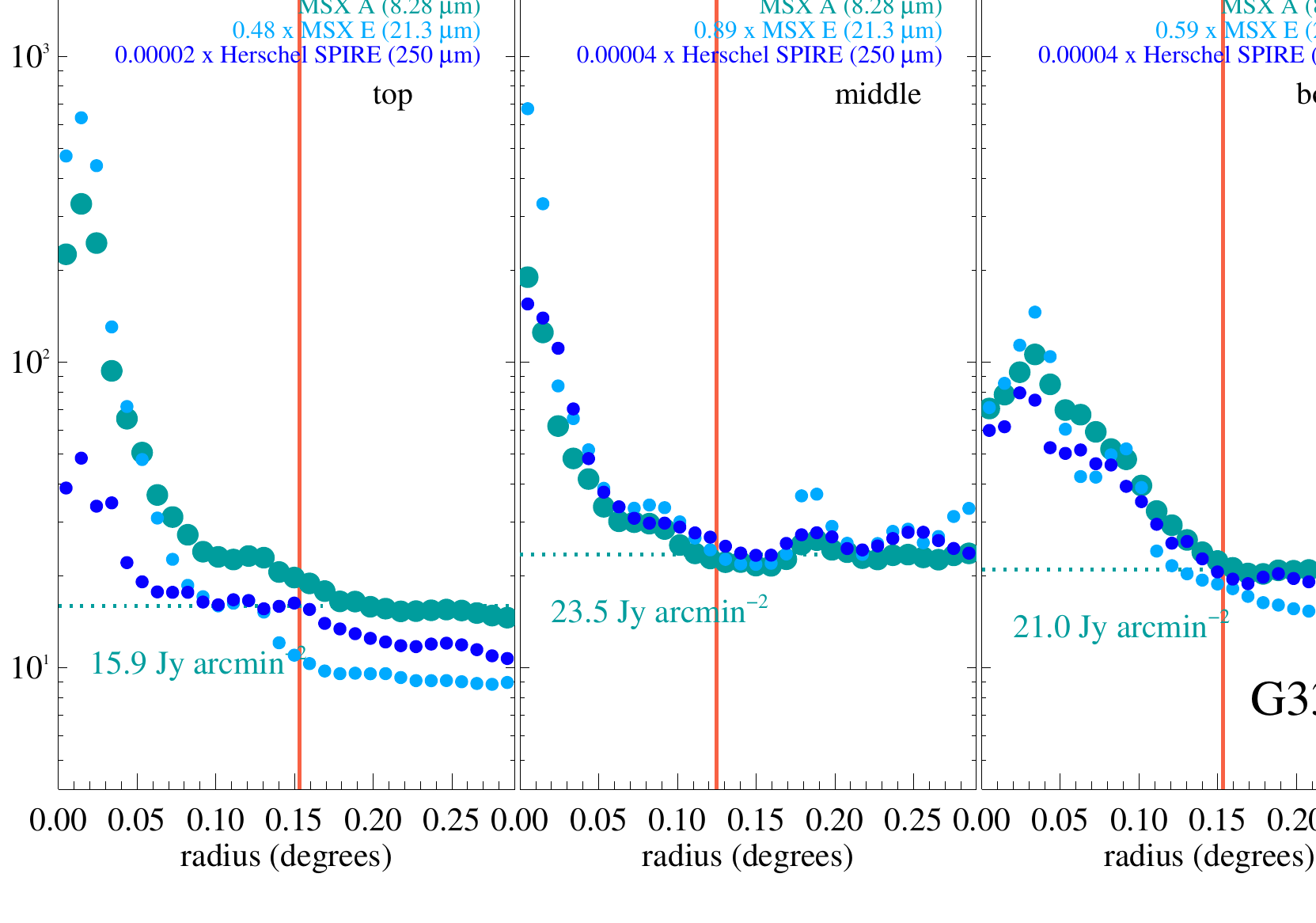}}
\figsetgrpnote{The RGB-rendered finding chart and surface brightness profile for G333. Blue is \msx A (8.28 \micron), green is \msx E (21.3 \micron), and red is \herschel SPIRE 250 \micron.}
\figsetgrpend

\figsetgrpstart
\figsetgrpnum{1.22}
\figsetgrptitle{W43 Finding Chart and Surface Brightness Profile}
\figsetplot{\includegraphics[width=0.31\linewidth,clip,trim=1.2cm 12.4cm 3.5cm 3.7cm]{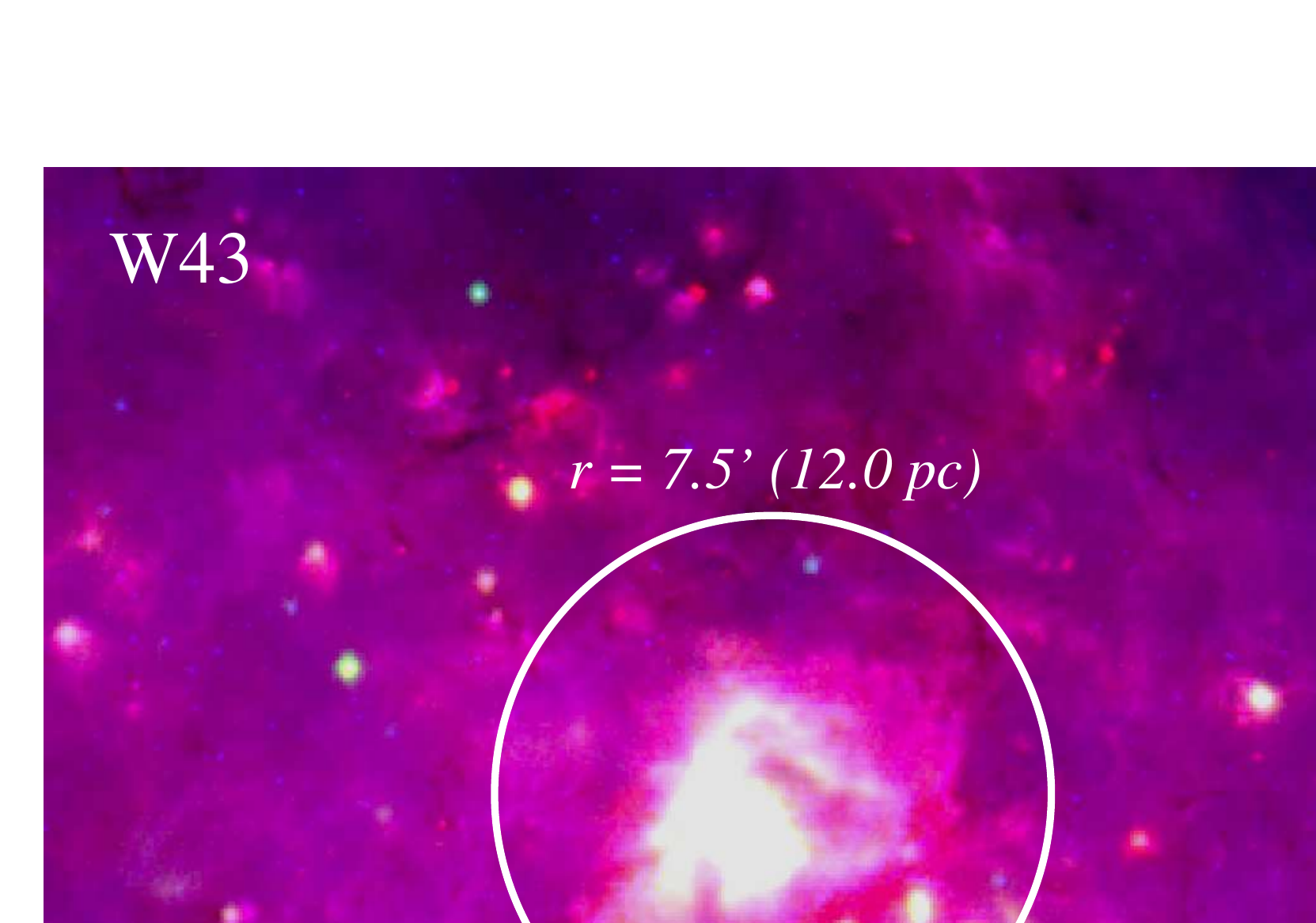}}
\figsetplot{\includegraphics[width=0.31\linewidth,clip,trim=1.2cm 12.4cm 3.5cm 3.7cm]{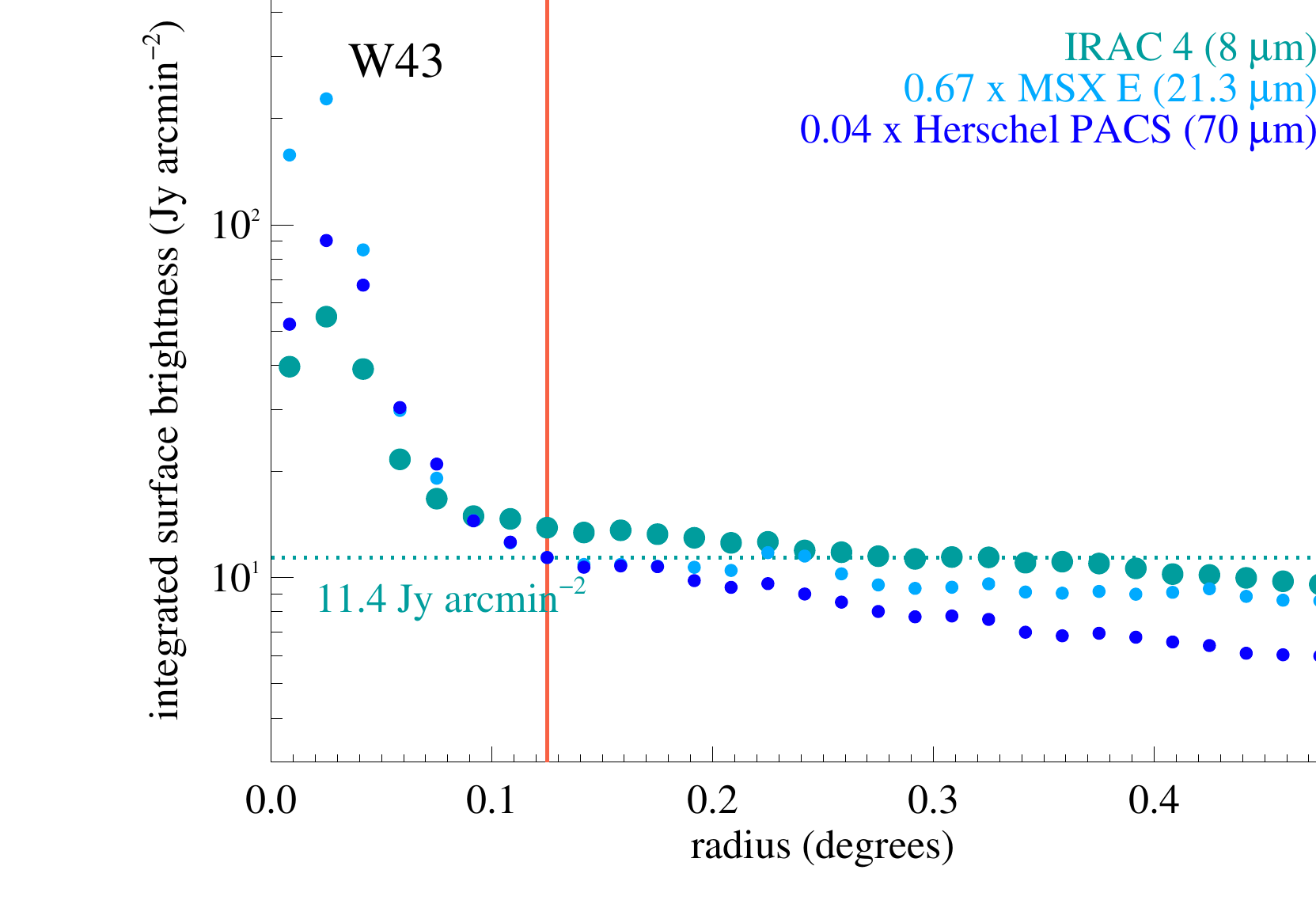}}
\figsetgrpnote{The RGB-rendered finding chart and surface brightness profile for W43. Blue is \spitzer IRAC 4 (8 \micron), green is \msx E (21.3 \micron), and red is \herschel PACS 70 \micron.}
\figsetgrpend

\figsetgrpstart
\figsetgrpnum{1.23}
\figsetgrptitle{RCW49 Finding Chart and Surface Brightness Profile}
\figsetplot{\includegraphics[width=0.31\linewidth,clip,trim=1.2cm 12.4cm 3.5cm 3.7cm]{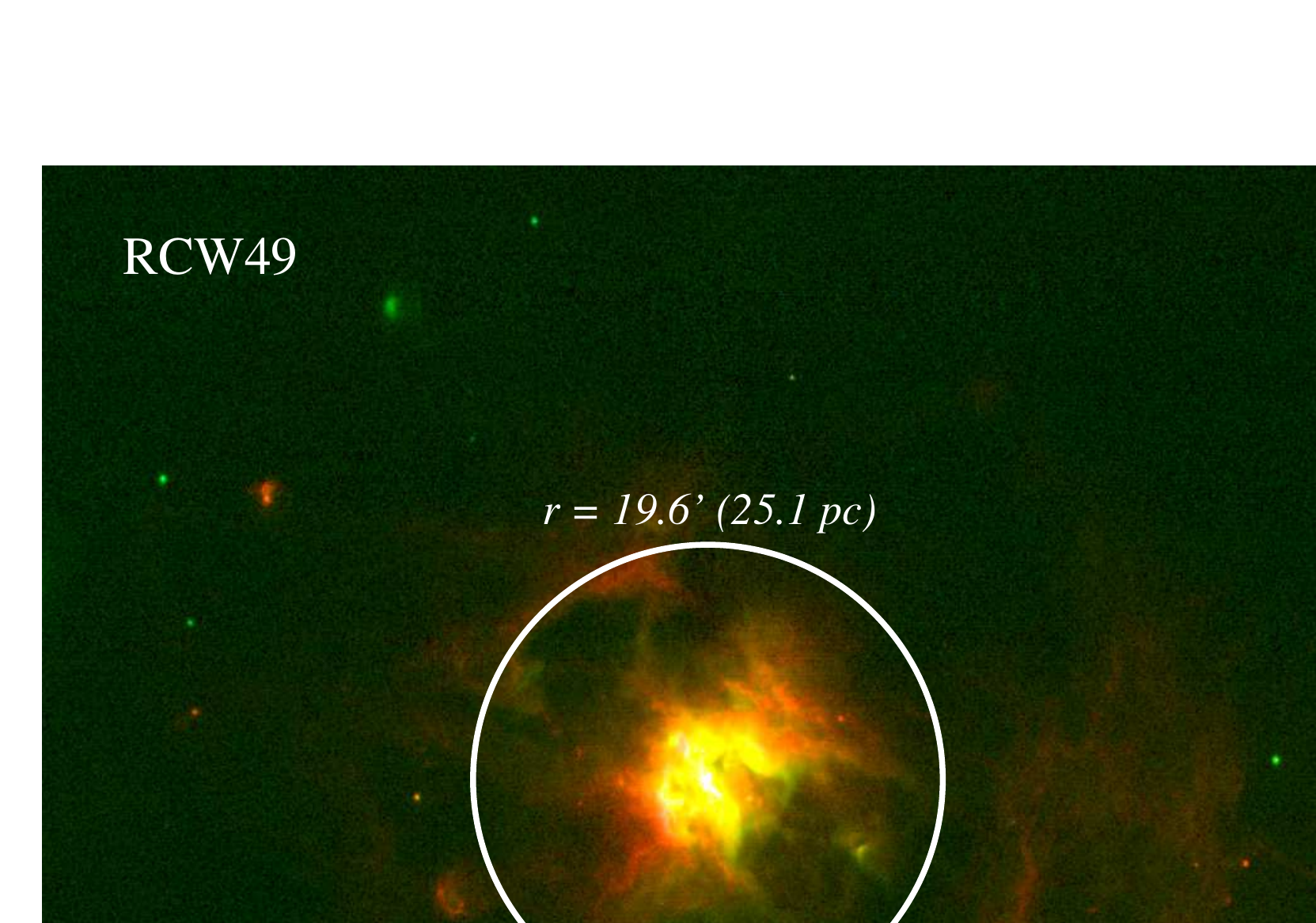}}
\figsetplot{\includegraphics[width=0.31\linewidth,clip,trim=1.2cm 12.4cm 3.5cm 3.7cm]{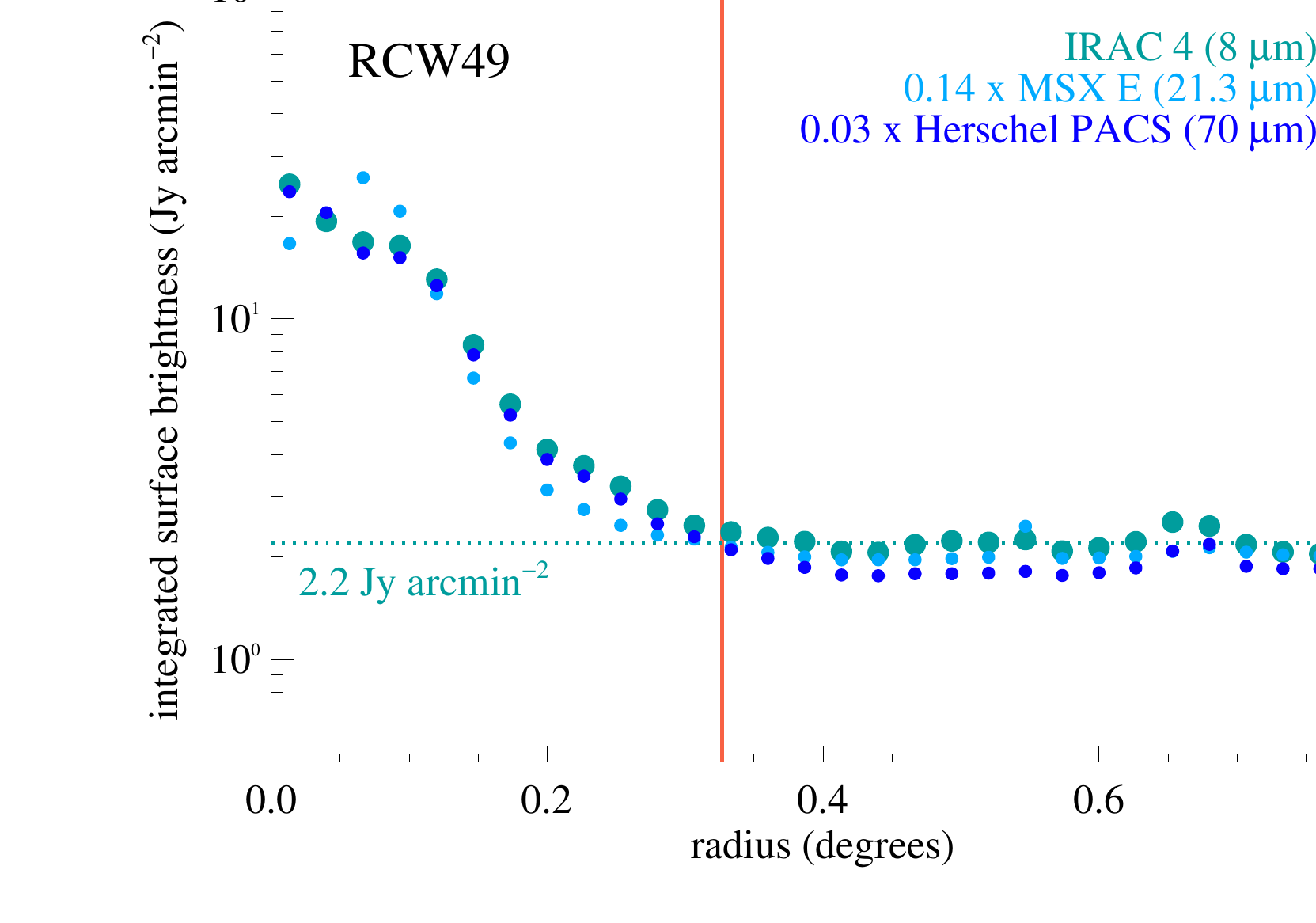}}
\figsetgrpnote{The RGB-rendered finding chart and surface brightness profile for RCW49. Blue is \spitzer IRAC 4 (8 \micron), green is \msx E (21.3 \micron), and red is \herschel PACS 70 \micron.}
\figsetgrpend

\figsetgrpstart
\figsetgrpnum{1.24}
\figsetgrptitle{G305 Finding Chart and Surface Brightness Profile}
\figsetplot{\includegraphics[width=0.31\linewidth,clip,trim=1.2cm 12.4cm 3.5cm 3.7cm]{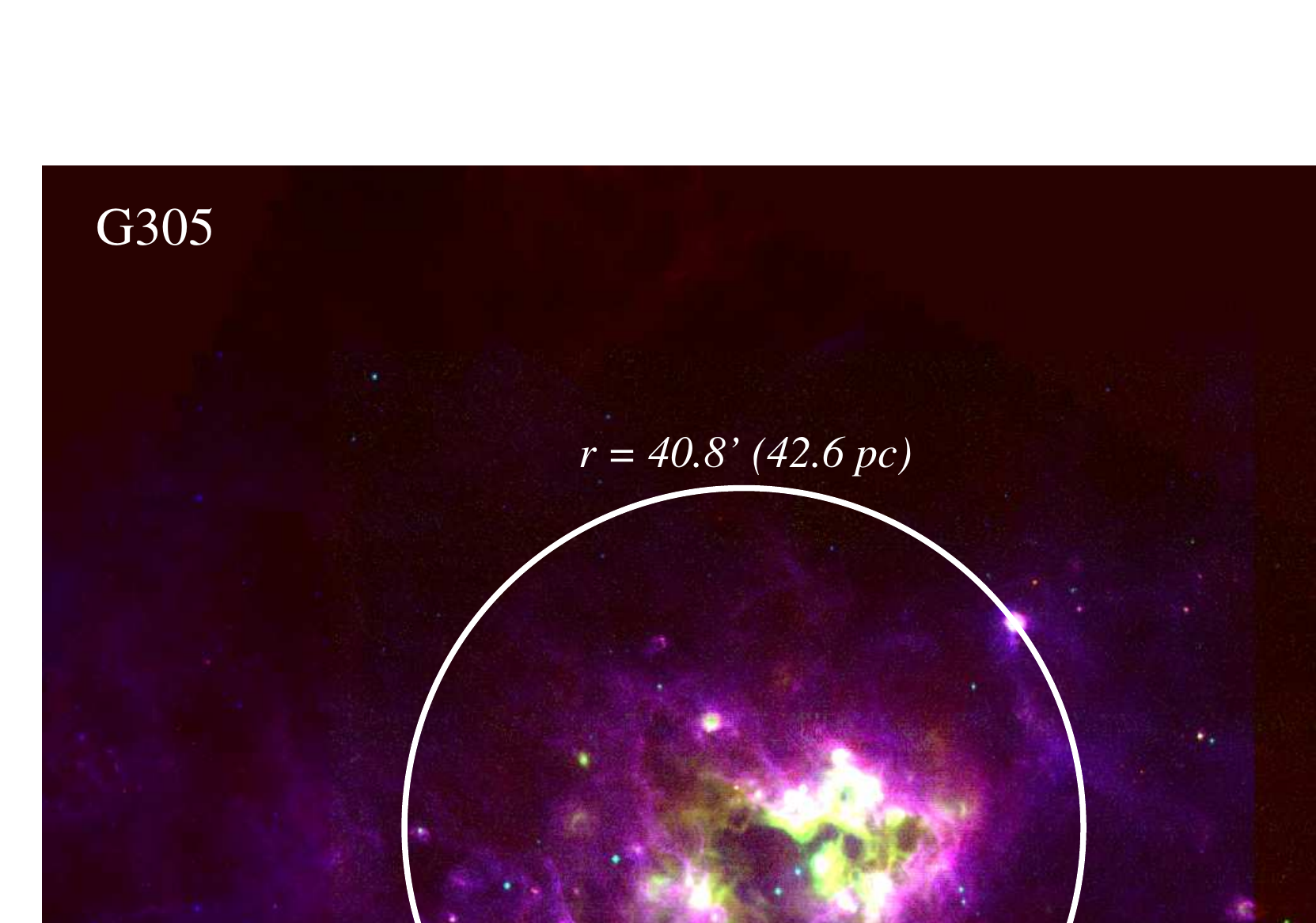}}
\figsetplot{\includegraphics[width=0.31\linewidth,clip,trim=1.2cm 12.4cm 3.5cm 3.7cm]{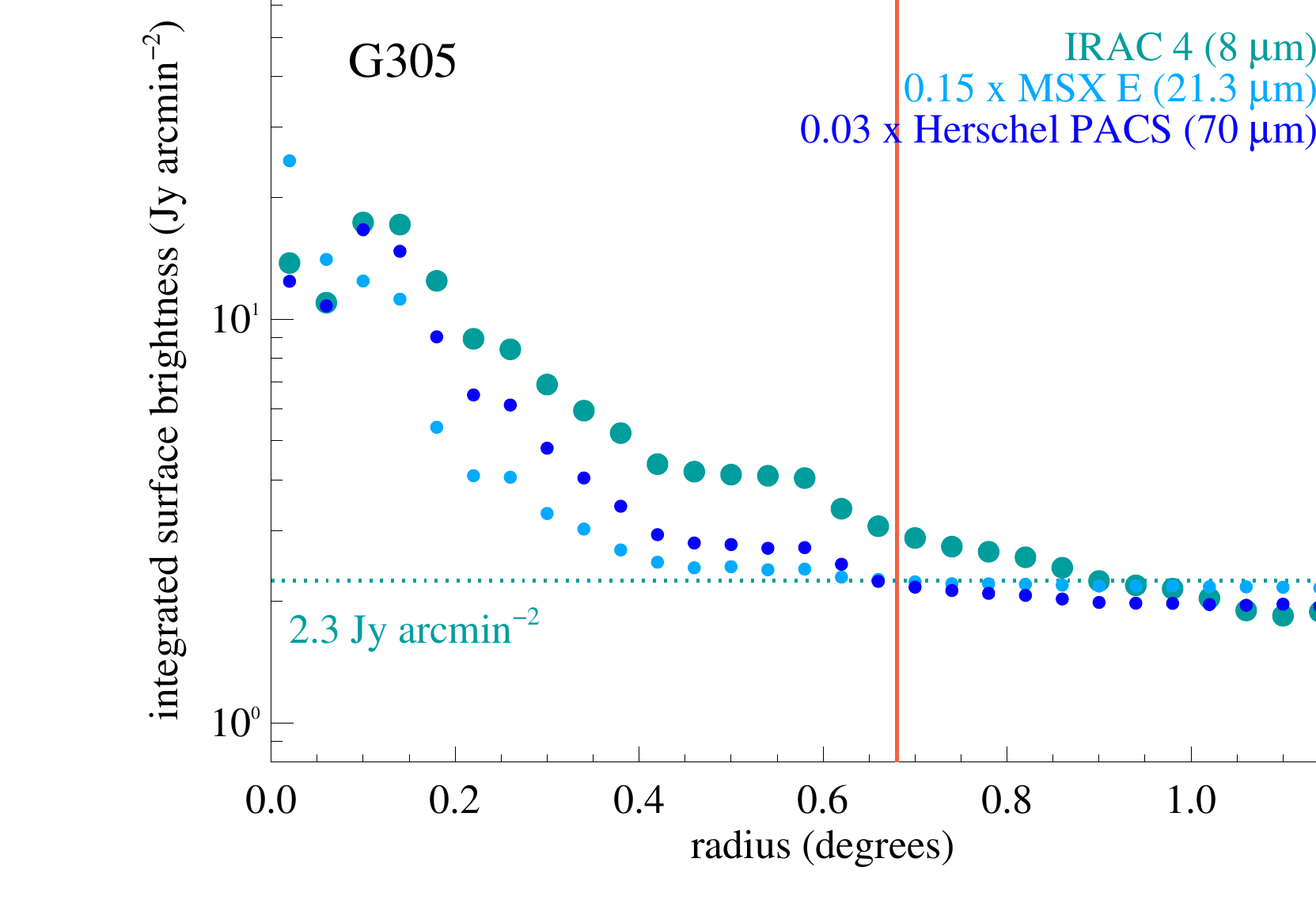}}
\figsetgrpnote{The RGB-rendered finding chart and surface brightness profile for G305. Blue is \spitzer IRAC 4 (8 \micron), green is \msx E (21.3 \micron), and red is \herschel PACS 70 \micron.}
\figsetgrpend

\figsetgrpstart
\figsetgrpnum{1.25}
\figsetgrptitle{W49A Finding Chart and Surface Brightness Profile}
\figsetplot{\includegraphics[width=0.31\linewidth,clip,trim=1.2cm 12.4cm 3.5cm 3.7cm]{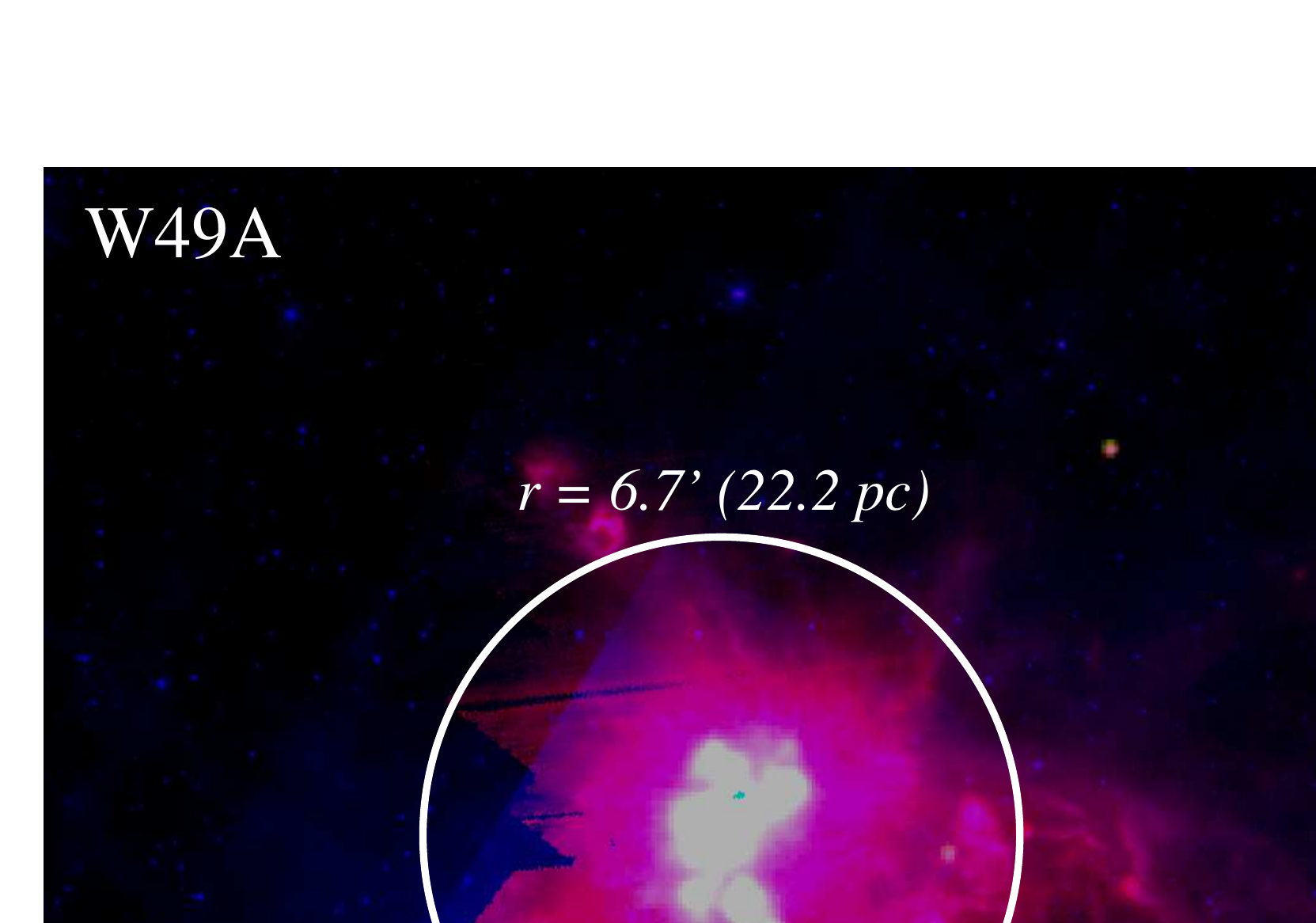}}
\figsetplot{\includegraphics[width=0.31\linewidth,clip,trim=1.2cm 12.4cm 3.5cm 3.7cm]{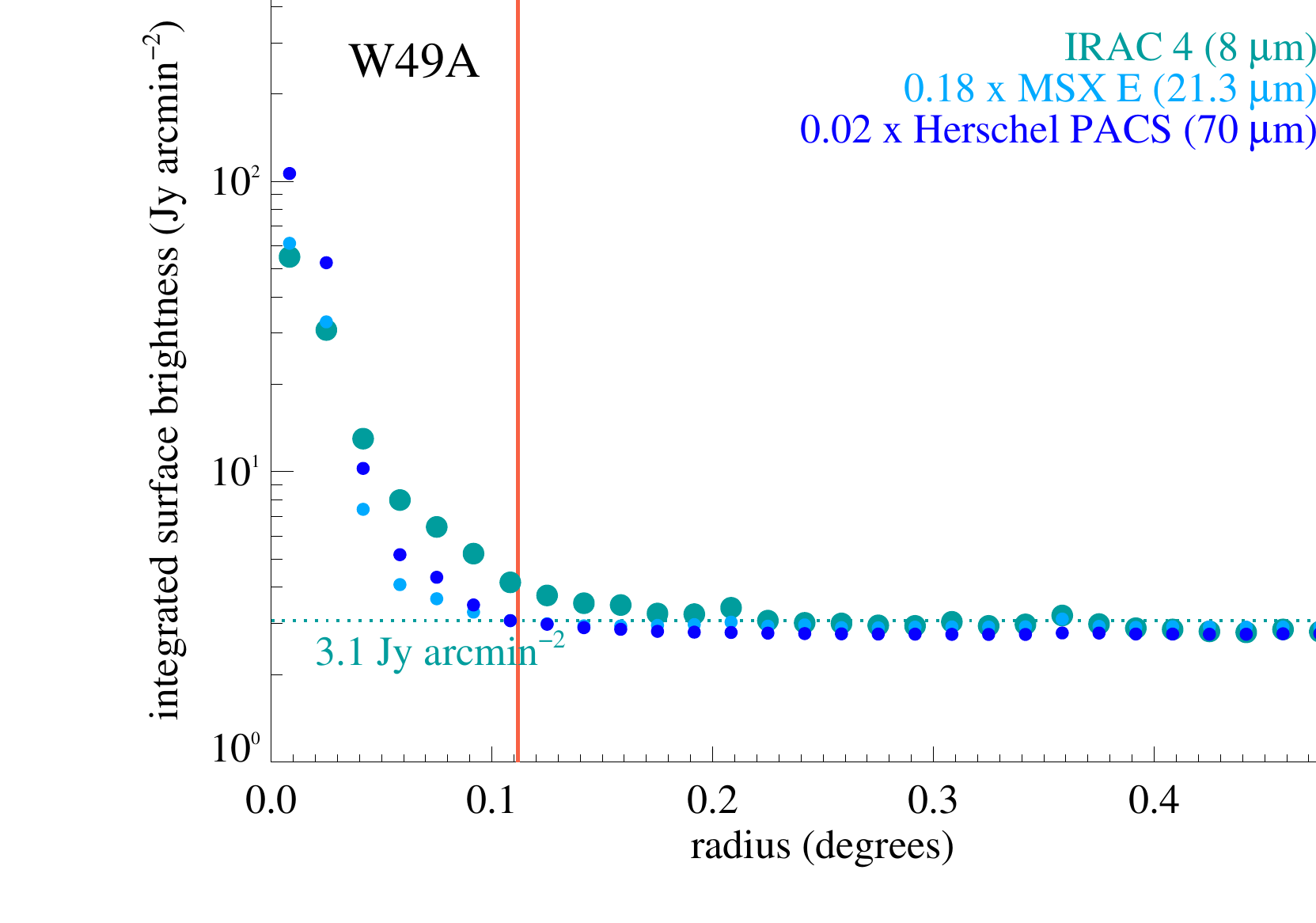}}
\figsetgrpnote{The RGB-rendered finding chart and surface brightness profile for W49A. Blue is \spitzer IRAC 4 (8 \micron), green is \msx E (21.3 \micron), and red is \herschel PACS 70 \micron.}
\figsetgrpend

\figsetgrpstart
\figsetgrpnum{1.26}
\figsetgrptitle{Carina Nebula Finding Chart and Surface Brightness Profile}
\figsetplot{\includegraphics[width=0.31\linewidth,clip,trim=1.2cm 12.4cm 3.5cm 3.7cm]{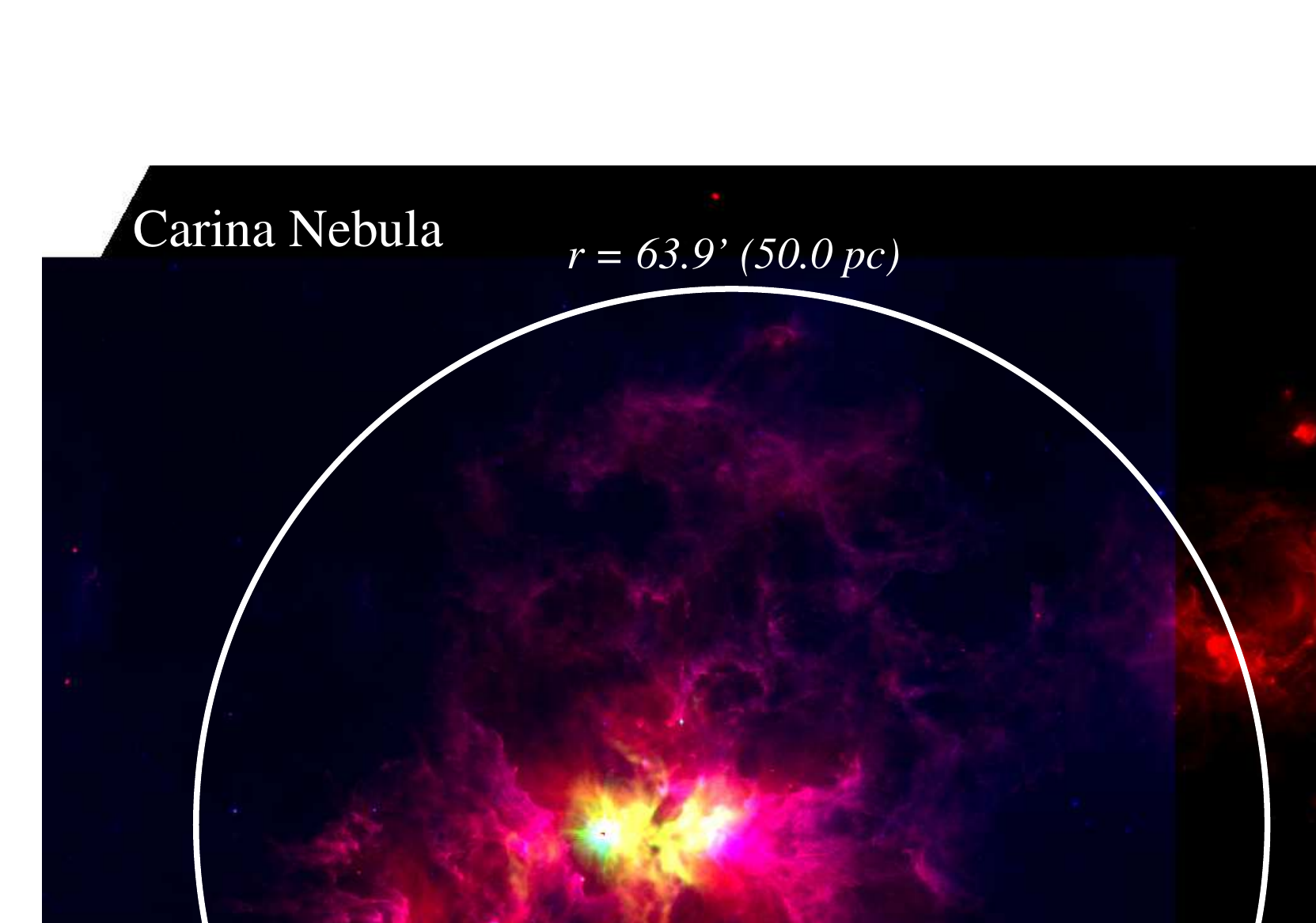}}
\figsetplot{\includegraphics[width=0.31\linewidth,clip,trim=1.2cm 12.4cm 3.5cm 3.7cm]{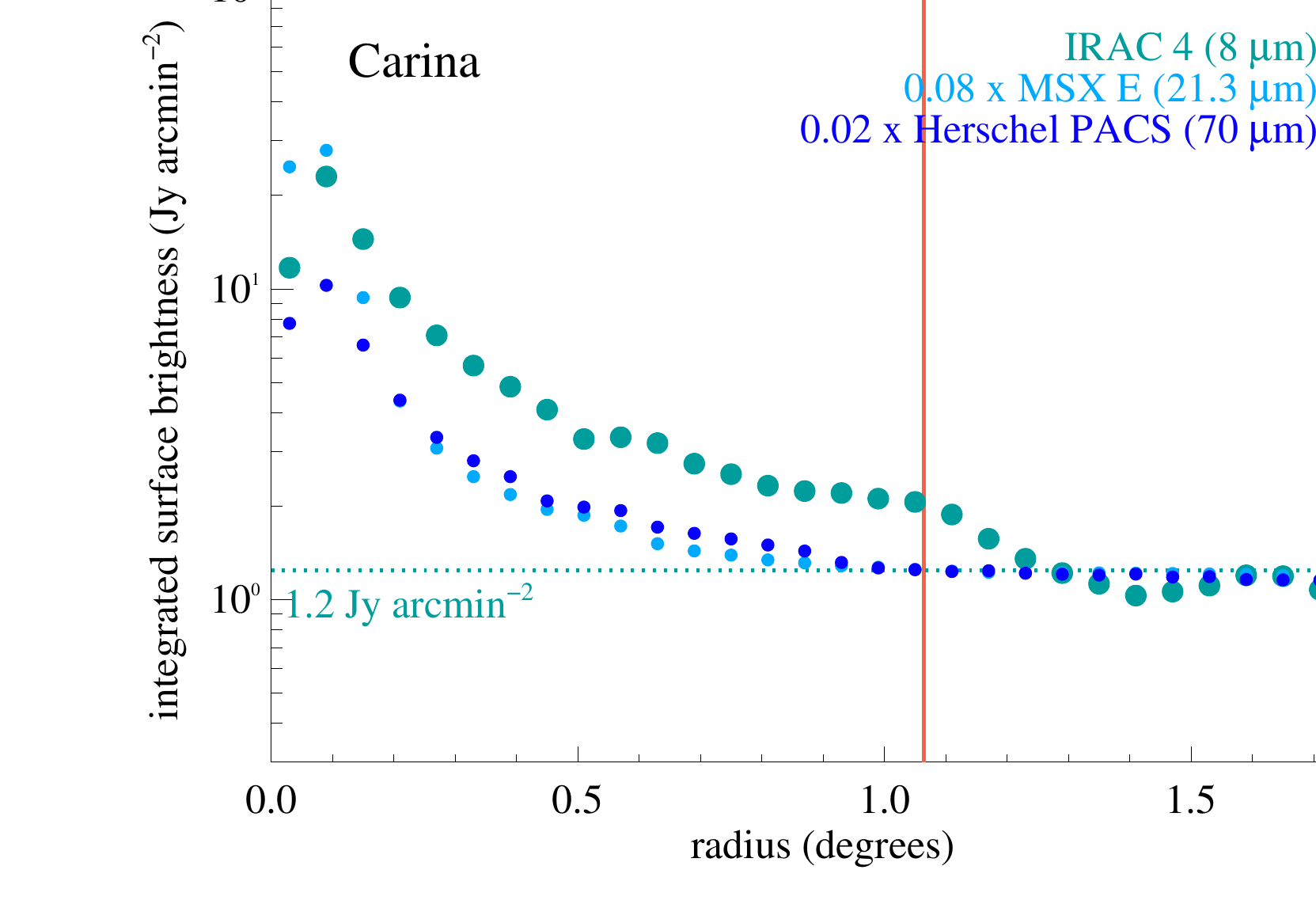}}
\figsetgrpnote{The RGB-rendered finding chart and surface brightness profile for the Carina Nebula. Blue is \spitzer IRAC 4 (8 \micron), green is \msx E (21.3 \micron), and red is \herschel PACS 70 \micron.}
\figsetgrpend
 
\figsetgrpstart
\figsetgrpnum{1.27}
\figsetgrptitle{W51A Finding Chart and Surface Brightness Profile}
\figsetplot{\includegraphics[width=0.31\linewidth,clip,trim=1.2cm 12.4cm 3.5cm 3.7cm]{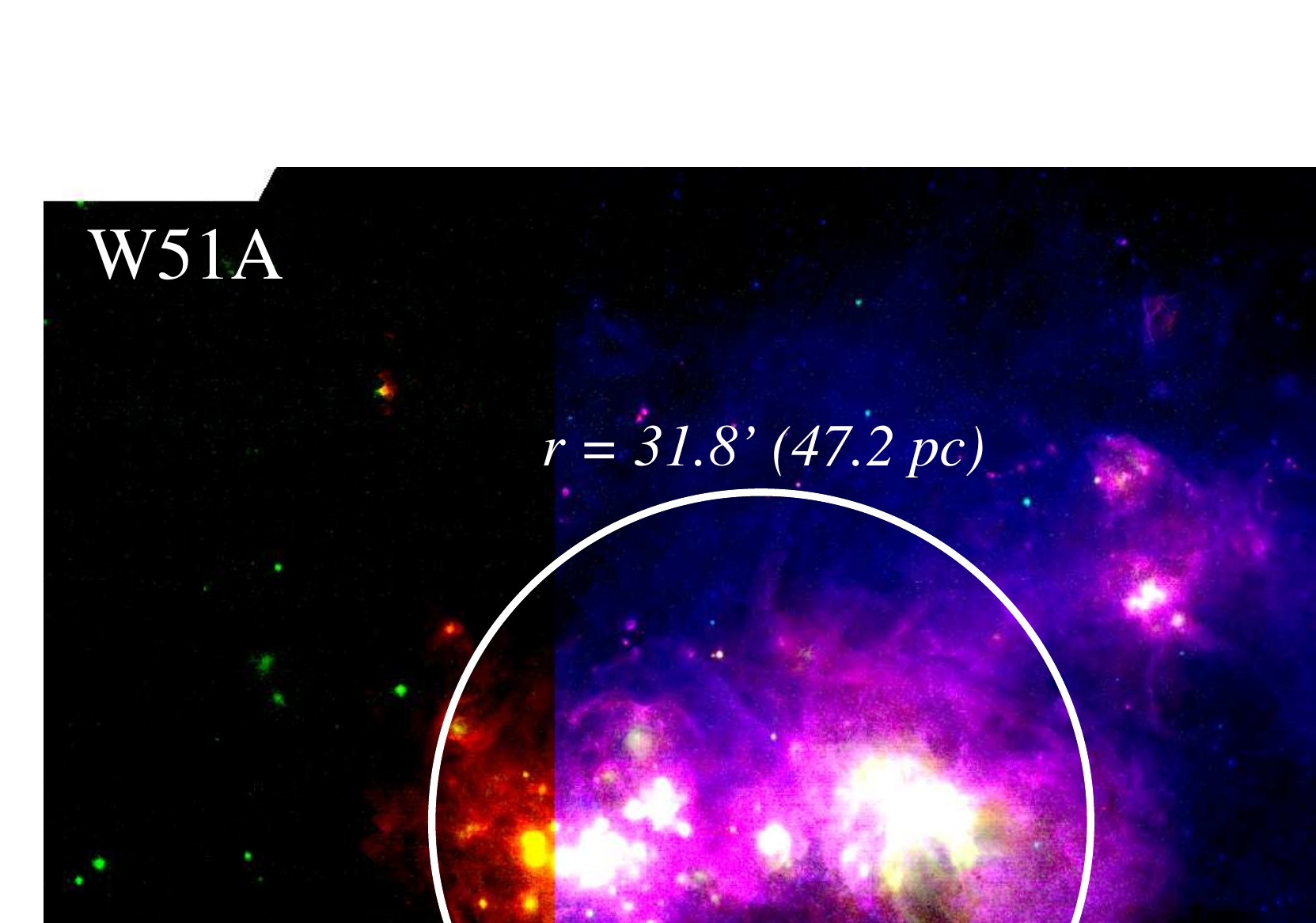}}
\figsetplot{\includegraphics[width=0.31\linewidth,clip,trim=1.2cm 12.4cm 3.5cm 3.7cm]{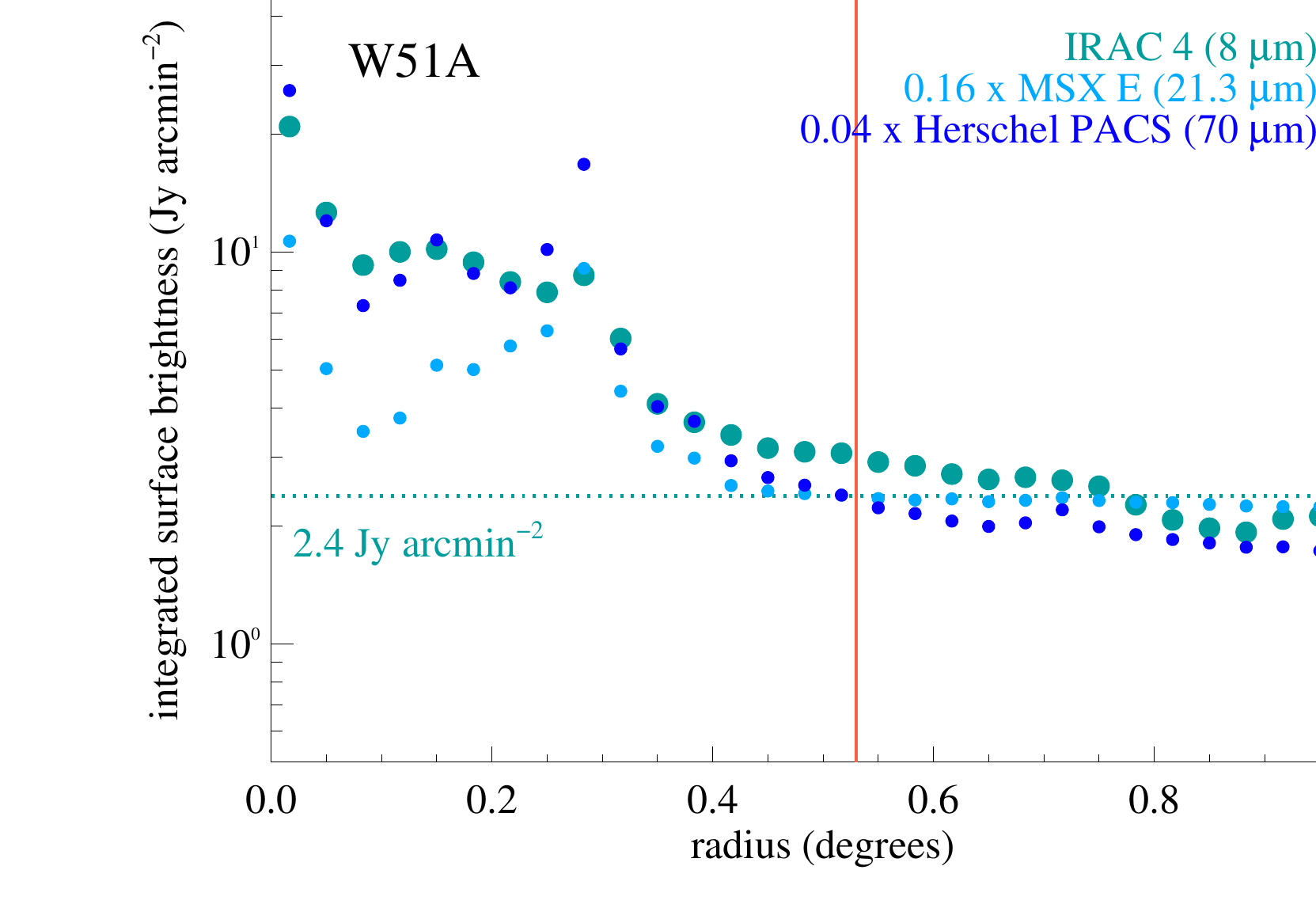}}
\figsetgrpnote{The RGB-rendered finding chart and surface brightness profile for W51A. Blue is \spitzer IRAC 4 (8 \micron), green is \msx E (21.3 \micron), and red is \herschel PACS 70 \micron.}
\figsetgrpend

\figsetgrpstart
\figsetgrpnum{1.28}
\figsetgrptitle{NGC 3603 Finding Chart and Surface Brightness Profile}
\figsetplot{\includegraphics[width=0.31\linewidth,clip,trim=1.2cm 12.4cm 3.5cm 3.7cm]{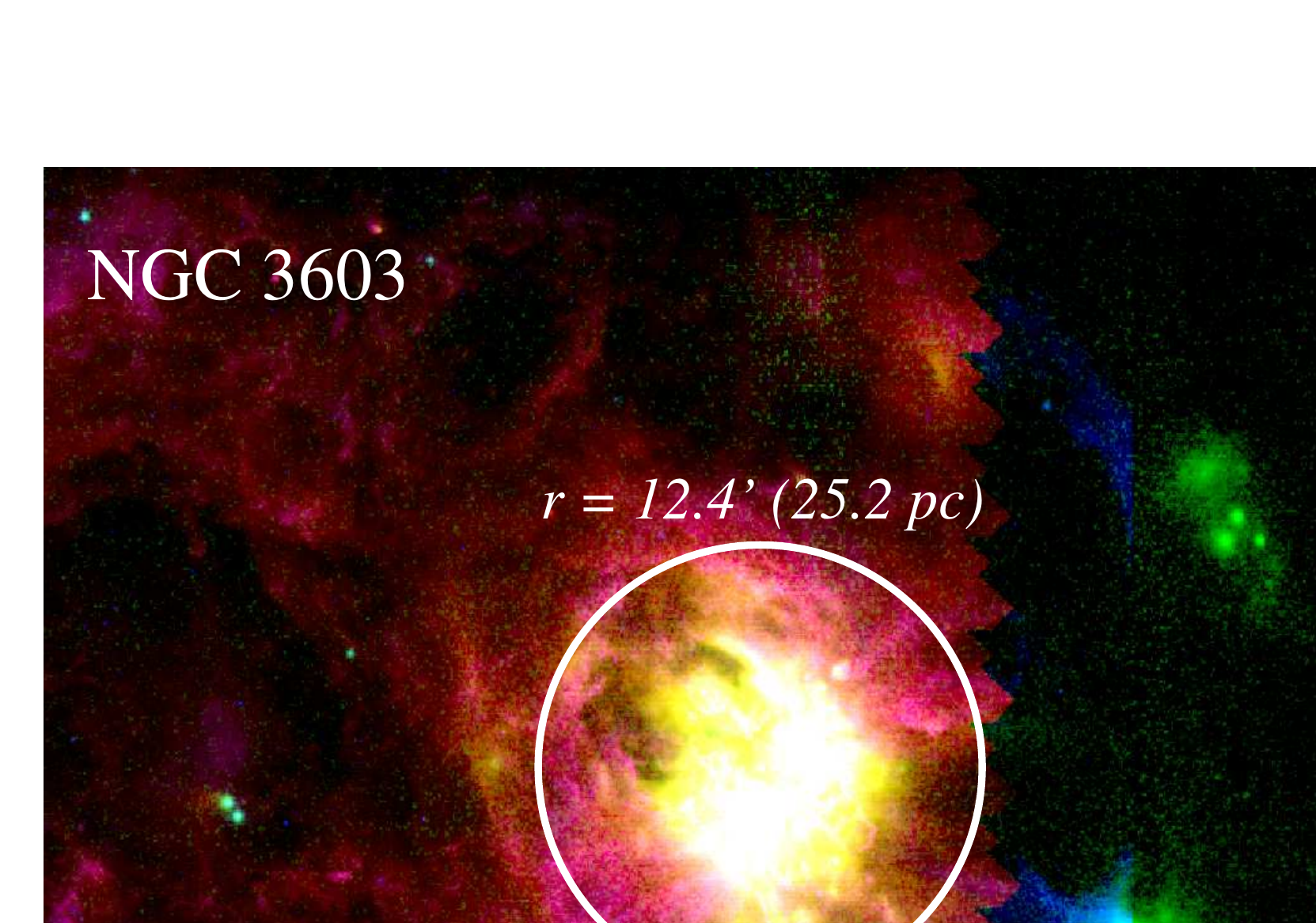}}
\figsetplot{\includegraphics[width=0.31\linewidth,clip,trim=1.2cm 12.4cm 3.5cm 3.7cm]{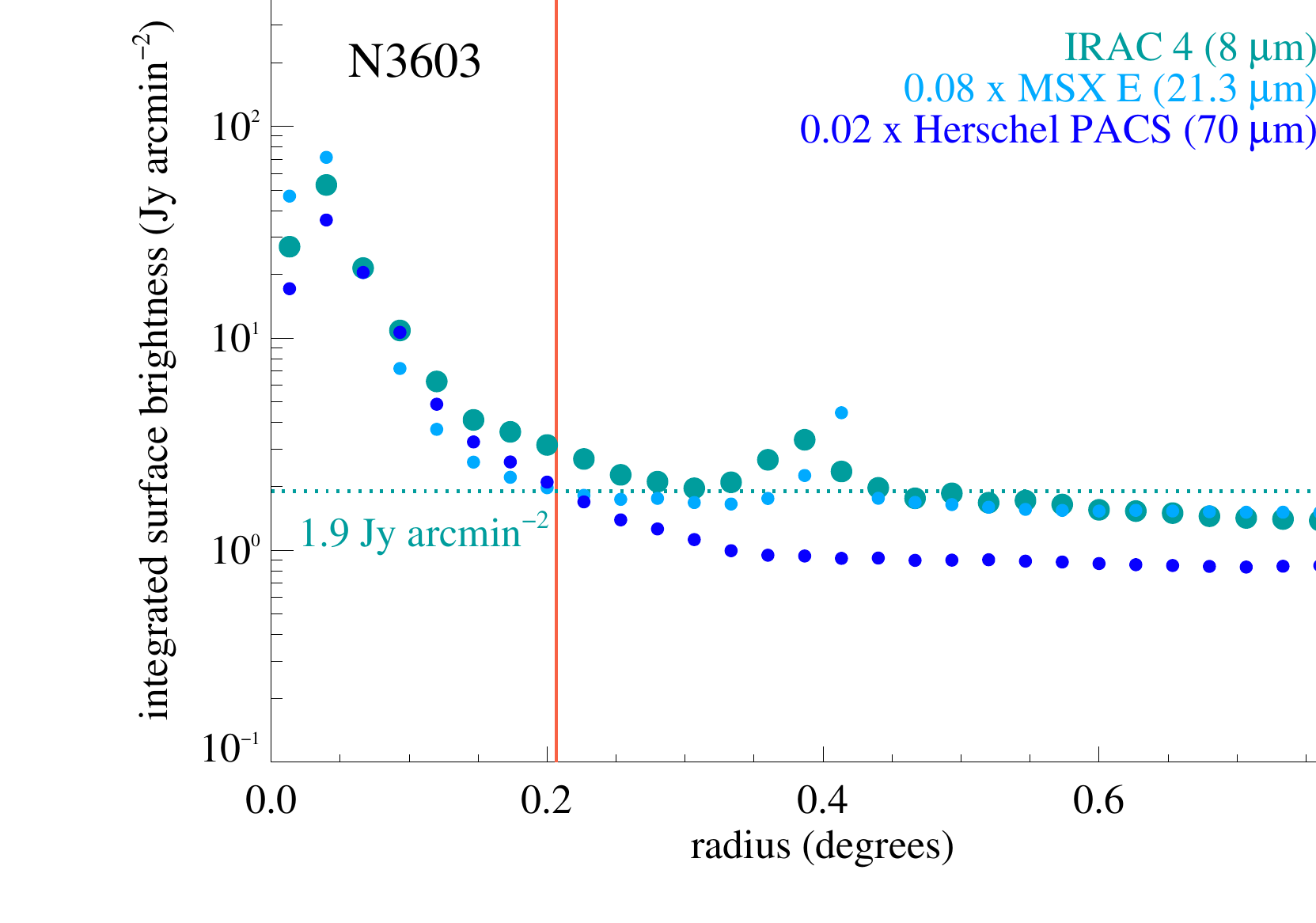}}
\figsetgrpnote{The RGB-rendered finding chart and surface brightness profile for NGC 3603. Blue is \spitzer IRAC 4 (8 \micron), green is \msx E (21.3 \micron), and red is \herschel PACS 70 \micron.}
\figsetgrpend
\figsetend

\begin{figure}[htp]
	\begin{tabular}{c}
		\includegraphics[width=0.95\linewidth,clip,trim=1cm 5cm 1cm 5cm]{fc_Orion.pdf} \\
		\includegraphics[width=1\linewidth,clip,trim=3.1cm 13.1cm 2.3cm 3.3cm]{sb_Orion.pdf}
	\end{tabular}
	\caption{{\it Top:} The RGB image (using a logarithmic stretch function) of the Orion Nebula, with the extraction aperture shown in white. Blue is \spitzer IRAC 4 (8.0 \micron), green is \msx E (21.3 \micron), and red is \herschel PACS 70 \micron. {\it Bottom:} Surface brightness profiles from \spitzer IRAC [8.0], \msx E, and PACS 70 \micron\xspace. The vertical red line indicates the outermost radius from within which the SEDs for the region were extracted. The horizontal dotted line indicates the background flux level. The complete figure set (56 images) is available in the online journal.}
\end{figure}\label{figure:surface_brightness}

In many cases, the 8 \micron\xspace aperture radius was smaller than the \planck FWHM (especially at 100 GHz and 70 GHz), which would have led to a potentially significant loss of radio flux. We therefore used {\it either} the IR-derived aperture or the \planck FWHM at each frequency, whichever was larger, to compute the radio flux. For some MSFRs in crowded regions, source confusion (particularly at lower frequencies) was a serious issue -- we only measure the radio flux in the frequency range where the MSFR is clearly resolved. The aperture size $r_{\rm ap}$ used to compute the radio flux for each region is given in Appendix~\ref{appendix:supplemental_figs}.

Data obtained from the various missions used in this work are reported in different units (for example, \herschel/PACS images are calibrated in units of Jy pixel$^{-1}$, while SPIRE images are in MJy sr$^{-1}$). We integrated the intensity images over the apertures defined above in MJy sr$^{-1}$ before converting each value to a flux density $S_0$ (Jy). The background-subtracted flux densities were calculated as
	
\begin{equation}
S = S_0 - B n_{\rm pix},
\end{equation}
	
\noindent where $n_{\rm pix}$ is the number of pixels contained within the source aperture and $B$ is the background level. The uncertainties are estimated as
	
\begin{equation}
\frac{\delta S}{S} = \frac{Bn_{\rm pix}}{S_0}.
\end{equation}
		
Additional sources of systematic error affect the absolute diffuse flux calibrations of IRAC images. \citet{Cohen+07} compared the IRAC 8 \micron\xspace and \msx 8.3 \micron\xspace for a sample of 43 Galactic \HII regions and found, correcting for the difference in bandpasses, that the present calibration of the IRAC 8 \micron\xspace band tends to overestimate diffuse fluxes by 36\%. This discrepancy is attributed to scattered light inside the camera, likely affecting the IRAC 5.8 \micron\xspace band as well. Aperture correction factors have been estimated for all four bands; we adopt the SSC ``infinite-aperture'' correction factors\footnote{See \url{http://ssc.spitzer.caltech.edu/irac/calib/extcal}} and multiply our flux densities at 3.6, 4.5, 5.8, and 8.0 \micron\xspace by factors of 0.91, 0.94, 0.73, and 0.74, respectively. All of our measured flux densities, aperture central coordinates, and adopted radio apertures are reported in Table~\ref{table:all_photometry} in the Appendix.

\section{The Spectral Energy Distributions}\label{section:SED_model_description}
	Our SED model contains three basic components: a set of dust models to describe the ``warm'' dust component ($\sim$3--100 \micron), a ``cool'' blackbody component to describe the far-IR ($\sim$20 K, at $\sim$100--500 \micron) observations, and the radio continuum:
	
	\begin{equation}
	S_{\nu} = S_{\rm dust} + S_{\rm blackbody} + S_{\rm power law}.
	\end{equation}

\noindent We discuss each component in detail below.
	
	\subsection{The ``Warm'' Dust Model}
We employ the dust emissivity models of \citet{Draine+Li07}, using the Galactic ``MW'' grain size distribution models \citep{Weingartner+01}. These models assume a canonical extinction law $A_V/E(B-V)=3.1$ mag. This extinction law may underestimate the dust emissivity at longer wavelengths (e.g., where the colder dust, described below, begins to dominate the SED). For this reason, we do not attempt to constrain the dust emissivity or dust mass using the \citet{Draine+Li07} models; the primary purpose of the SED modeling is to obtain accurate IR luminosities of the MSFRs. The dust is assumed to be a mixture of amorphous silicate and graphitic grains, heated by starlight, with the smallest carbonaceous grains having the physical properties of polycyclic aromatic hydrocarbon (PAH) molecules. The size distributions of these particles are chosen to reproduce the wavelength-dependent extinction in the Milky Way. The silicate and carbonaceous content of the dust grains was constrained by observations of the gas phase depletions in the interstellar medium \citep{Weingartner+01}. The PAH abundance in each model is characterized by the index $q_{\rm PAH}$, defined to be the percentage of the total grain mass contributed by PAHs containing less than 10$^3$ C atoms, which can range from 0.46\% to 4.6\%.

In addition to the physical dust mixture, the models also specify the intensity of the radiation field that is heating the dust grains. The IR emission spectrum is relatively insensitive to the detailed spectrum of the $h\nu<13.6$ eV photons, and the \citet{Draine+Li07} dust models simply adopt the spectrum of the local interstellar radiation field (ISRF). The specific energy density of starlight is therefore taken to be
	
\begin{equation}
u_{\nu} = U u_{\nu}^{\rm (MMP83)},
\end{equation}
	
\noindent where $u_{\nu}^{\rm (MMP83)}$ is the specific energy density estimated by \citet{Mathis+83} for the local Galactic ISFR and $U$ is a dimensionless scale factor. In order to account for the range of starlight intensities that may be present in MSFRs, we parameterize the starlight as the sum of two contributions: one describing the radiation field due to the central ionizing cluster, assumed to be a delta function where $U_{\rm min,1} = U_{\rm max,1} = U_1$, and the other describing a range of stellar intensities ranging from $U_{\rm min,2}$ to $U_{\rm max,2}$. The second contribution allows the stellar radiation field to decrease with increasing distance from the principal ionizing OB star(s) and as a result of attenuation by intervening dust. The flux density of the total warm dust model used in our fits is therefore given by

\begin{equation}
f_{\rm dust}(\lambda) = N_{\rm dust} \left[\gamma f_{\rm dust,1}\left( \lambda \right) + (1-\gamma) f_{\rm dust,2}\left( \lambda \right) \right],
\end{equation}

\noindent where $f_{\rm dust,1}(\lambda)$ is the $\delta$-function radiation field and $f_{\rm dust,2}$ is the radiation field described by a range of stellar intensities. The total warm dust model is defined by the PAH fraction of each component ($q_{\rm PAH,1}$ and $q_{\rm PAH,2}$), the minimum and maximum stellar radiation fields experienced by component two ($U_{\rm min,2}$ and $U_{\rm max,2}$), the radiation field experienced by component one ($U_1$), the fraction of flux density emitted by each component ($\gamma$), and a normalization constant ($N_{\rm dust}$). Typically, $U_{\rm min,2}$ spans 0.1--1.00 while $U = U_{\rm max,2}$ = 10$^3$--10$^5$, and $\gamma$ is small ($\sim$10$^{-5}$). For regions without complete \spitzer coverage, this dust component is almost completely unconstrained. In these cases, we utilize a single dust component and report only $U_1$ and $q_{\rm PAH}$.

	\subsection{The ``Cool'' Blackbody}
	A single-temperature blackbody modified by an emissivity law proportional to $\lambda^{-\beta}$ is used to fit the cool ($\sim$20--30 K) dust component of the MSFRs, captured primarily by the SPIRE 250 \micron, 350 \micron, and 500 \micron\xspace channels. We refer to this component as the ``cool'' blackbody to differentiate it from ``cold'' ($\leq$10 K) dust in the ISM. Laboratory studies of interstellar dust analogs have found that $\beta\sim1-2$ for carbonaceous grains \citep{Mennella+95,Zubko+96,Jager+98} and $\beta\sim$2 for silicate grains \citep{Mennella+98,Boudet+05,Coupeaud+11} at FIR wavelengths. The effective value of $\beta$ for interstellar dust depends on the interstellar dust mixture and the interstellar radiation field. We assume $\beta$ = 1.5, consistent with observational constraints from SPIRE \citep[e.g.,][]{Dunne+Eales01,Paradis+09,Gordon+10,Gordon+14,Skibba+11}. The inferred dust temperatures depend marginally on the assumed emissivity, with $\beta$ = 2 yielding temperatures systematically lower by a few degrees \citep{Bendo+03}. With $\beta$ fixed, the blackbody component of our SED model is defined only by the dust temperature ($T_{\rm BB}$) and a normalization component ($N_{\rm BB}$).
	
We note that the dust opacity and total dust mass will depend on the normalization components $N_{\rm BB}$ and $N_{\rm dust}$. Due to the uncertainties of the dust properties, we do not attempt to estimate the dust mass for any of the MSFRs. The normalizations are used only for estimating the total IR luminosity of each region (see below, Section~\ref{section:SED_fit_results}).

	\subsection{The Radio Continuum}
	The nebular radio emission from MSFRs \citep[as well as entire star-forming galaxies;][]{Deeg+93} originates from two principal mechanisms: thermal bremsstrahlung (free-free) emission and non-thermal synchrotron radiation from supernovae (SNe). Both free-free and synchrotron radiation produce power law radio continua, with a spectral index $\alpha$ defined by

\begin{equation}
\alpha = \frac{d\text{log}S_{\nu}}{d\text{log}\nu}.
\end{equation}
	
\noindent We hence adopt the sign convention for which negative values of $\alpha$ indicate decreasing flux density with increasing frequency. Optically thin free-free emission is characterized by $\alpha=-0.1$, while non-thermal synchrotron emission typically yields $\alpha=-0.5$ \citep[e.g., see][and references therein]{Klein+88,Carlstrom+91}. For regions where we are able to estimate the radio flux at all four \planck frequencies, both the power law spectral index $\alpha$ and the power law normalization are free parameters in our fit. For regions where source confusion is an issue, and only one or two radio flux measurements are available, we assume $\alpha=-0.1$ (corresponding to a pure thermal continuum) and only fit the normalization component.

The IRAC [4.5] band, while free of strong PAH emission bands, contains the H~I Br$\alpha$ recombination line at 4.05 \micron, a potentially strong emission feature in \HII regions. Following the method of \citet{Povich+07} we use the thermal continuum flux density from the \planck radio observations to calculate the contribution of the Br$\alpha$ line to the IRAC [4.5] flux density, which is typically ${\sim}1$--20\%. We then increase the model-predicted 4.5 \micron\xspace flux by this amount prior to fitting the SEDs.

	\subsection{Performing the Fit}
Our model SEDs are well-sampled in wavelength, whereas our observed SEDs are not. We therefore integrate the model SED flux density ($S_{\nu}$) over the (broad) response functions for each filter using
	
\begin{equation}
S_{\rm band} = \frac{\int S_{\nu} R_E\left( \nu \right) d\nu}{\int \left(\nu_0 / \nu \right)^{-1} R_E\left( \nu \right) d\nu}.
\end{equation}

\noindent $R_E(\nu)$ is the response function for each filter\footnote{Filter profiles were obtained from the SVO Filter Profile Service, \url{http://svo2.cab.inta-csic.es/theory/fps/}.} used in our SED modeling, and $\nu_0$ is the central frequency of the bandpass. For simplicity, we assume a $S(\nu) = \nu^{-1}$ reference spectrum \citep[e.g., as used in the calibration of \herschel PACS and SPIRE, where the SED peaks; ][]{Gordon+14}. Although the shape of the underlying \HII spectrum is frequency-dependent, the individual filter bandpasses are significantly narrower than the full SED; thus, changes the reference spectrum will change the model flux in each filter by only a few percent.

We use the IDL routine \texttt{mpfitfun} \citep{Markwardt09} to fit the observed fluxes to the model fluxes, and the model yielding the lowest $\chi^2$ per degree of freedom (\chisqr) is selected as the best-fit model. Our model is applied to a grid of possible dust model combinations defined by $[{q_{\rm PAH,1}, q_{\rm PAH,2}, U_{\rm min,2}, U_{\rm max,2}, U}$]. The best-fit values of $\gamma$, $N_{\rm dust}$, $T_{\rm BB}$, $N_{\rm BB}$, $\alpha$, $f_{\rm Br\alpha}$, and $N_{\rm PL}$ are determined for each parameter set; these are the parameters which determine the number of degrees of freedom for each model.

	\subsection{Results of SED Fitting}\label{section:SED_fit_results}
We calculate the bolometric luminosity $L_{\rm TIR}$ of each \HII region by integrating our best-fit model over the wavelength range probed by our IR photometry (3.6--500 \micron), assuming the distance to each \HII region listed in Table~\ref{table:targets}. We also integrate over the model-predicted fluxes from the warm dust component only, and compute the fraction of the bolometric luminosity that is emitted by the warm dust component ($f_{\rm bol}$).

To examine the robustness of the dust model parameters, we additionally examined the model with the second-lowest \chisqr (presented in Table~\ref{table:SED_global_2nd_best} in Appendix~\ref{appendix:supplemental_figs}). In general, the differences in \chisqr between the best and second-best models ($\Delta \chi^2$) are small, and the bolometric luminosities inferred from the two models are within 1$\sigma$ of one another for all MSFRs. Only three MSFRs have $\Delta \chi^2 \geq 0.5$ (the Eagle Nebula, the Carina Nebula, and NGC~3603). The radio continua are consistent with thermal emission for all MSFRs for which it could be measured.

The intensity of the thermal radio continuum is proportional to the number of apparent Lyman continuum photons $N_C^{\prime}$. The Lyman continuum photon flux required to maintain ionization is given by
	
	\begin{equation}
	\begin{split}
	N_C^{\prime} = 7.489 \times10^{46} \left( \frac{\nu}{\text{GHz}} \right)^{0.1} \left( \frac{T_e}{10^4 \text{ K}} \right)^{-0.5} \left( \frac{S_{\nu}}{\text{Jy}} \right) \\
	\times \left( \frac{D}{\text{kpc}} \right)^2 \text{ ph s}^{-1},
	\end{split}
	\end{equation}
	
\noindent where $S_{\nu}$ is the (thermal) continuum flux density measured by \planck, and $D$ is the distance to the source (Table~\ref{table:targets}). We assume an electron temperature of $T_e=10^4$ K \citep{Subrahmanyan+Goss96}. The best-fit parameters for all MSFRs in our sample are summarized in Table~\ref{table:SED_global}, sorted by increasing bolometric luminosity. The table also includes the circular aperture radius used in our photometry analysis, along with the corresponding physical size of the region (note: the precise central coordinates of our apertures are given in Table~\ref{table:all_photometry} in Appendix~\ref{appendix:supplemental_figs}). Figure~\ref{figure:SED_examples} shows the SED and best-fit model for the Orion Nebula; the remaining SEDs are shown in Figures~\ref{figure:SED_examples}.1 through \ref{figure:SED_examples}.28 in the online figure set.

\figsetstart
\figsetnum{2}
\figsettitle{The global SEDs with the best-fit models superimposed.}

\figsetgrpstart
\figsetgrpnum{2.1}
\figsetgrptitle{Flame Nebula SED}
\figsetplot{\includegraphics[width=0.31\linewidth,clip,trim=1.2cm 12.4cm 3.5cm 3.7cm]{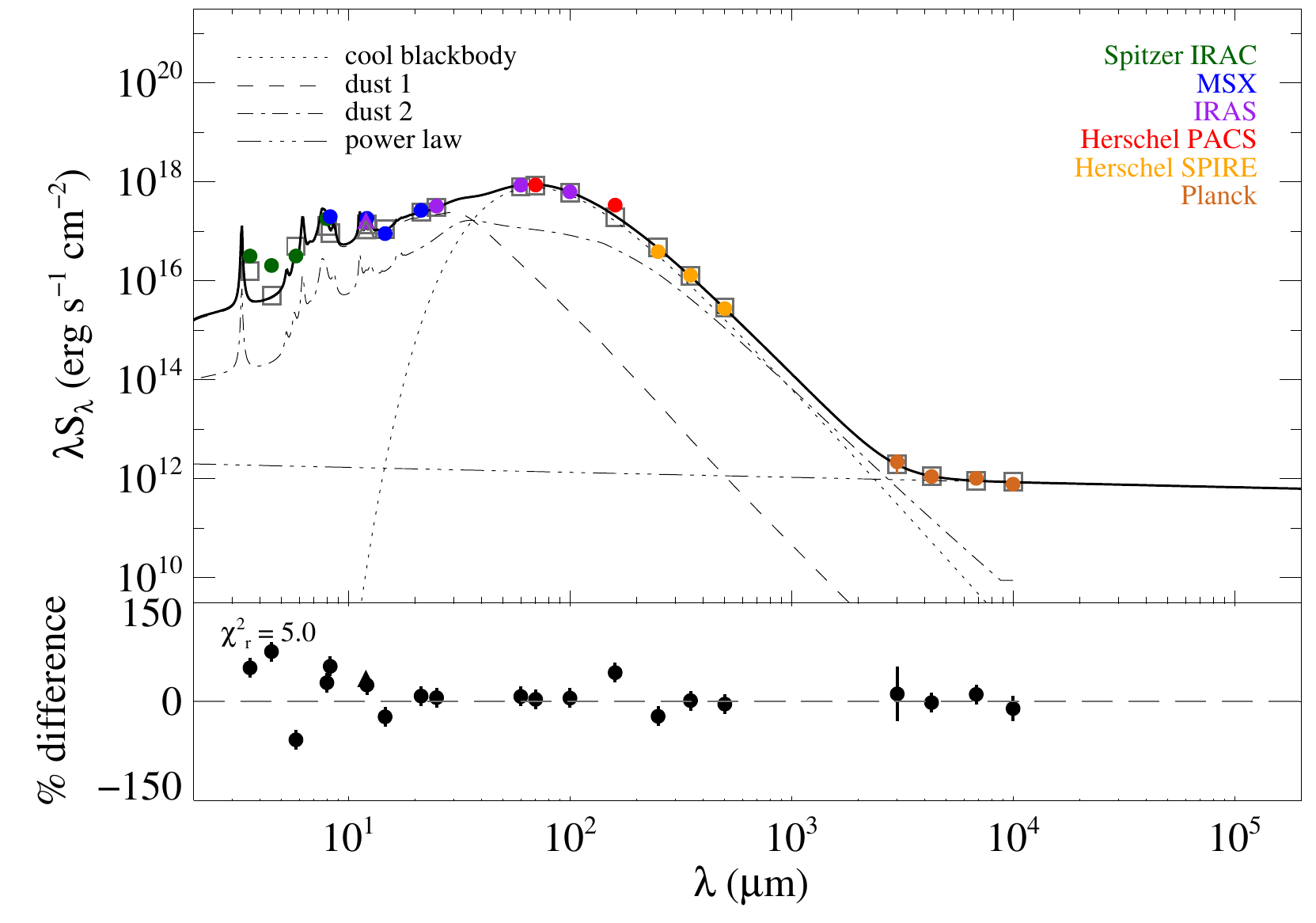}}
\figsetgrpnote{The SED and best-fit model for the Flame Nebula}
\figsetgrpend

\figsetgrpstart
\figsetgrpnum{2.2}
\figsetgrptitle{W40 SED}
\figsetplot{\includegraphics[width=0.31\linewidth,clip,trim=1.2cm 12.4cm 3.5cm 3.7cm]{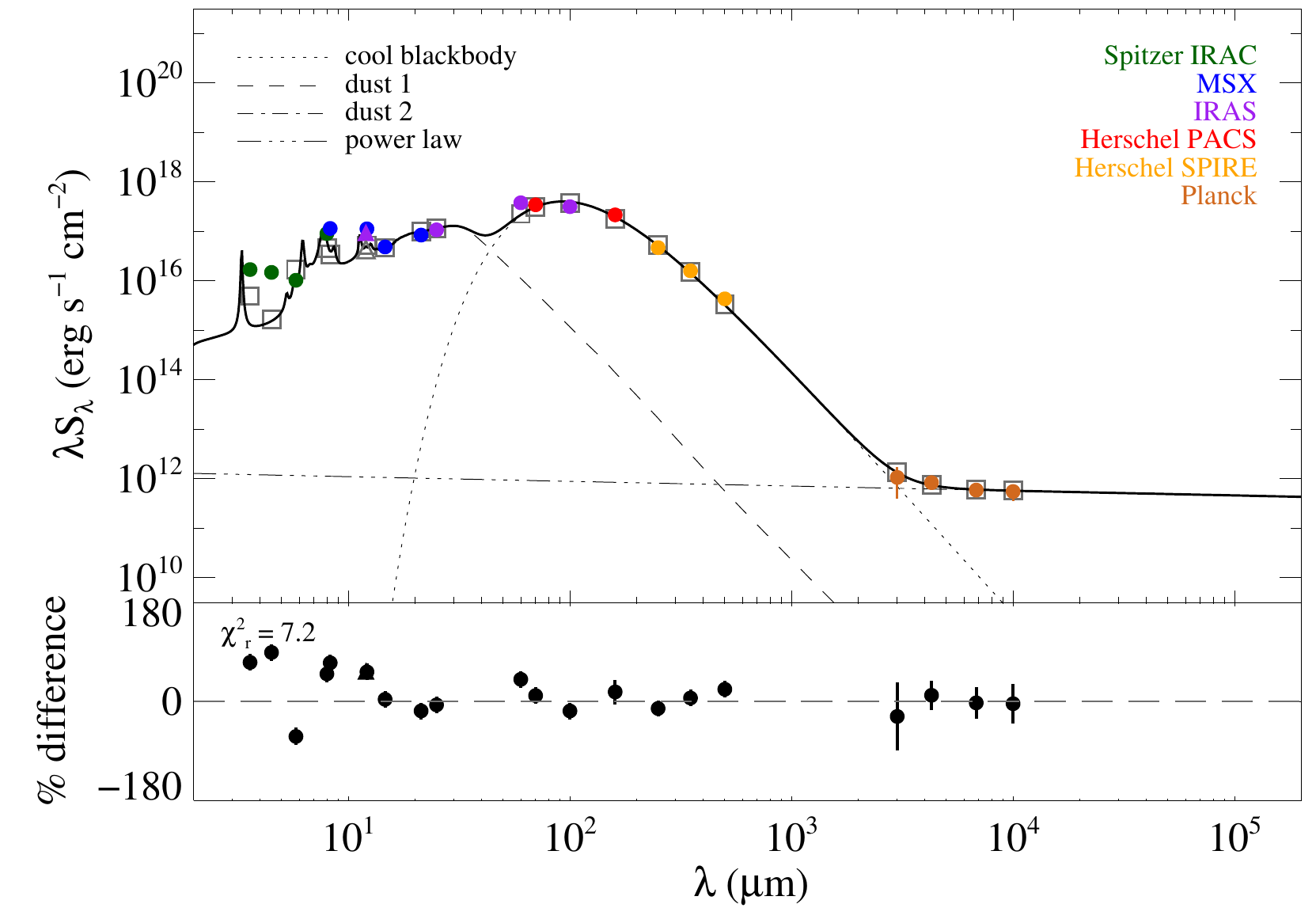}}
\figsetgrpnote{The SED and best-fit model for W40}
\figsetgrpend

\figsetgrpstart
\figsetgrpnum{2.3}
\figsetgrptitle{Westerlund 1 SED}
\figsetplot{\includegraphics[width=0.31\linewidth,clip,trim=1.2cm 12.4cm 3.5cm 3.7cm]{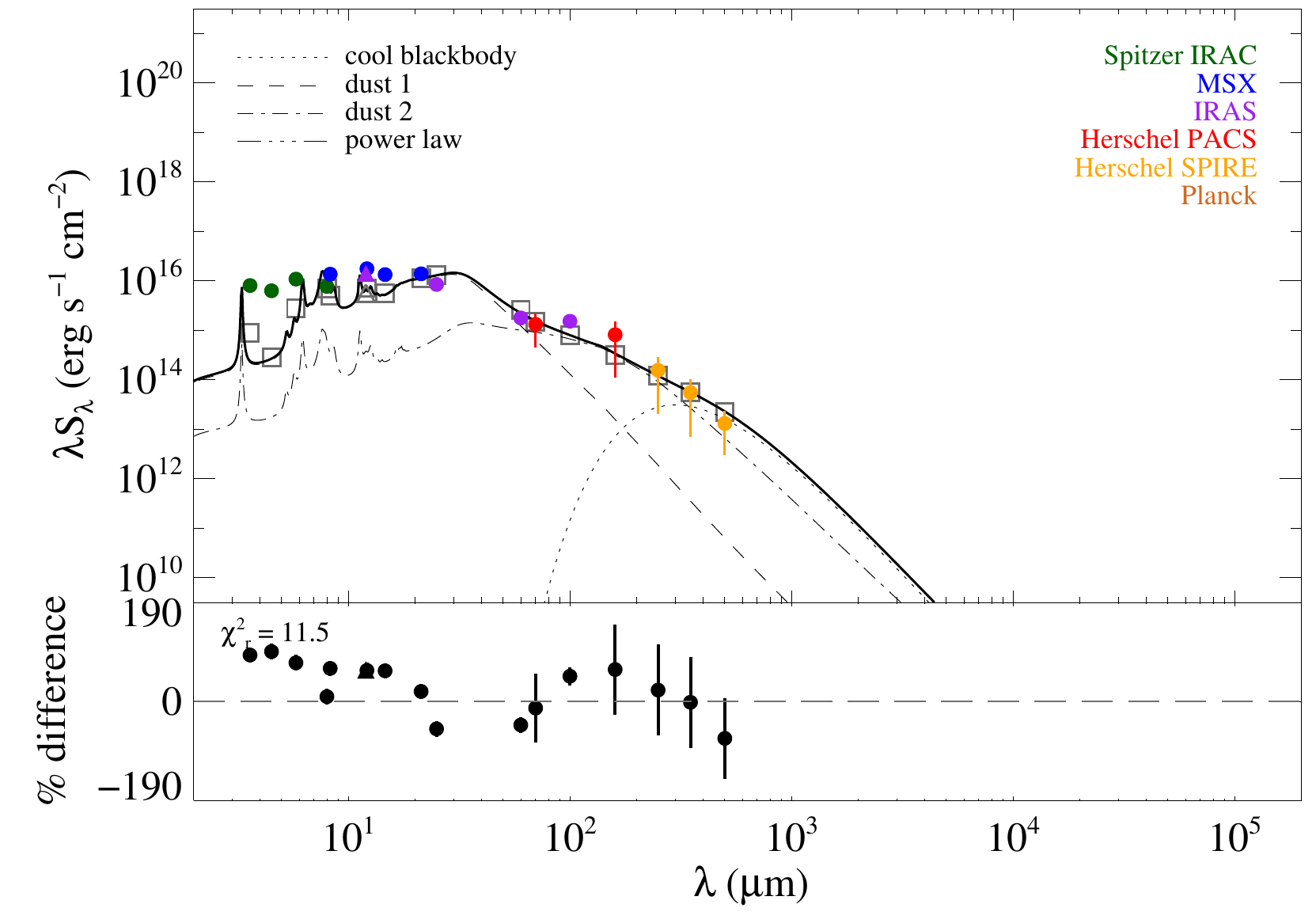}}
\figsetgrpnote{The SED and best-fit model for Wd 1}
\figsetgrpend

\figsetgrpstart
\figsetgrpnum{2.4}
\figsetgrptitle{RCW36 SED}
\figsetplot{\includegraphics[width=0.31\linewidth,clip,trim=1.2cm 12.4cm 3.5cm 3.7cm]{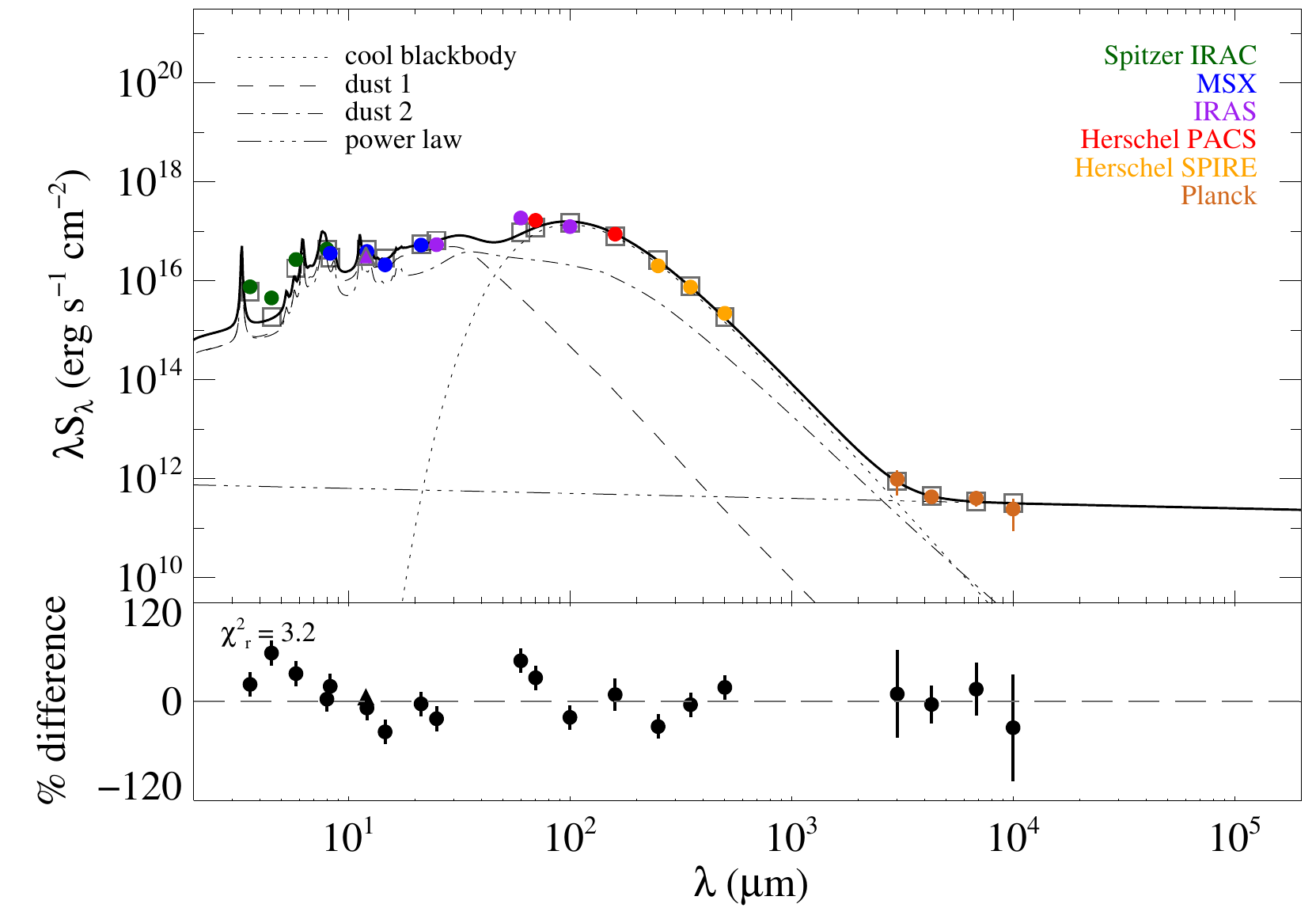}}
\figsetgrpnote{The SED and best-fit model for RCW36}
\figsetgrpend

\figsetgrpstart
\figsetgrpnum{2.5}
\figsetgrptitle{Berkeley 87 SED}
\figsetplot{\includegraphics[width=0.31\linewidth,clip,trim=1.2cm 12.4cm 3.5cm 3.7cm]{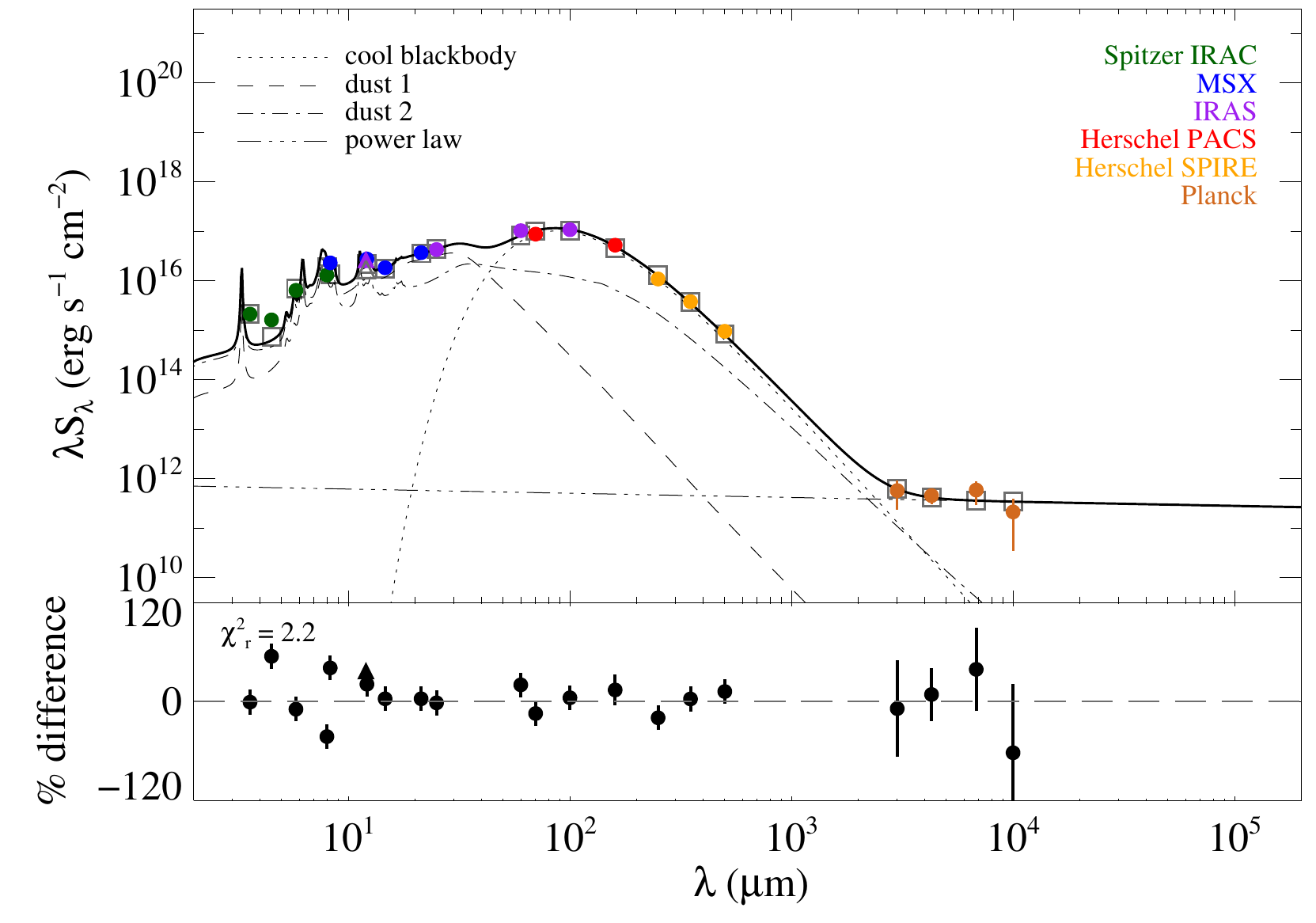}}
\figsetgrpnote{The SED and best-fit model for Berkeley 87}
\figsetgrpend

\figsetgrpstart
\figsetgrpnum{2.6}
\figsetgrptitle{Orion Nebula SED}
\figsetplot{\includegraphics[width=0.31\linewidth,clip,trim=1.2cm 12.4cm 3.5cm 3.7cm]{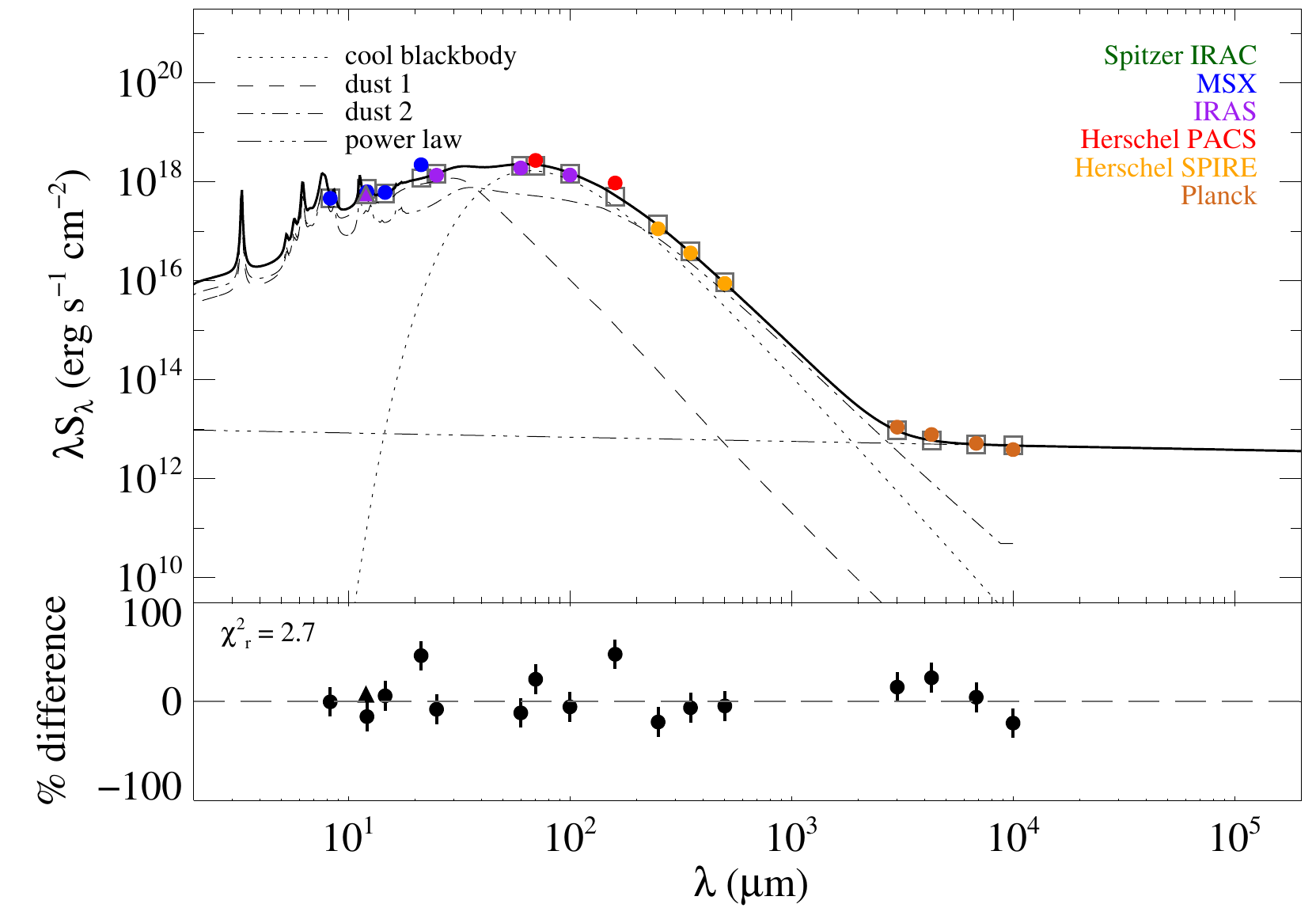}}
\figsetgrpnote{The SED and best-fit model for the Orion Nebula}
\figsetgrpend

\figsetgrpstart
\figsetgrpnum{2.7}
\figsetgrptitle{Lagoon Nebula SED}
\figsetplot{\includegraphics[width=0.31\linewidth,clip,trim=1.2cm 12.4cm 3.5cm 3.7cm]{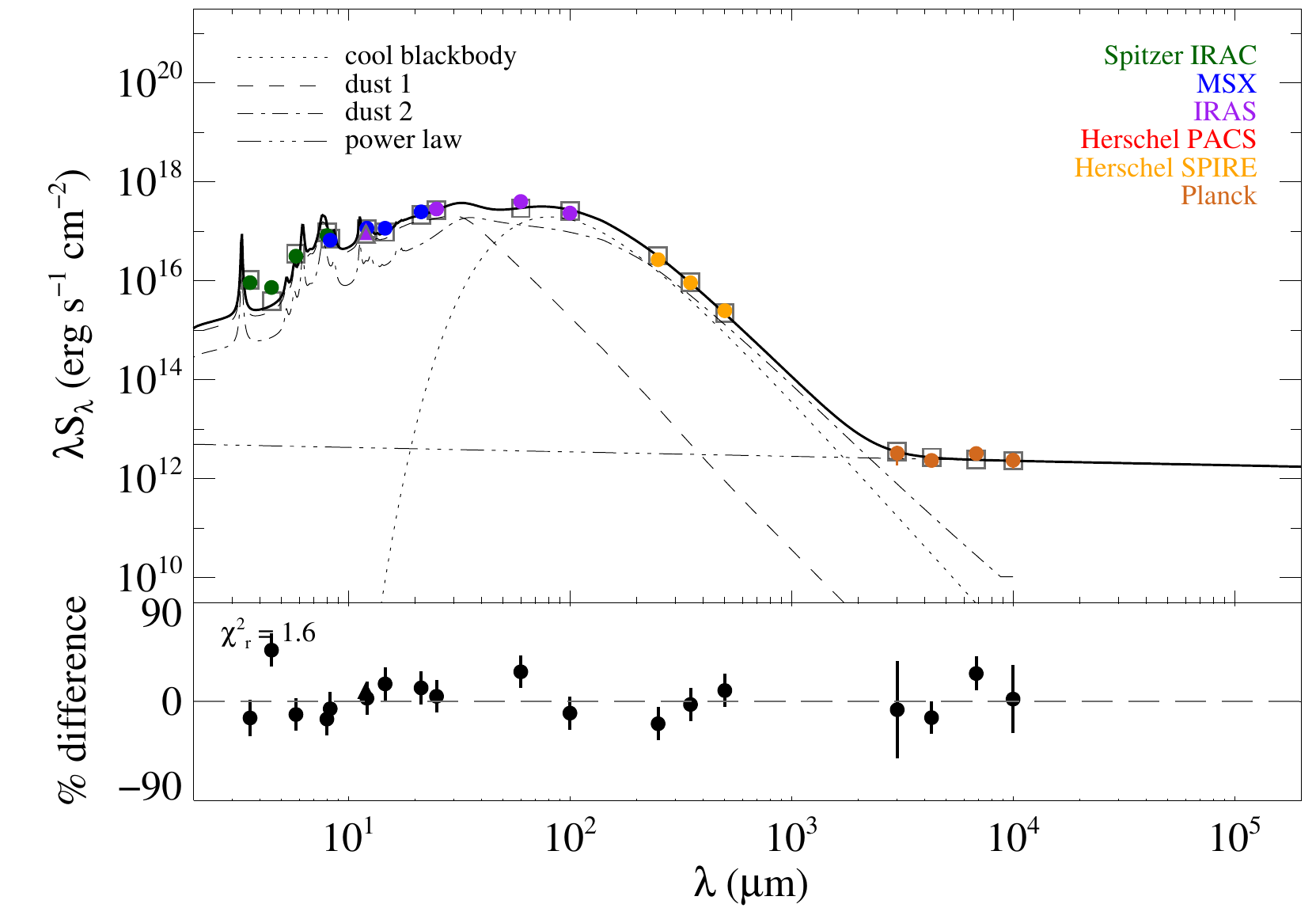}}
\figsetgrpnote{The SED and best-fit model for the Lagoon Nebula}
\figsetgrpend

\figsetgrpstart
\figsetgrpnum{2.8}
\figsetgrptitle{Trifid Nebula SED}
\figsetplot{\includegraphics[width=0.31\linewidth,clip,trim=1.2cm 12.4cm 3.5cm 3.7cm]{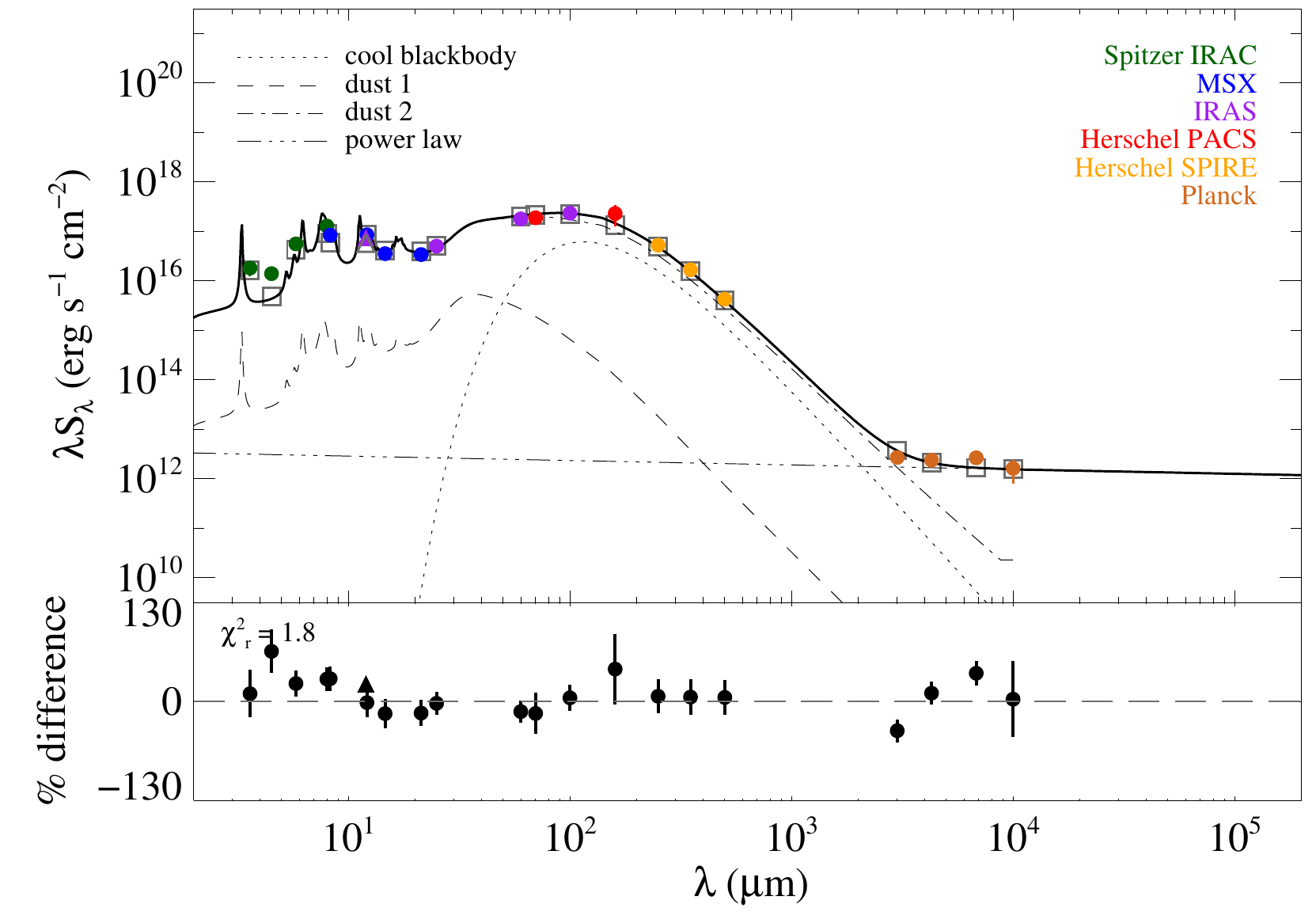}}
\figsetgrpnote{The SED and best-fit model for the Trifid Nebula}
\figsetgrpend

\figsetgrpstart
\figsetgrpnum{2.9}
\figsetgrptitle{W42 SED}
\figsetplot{\includegraphics[width=0.31\linewidth,clip,trim=1.2cm 12.4cm 3.5cm 3.7cm]{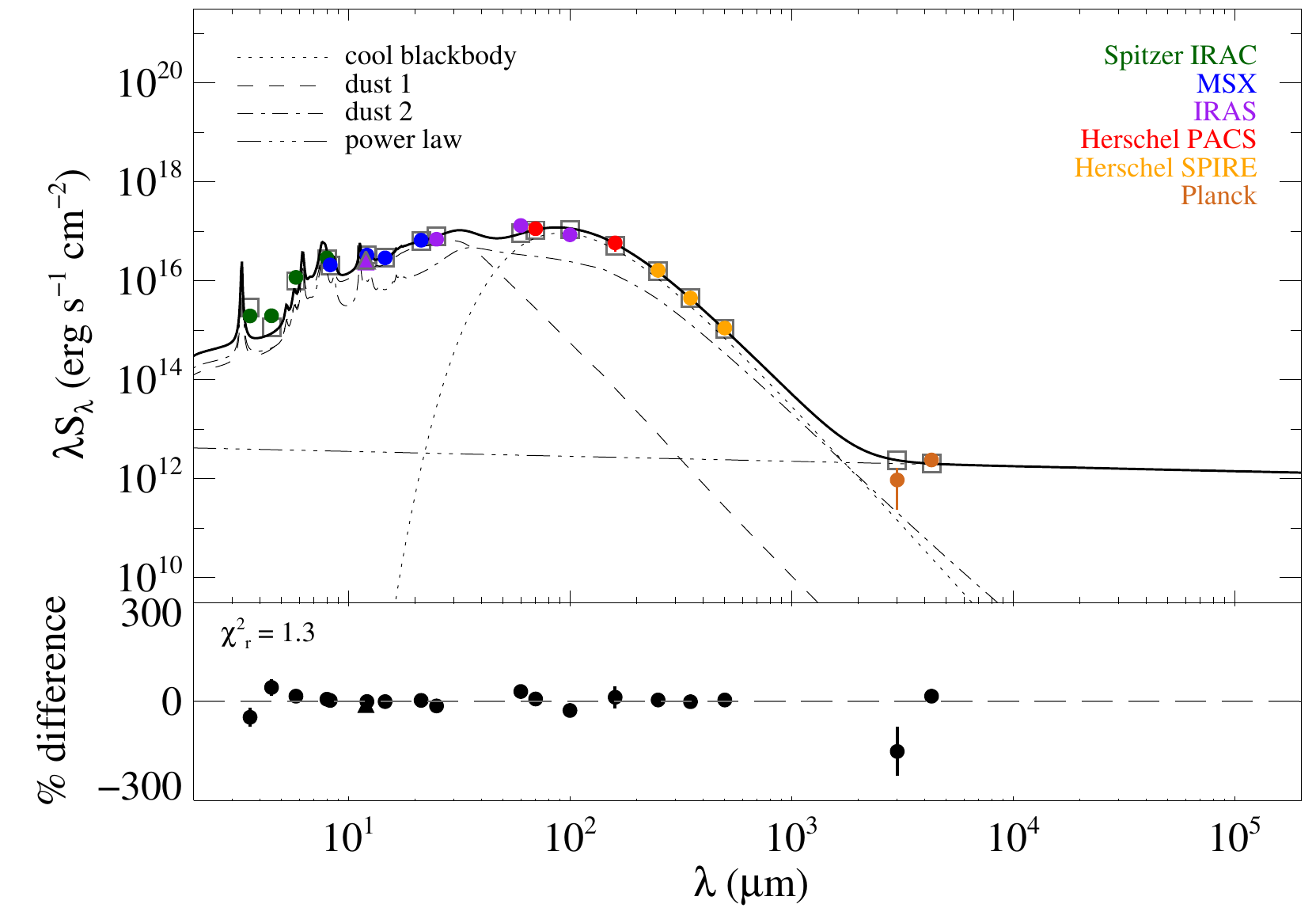}}
\figsetgrpnote{The SED and best-fit model for W42}
\figsetgrpend

\figsetgrpstart
\figsetgrpnum{2.10}
\figsetgrptitle{NGC 7538 SED}
\figsetplot{\includegraphics[width=0.31\linewidth,clip,trim=1.2cm 12.4cm 3.5cm 3.7cm]{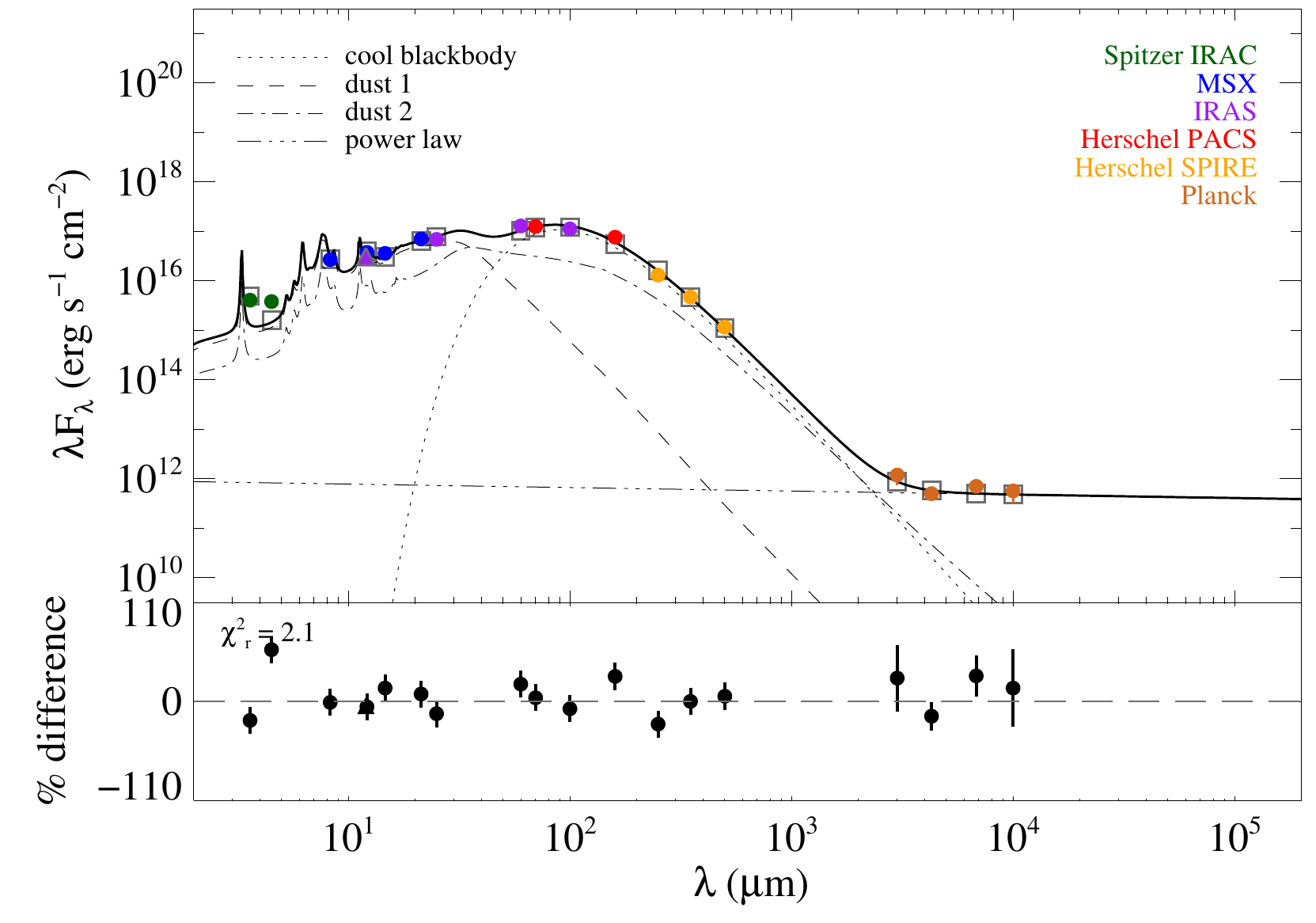}}
\figsetgrpnote{The SED and best-fit model for NGC 7538}
\figsetgrpend

\figsetgrpstart
\figsetgrpnum{2.11}
\figsetgrptitle{W4 SED}
\figsetplot{\includegraphics[width=0.31\linewidth,clip,trim=1.2cm 12.4cm 3.5cm 3.7cm]{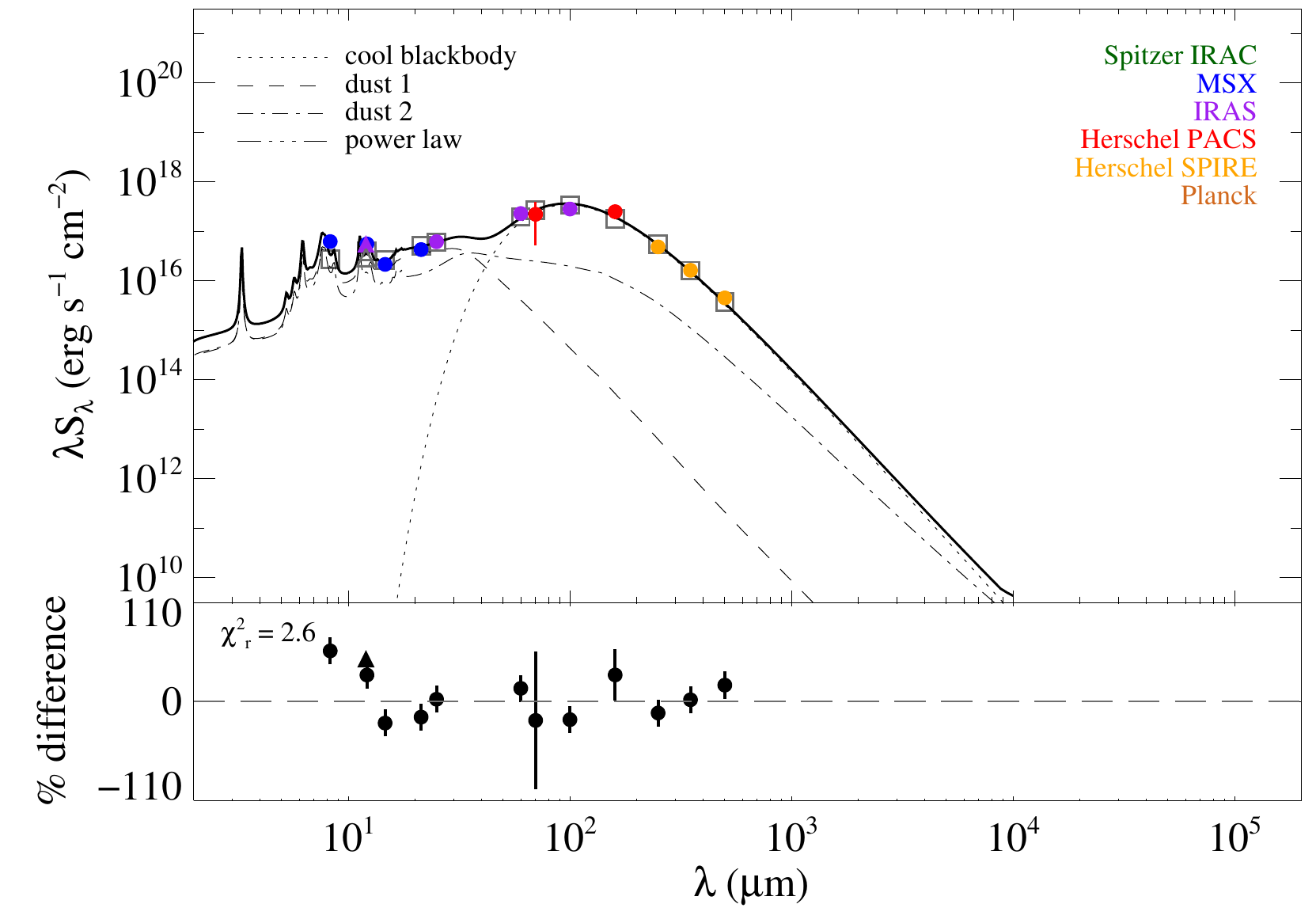}}
\figsetgrpnote{The SED and best-fit model for W4}
\figsetgrpend

\figsetgrpstart
\figsetgrpnum{2.12}
\figsetgrptitle{Eagle Nebula SED}
\figsetplot{\includegraphics[width=0.31\linewidth,clip,trim=1.2cm 12.4cm 3.5cm 3.7cm]{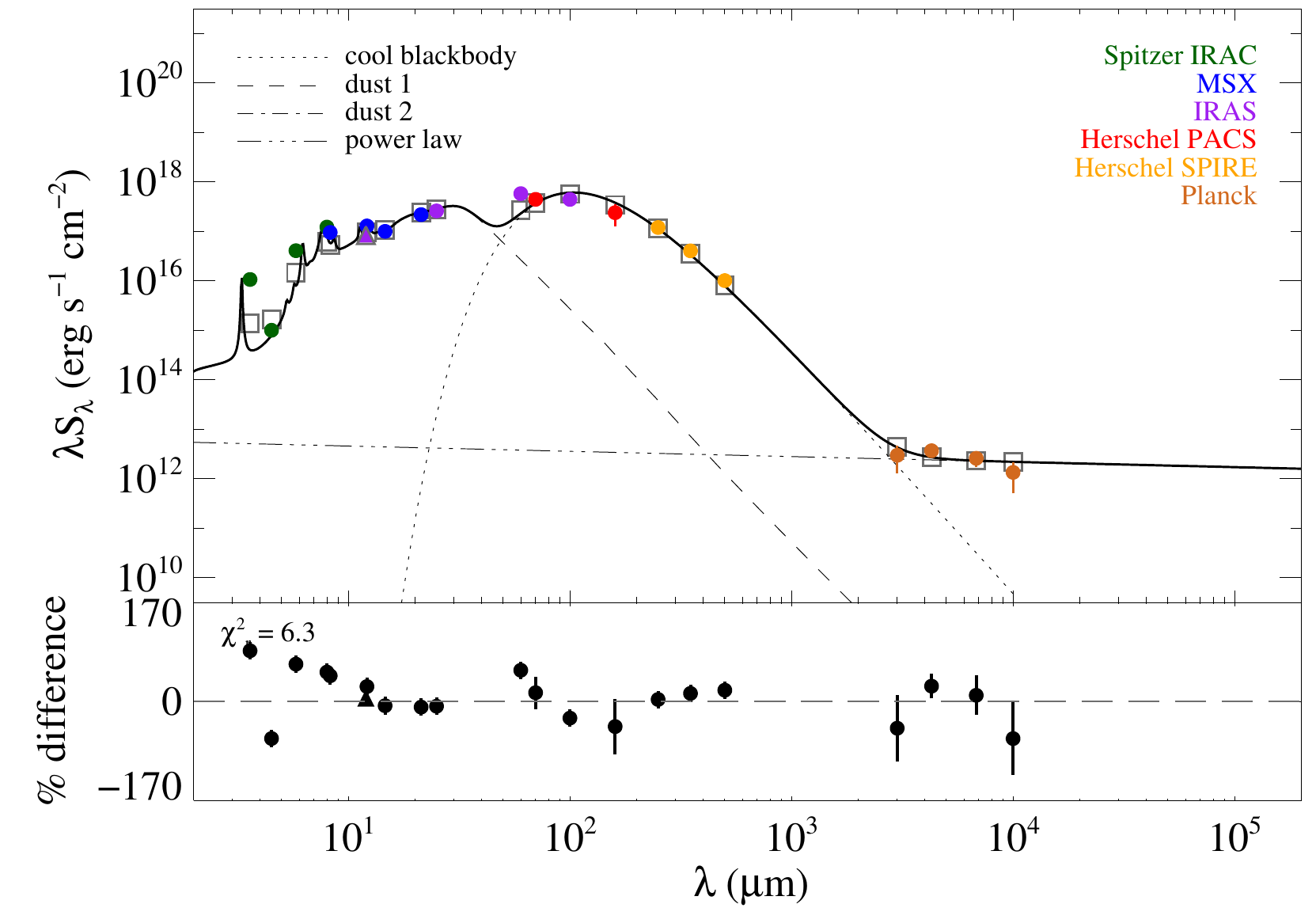}}
\figsetgrpnote{The SED and best-fit model for the Eagle Nebula}
\figsetgrpend

\figsetgrpstart
\figsetgrpnum{2.13}
\figsetgrptitle{W33 SED}
\figsetplot{\includegraphics[width=0.31\linewidth,clip,trim=1.2cm 12.4cm 3.5cm 3.7cm]{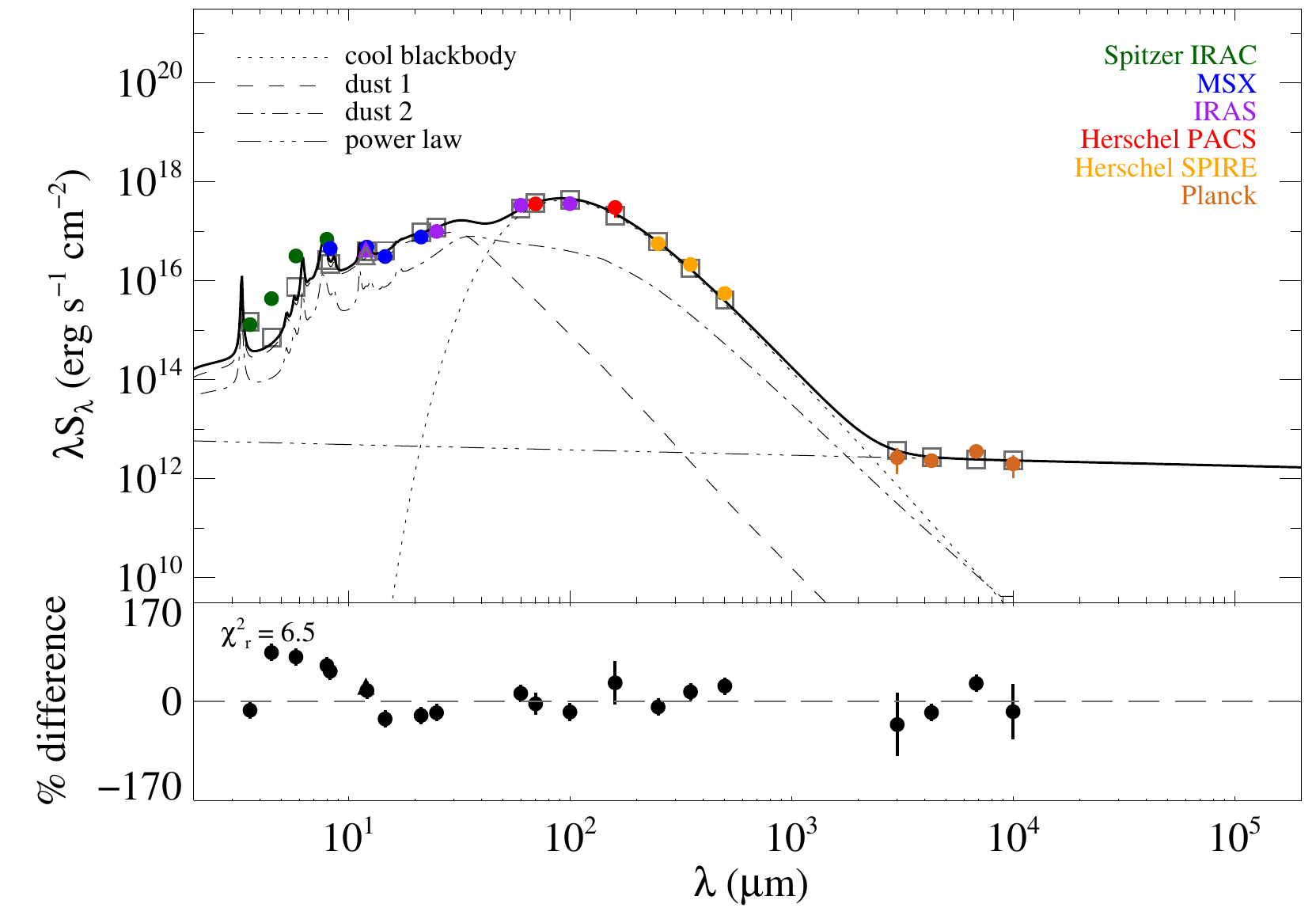}}
\figsetgrpnote{The SED and best-fit model for W33}
\figsetgrpend

\figsetgrpstart
\figsetgrpnum{2.14}
\figsetgrptitle{RCW38 SED}
\figsetplot{\includegraphics[width=0.31\linewidth,clip,trim=1.2cm 12.4cm 3.5cm 3.7cm]{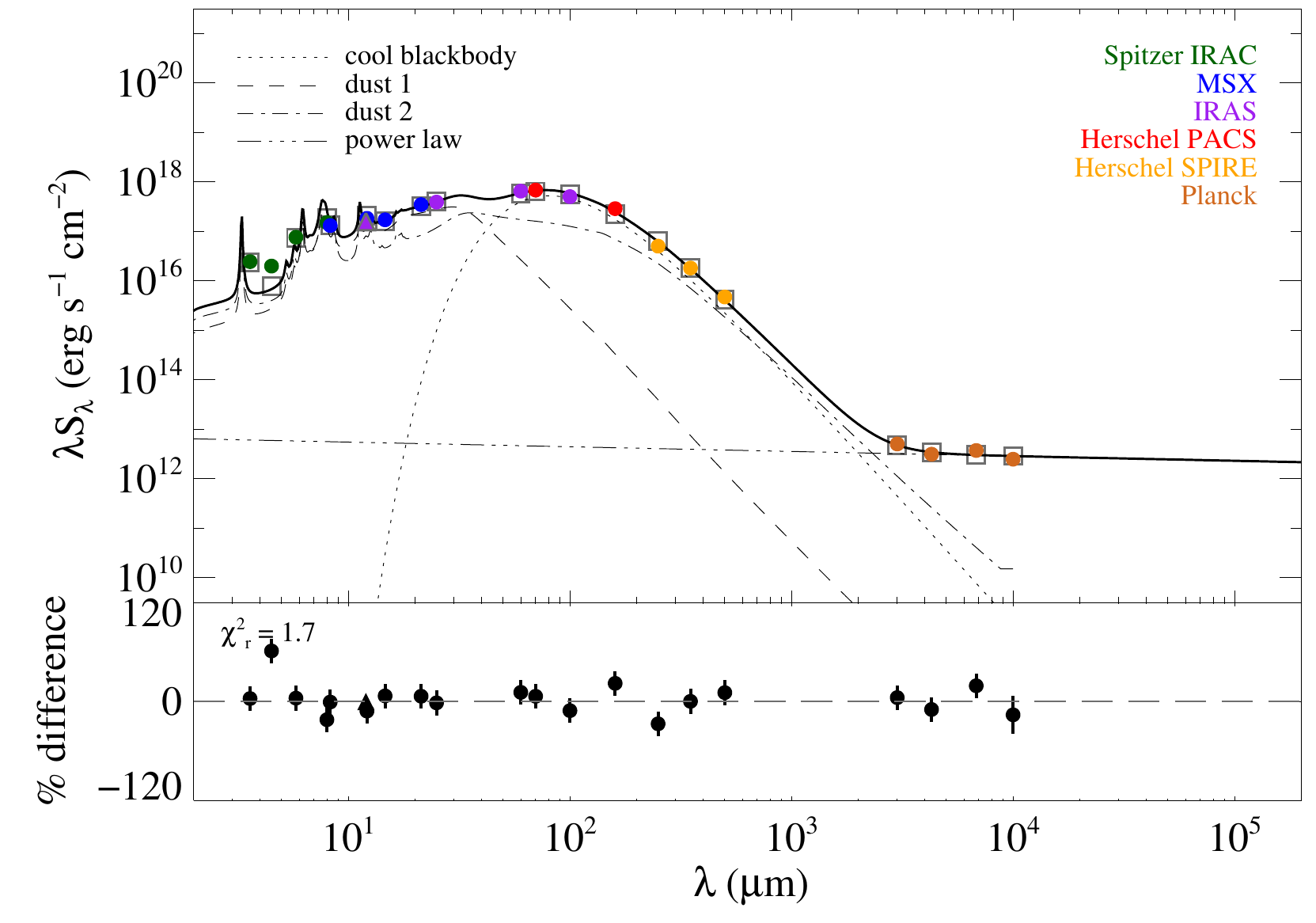}}
\figsetgrpnote{The SED and best-fit model for RCW38}
\figsetgrpend

\figsetgrpstart
\figsetgrpnum{2.15}
\figsetgrptitle{W3 SED}
\figsetplot{\includegraphics[width=0.31\linewidth,clip,trim=1.2cm 12.4cm 3.5cm 3.7cm]{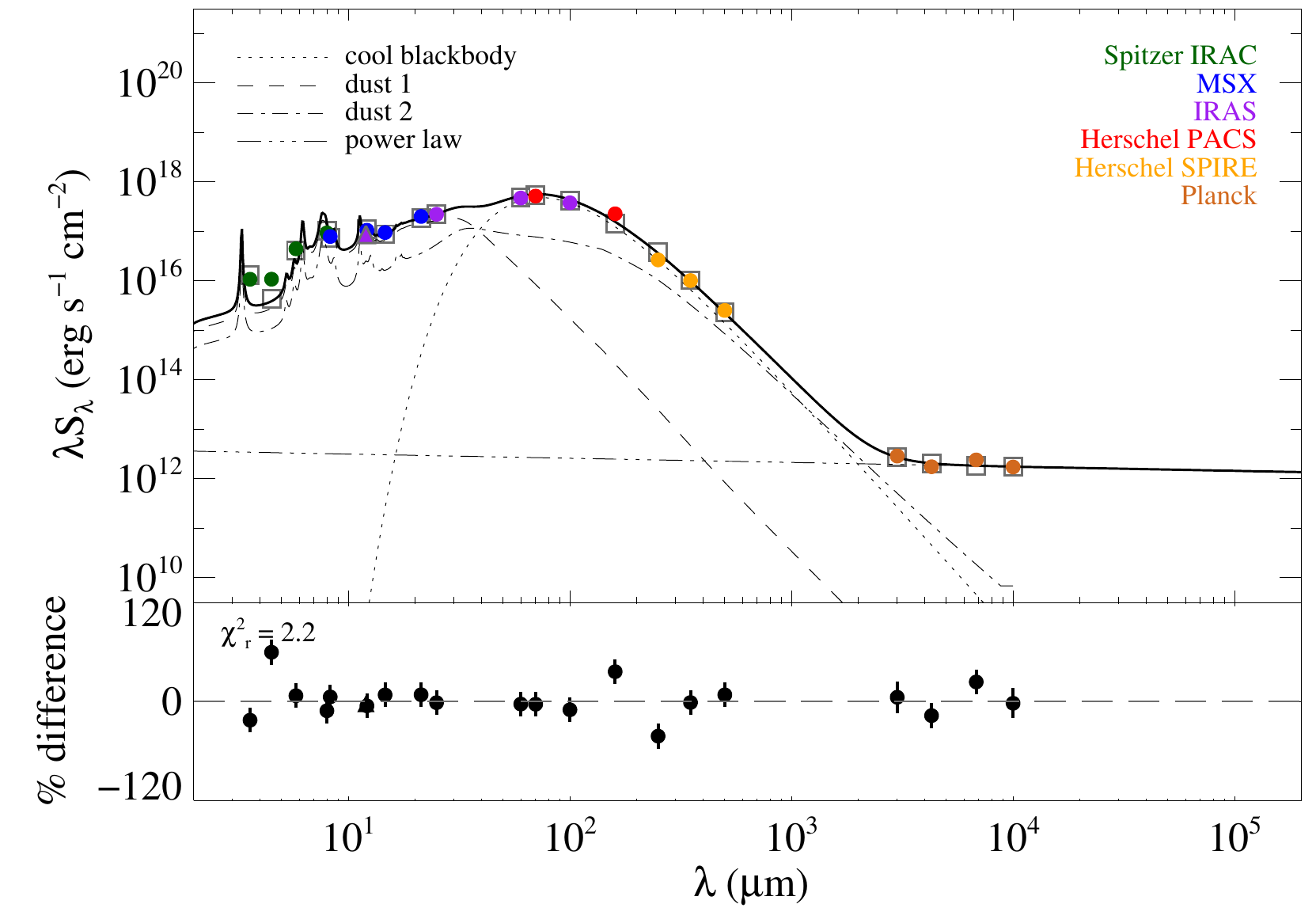}}
\figsetgrpnote{The SED and best-fit model for W3}
\figsetgrpend

\figsetgrpstart
\figsetgrpnum{2.16}
\figsetgrptitle{NGC 3576 SED}
\figsetplot{\includegraphics[width=0.31\linewidth,clip,trim=1.2cm 12.4cm 3.5cm 3.7cm]{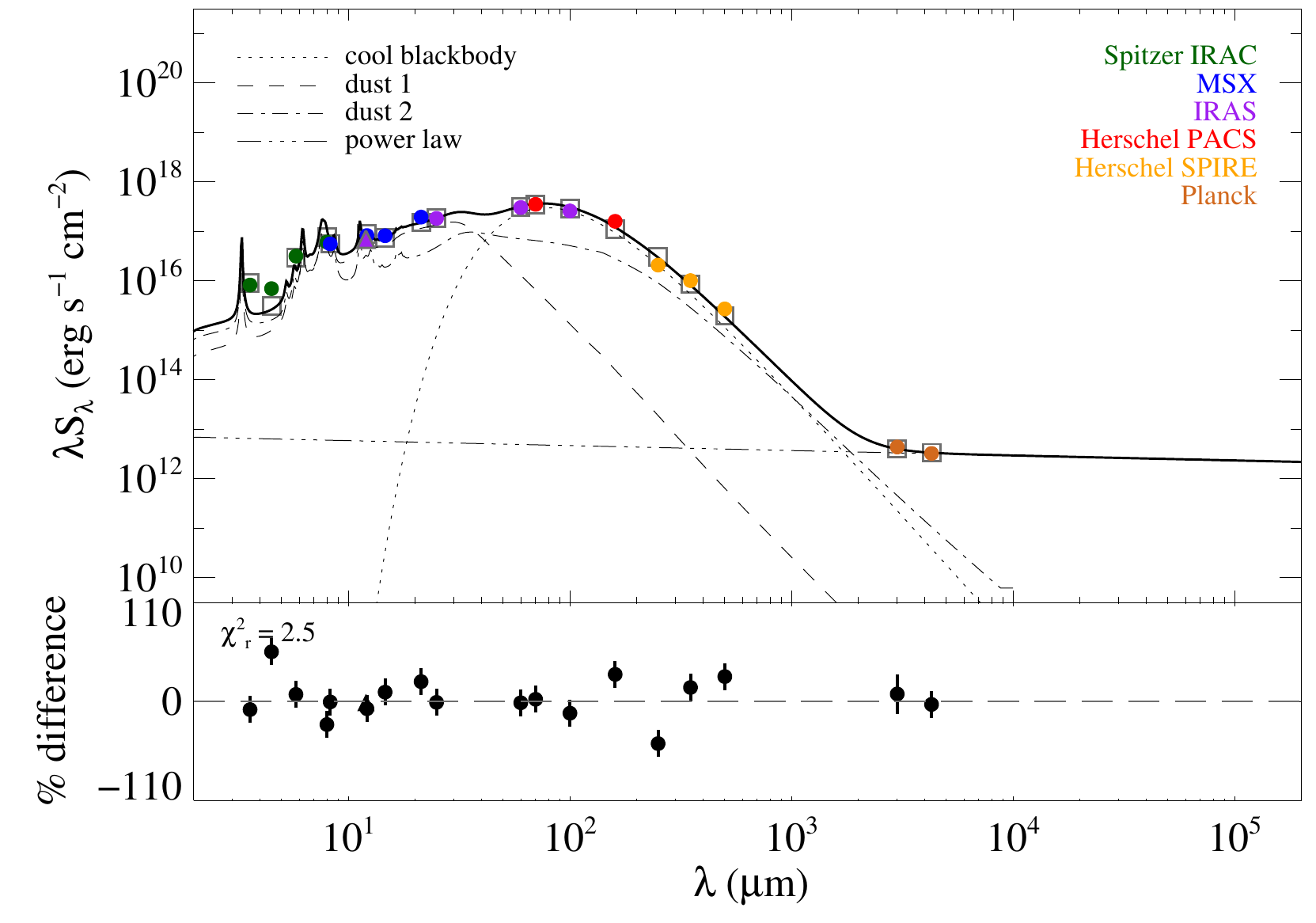}}
\figsetgrpnote{The SED and best-fit model for NGC 3576}
\figsetgrpend

\figsetgrpstart
\figsetgrpnum{2.17}
\figsetgrptitle{NGC 6334 SED}
\figsetplot{\includegraphics[width=0.31\linewidth,clip,trim=1.2cm 12.4cm 3.5cm 3.7cm]{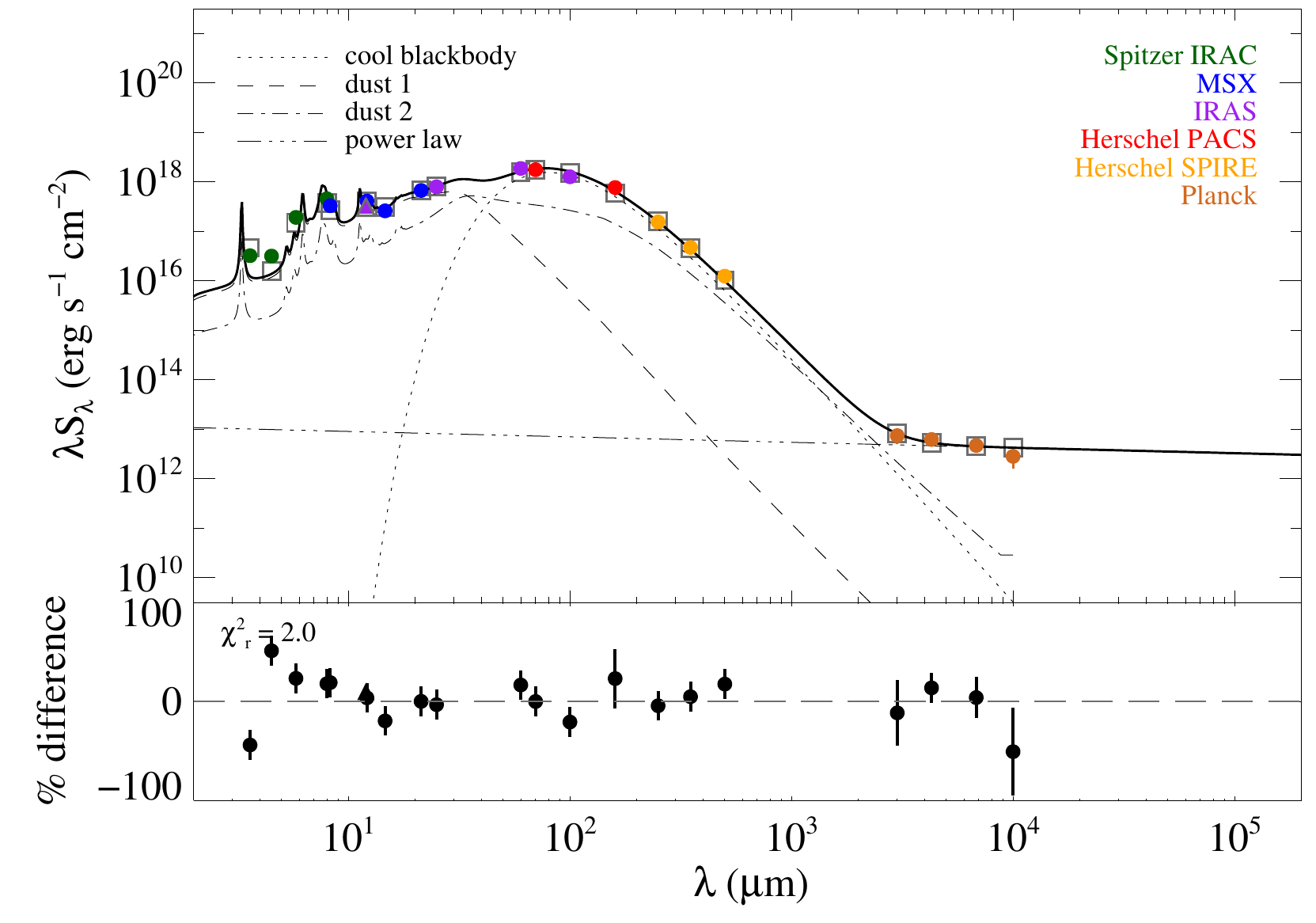}}
\figsetgrpnote{The SED and best-fit model for NGC 6334}
\figsetgrpend

\figsetgrpstart
\figsetgrpnum{2.18}
\figsetgrptitle{G29.96-0.02 SED}
\figsetplot{\includegraphics[width=0.31\linewidth,clip,trim=1.2cm 12.4cm 3.5cm 3.7cm]{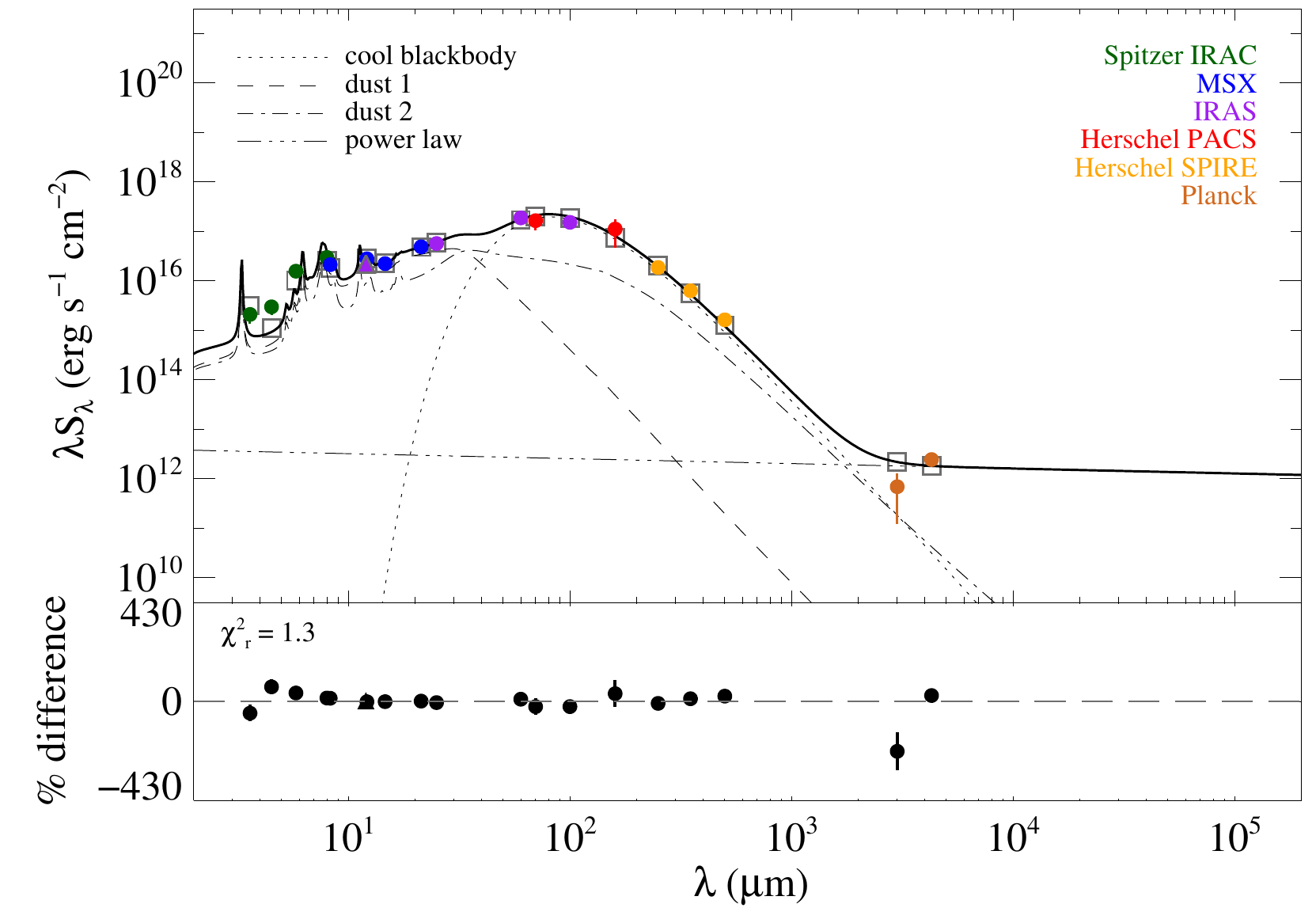}}
\figsetgrpnote{The SED and best-fit model for G29.96-0.02}
\figsetgrpend

\figsetgrpstart
\figsetgrpnum{2.19}
\figsetgrptitle{NGC 6357 SED}
\figsetplot{\includegraphics[width=0.31\linewidth,clip,trim=1.2cm 12.4cm 3.5cm 3.7cm]{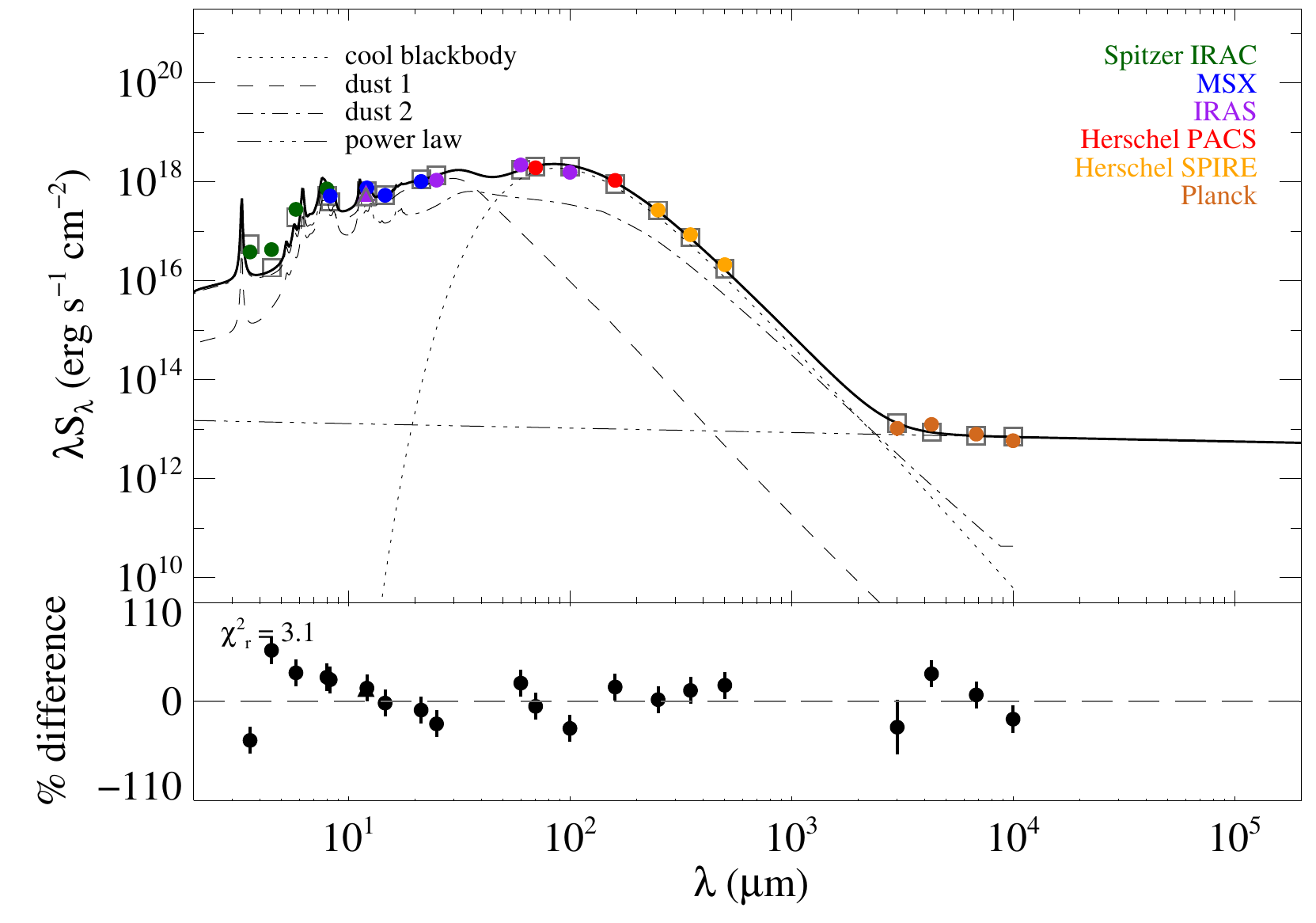}}
\figsetgrpnote{The SED and best-fit model for NGC 6357}
\figsetgrpend

\figsetgrpstart
\figsetgrpnum{2.20}
\figsetgrptitle{M17 SED}
\figsetplot{\includegraphics[width=0.31\linewidth,clip,trim=1.2cm 12.4cm 3.5cm 3.7cm]{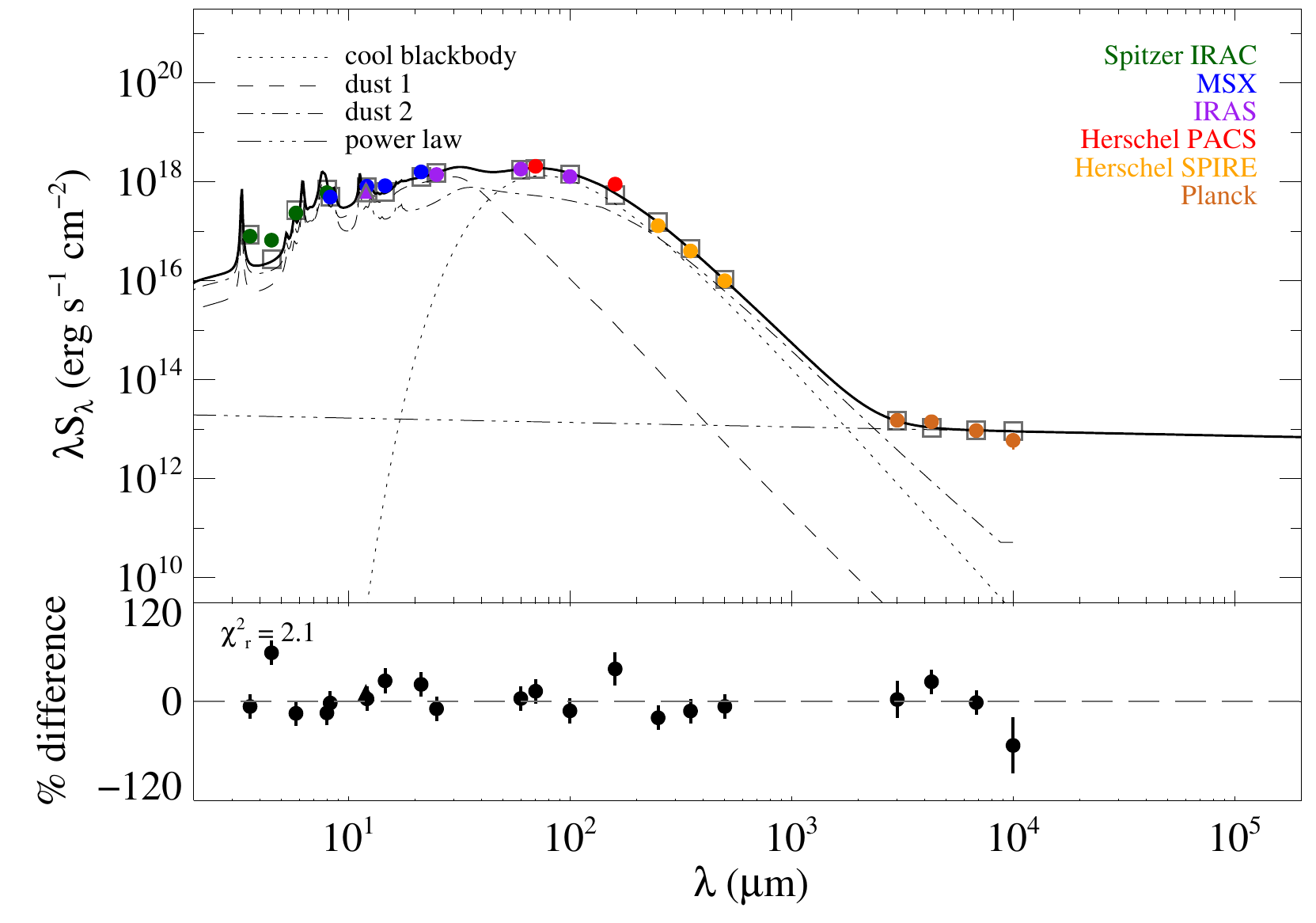}}
\figsetgrpnote{The SED and best-fit model for M17}
\figsetgrpend

\figsetgrpstart
\figsetgrpnum{2.21}
\figsetgrptitle{G333 SED}
\figsetplot{\includegraphics[width=0.31\linewidth,clip,trim=1.2cm 12.4cm 3.5cm 3.7cm]{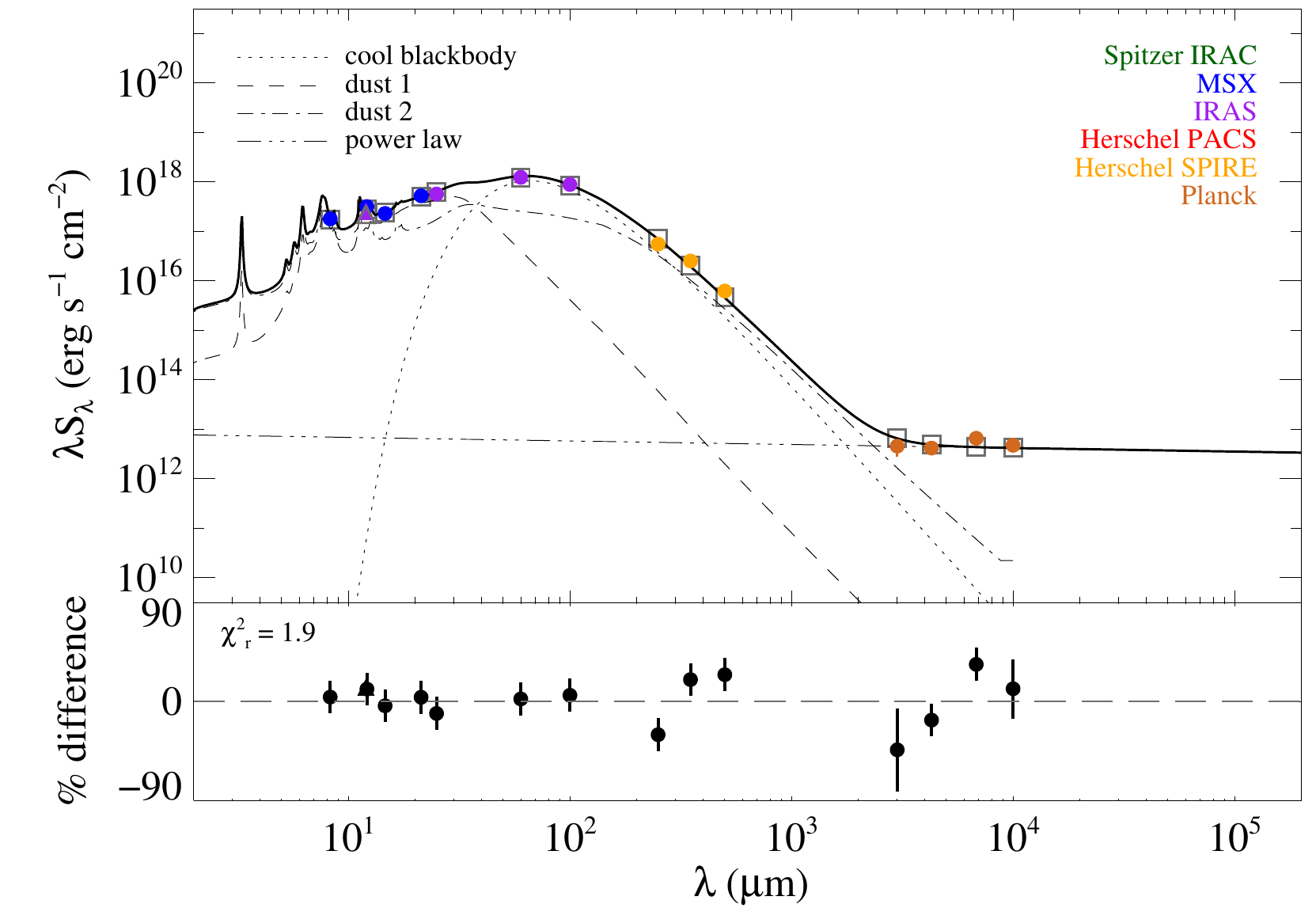}}
\figsetgrpnote{The SED and best-fit model for G333}
\figsetgrpend

\figsetgrpstart
\figsetgrpnum{2.22}
\figsetgrptitle{W43 SED}
\figsetplot{\includegraphics[width=0.31\linewidth,clip,trim=1.2cm 12.4cm 3.5cm 3.7cm]{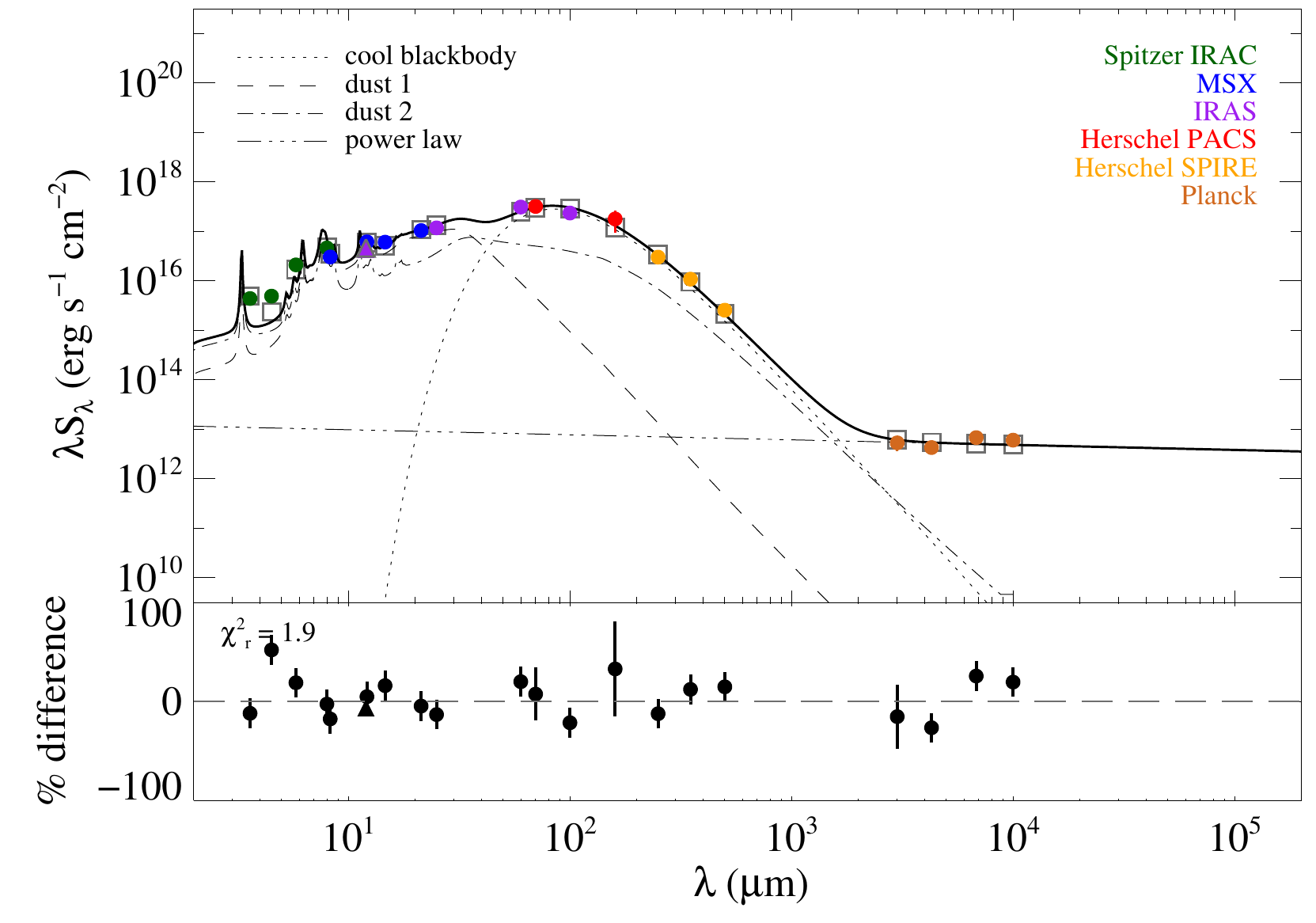}}
\figsetgrpnote{The SED and best-fit model for W43}
\figsetgrpend

\figsetgrpstart
\figsetgrpnum{2.23}
\figsetgrptitle{RCW49 SED}
\figsetplot{\includegraphics[width=0.31\linewidth,clip,trim=1.2cm 12.4cm 3.5cm 3.7cm]{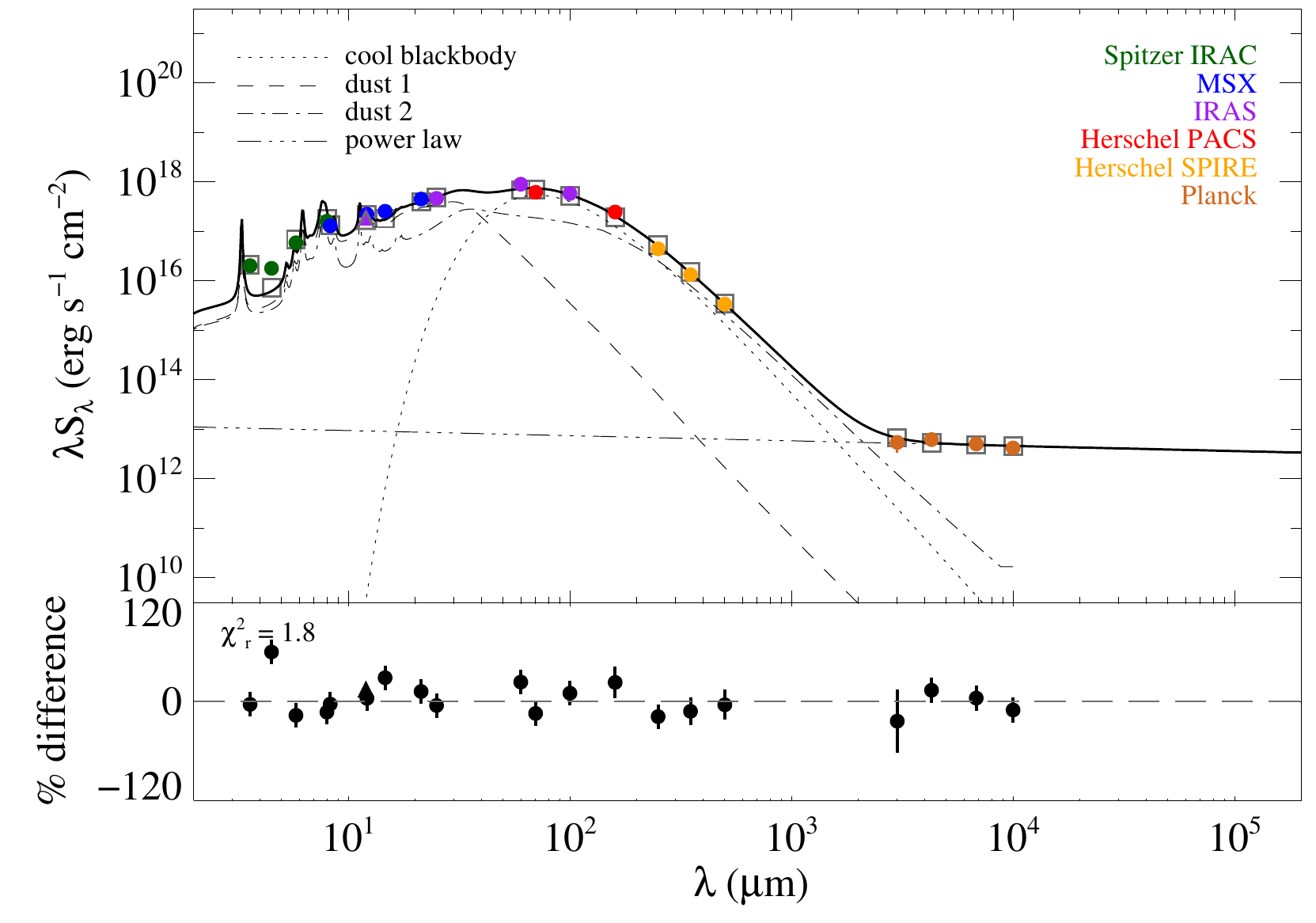}}
\figsetgrpnote{The SED and best-fit model for RCW49}
\figsetgrpend

\figsetgrpstart
\figsetgrpnum{2.24}
\figsetgrptitle{G305 SED}
\figsetplot{\includegraphics[width=0.31\linewidth,clip,trim=1.2cm 12.4cm 3.5cm 3.7cm]{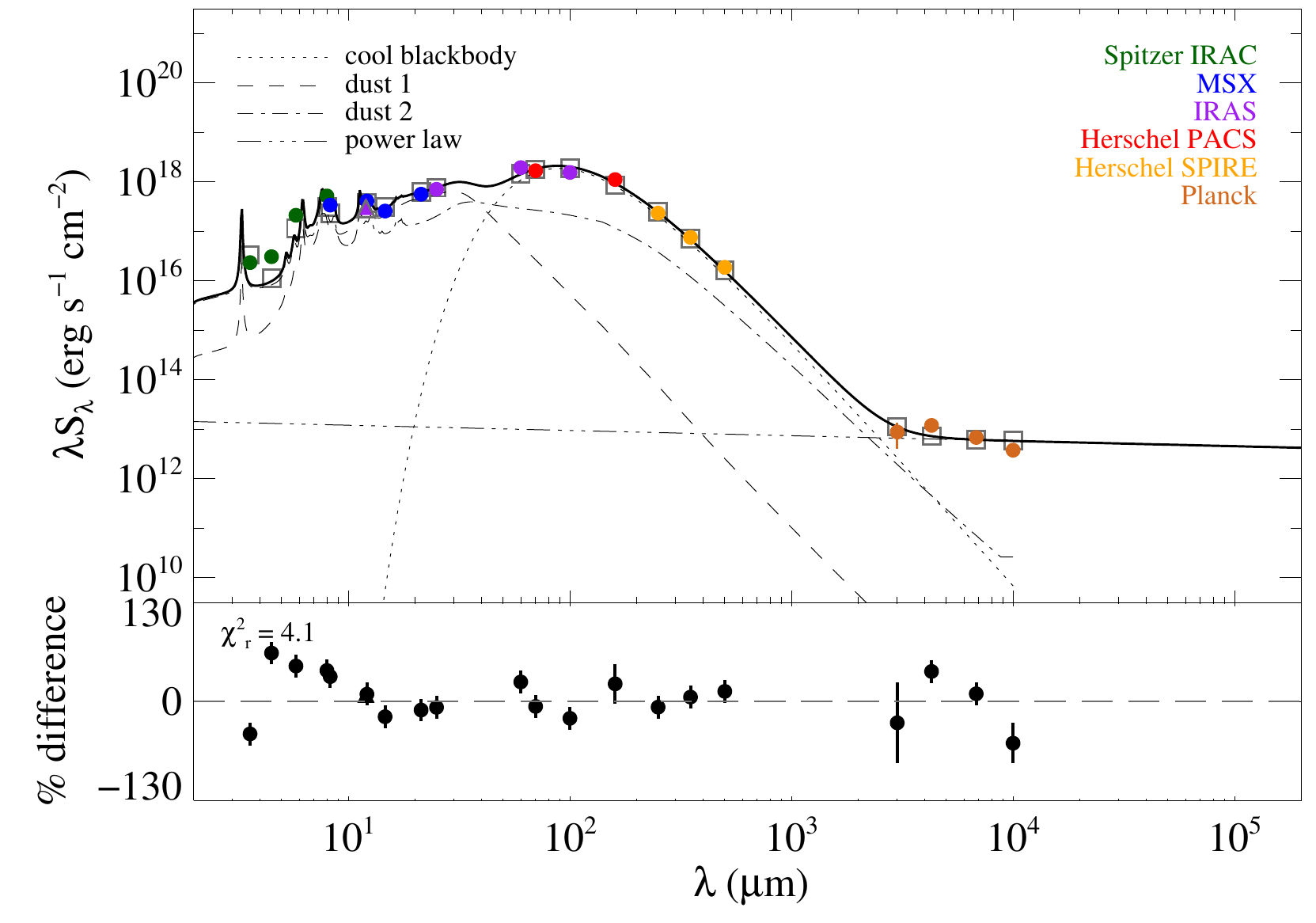}}
\figsetgrpnote{The SED and best-fit model for G305}
\figsetgrpend

\figsetgrpstart
\figsetgrpnum{2.25}
\figsetgrptitle{W49A SED}
\figsetplot{\includegraphics[width=0.31\linewidth,clip,trim=1.2cm 12.4cm 3.5cm 3.7cm]{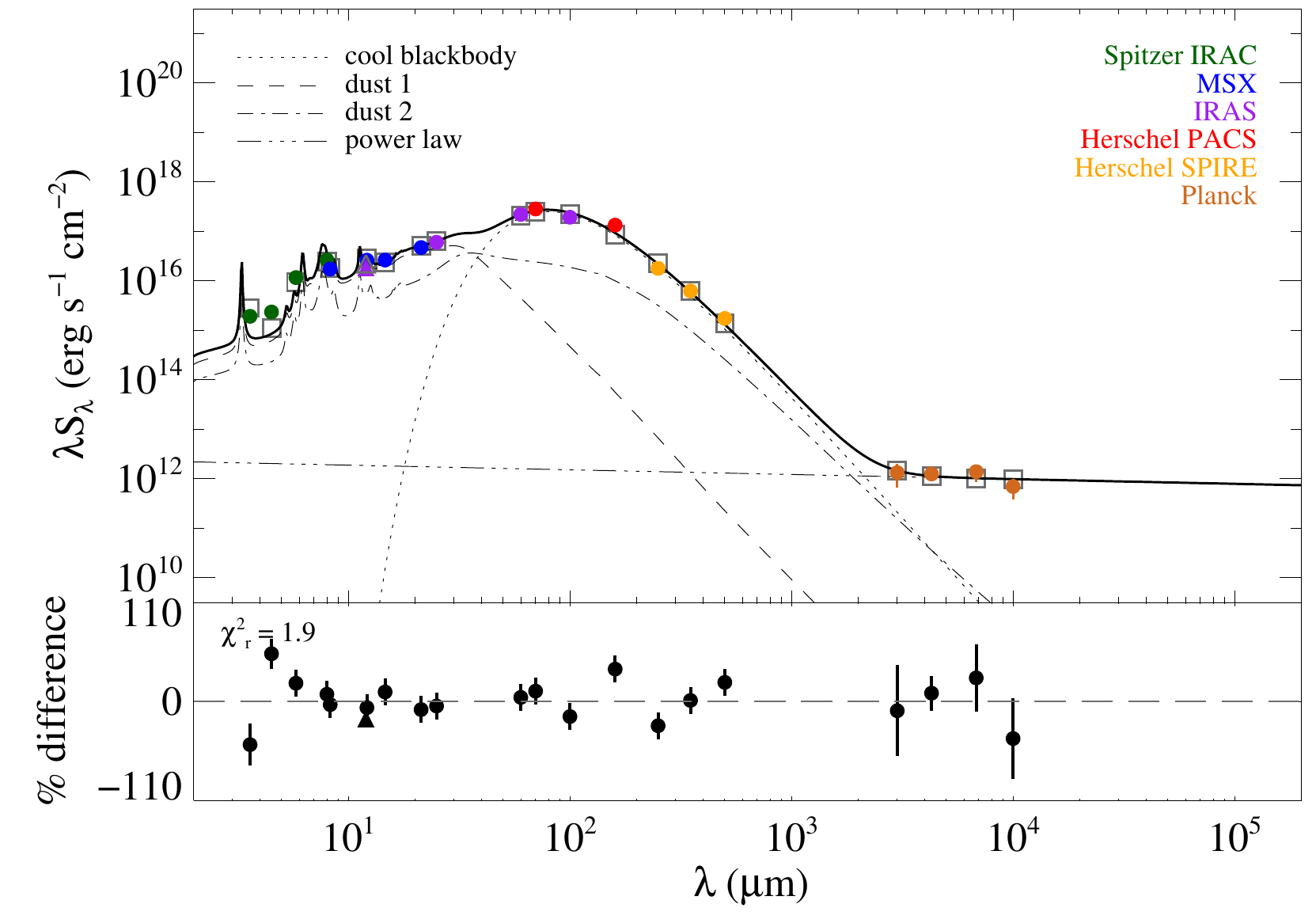}}
\figsetgrpnote{The SED and best-fit model for W49A}
\figsetgrpend

\figsetgrpstart
\figsetgrpnum{2.26}
\figsetgrptitle{Carina Nebula SED}
\figsetplot{\includegraphics[width=0.31\linewidth,clip,trim=1.2cm 12.4cm 3.5cm 3.7cm]{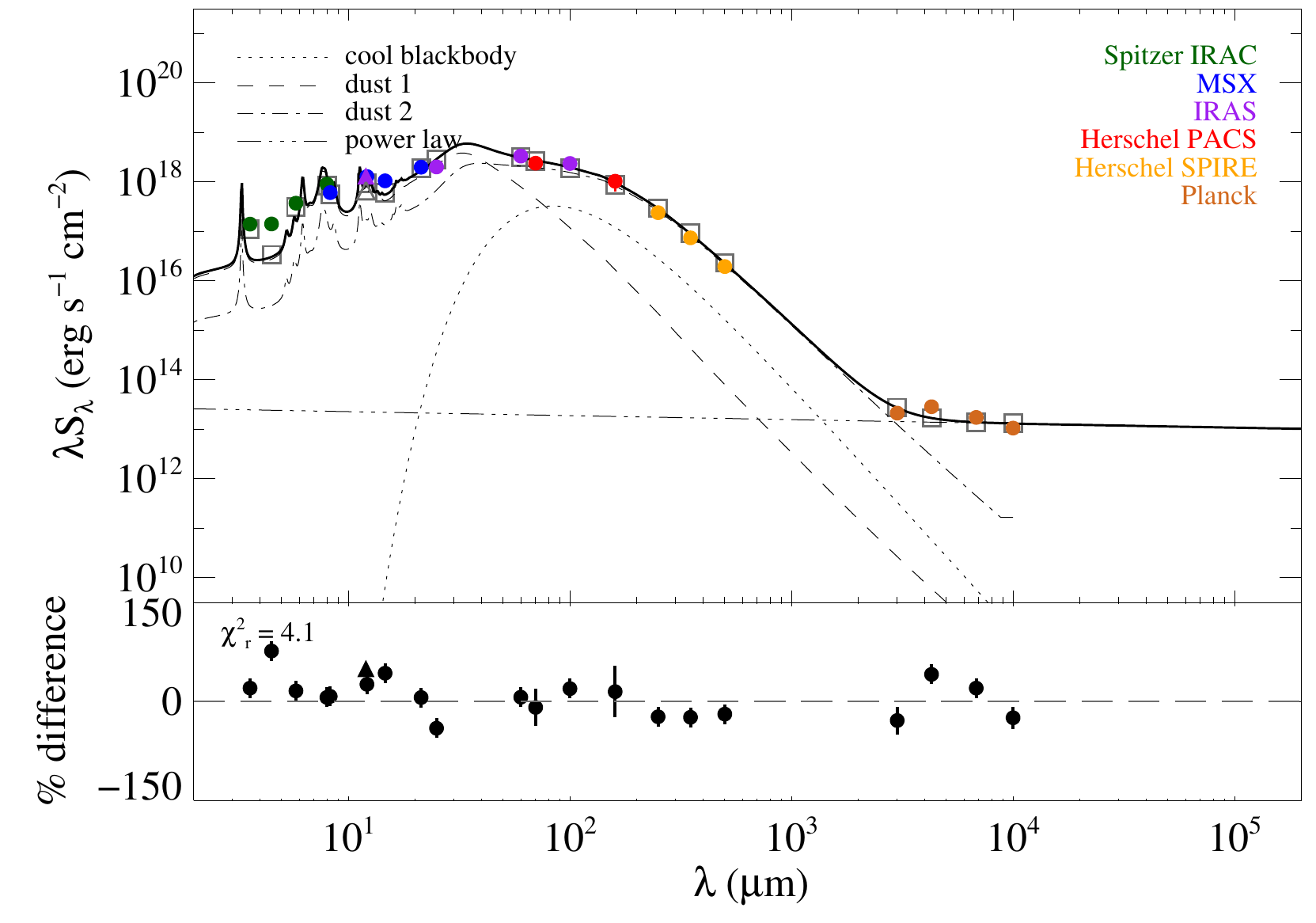}}
\figsetgrpnote{The SED and best-fit model for the Carina Nebula}
\figsetgrpend
 
\figsetgrpstart
\figsetgrpnum{2.27}
\figsetgrptitle{W51A SED}
\figsetplot{\includegraphics[width=0.31\linewidth,clip,trim=1.2cm 12.4cm 3.5cm 3.7cm]{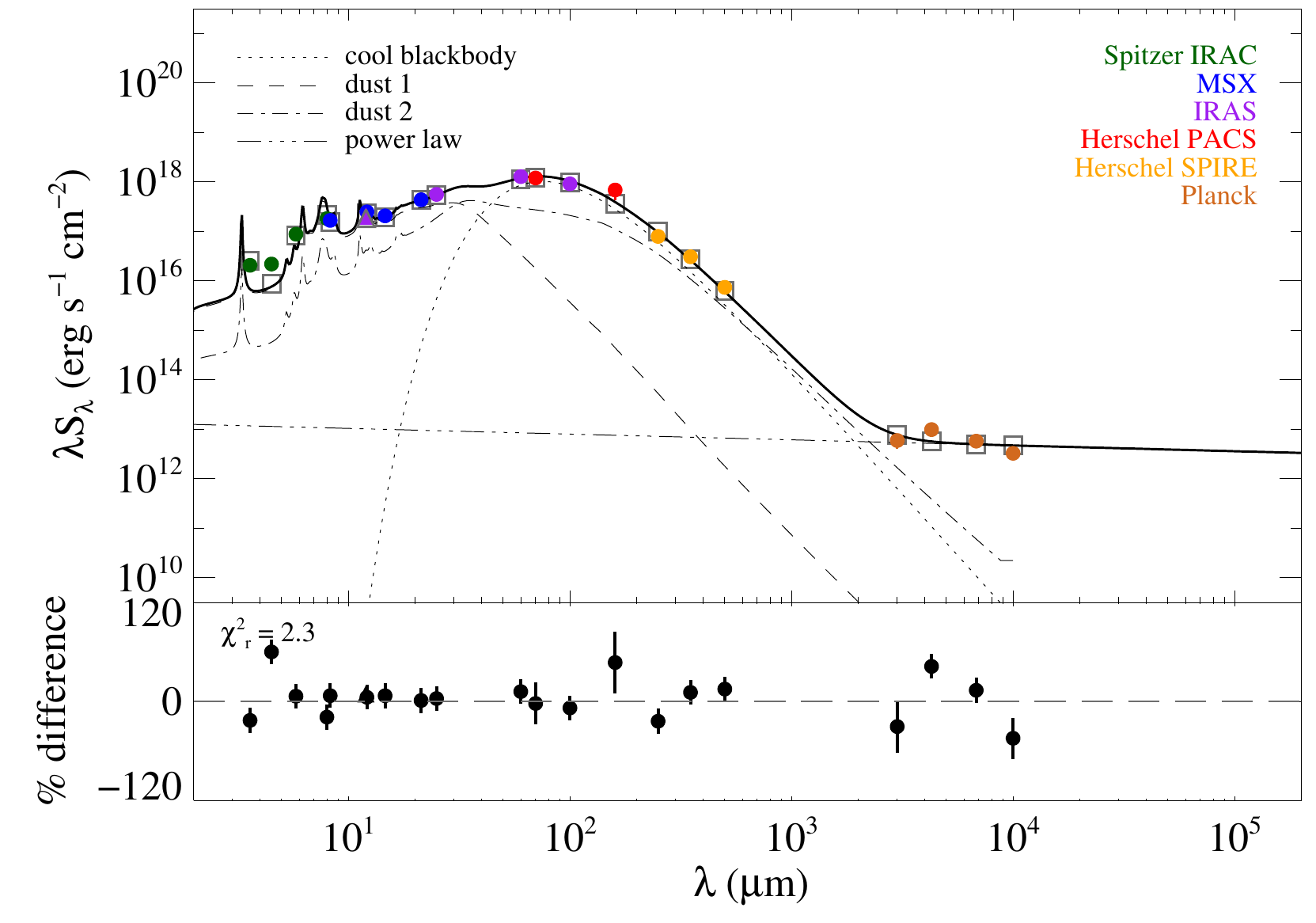}}
\figsetgrpnote{The SED and best-fit model for W51A}
\figsetgrpend

\figsetgrpstart
\figsetgrpnum{2.28}
\figsetgrptitle{NGC 3603 SED}
\figsetplot{\includegraphics[width=0.31\linewidth,clip,trim=1.2cm 12.4cm 3.5cm 3.7cm]{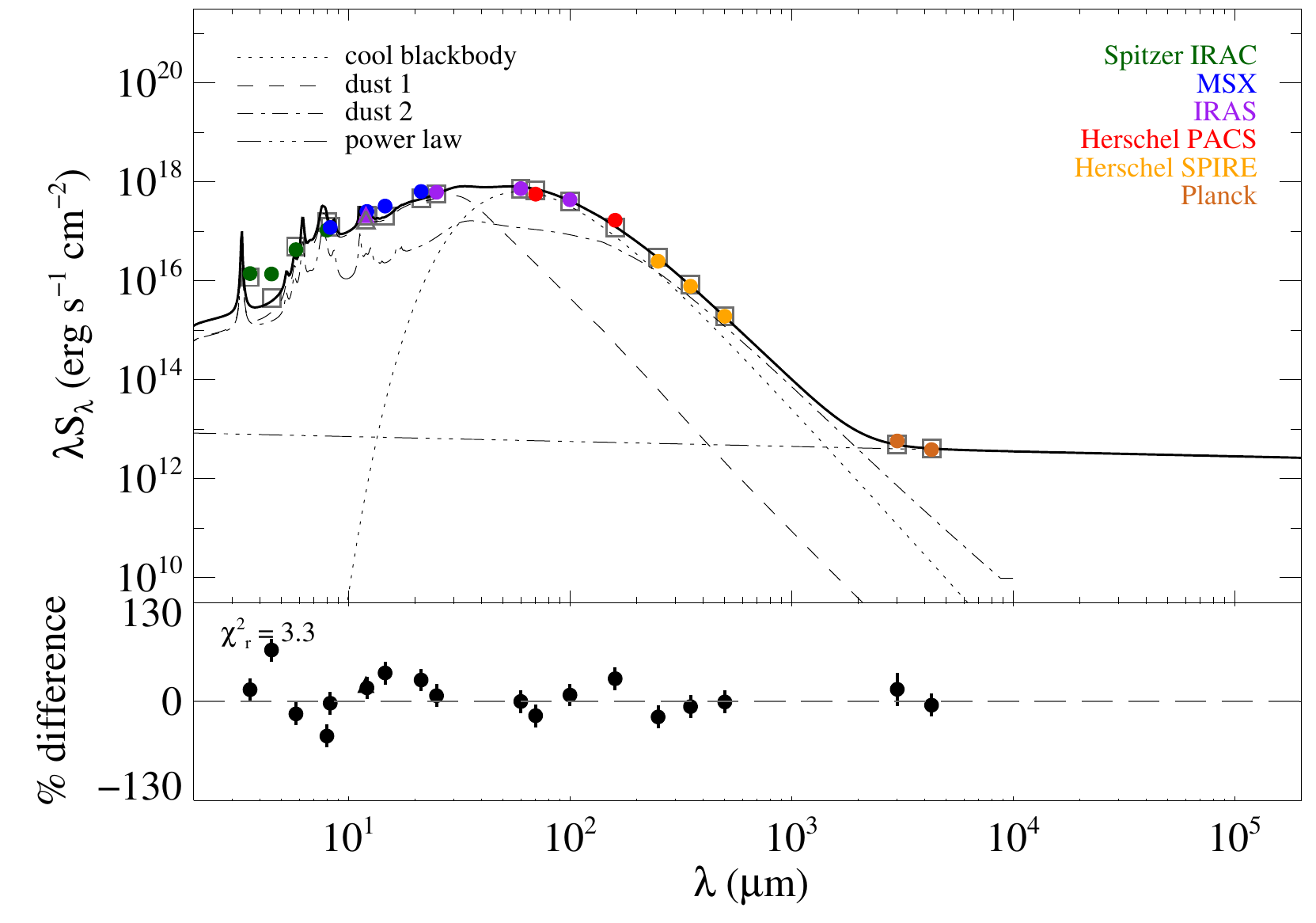}}
\figsetgrpnote{The SED and best-fit model for NGC 3603}
\figsetgrpend
\figsetend

\begin{figure}[htp]
\centering
\includegraphics[width=1\linewidth,clip,trim=1.2cm 12.4cm 3.5cm 3.7cm]{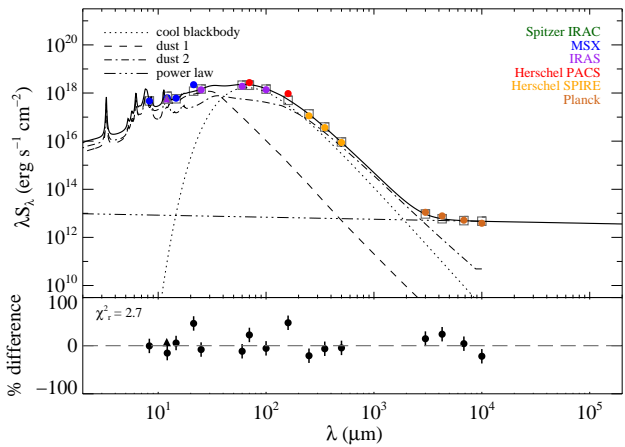}
	\caption{The 3.6 \micron\ -- 10 mm SED for the Orion Nebula, with the best-fit model superimposed. The complete figure set (28 images) is available in the online journal.}
\end{figure}\label{figure:SED_examples}

\begin{figure}
	\includegraphics[width=1\linewidth,clip,trim=1cm 12.5cm 2.5cm 3cm]{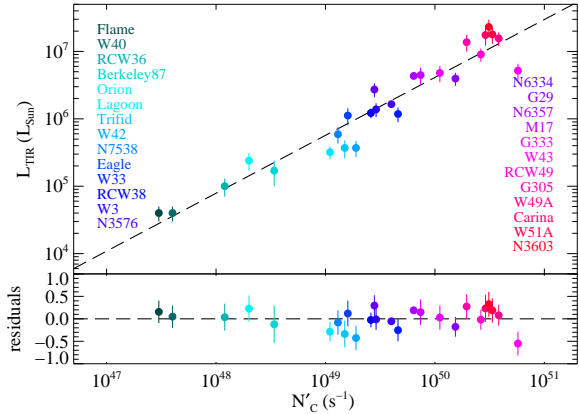}
	\caption{Dust-processed bolometric luminosity $L_{\rm TIR}$ as a function of $N_C^{\prime}$ from the \planck radio continuum. The dashed black line shows the best-fit relation (log$L_{\rm TIR}=(-36.33\pm2.53) + (0.86\pm0.05)$ log$N_C^{\prime}$). Only regions with $N_C^{\prime}$ measurements from \planck are used in the fit.}
\end{figure}\label{figure:Lbol_vs_Ncprime}

\begin{table*}[ht]\setlength{\tabcolsep}{2.5pt}
\centering
\scriptsize 
\caption{Global SED Model Fits}
\begin{tabular}{ccccccccccccccc}
\hline \hline
Name	& aperture			& $U_1$			& $q_{\rm PAH,1}$	& $U_{\rm min,2}$	& $U_{\rm max,2}$	& $q_{\rm PAH,2}$	& $f_{\rm bol}$	& 1-$\gamma$		& $L_{\rm TIR}$$^a$		& $T_{\rm BB}$		& $\alpha$		& $f_{\rm Br\alpha}$	& $N_C^{\prime}$	& \chisqr	\\ 
		& radius 			&				& (\%)			& 				&				& (\%)			& (\%)		& (10$^{-5}$)		& (10$^6$ L$_{\odot}$)	& (K)				&				& (\%)			& ($10^{49}$)		&		\\
 (1)		& (2)				& (3)	 			& (4)	 			& (5)		 		& (6)				& (7)				& (8)			& (9)				& (10)				& (11)			& (12)			& (13)			& (14)			& (15)	\\
\hline
Flame			& 15\farcm9 (1.5 pc)		& 10$^5$		& 4.58	& 0.50		& 10$^5$		& 0.47	& 41		& 7.5$\pm$0.9		& 0.04$\pm$0.01	& 34.9$\pm$9.3	& -0.10$\pm$0.01	& 2	& 0.03$\pm$0.01	& 5.0		\\
W40				& 22\farcm4 (3.3 pc)		& 10$^5$		& 3.19	& \nodata		& \nodata		& \nodata	& 26		& \nodata			& 0.04$\pm$0.01	& 25.8$\pm$1.3	& -0.09$\pm$0.01	& 2	& 0.04$\pm$0.01	& 7.2		\\
Wd~1$^e$		& 2\farcm8 (3.2 pc)		& 10$^5$		& 4.58	& 1.00		& 10$^5$		& 3.19	& 100	& 84.1$\pm$9.9	& 0.09$\pm$0.04	& 8.0$\pm$7.5		& \nodata			& \nodata	& \nodata		& 11.5	\\
RCW36			& 16\farcm4 (5.2 pc)		& 10$^5$		& 4.58	& 0.50		& 10$^5$		& 4.58	& 49		& 5.2$\pm$0.3		& 0.10$\pm$0.03	& 24.2$\pm$2.4	& -0.10$\pm$0.02	& 3	& 0.12$\pm$0.02	& 3.2		\\
Berkeley~87		& 8\farcm4 (4.3 pc)		& 10$^5$		& 1.12	& 0.50		& 10$^5$		& 4.58	& 43		& 5.7$\pm$0.9		& 0.17$\pm$0.07	& 27.4$\pm$4.1	& -0.09$\pm$0.01	& 11	& 0.34$\pm$0.17	& 2.2		\\
Orion			& 22\farcm6 (2.7 pc)		& 10$^5$		& 2.50	& 0.50		& 10$^5$		& 3.90	& 61		& 5.8$\pm$0.9		& 0.24$\pm$0.07	& 36.2$\pm$3.0	& -0.09$\pm$0.01	& 26	& 0.2$\pm$0.1		& 2.7		\\
Lagoon$^d$		& 13\farcm0 (4.4 pc)		& 10$^5$		& 3.19	& 0.50		& 10$^5$		& 1.12	& 72		& 5.0$\pm$0.2		& 0.32$\pm$0.07	& 29.6$\pm$2.2	& -0.09$\pm$0.01	& 17	& 1.1$\pm$0.2		& 1.6		\\
Trifid				& 16\farcm9 (7.7 pc)		& 10$^3$		& 2.50	& 0.50		& 10$^3$		& 4.58	& 87		& 5.0$\pm$0.8		& 0.37$\pm$0.11	& 20.9$\pm$3.1	& -0.09$\pm$0.01	& 6	& 1.5$\pm$0.4		& 1.8		\\
W42				& 5\farcm2 (3.3 pc)		& 10$^5$		& 1.77	& 0.50		& 10$^5$		& 2.50	& 60		& 5.5$\pm$0.8		& 0.37$\pm$0.10	& 26.4$\pm$0.8	& -0.10 (fixed)		& 25	& 1.9$\pm$0.1		& 1.3		\\
NGC~7538$^b$ 	& 8\farcm0 (6.2 pc)		& 10$^5$		& 4.58	& 0.50		& 10$^5$		& 1.12	& 58		& 6.3$\pm$0.2		& 0.59$\pm$0.16	& 27.0$\pm$4.0	& -0.07$\pm$0.01	& 7	& 1.3$\pm$0.1		& 2.1		\\
W4$^{c,e}$		& 55\farcm7 (36.3 pc)	& 10$^5$		& 4.58	& 0.50		& 10$^5$		& 4.58	& 26		& 5.0$\pm$0.3		& 0.77$\pm$0.07	& 24.8$\pm$1.2	& \nodata			& \nodata	& \nodata		& 2.6		\\
Eagle			& 20\farcm5 (10.2 pc)	& 10$^5$		& 0.47	& \nodata		& \nodata		& \nodata	& 53		& \nodata			& 1.12$\pm$0.32	& 22.1$\pm$1.8	& -0.10$\pm$0.02	& 86	& 1.6$\pm$0.4		& 7.5		\\
W33				& 12\farcm0 (8.4 pc)		& 10$^5$		& 1.12	& 0.50		& 10$^4$		& 0.47	& 35		& 4.9$\pm$0.4		& 1.18$\pm$0.29	& 25.5$\pm$1.8	& -0.09$\pm$0.01	& 39	& 4.6$\pm$0.8		& 6.2		\\
RCW38			& 21\farcm2 (10.5 pc)	& 10$^5$		& 2.50	& 0.50		& 10$^5$		& 3.90	& 59		& 5.0$\pm$0.1		& 1.22$\pm$0.20	& 29.9$\pm$1.2	& -0.09$\pm$0.01	& 8	& 2.6$\pm$0.1		& 1.7		\\
W3				& 13\farcm4 (8.5 pc)		& 10$^5$		& 3.90	& 0.50		& 10$^5$		& 2.50	& 45		& 7.1$\pm$1.1		& 1.38$\pm$0.32	& 32.6$\pm$1.1	& -0.08$\pm$0.01	& 7	& 2.9$\pm$0.2		& 2.2		\\
NGC~3576	 	& 12\farcm9 (10.4 pc)	& 10$^5$		& 1.77	& 0.50		& 10$^5$		& 3.90	& 52		& 5.8$\pm$0.4		& 1.65$\pm$0.12	& 30.6$\pm$0.7	& -0.10 (fixed)		& 13	& 4.0$\pm$0.1		& 2.5		\\
NGC~6334		& 19\farcm8 (9.4 pc)		& 10$^5$		& 4.58	& 0.50		& 10$^5$		& 1.12	& 50		& 6.1$\pm$0.3		& 2.72$\pm$0.63	& 30.2$\pm$1.4	& -0.11$\pm$0.02	& 7	& 2.8$\pm$0.3		& 2.0		\\
G29.96--0.02		& 9\farcm7 (17.5 pc)		& 10$^5$		& 3.19	& 0.50		& 10$^5$		& 2.50	& 37		& 4.6$\pm$0.2		& 3.96$\pm$0.88	& 29.3$\pm$1.0	& -0.10 (fixed)		& 14	& 15.4$\pm$0.4	& 1.3		\\
NGC~6357 		& 25\farcm8 (13.4 pc)	& 10$^5$		& 0.47	& 0.50		& 10$^5$		& 4.58	& 55		& 6.0$\pm$0.5		& 4.33$\pm$0.34	& 27.5$\pm$1.7	& -0.09$\pm$0.01	& 7	& 6.4$\pm$0.2		& 3.1		\\
M17				& 23\farcm0 (12.2 pc)	& 10$^5$		& 1.77	& 0.50		& 10$^5$		& 4.58	& 67		& 5.8$\pm$0.2		& 4.46$\pm$1.28	& 32.1$\pm$4.4	& -0.09$\pm$0.01	& 5	& 7.4$\pm$0.6		& 2.1		\\ \cline{2-2}
				& 9\farcm2 (7.0 pc),		&			&		&			&			&		&		&				&				&				&				&	&				&		\\
G333			& 7\farcm5 (5.7 pc),		& 10$^5$		& 0.47	& 0.50		& 10$^5$		& 3.90	& 51		& 4.9$\pm$0.7		& 4.80$\pm$1.29	& 36.3$\pm$1.9	& -0.07$\pm$0.01	& 18	& 11.1$\pm$0.2		& 1.9		\\
				& 9\farcm2 (7.0 pc)		&			&		&			&			&		&		&				&				&				&				&	&				&		\\ \cline{2-2}
W43				& 7\farcm5 (12.0 pc)		& 10$^5$		& 1.12	& 0.50		& 10$^5$		& 3.19	& 46		& 5.5$\pm$0.8		& 5.19$\pm$1.37	& 28.3$\pm$0.9	& -0.10$\pm$0.02	& 51	& 57.4$\pm$9.5	& 1.9		\\
RCW49			& 19\farcm6 (25.1 pc)	& 10$^5$		& 2.50	& 0.50		& 10$^5$		& 2.50	& 62		& 5.8$\pm$0.3		& 9.02$\pm$2.03	& 33.7$\pm$1.1	& -0.10$\pm$0.02	& 10	& 26.3$\pm$3.0	& 1.8		\\
G305			& 40\farcm8 (42.6 pc)	& 10$^5$		& 0.47	& 0.50		& 10$^5$		& 4.58	& 42		& 5.4$\pm$0.8		& 13.73$\pm$3.74	& 26.9$\pm$1.7	& -0.11$\pm$0.02	& 8	& 19.5$\pm$3.5	& 4.1		\\
W49A			& 6\farcm7 (22.2 pc)		& 10$^5$		& 3.19	& 0.50		& 10$^5$		& 1.77	& 31		& 6.3$\pm$0.9		& 15.61$\pm$3.63	& 29.9$\pm$1.1	& -0.09$\pm$0.01	& 23	& 38.3$\pm$9.2	& 1.9		\\
Carina			& 63\farcm9 (50.0 pc)	& 10$^4$		& 3.19	& 0.50		& 10$^4$		& 0.47	& 96		& 46.5$\pm$1.9	& 17.51$\pm$5.31	& 28.7$\pm$4.3	& -0.08$\pm$0.01	& 5	& 29.0$\pm$3.1	& 4.1		\\
W51A			& 31\farcm8 (47.2 pc)	& 10$^5$		& 4.58	& 0.50		& 10$^5$		& 0.47	& 51		& 4.7$\pm$0.3		& 17.88$\pm$4.86	& 31.9$\pm$4.8	& -0.12$\pm$0.02	& 9	& 33.5$\pm$5.2	& 2.3		\\
NGC~3603		& 12\farcm4 (25.2 pc)	& 10$^5$		& 1.12	& 0.50		& 10$^5$		& 2.50	& 56		& 12.6$\pm$1.1	& 23.10$\pm$6.40	& 40.6$\pm$6.1	& -0.10 (fixed)		& 8	& 31.1$\pm$0.9	& 3.3		\\
\hline \hline
\end{tabular}
\label{table:SED_global}
\tablecomments{$^a$In this and subsequent tables, MSFRs are listed in order of increasing $L_{\rm TIR}$.  $^b$Missing \spitzer [5.8] and [8.0] observations. $^c$\spitzer observations not used in fit due to incomplete coverage of the region. $^d$Missing \herschel PACS observations. $^e$Missing or insufficient radio emission in \planck.}
\end{table*}

In Figure~\ref{figure:Lbol_vs_Ncprime} we plot $L_{\rm TIR}$ against $N_C^{\prime}$. Previous studies have utilized the relationship between the luminosity at 24 \micron\xspace and $N_C^{\prime}$ as a foundation for calibrations of the extragalactic, mid-IR SFR determinations \citep{Calzetti+07, Chomiuk+Povich11,VEH16}. We find a sub-linear correlation, with log$L_{\rm TIR}=(-36.33\pm2.53) + (0.86\pm0.05)$ log$N_C^{\prime}$. This is consistent with sub-linear correlation between radio and MIR tracers of star formation found by \citet{VEH16}.

The cool blackbody components of the MSFRs in our sample have an average temperature $\langle T_{\rm BB}\rangle= 28.6\pm6.0$~K. This temperature range is consistent with the galaxy-wide SED modeling results found in the KINGFISH survey \citep{Hunt+15}.

\section{\HII Region Reprocessing of Starlight and Ionizing Photons}\label{section:reprocessing}
We compiled the known massive stellar content (stars with spectral types earlier than B2) of each region and estimated the Lyman continuum photon rate ($N_C$) and bolometric luminosity ($L_{\star}$) produced by the stars in each region. We used the models of \citep{Martins+05b} to estimate $N_C$ for the cataloged OB population in each region. This grid covers the log $g$--$T_{\rm eff}$ plane for O- and early-B stars, and includes non-LTE treatment and line-blanketing. We used the observed spectral type of each massive star (B2 or earlier) to assign a corresponding $L_{\star}$ and $N_C$, summarized in Table~\ref{table:stellar_models}. For Wolf-Rayet stars, we adopt the luminosities and ionizing photon rates provided in \citet[][their Table~2]{Crowther07}. A detailed discussion of each region, including references to the previously catalogued stellar content or assumptions made regarding the spectral types, is presented in Appendix~\ref{appendix:region_discussion}.

\begin{table*}[ht]
\centering
\caption{Adopted Stellar Parameters by Spectral Type}
\begin{tabular}{ccccccccccccccc}
\hline \hline
		& \multicolumn{2}{c}{Class V}	&& \multicolumn{2}{c}{Class III}		&& \multicolumn{2}{c}{Class I}		&&  \multicolumn{2}{c}{Wolf Rayet (WN+)}		&&  \multicolumn{2}{c}{Wolf Rayet (WC+)}		\\ \cline{2-3} \cline{5-6} \cline{8-9} \cline{11-12} \cline{14-15}
Spectral	& log $L_{\star}$ 	& log $N_C$ 	&&  log $L_{\star}$ 		& log $N_C$	&&  log $L_{\star}$	& log $N_C$ 		&&  log $L_{\star}$	& log $N_C$	&&  log $L_{\star}$	& log $N_C$	\\
Type		& (\Lsun)		& (s$^{-1}$)	&& (\Lsun)		& (s$^{-1}$)	&& (\Lsun)	& (s$^{-1}$)		&& (\Lsun)	& (s$^{-1}$)	&& (\Lsun)	& (s$^{-1}$)	\\
(1)		& (2)			& (3)			&& (4)			& (5)			&& (6)		& (7)				&& (8)		& (9)			&& (10)		& (11)		\\
\hline
O3		& 5.84	& 49.64	&& 5.96	& 49.77	&& 5.99	& 49.78	&& 5.34	& 49.20	&& ...	& ...		\\
O3.5		& 5.76	& 49.54	&& 5.91	& 49.71	&& 5.96	& 49.74	&& ...	& ...		&& ...	& ...		\\
O4		& 5.67	& 49.44	&& 5.85	& 49.64	&& 5.93	& 49.70	&& 5.30	& 49.20	&& 5.54	 & 49.40	\\
O4.5		& 5.58	& 49.33	&& 5.70	& 49.56	&& 5.90	& 49.66	&& ...	& ...		&& ...	& ...		\\
O5		& 5.49	& 49.22	&& 5.73	& 49.48	&& 5.87	& 49.62	&& 5.20	& 49.00	&& 5.10	 & 48.90	\\
O5.5		& 5.41	& 49.10	&& 5.67	& 49.40	&& 5.84	& 49.58	&& ...	& ...		&& ...	& ...		\\
O6		& 5.32	& 48.99	&& 5.61	& 49.32	&& 5.81	& 49.52	&& 5.20	 & 49.10	&& 5.06	 & 48.90	\\
O6.5		& 5.23	& 48.88	&& 5.54	& 49.23	&& 5.78	& 49.46	&& ...	& ...		&& ...	& ...		\\
O7		& 5.14	& 48.75	&& 5.48	& 49.13	&& 5.75	& 49.41	&& 5.54	 & 49.40	&& 5.34	 & 49.10	\\
O7.5		& 5.05	& 48.61	&& 5.42	& 49.01	&& 5.72	& 49.31	&& ...	& ...		&& ...	& ...		\\
O8		& 4.96	& 48.44	&& 5.35	& 48.88	&& 5.68	& 49.25	&& 5.38	 & 49.10	&& 5.14	 & 49.00	\\
O8.5		& 4.86	& 48.27	&& 5.28	& 48.75	&& 5.65	& 49.19	&& ...	& ...		&& ...	& ...		\\
O9		& 4.77	& 48.06	&& 5.21	& 48.65	&& 5.61	& 49.11	&& 5.70	 & 48.90	&& 4.94	 & 48.60	\\
O9.5		& 4.68	& 47.88	&& 5.15	& 48.42	&& 5.57	& 49.00	&& ...	& ...		&& ...	& ...		\\
B0		& 4.57	& 47.70	&& 5.08	& 48.28	&&  ...	& ...		&& ...	& ...		&& ...	& ...		\\
B0.5		& 4.47	& 47.50	&& 5.00	& 48.10	&&  ...	& ...		&& ...	& ...		&& ...	& ...		\\
B1		& 4.37	& 47.28	&& 4.93	& 47.90	&&  ...	& ...		&& ...	& ...		&& ...	& ...		\\
B1.5		& 4.28	& 47.05	&& 4.86	& 47.68	&& ...	& ...		&& ...	& ...		&& ...	& ...		\\
B2		& 4.19	& 46.80	&& 4.78	& 47.44	&& ...	& ...		&& ...	& ...		&& ...	& ...		\\
\hline \hline
\end{tabular}
\label{table:stellar_models}
\end{table*}

In Table~\ref{table:energy_budget} we summarize the expected ($N_C$) and ($L_{\star}$) in each MSFR, the ionizing photon flux estimated from the \planck radio observations ($N_C^{\prime}$), the bolometric luminosity ($L_{\rm TIR}$) measured by our SED fitting, and the ratios $N_C^{\prime}/N_C$ and $L_{\rm TIR}/L_{\star}$ for each region. The spectral types of the ionizing stellar populations are, of course, not known with perfect accuracy. To assign uncertainties to $N_C$ and $L_{\star}$, we assume that the cataloged spectral type of each star may differ by up to one type from the true value; e.g., an O6V star may be as late as an O7V or as early as an O5V. For each star in the MSFR, we randomly select a spectral type that can be the same as the reported spectral type, or a half- or full-spectral type earlier or later than the reported spectral type. We then re-compute $N_C$ and $L_{\star}$ for the region. This process is repeated 500 times for each MSFR, yielding distributions of plausible $N_C$ and $L_{\star}$ values for each region. The standard deviations of these distributions are then used as the uncertainties on $N_C$ and $L_{\star}$ reported in Table~\ref{table:energy_budget}; typically, the means of these distributions agree with the values computed using the cataloged spectral types.

\begin{table*}[ht]
\footnotesize
\centering
\caption{Lyman Continuum Rates and $L_{\star}$ vs. $L_{\rm TIR}$}
\begin{tabular}{cccccccc}
\hline \hline
			& \multicolumn{2}{c}{OB Stars}								&& \multicolumn{2}{c}{SED Model/Stellar Population}		&			&				\\ \cline{2-3} \cline{5-6}
Region		& $N_C$ (10$^{49}$ s$^{-1}$) 	& $L_{\star}$ (10$^6$ $L_{\odot}$) 	&& $N_C^{\prime}/N_C$			& $L_{\rm TIR}/L_{\star}$	& f$_{\rm esc}$	& $f_{C,\rm abs}$ 	\\
(1)			& (2)						& (3)							&& (4)						& (5)					& (6)			& (7)				\\
\hline
Flame$^a$		& 0.15$\pm$0.07	& 0.09$\pm$0.02	&& 0.20$\pm$0.11		& 0.44$\pm$0.15	& 0.56$\pm$0.19	& 0.24$\pm$0.16	\\		
W40				& 0.09$\pm$0.05	& 0.07$\pm$0.01	&& 0.44$\pm$0.27		& 0.57$\pm$0.16	& 0.43$\pm$0.12	& 0.13$\pm$0.09	\\		
Wd~1                 	& 6.33$\pm$4.11	& 2.64$\pm$0.65 	&& \nodata			& 0.03$\pm$0.02 	& $>$0.32			& \nodata		 	\\		
RCW~36			& 0.19$\pm$0.09	& 0.11$\pm$0.02	&& 0.63$\pm$0.32		& 0.91$\pm$0.32	& 0.09$\pm$0.03	& 0.28$\pm$0.17	\\		
Berkeley~87		& 1.86$\pm$0.72	& 0.72$\pm$0.15	&& 0.18$\pm$0.11		& 0.24$\pm$0.11	& $>$0.41			& 0.06$\pm$0.05 	\\		
Orion			& 0.77$\pm$0.23	& 0.29$\pm$0.04	&& 0.26$\pm$0.15		& 0.83$\pm$0.27	& 0.17$\pm$0.06	& 0.57$\pm$0.38	\\		
Lagoon$^b$		& 3.71$\pm$2.09	& 1.14$\pm$0.27	&& 0.30$\pm$0.18		& 0.28$\pm$0.09	& 0.72$\pm$0.23	& 0		 		 \\		
Trifid$^c$			& 0.58$\pm$0.22	& 0.16$\pm$0.04	&& 2.59$\pm$1.20		& 2.31$\pm$0.90	& \nodata			& \nodata			\\		
W42				& 1.71$\pm$0.54	& 0.35$\pm$0.08	&& 1.11$\pm$0.36		& 1.06$\pm$0.38	& \nodata			& 0				\\		
NGC~7538$^a$	& 4.74$\pm$0.89	& 1.01$\pm$0.13	&& 0.27$\pm$0.05		& 0.58$\pm$0.17	& 0.42$\pm$0.12	& 0.31$\pm$0.10 	\\		
W4$^b$			& 11.50$\pm$2.12	& 2.60$\pm$0.28	&& \nodata			& 0.30$\pm$0.03	& 0.70$\pm$0.07	& \nodata			\\		
Eagle$^b$		& 4.31$\pm$2.88	& 2.01$\pm$0.41	&& 0.37$\pm$0.26		& 0.56$\pm$0.20	& 0.44$\pm$0.16	& 0.19$\pm$0.15	\\		
W33				& 10.78$\pm$1.30	& 1.98$\pm$0.16	&& 0.43$\pm$0.09		& 0.60$\pm$0.16	& 0.40$\pm$0.11	& 0.17$\pm$0.06 	\\		
RCW~38$^{a,b}$	& 3.50$\pm$1.36	& 0.76$\pm$0.18	&& 0.74$\pm$0.29		& 1.61$\pm$0.46	& \nodata			& \nodata			\\		
W3$^b$			& 5.87$\pm$2.37	& 1.68$\pm$0.31	&& 0.49$\pm$0.20		& 0.82$\pm$0.24	& 0.18$\pm$0.05	& 0.30$\pm$0.17 	\\		
NGC~3576$^c$	& 2.71$\pm$1.26	& 0.88$\pm$0.16	&& 1.48$\pm$0.69		& 1.88$\pm$0.37	& \nodata			& \nodata		 	\\		
NGC~6334$^{b,c}$	& 4.24$\pm$1.74	& 1.17$\pm$0.27	&& 0.66$\pm$0.28		& 2.32$\pm$0.76 	& \nodata			& \nodata			\\		
G29.96--0.02$^c$	& 4.17$\pm$0.51	& 0.74$\pm$0.06	&& 3.69$\pm$0.46		& 5.35$\pm$1.24	& \nodata			& \nodata 			\\		
NGC~6357$^b$	& 32.95$\pm$3.41	& 7.13$\pm$0.53	&& 0.19$\pm$0.02		& 0.61$\pm$0.07 	& 0.39$\pm$0.04	& 0.42$\pm$0.07	\\ 		
M17$^b$			& 22.39$\pm$4.08	& 5.89$\pm$0.54	&& 0.33$\pm$0.07		& 0.76$\pm$0.23	& 0.24$\pm$0.07	& 0.43$\pm$0.16 	\\		
G333$^{a,c}$		& 8.30$\pm$1.26	& 1.55$\pm$0.18	&& 1.33$\pm$0.20		& 1.22$\pm$0.36	& \nodata			& \nodata		 	\\		
W43				& 39.70$\pm$1.95	& 7.10$\pm$0.25	&& 1.44$\pm$0.25		& 0.73$\pm$0.19	& 0.27$\pm$0.07	& \nodata		 	\\		
RCW~49			& 56.24$\pm$2.82	& 10.38$\pm$0.42	&& 0.47$\pm$0.06		& 0.87$\pm$0.20	& 0.13$\pm$0.03	& 0.40$\pm$0.11 	\\		
G305$^c$			& 29.31$\pm$2.34	& 6.70$\pm$0.34	&& 0.67$\pm$0.13		& 2.05$\pm$0.57	& \nodata			& \nodata	 		\\		
W49A$^c$		& 60.89$\pm$3.18	& 10.53$\pm$0.45	&& 0.63$\pm$0.15		& 1.48$\pm$0.35	& \nodata			& \nodata			\\		
Carina			& 93.79$\pm$6.43	& 22.76$\pm$0.97	&& 0.31$\pm$0.04		& 0.77$\pm$0.24	& 0.23$\pm$0.07	& 0.46$\pm$0.16 	\\		
W51A$^{c,d}$		& 42.95$\pm$2.81	& 9.15$\pm$0.43	&& 0.78$\pm$0.13		& 1.95$\pm$0.28	& \nodata			& \nodata		 	\\		
NGC~3603		& 137.08$\pm$5.02	& 23.03$\pm$0.74	&& 0.23$\pm$0.02		& 1.00$\pm$0.28	& $<$0.28			& 0.77$\pm$0.23 	\\		
\hline \hline
\end{tabular}\label{table:energy_budget}
\tablecomments{$^a$Assumptions made about the spectral type(s) of one or more stars in the region. $^b$Includes candidate OB members from \citet{Povich+17}. $^c$Stellar content likely incomplete. $^d$Significant distance uncertainty.}
\end{table*}  

\begin{figure*}[htp]
\centering
\includegraphics[width=0.45\linewidth,clip,trim=1.5cm 12.5cm 2cm 3cm]{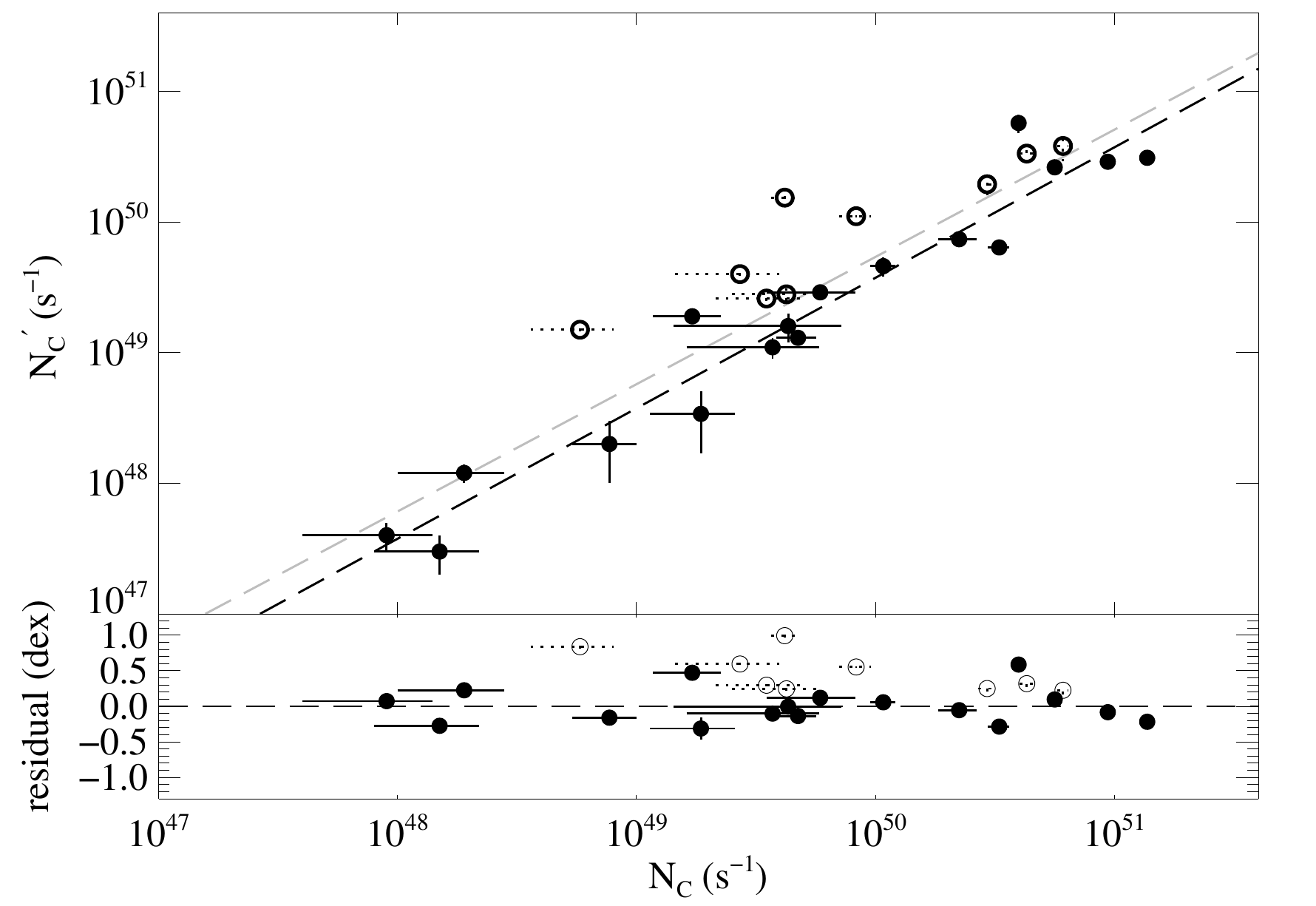} \quad
\includegraphics[width=0.45\linewidth,clip,trim=1.5cm 12.5cm 2cm 3cm]{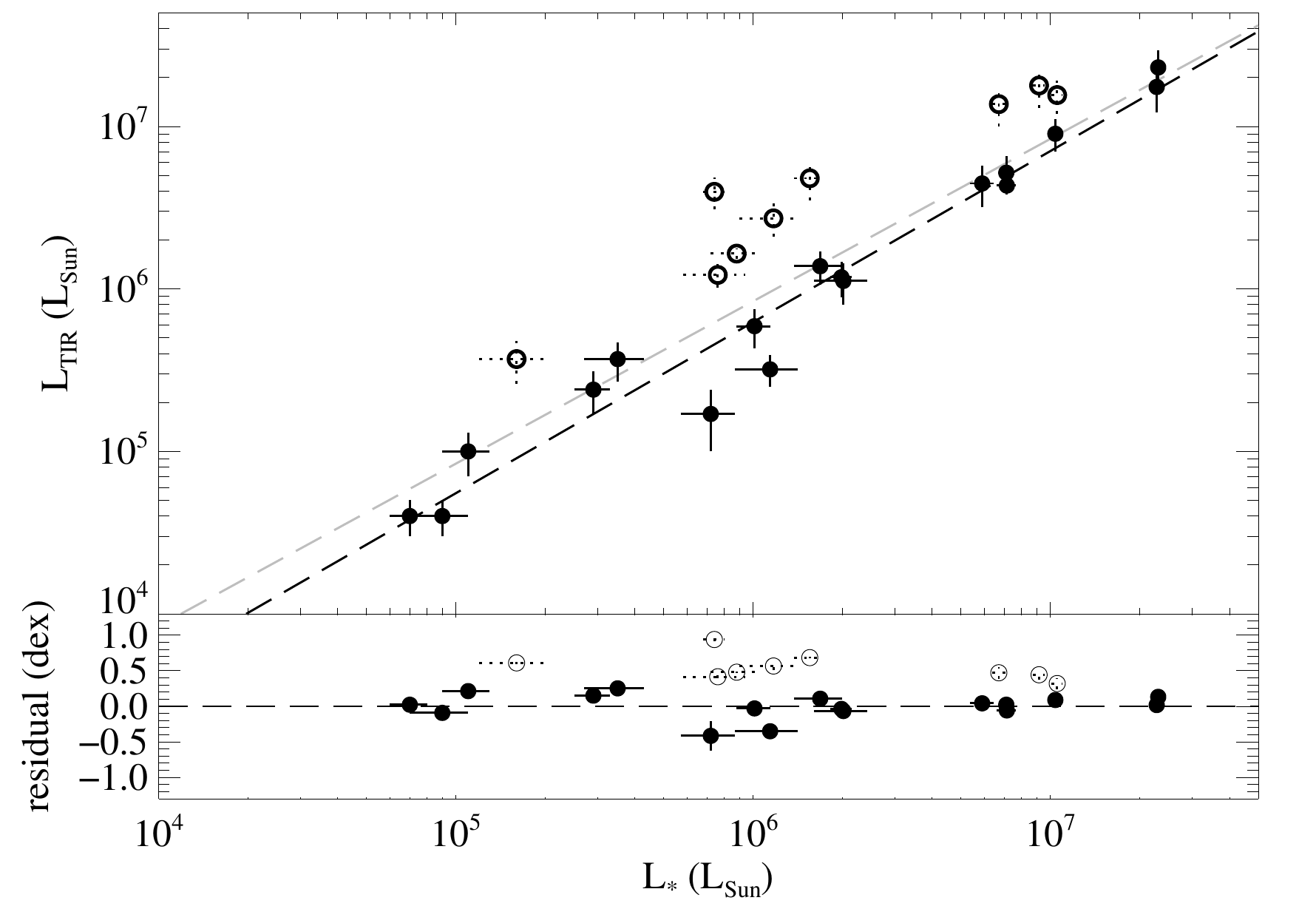}
\caption{{\it Left}: The Lyman continuum photon rate from radio observations ($N_C^{\prime}$) as a function of the stellar content ($N_C$). Excluding the MSFRs plotted as open circles with dotted-line error bars (see text for details), $\log{N_C^{\prime}} = (-0.40\pm0.80) + (1.00\pm0.07)\log{N_C}$ (black dashed line). Using all MSFRs, the slope becomes 0.97$\pm$0.08 and the $y$-intercept is 1.00$\pm$0.76. {\it Right}: The bolometric luminosity ($L_{\rm TIR}$) as a function of the stellar content ($L_{\star}$). The fit to the filled circle points gives $\log{L_{\rm TIR}} = (-0.52\pm0.35) + (1.04\pm0.06)\log{L_{\star}}$ (black dashed line). Using all MSFRs, the slope becomes $1.00\pm0.12$ and the $y$-intercept becomes -0.07$\pm$0.73 (gray dashed line). The bottom panels show the scatter about the best-fit relationships.}
\label{figure:SED_vs_OB}
\end{figure*}

In Figure~\ref{figure:SED_vs_OB} we plot $N_C^\prime$ and $L_{\rm TIR}$ derived from our SED fits against against $N_C$ and $L_\star$, respectively, from the massive stellar content in all 28 MSFRs. Excluding MSFRs that lack secure distances, well-characterized massive stellar populations, or measurable $N_C^{\prime}$ from the \planck observations, the best-fit relationship for the Lyman continuum photon rates yields a linear slope, $\log{N_C^{\prime}} = (-0.40\pm0.80) + (1.00\pm0.07)\log{N_C}$, as does the relationship for bolometric luminosities, with $\log{L_{\rm TIR}} = (-0.52\pm0.35) + (1.04\pm0.06)\log{L_{\star}}$. Including all MSFRs does not significantly change the power-law slopes of these relationships.

Dust can absorb Lyman continuum photons before they contribute to the ionization of \HII regions \citep{McKee+Williams97}, reducing $N_C^\prime/N_C$ while contributing to $L_{\rm TIR}/L_\star$. Previous studies have found that $\sim$20-50\% of UV photons produced by massive stars in the Milky Way and Local Group galaxies are absorbed by dust in surrounding \HII regions \citep[see][and references therein]{Inoue01,Inoue+01}. Similar studies of external galaxies have suggested 30--70\% of the emitted UV photons escape from \HII regions and interact with the ISM \citep[e.g.,][]{Oey+Kennicutt97,Zurita+02,Giammanco+05,Pellegrini+12}. 

Our SED modeling results allow us to estimate both the fraction of stellar luminosity that escapes the MSFR and the fraction of Lyman continuum photons absorbed by dust within the \HII regions. Strong density inhomogeneities in the \HII regions and surrounding PDRs create low-density pathways through which Lyman continuum and longer-wavelength photons may reach the diffuse ISM without first being absorbed by local dust or gas associated with the MSFR. UV photons carry the bulk of the emitted stellar luminosity and have characteristically large interaction cross-sections with both dust and gas. The average hydrogen gas density $n_H$ of the diffuse ISM is typically lower than that within a young \HII region by a factor of $10^{-3}$. Since the Str{\"o}mgren radius is proportional to $n_H^{-2/3}$ \citep{Stromgren39}, Lyman continuum photons that manage to escape MSFRs can ionize regions 100 times larger than their parent \HII regions. The largest \HII regions in our sample have diameters of tens of pc, hence their escaped Lyman continuum photons contribute to the ionization of the warm ionized medium, with its ${\sim}1$~kpc scale height \citep{Haffner+03}.

The fraction of stellar luminosity escaping from each MSFR is simply

\begin{equation}\label{equation:fesc}
  f_{\rm esc} = 1 - L_{\rm TIR}/L_{\star},
\end{equation}

\noindent which we calculate for the 18 MSFRs with well-characterized massive stellar populations, well-constrained distances (hence excluding the 8 regions marked with a $c$ or $d$ in column 1 of Table~\ref{table:energy_budget}; values of $f_{\rm esc}$ are presented in column 6), and measurable $N_C^{\prime}$. For these regions we find an average $\langle L_{\rm TIR}/L_{\star}\rangle=0.74\pm0.22$, so approximately three-quarters of the emitted stellar luminosity is absorbed and reprocessed by the \HII regions and surrounding PDRs and one-quarter escapes into the diffuse ISM. The average ratio of Lyman continuum photon rate emitted by the massive stars to ionizing photon rate measured from the \planck thermal radio continuum is $\langle N_C^{\prime}/N_C\rangle=0.47\pm0.24$. In other words, we find that only $\sim$50\% of UV photons emitted by massive stars contribute to the ionization of their surrounding \HII regions, consistent with \citet{Inoue01}. Ionizing continuum photons are lost to the \HII regions due to the combination of dust absorption and escape, hence we define the fraction ($f_{C,\rm abs}$) of Lyman continuum photons absorbed by dust within \HII regions using

\begin{equation}
  f_{C,\rm abs} + f_{\rm esc} = (N_{C,\rm abs} + N_{C,\rm esc})/N_C = 1 - N_C^\prime/N_C.
\end{equation}

\noindent Substituting Equation~\ref{equation:fesc} into the above and rearranging terms, we have

\begin{equation}
  f_{C,\rm abs} = L_{\rm TIR}/L_{\star} - N_C^\prime/N_C.
\end{equation}

We have tacitly assumed that $f_{\rm esc}$ does not differ significantly between Lyman continuum and longer-wavelength UV photons. Values of $f_{C,\rm abs}$ are reported in column 7 of Table~\ref{table:energy_budget}. The uncertainties are relatively large, and $f_{C,\rm abs}$ falls within $1\sigma$ of zero for roughly 20\% (3/17) of \HII regions for which it could be calculated; for these regions we report upper limits only.

In Table~\ref{table:subgroup_mean} we divide our MSFR into subgroups based on luminosity and then compute the average values of $N_C^\prime/N_C$, $L_{\rm TIR}/L_\star$, and $f_{C,\rm abs}$ for each subgroup. The luminosity subgroups were defined to categorize the ionizing stellar population as follows: those with a fully-populated upper stellar initial mass function (IMF) containing multiple O2/O3 stars plus Wolf-Rayet stars ($\log{L_{\rm TIR}/L_\sun}\ge 6.8$ or $\log{L_{\star}/L_\sun}\ge 6.8$), those ionized by the equivalent of a single O6 or later-type star ($\log{N_C/{\rm s}^{-1}}< 49$), and the intermediate case of \HII regions ionized by one or more early O stars but may still be influenced by stochastic sampling of the upper IMF \citep[see][]{Kennicutt+12}. Only regions with reasonably secure distances, well-catalogued stellar populations, and for which we were able to calculate non-zero values of $f_{C,\rm abs}$ (e.g., half of our sample) were used for this analysis.

\begin{table*}[htp]
\centering
\caption{Mean Fractions of Starlight Reprocessed by Dust in MSFRs}
\begin{tabular}{ccccccc}
\hline \hline
MSFR Subgroup	& $N$	& $\langle \log{(N_C^{\prime}/{\rm s}^{-1})}\rangle$ 	& $\langle \log{(L_{\rm TIR}/L_\sun)}\rangle$	& $\langle N_C^{\prime}/N_C\rangle$	& $\langle L_{\rm TIR}/L_{\star}\rangle$ & $\langle f_{C,\rm abs}\rangle$	\\
(1)	& (2)	& (3)	& (4)	& (5)	& (6)	& (7)		\\
\hline
Fully-populated upper IMF		& 4	& 50.295	& 7.050	& 0.30 	& 0.82	& 0.51	\\
Some early O stars	       		& 6	& 49.307	& 5.985	& 0.35	& 0.59	& 0.24	\\
Single O6 or later	        		& 4	& 47.865	& 4.896	& 0.38	& 0.69	& 0.31	\\
\hline
All MSFRs		        		& 14	& 49.177	& 5.978	& 0.34	& 0.68	& 0.34 	\\
\hline \hline
\end{tabular}\label{table:subgroup_mean}
\end{table*}

\begin{figure}
\centering
\includegraphics[width=0.9\linewidth,clip,trim=2.5cm 13cm 2cm 3cm]{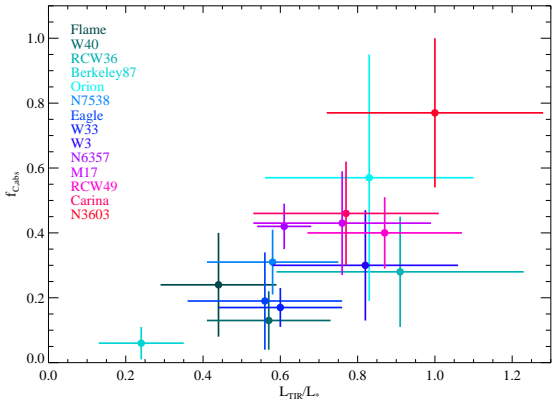} 
\caption{The fraction of Lyman continuum photons absorbed by dust within each \HII region ($f_{C,\rm abs}$) as a function of the luminosity ratio $L_{\rm TIR}/L_{\star}$.}
\label{figure:Lfrac_vs_fcabs}
\end{figure}

The main result highlighted in Table~\ref{table:subgroup_mean} is that the high-luminosity regions with fully-populated upper IMFs lose 50\% of their Lyman continuum photons to dust absorption that would otherwise contribute to ionizing their \HII regions, which is significantly above the average $f_{C,\rm abs}=30\%$ for the entire sample. This result agrees very well with the predictions of \citet{McKee+Williams97}, who calculated that the fraction of ionizing photons absorbed by dust increases with ionizing photon rate in \HII regions. This trend was caused by the higher ionization fraction lowering the absorption cross-section of the gas toward ionizing photons.

In Figure~\ref{figure:Lfrac_vs_fcabs}, we plot $f_{C,\rm abs}$ as a function of the luminosity ratio $L_{\rm TIR}/L_{\star}$ (the regions included in Figure~\ref{figure:Lfrac_vs_fcabs} are the same regions used to compute averages in Table~\ref{table:subgroup_mean}). Regions with higher fractions of their stellar luminosities reprocessed by dust show higher fractions of Lyman continuum photon absorption.

Given the extreme feedback effects produced by the radiation and stellar winds of the most massive young clusters, one might expect that dust grains would be more efficiently destroyed or evacuated from more luminous regions. \citet{Everett+Churchwell10} found that dust must be continually replenished within \HII regions to produce the observed 24~\micron\ emission. \citet{McKee+Williams97} did not address the question of whether dust properties vary among different \HII regions. We find no significant depletion of dust among the more luminous Galactic \HII regions in our sample, indeed, our results seem to imply the opposite, that more luminous \HII regions are somehow dustier than less-luminous ones.

More massive clusters are formed from the densest clumps within the most massive giant molecular clouds, and the high gravitational potential combined with the self-shielding effects of very dense, dusty gas could preserve massive reservoirs of cold dust within filaments, pillars, and globules, that are in close proximity to or even surrounded by ionized gas \citep[see, e.g.][]{Dale+Bonnell11}. Photoevaporation of this cold, dusty gas could hence provide a readily-available source of dust replenishment for luminous \HII regions. This is consistent with the observed morphologies of dusty, giant \HII regions, which feature large cavities filled with hot, low-density X-ray-emitting plasma surrounded by relatively thin shells traced by both the brightest mid-IR and radio emission \citep{Townsley+03,Povich+07,Townsley+11b}. In our sample, M17, W43, and NGC 3603 exemplify this morphology.

Part of this trend is likely be due to selection bias in our sample; after all we have targeted IR-bright Galactic MSFRs, not a representative sample of all Galactic MSFRs. The most massive young clusters are located at relatively large heliocentric distances in the bright, crowded reaches of the Galactic midplane, so very massive clusters that have been cleared of dust (such as Wd~1) are difficult to identify. But our sample was constructed so that the selection biases should be similar to those of a spatially-resolved, IR observation of a nearby disk galaxy targeting the brightest compact ``knots'' of IR emission \citep[e.g.,][]{Calzetti+07}. We have thus included a representative sample of regions that are bright in the IR and sufficiently isolated (none are within 5~kpc of the Galactic center) that they would stand out among the brightest compact IR sources in to an external observer of the Milky Way.

\section{Monochromatic Luminosities and Predicted Star Formation Rates}\label{section:lum_and_SFRs}
Monochromatic luminosities at various IR wavelengths have been developed as more convenient substitutes for $L_{\rm TIR}$ to measure SFRs in galaxies, generally calibrated against extinction-corrected H$\alpha$ emission as a proxy for the ionizing photon rate. In this section we analyze the behavior, among our sample of Galactic MSFRs, of three monochromatic luminosities that have been widely investigated in the extragalactic context. Monochromatic luminosities reported here are measured from the SED model luminosities convolved with the relevant instrumental filter bandpass, which may differ from the flux density measured directly from aperture photometry in that bandpass. In the case of the 24~\micron\ luminosity the direct measure is not available, as our MSFRs usually saturate the MIPS 24~\micron\ images. Saturation frequently affects the IRAC 8~\micron\ flux densities in bright regions as well.

At shorter wavelengths (3~\micron~$\la\lambda\la 20$~\micron), IR emission from MSFRs is dominated by the emission features of PAHs and the warm (${>}150$~K) dust continuum. Not surprisingly, short-wavelength tracers such as the monochromatic 8 \micron\xspace luminosity ($L_8$) are inaccurate, showing large degrees of variability with respect to metallicity \citep[e.g.,][]{Madden00,Madden+06,Engelbracht+05,Engelbracht+08,Draine+Li07,Galliano+08,Gordon+08,MunozMateos+09,Marble+10} and the shape of the SEDs \citep{Dale+05,Calzetti+07}. For these reasons, $L_8$ is not generally considered a reliable SFR tracer. The strong ionizing radiation fields of early-type stars are extremely efficient at destroying PAHs in their vicinity, thereby reducing their line strength and relative contribution to $L_8$. Thus, $L_8$ may be a better tracer of the B-star population in a region than the overall SFR \citep{Peeters+04}. 

The monochromatic 24 \micron\xspace luminosity ($L_{24}$) is often utilized as a SFR tracer \citep{Calzetti+07}. Although the $L_{\rm 24}$/SFR ratio is reasonably consistent on local scales, it is systematically higher when applied to starburst galaxies or ULRIGs \citep{Calzetti+05}. In this intermediate wavelength range ($\sim$20--60 \micron), warm dust ($\sim$50 K) emission transitions from being dominated by stochastically heated small grains to being dominated by larger grains in thermal equilibrium. Thus, variations in $L_{24}$ are related to the shapes of the observed SED of the star-forming region, and hence should be sensitive to the radiation field strength of the ionizing stars and to the dust temperature.

On ${\sim}500$~pc scales in spatially-resolved observations of nearby galaxies, \citet{Calzetti+07} found a sublinear relationship between the SFR and $L_{24}$ (their Equation 9),

 \begin{equation}
\frac{\text{SFR}_{24}}{M_{\odot}~\text{yr}^{-1}} = \left(1.3\pm0.2\right)\times10^{-38} \left(\frac{L_{24}}{\text{erg~s}^{-1}}\right)^{\left(0.89\pm0.02\right)},
 \end{equation}

\noindent derived over a luminosity range of $3\times 10^{6} \leq (L_{24}/L_{\sun}) \leq 10^{11}$.

The non-linearity of this correlation is characteristic of this tracer \citep{AlonsoHerrero+06,PerezGonzalez+06,Calzetti+07,Relano+07,Murphy+11}. Proposed explanations for this trend invoke increasing dust opacity in star-forming regions and/or increasing mean dust temperature with increasing $L_{24}$.

Longward of $\sim$60 \micron, the emission from star-forming regions is dominated by thermal emission from larger dust grains at $\sim$20 K (typically referred to as the ``cool'' or ``cold'' dust component, and represented in our SED model by the cold blackbody component). Although heating from lower-mass (and potentially older) stars contributes more at these cold temperatures than at shorter wavelengths, the 70 \micron\xspace luminosity has been found to be an accurate monochromatic SFR indicator \citep{Dale+05}. The relationship between SFR and $L_{70}$ is linear; using a sample of over 500 star-forming regions, \citet{Li+10} found 

\begin{equation}\label{eq:SFRcalconst}
\frac{\text{SFR}_{70}}{M_{\odot}~\text{yr}^{-1}} = c_{70}\times 10^{-43} \left(\frac{L_{70}}{\text{erg s}^{-1}}\right),
\end{equation}

\noindent with calibration constant $c_{70}=0.94$, over 1 kpc scales for $10^{7} \la (L_{70}/L_{\sun}) \la 10^{10}$ (their Equation 4). The formal uncertainty reported for the calibration constant was ${\sim}2\%$. \citet{Lawton+10} found a similar relationship for dust-obscured \HII regions in the Large and Small Magellanic Clouds. 

In Figure~\ref{figure:Lbol_Lmon} we plot the ratio of these monochromatic luminosities to $L_{\rm TIR}$ against $L_{\rm TIR}$ for each of the MSFRs in our sample. On average, we find $L_8$ comprises 17$\pm$5\% of the bolometric luminosity of the regions in our sample, comparable to previous studies of the $L_8$-SFR relationship in other metal-rich, star forming galaxies \citep{Crocker+12,Treyer+10,Elbaz+11}. The 24~\micron\ luminosity accounts for 25$\pm$7\% of the bolometric luminosity, and 52$\pm$10\% of the bolometric luminosity is emitted at 70~\micron. The peak of the IR SEDs almost always falls near or within the PACS 70~\micron\ band (with the notable exception of Wd 1, indicated by an open circle in Figure~\ref{figure:Lbol_Lmon}). 

\begin{figure}
	\includegraphics[width=1\linewidth,clip,trim=1.5cm 12.5cm 2cm 3cm]{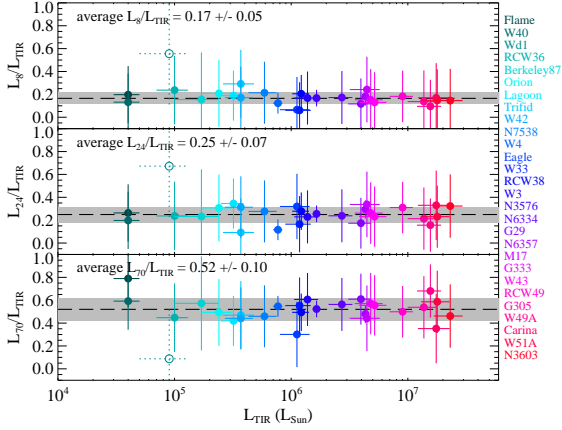}
\caption{The fraction of the monochromatic luminosities to total luminosity as a function of total luminosity derived from the global SED fitting. The gray shaded region in each panel shows the 1$\sigma$ uncertainty in the average fraction. Wd 1 (open circle and dotted-line error bars) has been excluded from the fit.}
\end{figure}\label{figure:Lbol_Lmon}

We do not observe a trend consistent with increasing dust temperature enhancing $L_{24}$ as $L_{\rm TIR}$ (or, implicitly, SFR) increases. Our SED modeling similarly reveals no correlation between $f_{\rm bol}$ (the fraction of luminosity in the warm dust component) and $L_{\rm TIR}$ (Table~\ref{table:SED_global}). One possible explanation could be that increased feedback in more luminous regions tends to expel or destroy dust from the immediate vicinity of the early-type stars, hence the remaining dust persists at larger distances from the ionizing cluster(s), maintaining cooler equilibrium temperatures than would be predicted by models that increase the radiation field without changing the spatial distribution of dust. While the upper end of our sampled luminosities overlaps with the luminosity range studied by \citet{Calzetti+07}, we cannot rule out the possibility the $L_{24}$--$L_{\rm TIR}$ relation steepens at higher luminosities, so there is no evident tension between our results and the extragalactic calibrations.

\subsection{Luminosity and the PAH Fraction}
The \citet{Draine+Li07} dust models may provide insight into the physical nature of the dust present in each star-forming region. Of particular interest is the PAH fraction required to best-fit the short-wavelength \spitzer fluxes in the MSFR SEDs. We calculate a luminosity-weighted average PAH fraction $\langle q_{\rm PAH} \rangle$ from our best-fit SED model (e.g., with the value of $q_{\rm PAH}$ of each dust component weighted by the luminosity produced by that component); uncertainties on $\langle q_{\rm PAH} \rangle$ are derived from the uncertainties in the luminosities of each dust component. 

We observe marginal evidence for a correlation between the average $q_{\rm PAH}$ and $L_{\rm TIR}$; brighter regions have systematically lower average $q_{\rm PAH}$ values. Figure~\ref{figure:av_qpah} shows the average $q_{\rm PAH}$ as a function of $L_{\rm TIR}$ for each of the three luminosity categories listed in Table~\ref{table:subgroup_mean}, as well as for the entire MSFR sample. Regions with fully-populated upper IMFs exhibit the lowest $\langle q_{\rm PAH} \rangle$ values, 2.2$\pm$0.5\%, compared to regions with some early O-stars ($\langle q_{\rm PAH} \rangle$ = 2.8$\pm$0.9\%) or a single O6-type or later ($\langle q_{\rm PAH} \rangle$ = 3.5$\pm$1.0\%). We caution that the inferred PAH fractions depend on the \citet{Draine+Li07} dust models, which assume a canonical extinction law ($R_V=3.1$ mag) that may underestimate the dust emissivity at longer wavelengths; the absolute $\langle q_{\rm PAH} \rangle$ values, therefore, may depend on choice of extinction law. The best-fit relationship between $\langle q_{\rm PAH} \rangle$ and $L_{\rm TIR}$ using only the average quantities for the three MSFR subgroups (the black dashed line in Figure~\ref{figure:av_qpah}) is given by

\begin{equation}
\langle q_{\rm PAH} \rangle = (6.8\pm1.4) - (0.7\pm0.2) \text{log }L_{\rm TIR}~(\%).
\end{equation}

\noindent There is no significant difference in the fit parameters when all MSFRs are used. This correlation may arise from weaker radiation fields produced by late O-type stars being inefficient at destroying a large percentage of PAH molecules. 

\begin{figure}
	\includegraphics[width=1\linewidth,clip,trim=3cm 13cm 2cm 3cm]{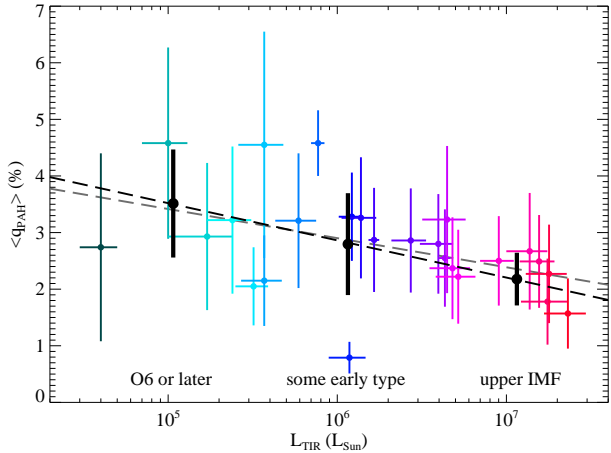}
	\caption{The average $q_{\rm PAH}$ of MSFRs in our sample sorted by luminosity subgroup; colors are the same as in previous figures. Regions with a fully-populated upper IMF have the lowest median $q_{\rm PAH}$ at (2.2$\pm$0.5)\%, compared with intermediate luminosity regions (e.g., those powered by some early O stars; 2.8$\pm$0.9\%) and fainter regions (e.g., those powered by a single O6 star or later; 3.5$\pm$1.0\%). The black dashed line shows the best-fit relationship between $\langle q_{\rm PAH} \rangle$ and $L_{\rm TIR}$ using only the average quantities for the three MSFR subgroups; the gray dashed line shows the relationship derived from all MSFRs. }
\end{figure}\label{figure:av_qpah}

\subsection{Calibrating SFR Tracers Against Ionizing Photon Rates of Cataloged Massive Stellar Populations}
The Galaxy offers the advantage that the ionizing stellar populations in MSFRs can be resolved at the level of individual OB stars, with unresolved binary/multiple systems revealed through spectroscopy. We can therefore calibrate SFRs directly against the Lyman continuum photon rate $N_C$ expected from the cataloged OB populations (Tables~\ref{table:stellar_models} and \ref{table:energy_budget}). Using the Starburst99 population synthesis code \citep{Leitherer+99} and assuming a Kroupa initial mass function \citep{K+W03} the relationship between $N_C$ and SFR is\footnote{Many studies still use an older calibration based on a Salpeter IMF from \citet{Kennicutt98}, which increases derived SFRs by a factor of 1.44  assuming a minimum stellar mass of 0.1~\Msun.}

\begin{equation}\label{eq:SFR(LyC)}
\frac{\text{SFR$_{\rm LyC}$}}{M_{\odot}~\text{yr}^{-1}} = (7.5 \times10^{-54})\frac{N_C} {\text{s}^{-1}}
\end{equation}

\noindent \citep{Kennicutt+09,Chomiuk+Povich11}. \citet{Chomiuk+Povich11} caution that this relationship likely underestimates the SFR for the very young MSFRs in our sample. SFR$_{\rm LyC}$ is the {\it continuous} SFR required to maintain ``steady state'' conditions within the MSFR in which the ionizing star birth rate equals the death rate. The lifetime of the ionizing stars is therefore assumed in this relationship, and this timescale is likely to be longer than the duration of star formation activity in many of the MSFRs in our sample.

Values for SFR$_{\rm LyC}$ in each MSFR computed using Equation~\ref{eq:SFR(LyC)} are reported in Table~\ref{table:model_luminosity_SFRs}; for the six regions where we have determined that an incomplete massive stellar census (e.g., where the principal ionizing star or stars have not yet been identified, marked with a $c$ in Table~\ref{table:energy_budget}) we report lower limits. We then assume SFR$_{\rm LyC}$ for the left-hand side of Equation~\ref{eq:SFRcalconst} and substitute $L_{\rm TIR}$, $L_{24}$, or $L_{70}$ to calculate the calibration constants $c_{\rm TIR}$, $c_{24}$, and $c_{70}$, respectively. We additionally derive a radio calibration constant $c_{\rm radio}$ from the \planck radio observations by substituting $N_C^{\prime}$ into the right-hand side of Equation~\ref{eq:SFR(LyC)}. Excluding the eight MSFRs with incompletely cataloged stellar populations or very uncertain distances (e.g., those marked with a $c$ or $d$ in Table~\ref{table:energy_budget}) and Wd~1, we derive $c_{\rm TIR}=1.2\pm0.7$, $c_{24}=3.7\pm2.4$, $c_{70}=2.7\pm1.4$, and $c_{\rm radio}=21.8\pm11.4$ averaged across 19 Galactic MSFRs. We also computed median values for each calibration constant and found that they did not differ significantly from the mean values.

Using these calibration constants, we computed SFR$_{\rm TIR}$ (analogous to the \citealp{Kennicutt+09}  ``total IR'' SFR tracer), SFR$_{24}$, SFR$_{70}$, and SFR$_{\rm radio}$ for all MSFRs in our sample. These various SFRs are presented in Table~\ref{table:model_luminosity_SFRs} along with the monochromatic 8, 24, and 70~\micron\ fluxes and luminosities. Comparisons between each of these IR SFR indicators and SFR$_{\rm LyC}$ are shown in Figure~\ref{figure:SFRs}. Not surprisingly, Wd~1 is a clear outlier, with a predicted SFR$_{\rm TIR}$ that is only $\sim$6\% SFR$_{\rm LyC}$. Figure~\ref{figure:SFRs} shows that all three IR indicators begin to overestimate the SFR for SFR$_{\rm LyC}\lesssim10^{-4}$ \Msun yr$^{-1}$, but the SFR derived from the radio observations does not.

\begin{table*}
\centering
\setlength\tabcolsep{2 pt}
\tiny 
\caption{Monochromatic Luminosities and SFRs}
\begin{tabular}{ccccccccccccccccc}
\hline \hline
		&  & &	& \multicolumn{3}{c}{8~\micron}			&& \multicolumn{4}{c}{24~\micron}						&& \multicolumn{4}{c}{70 \micron}										\\ \cline {5-7} \cline{9-12} \cline{14-17}
Name	& SFR$_{\rm LyC}$  & SFR$_{\rm TIR}$ & SFR$_{\rm radio}$	& $S_{\nu}$ 	& $L_{8}$		& $\frac{L_{8}}{L_{\rm TIR}}$	&& $S_{\nu}$ 	& $L_{24}$		& SFR$_{24}$ 	& $\frac{L_{24}}{L_{\rm TIR}}$	&& $S_{\nu}$	& $L_{70}$		& SFR$_{70}$ 		& $\frac{L_{70}}{L_{\rm TIR}}$	\\

		&($10^{-3} \frac{M_{\sun}}{\rm yr}$) &($10^{-3} \frac{M_{\sun}}{\rm yr}$) &($10^{-3} \frac{M_{\sun}}{\rm yr}$) & (Jy)		& ($10^{39}\frac{\rm erg}{\rm s}$)		& 	&& (Jy)		& ($10^{39}\frac{\rm erg}{\rm s}$)		& ($10^{-3} \frac{M_{\sun}}{\rm yr}$) &	&& (Jy)		& ($10^{39}\frac{\rm erg}{\rm s}$)		& ($10^{-3} \frac{M_{\sun}}{\rm yr}$)	&	\\
(1)		& (2)				& (3)	 		& (4)	 			& (5)				& (6)		& (7)				&& (8)				& (9)			& (10) & (11) && (12) & (13) 	& (14)	\\
\hline
Flame					& 0.011$\pm$0.005		& 0.018$\pm$0.004 		& 0.007$\pm$0.003	& 5,170   	& 0.03 	& 0.14	&& 22,840		& 0.04 	& 0.02   	& 0.65	&& 206,760	& 0.12	& 0.03  	& 0.08 \\
W40						& 0.007$\pm$0.005 		& 0.018$\pm$0.004 		& 0.009$\pm$0.005	& 1,740   	& 0.02 	& 0.12	&& 8,820		& 0.03 	& 0.01	& 0.19	&& 72,120		& 0.09	& 0.02  	& 0.54 \\ 
Wd~1					& 0.48$\pm$0.26  		& 0.04$\pm$0.02	 	& \nodata			& 280    	& 0.19 	& 0.53	&& 1,000		& 0.23 	& 0.08	& 0.63      && 360		& 0.03 	& 0.01 	& 0.08 \\
RCW~36					& 0.014$\pm$0.006 		& 0.05$\pm$0.01 		& 0.026$\pm$0.014	& 1,720   	& 0.09 	& 0.23	&& 4,930  		& 0.09 	& 0.03	& 0.22	&& 27,640		& 0.17	& 0.05 	& 0.42 \\
Berkeley~87				& 0.14$\pm$0.02		& 0.08$\pm$0.03	  	& 0.074$\pm$0.039 	& 750	& 0.10 	& 0.15	&& 3,310		& 0.15 	& 0.06	& 0.22	&& 24,040		& 0.37	& 0.10	& 0.65 \\
Orion					& 0.059$\pm$0.023 		& 0.11$\pm$0.03 		& 0.044$\pm$0.023	& 25,860	& 0.19 	& 0.21	&& 110,710	& 0.28	& 0.10	& 0.31	&& 520,060	& 0.45	& 0.12	& 0.49 \\
Lagoon					& 0.28$\pm$0.15		& 0.14$\pm$0.03 		& 0.24$\pm$0.13	& 3,800	& 0.23 	& 0.19	&& 20,380		& 0.42	& 0.15	& 0.34  	&& 73,410		& 0.51	& 0.14	& 0.42 \\
Trifid						& 0.044$\pm$0.005 		& 0.17$\pm$0.05		& 0.33$\pm$0.17	& 3,730 	& 0.41 	& 0.29 	&& 3,500		& 0.13	& 0.05	& 0.09 	&& 52,010		& 0.66	& 0.18 	& 0.46 \\
W42						& 0.13$\pm$0.04 		& 0.17$\pm$0.05	 	& 0.41$\pm$0.22	& 1,090	& 0.24 	& 0.16	&& 6,050		& 0.44	& 0.16	& 0.30  	&& 25,080		& 0.62	& 0.17	& 0.43 \\
NGC~7538				& 0.36$\pm$0.11		& 0.26$\pm$0.07	 	& 0.28$\pm$0.15	& 1,540    	& 0.48 	& 0.21	&& 5,920		& 0.62	& 0.23	& 0.27  	&& 28,640		& 1.03	& 0.27	& 0.46 \\
W4						& 0.86$\pm$0.28		& 0.34$\pm$0.03 		& \nodata			& 1,600	& 0.36 	& 0.12	&& 4,560		& 0.34 	& 0.13	& 0.12	&& 62,330		& 1.60	& 0.43	& 0.55 \\
Eagle					& 0.37$\pm$0.25		& 0.50$\pm$0.14	 	& 0.35$\pm$0.18	& 2,110	& 0.28 	& 0.06	&& 31,110  	& 1.36	& 0.50	& 0.32 	&& 85,840		& 1.28	& 0.43 	& 0.30 \\
W33						& 0.82$\pm$0.18		& 0.53$\pm$0.13 		& 1.00$\pm$0.52	& 1,050	& 0.27 	& 0.06	&& 8,630		& 0.74	& 0.27	& 0.16	&& 84,340		& 2.48	& 0.66	& 0.55 \\
RCW~38					& 0.42$\pm$0.19		& 0.54$\pm$0.09	 	& 0.57$\pm$0.30	& 7,380 	& 0.95 	& 0.21 	&& 30,030		& 1.29	& 0.47	& 0.44  	&& 154,970	& 2.29	& 0.61 	& 0.50 \\
W3						& 0.44$\pm$0.27		& 0.62$\pm$0.14 		& 0.63$\pm$0.33	& 4,160	& 0.88 	& 0.17 	&& 16,950		& 1.20	& 0.44	& 0.23  	&& 130,710	& 3.18	& 0.85	& 0.60 \\
NGC~3576\tablenotemark{a}	& ${>}0.20$			& 0.74$\pm$0.05 		& 0.87$\pm$0.46	& 3,060 	& 1.05 	& 0.17	&& 14,400		& 1.60	& 0.59	& 0.41	 && 87,340	& 3.28	& 0.87 	& 0.54 \\
NGC~6334\tablenotemark{a}	& $>$0.65				& 1.21$\pm$0.28	 	& 0.61$\pm$0.32	& 14,990	& 1.78 	& 0.17	&& 62,090		& 2.46	& 0.90	& 0.24	&& 427,740	& 5.82	& 1.55	& 0.56 \\
G29.96--0.02\tablenotemark{a}	& ${>}0.32$ 			& 1.77$\pm$0.39 		& 3.36$\pm$1.75	& 1,020   	& 1.76 	& 0.12     	&& 4,550		& 2.61	& 0.96	& 0.88	&& 46,620		& 9.17	& 2.44	& 0.62 \\
NGC~6357				& 2.47$\pm$0.26		& 1.93$\pm$0.15		& 1.40$\pm$0.73	& 21,050	& 2.98	& 0.18	&& 100,360	& 4.85	& 1.78	& 0.29	&& 460,600	& 7.89	& 2.10	& 0.45 \\
M17						& 1.99$\pm$0.21 		& 1.99$\pm$0.57 		& 1.62$\pm$0.84	& 27,690 	& 4.10 	& 0.24     	&& 115,980	& 5.72	& 2.10	& 0.34	&& 441,940	& 7.49	& 1.99 	& 0.44 \\
G333					& 1.03$\pm$0.26		& 2.14$\pm$0.58 		& 2.42$\pm$1.26	& 9,190 	& 2.78 	& 0.15 	&& 47,350		& 4.77	& 1.75	& 0.26 	&& 300,920	& 10.40	& 2.76	& 0.57 \\
W43						& 3.01$\pm$0.62		& 2.32$\pm$0.61	 	& 12.53$\pm$6.53	& 1,890  	& 2.56 	& 0.13	&& 10,100		& 4.55	& 1.67	& 0.23    	&& 70,790		& 10.95	& 2.91	& 0.56 \\
RCW~49					& 5.41$\pm$1.23 		& 4.03$\pm$0.91	 	& 5.74$\pm$2.99	& 7,220   	& 6.25 	& 0.18	&& 36,840		& 10.63	& 3.89	& 0.31	&& 172,710	& 17.10	& 4.54	& 0.50 \\
G305\tablenotemark{a}		& ${>}2.22$ 			& 6.13$\pm$1.67 		& 5.26$\pm$2.22	& 12,380	& 7.13 	& 0.14	&& 57,890		& 11.12	& 4.07	& 0.21	&& 425,800	& 28.07	& 7.46	& 0.54 \\
W49A\tablenotemark{a}		& ${>}4.62$ 			& 6.97$\pm$1.62	 	& 8.36$\pm$4.36	& 970	& 5.65 	& 0.10	&& 4,780		& 9.26	& 3.39	& 0.16	&& 60,760		& 40.38	& 10.73	& 0.68 \\
Carina					& 6.14$\pm$2.24 		& 7.81$\pm$2.37 		& 6.33$\pm$3.30	& 34,900	& 11.29 	& 0.17	&& 203,640	& 21.96	& 8.05	& 0.33	&& 632,930	& 23.42	& 6.22	& 0.35 \\
W51A\tablenotemark{a}		& ${>}3.43$ 			& 7.98$\pm$2.17	 	& 7.31$\pm$3.81	& 8,620  	& 10.03 	& 0.15	&& 40,450		& 15.68	& 5.74	& 0.23	&& 299,440	& 39.83	& 10.58	& 0.59 \\
NGC~3603				& 10.4$\pm$1.47 		& 10.31$\pm$2.86 		& 6.79$\pm$3.54	& 5,840  	& 12.79 	& 0.16	&& 38,870		& 28.39	& 10.40	& 0.34	&& 161,700	& 40.52	& 10.77 	& 0.49 \\
\hline \hline
\end{tabular}
\tablenotetext{a}{Regions for which massive stellar content remains incompletely cataloged; reported SFR$_{\rm LyC}$ is a lower limit.}
\end{table*}\label{table:model_luminosity_SFRs}

\begin{figure*}
\includegraphics[width=0.9\linewidth,clip,trim=1cm 11.8cm 2cm 6cm]{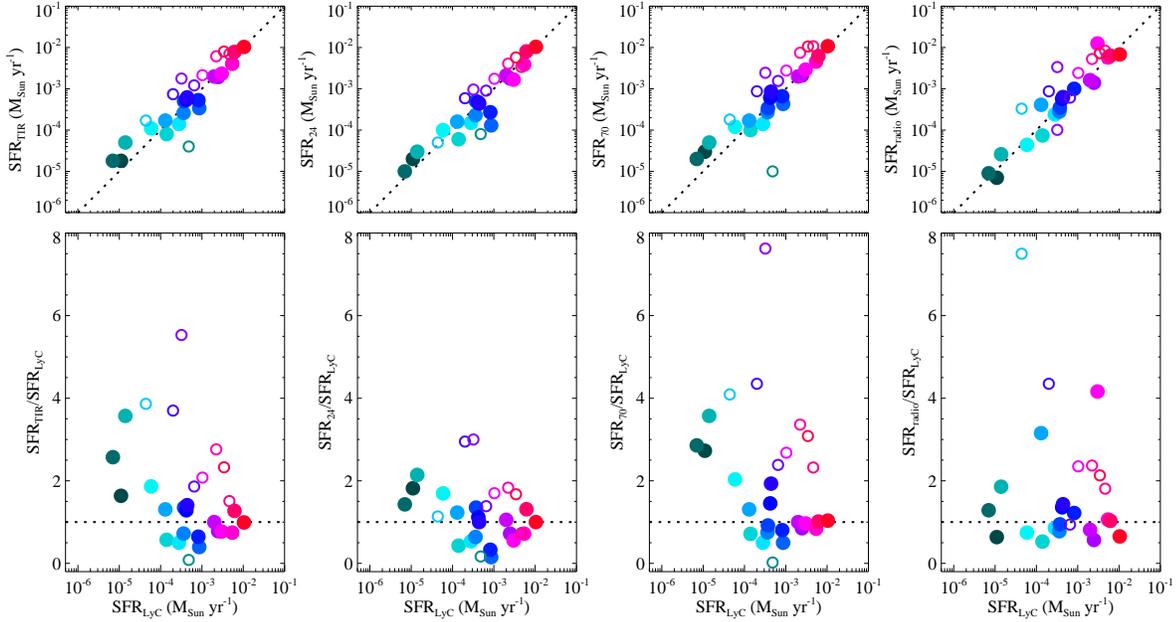}
\caption{The top panel shows comparisons (from left to right) of SFR$_{\rm TIR}$, SFR$_{24}$, SFR$_{70}$, and SFR$_{\rm radio}$ to SFR$_{\rm LyC}$. To bottom panel shows the ratio of the monochromatic luminosity-predicted SFR to SFR$_{\rm LyC}$, as a function of SFR$_{\rm LyC}$. The dotted lines show one-to-one correlations, not fits to the data. Regions with incompletely cataloged massive stellar populations or negligible obscuration are marked with open circles (as in Figure~\ref{figure:SED_vs_OB}) and excluded from the calculations of the calibration constants (see text). Colors are as in Figure~\ref{figure:Lbol_Lmon}.}
\end{figure*}\label{figure:SFRs}

\begin{figure}
\includegraphics[width=1\linewidth,clip,trim=1.5cm 11cm 7cm 2cm]{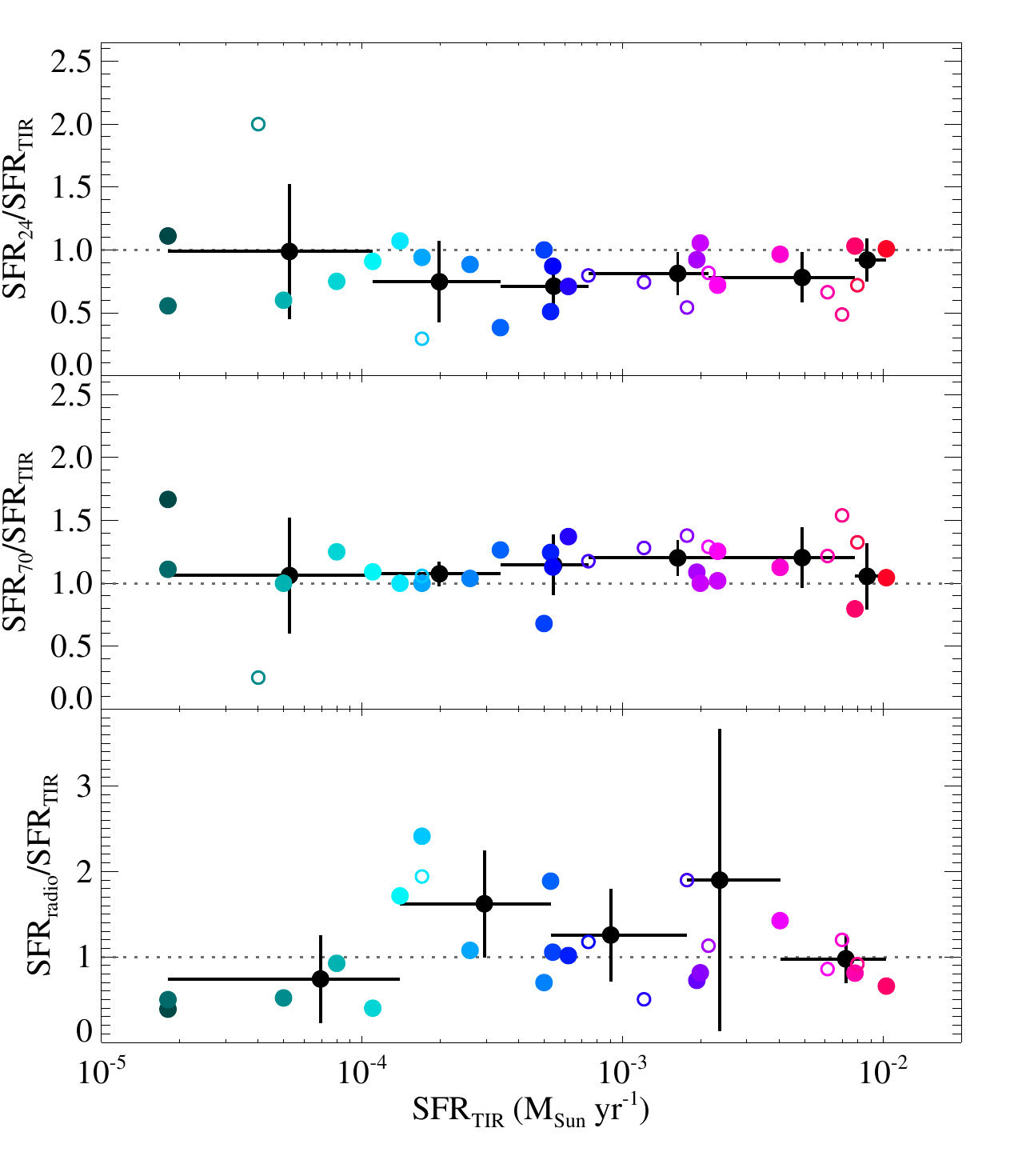}
	\caption{Monochromatic SFR indicators for 24~\micron\ ({\em top}), 70~\micron\ ({\em middle}), and the radio continuum ({\em bottom}) normalized to SFR$_{\rm TIR}$ measured from $L_{\rm TIR}$. Black points show average values. Symbols and colors are the same as in Figure~\ref{figure:SFRs}. The gray dotted line shows a perfect agreement between the monochromatic SFR and SFR$_{\rm TIR}$. }
\end{figure}\label{figure:SFRdiff_vs_Lbol}

In Figure~\ref{figure:SFRdiff_vs_Lbol} we compare each of the monochromatic SFR indicators directly with SFR$_{\rm TIR}$, which has the advantage of being independent of distance or knowledge of ionizing stellar content. Excluded regions in our sample include both very young, highly embedded \HII regions for which the massive stellar content is difficult to spectroscopically catalog, and older, unobscured regions that have largely dispersed their dust. Highly-embedded \HII regions are strong far-IR emitters due to their large fraction of cold dust that remains shielded from the nascent ionizing clusters, while unobscured regions have little to no warm dust remaining, although they may externally illuminate nearby cold, molecular cloud fragments.

\subsection{Comparison to Extragalactic Tracers of Dust-Obscured SFRs}
Using our value of $c_{24}$ yields SFR$_{24}$ estimates (as in Table~\ref{table:model_luminosity_SFRs} and Figure~\ref{figure:SFRs}) that are comparable to those derived from the sub-linear calibration from \cite[][their equation~9]{Calzetti+07}. Although there is considerable variation among individual MSFRs, the Galactic and resolved extragalactic tracers can be regarded as generally consistent. 

\citet{Calzetti+10} warn that SFR calibrations based on $L_{24}$ alone break down when applied to entire galaxies with $L_{24}<5\times 10^{43}$~erg~s$^{-1}$, a luminosity range that begins one order of magnitude above the brightest MSFRs in our sample. Our linear calibration constant $c_{24}$ is ${\sim}40$\% higher than the analogous extragalactic calibrations \citep[see references in][]{Calzetti+10}, albeit with large uncertainties. This is sensible because the average $L_{24}/L_{\rm TIR}\sim0.25$ (Figure~\ref{figure:Lbol_Lmon}) agrees with the upper end of the range of $L_{24}/L_{\rm TIR}$ measured for whole galaxies by \citet{Calzetti+10}. The mid-IR SEDs of our MSFRs thus resemble those of dusty, starburst galaxies, where the 24~\micron\ emission is completely dominated by heating from young stars. The average star-forming galaxy has a ratio that is lower by a factor of ${\sim}40$\%. Astrophysically, this discrepancy is explained by the increasing contribution from dust heated by older stellar populations to $L_{24}$ as SFR decreases, when the IR luminosity is measured using whole-galaxy apertures. 

The effect of increasing the aperture size over which the IR SEDs are measured becomes far more pronounced as the IR wavelength considered increases, because older stellar populations heat dust to lower temperatures than young stellar populations. Our 70 \micron\xspace calibration constant ($c_{70}$) is higher than the value measured by \citet{Calzetti+10} for whole galaxies by a factor of $\sim$5.5, a much greater discrepancy than we find for $c_{24}$. Meanwhile, our result that on average $L_{70}/L_{\rm TIR}=55\%$ is in excellent agreement with their measurements of star-forming galaxies.\footnote{Excluding luminous IR galaxies (LIRGs), which have systematically cooler dust temperatures and hence elevated $L_{70}$ and $L_{160}$ at the expense of $L_{24}$ compared to normal galaxies or most of our MSFRs.}

\citet{Li+13} explored the relationship between $c_{70}$ and physical aperture size and found that an adjustment to $c_{70}$ is required to ensure consistency of SFR$_{70}$ on different spatial scales. In Figure~\ref{figure:SFR_size}, we reproduce their Figure~9, including our significantly smaller, individual MSFRs (which have ${\sim}15$ pc typical physical aperture size). Our value of $c_{70}$, calibrated to SFR$_{\rm LyC}$ from the cataloged massive stellar populations within our MSFRs, is close to the value predicted by extrapolating the trend from the calibrations of \citet{Calzetti+10}, \citet{Li+10}, and \citet{Li+13} to smaller spatial scales.

Dust absorption produces a 50\% reduction in the ionizing photon rates compared to the production rates of Lyman continuum photons in the most luminous Galactic \HII regions. Such IR-luminous regions dominate the extragalactic calibrations of obscured star formation. This effect is potentially pernicious because it is very difficult to separate dust-absorbed from obscured star formation without knowing the ionizing stellar population. Indeed, the terms ``absorbed'' and ``obscured'' in this context are routinely used interchangeably in the extragalactic literature. Here we make the same distinction as did \citet{McKee+Williams97}. Lyman continuum photons ``absorbed'' by dust grains within \HII regions do not contribute to the ionization of the gas, hence this absorption reduces both the radio free-free and H$\alpha$ luminosity by the same factor. By contrast, ``obscuration'' refers to the effects of both absorption and scattering of photons below the Lyman limit by dust within or along the line-of-sight to an \HII region, which reduces the observed H$\alpha$ but does not affect the radio continuum. The empirical attenuation corrections typically applied to recombination-line studies of external galaxies account for the obscuration affecting visible/near-IR photons \citep[see the definition of ``attenuation'' provided by][]{Kennicutt+09}, but if they fail to completely correct for Lyman continuum photons absorbed by dust within the \HII regions the resulting SFR calibrations will underestimate the true obscured SFRs.

This dust absorption systematic is most important for smaller, IR-bright star-forming regions (such as our sample) and likely becomes insignificant for most galaxy-wide studies.\footnote{In the case of starburst galaxies or (U)LIRGs the potential impact is unclear. In the latter case it is generally preferable to assume 100\% obscured star-formation and hence base SFRs directly on $L_{\rm TIR}$ from population synthesis models, avoiding intermediate calibrations based on H$\alpha$.} For larger-scale galactic sub-regions and more evolved stellar populations, the contributions of unobscured \HII regions and Lyman continuum photons that have escaped from obscured \HII regions come to dominate the measured SFRs.

\begin{figure}
\includegraphics[width=0.9\linewidth,clip,trim=2.5cm 13.0cm 2.5cm 3cm]{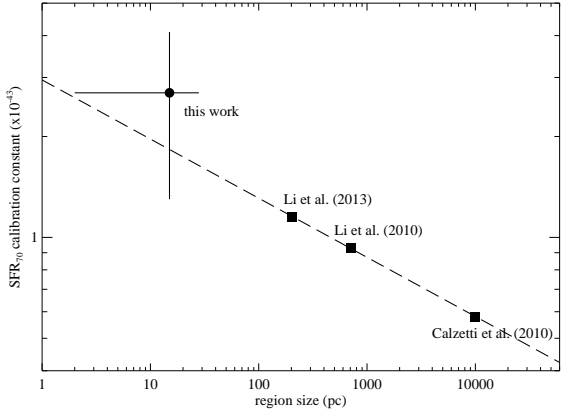}
\caption{The SFR$_{70}$ calibration constant $c_{70}$ as a function of physical size. The solid squares show measurements of the calibration constant from external galaxies and large, extragalactic star-forming regions. The solid squares show the predicted calibration constants for continuous star forming populations \citep{Li+10,Li+13}. The estimated average SFR$_{70}$ calibration constant for our regions (with an average size of $\sim$15 pc) is shown by the solid circle.}
\end{figure}\label{figure:SFR_size}

\section{Summary \& Conclusions}
We have presented a comprehensive study of the globally integrated IR-radio emission of 28 Galactic MSFRs. We fit the 3.6~\micron--10~mm spectral SEDs constructed from aperture photometry on \spitzer, \msx, \iras, and \herschel images plus \planck extended sources with models consisting of one or two \citet{Draine+Li07} dust components, one cold blackbody component, and a power-law continuum. From our SED model fits and adopted distances to each MSFR we derive the total IR luminosity $L_{\rm TIR}$ and ionizing photon rate $N_C^{\prime}$ required to maintain each radio \HII region. Our sampled MSFRs span three orders of magnitude in luminosity, ranging over $10^4~L_{\sun}\la L_{\rm TIR}\la 2\times 10^7~L_{\sun}$ in dust-reprocessed total infrared luminosity and $3\times 10^{47}~\rm{s}^{-1} \la N_C^{\prime}\la 5\times 10^{50}~\rm{s}^{-1}$ in ionizing photon rate required to maintain the observed radio \HII regions.

Modeling the IR+radio SED simultaneously offers considerable advantages over studying either the IR or radio emission alone. The free-free continuum is negligible at shorter mid-IR wavelengths. Although the true ionized gas spectrum departs from a pure power law at short wavelengths, this is unlikely to significant impact our results (e.g., the power law continuum contributes at most a few percent of the total flux at 3.6~\micron\; see Figure~\ref{figure:SED_examples}). However, the incorporation of the Br$\alpha$ emission-line flux at 4.5 \micron, constrained by the radio spectrum, has enabled an improved (although still not perfect) fit to the Spitzer [4.5] mid-IR band compared to models based on dust emission alone \citep{Stephens+14}.

We searched the literature to compile lists of the known massive stellar population in each MSFR to estimate the stellar bolometric luminosity ($L_{\star}$) and emitted Lyman continuum photon rate ($N_C$). We balance the ``energy budget'' in each MSFR in terms of the ratios $L_{\rm TIR}/L_{\star}$ and $N_C^{\prime}/N_C$. In 10/28 MSFRs the emergent dust-processed luminosity in the SED exceeds the bolometric luminosity input by the cataloged stars, leading us to conclude that the census of the massive stellar population is incomplete. 

Our main results are summarized as follows:
\begin{enumerate}
	\item A significant fraction ($f_{C,\rm{abs}}$) of Lyman continuum photons emitted by massive stars is absorbed by dust before contributing to the ionization of \HII regions. This absorption increases with bolometric luminosity; $f_{C,\rm{abs}}=34\%$ averaged across the 14 MSFRs for which it could be calculated and increases to $51\%$ averaged over the 4 most luminous MSFRs in our sample, which average $L_{\rm TIR}=10^7$~\Lsun\ each (Table~\ref{table:subgroup_mean}). This empirical result agrees well with the theoretical predictions of \citet{McKee+Williams97}, who calculated that the dust opacity in giant \HII regions increases with ionizing photon luminosity, reaching an average $\langle f_{C,\rm{abs}}\rangle=0.46$ for Galactic radio \HII regions with $N_{C}^{\prime}>1.5\times 10^{50}$~s$^{-1}$.
	\item We calculate an average PAH fraction from our dust models and find that it is systematically higher in regions that are powered by a single O6-type star or later, with lower PAH fractions observed in regions will fully populated upper IMFs. This radiation fields in these lower-luminosity \HII regions are relatively weak, and inefficient at destroying PAH molecules.
	\item We calibrate SFRs based on the monochromatic luminosities $L_{24}$ and $L_{70}$ from our SED models against the Lyman continuum photon rates of the cataloged massive stars in each region. We find that standard extragalactic calibrations of monochromatic SFRs based on population synthesis models are generally consistent with our values, although there is large variation among the 28 individual MSFRs in our sample. Our results are consistent with the \citet{Calzetti+07} 24~\micron\ calibration, and an extrapolation of the \citet{Li+13} 70~\micron\ SFR to the smaller size scales of the Galactic regions is broadly consistent our SFRs.
	\item The preferred monochromatic luminosity for measuring obscured SFRs is $L_{70}$, which captures, on average, $52\%$ of $L_{\rm TIR}$ in our regions, a result that is in excellent agreement with comparable extragalactic studies \citep[e.g.,][]{Calzetti+10}.
\end{enumerate}

SFR studies using Galactic radio \HII regions have long included corrections for Lyman continuum photons lost to dust absorption \citep{SBM78,Inoue+01,Murray+Rahman10,Lee+12}. Such corrections are typically not incorporated into extragalactic calibrations, as most H$\alpha$ emission observed on galaxy-wide scales originates from regions with negligible dust. Other SFR tracers, such as integrated UV emission, that do not rely on Lyman continuum photon rates avoid this issue entirely. However, dust absorption becomes significant for spatially-resolved studies of obscured star formation. While current, widely-used calibrations of obscured SFRs account for Lyman continuum photons that escape into the diffuse ISM by using a combination of recombination lines and IR broadband emission \citep[e.g][]{Calzetti+07,Kennicutt+09}, these calibrations could be biased toward low SFRs at the smallest spatial scales and/or highest dust obscurations. 

The IR and radio SFR calibrations presented in this work are preferred for application to Milky Way studies over the analogous extragalactic calibrations, given the orders-of-magnitude differences in timescales, physical sizes, and luminosities separating whole galaxies from individual Galactic star-forming regions. Systematics due to heating of dust by older stellar populations are most pronounced for the total IR or $L_{70}$ SFR tracers. The \citet{Calzetti+07} $L_{24}$ calibration, which was based on individual IR-bright knots in nearby galaxies, is most consistent with our $L_{24}$ calibration, and both appear to give reasonable results when applied to Galactic regions with sufficiently high IR luminosities \citep[see][whose sample of Galactic star-forming regions overlaps with the low-luminosity end of our sample]{VEH16}. Even within the Milky Way, our calibrations would likely break down when applied to star-forming clouds that are either too low-mass or too early in their evolution to have formed massive stars ionizing radio \HII regions \citep{V+E13,Povich+16}.

Thermal radio continuum has been relied upon over the past four decades to measure the total ionizing photon rate of the Milky Way and hence the Galactic SFR \citep[][and references therein]{Chomiuk+Povich11,Kennicutt+12}. We have demonstrated, across nearly three orders of magnitude in luminosity, that the average ionizing photon rate required to maintain the ionization of radio \HII regions is only one-third of the Lyman continuum photon rate emitted by the massive stellar content of these regions. It is therefore important to account for both the escape of Lyman continuum photons from compact radio \HII regions and their absorption by dust within the \HII regions to derive accurate SFRs or simply to infer the ionizing stellar populations within radio \HII regions. For example, the work by \citet{Murray+Rahman10} to measure the Galactic SFR using free-free emission measured by the {\it Wilkinson Microwave Anisotropy Probe ({\it WMAP})} used the calculations of \citet{McKee+Williams97} to correct their Lyman continuum photon rates for dust absorption. This absorption correction seems appropriate, and because {\it WMAP} measured free-free emission across Galactic scales the escape of ionizing photons would be negligible. For smaller, less-luminous \HII regions such as those studied by \citet{VEH16}, neither absorption nor escape of Lyman continuum photons can be safely neglected.

Our comparisons of Galactic and extragalactic SFR calibrations required that we assume a standard conversion of Lyman continuum photon rates to absolute SFR based on population synthesis models. While it is encouraging to see convergence between the IR nebular SFR tracers in the Galactic and extragalactic cases, \citet{Chomiuk+Povich11} warned that the assumed star formation timescale in this conversion is likely to be too long by a factor of a few compared to the actual duration of star formation in individual, IR-bright regions, hence such calibrations likely underestimate the true absolute SFRs. In future work we will measure SFRs directly from the spatially resolved low- and intermediate-mass stellar populations to provide a more direct, empirical SFR calibration for the IR and radio nebular tracers.

\acknowledgements
The authors would like to thank the referee Neal Evans for comments and suggestions that significantly improved this manuscript. The authors thank Karin Sandstrom for helpful discussions about SED modeling and Roberta Paladini for her assistance in obtaining the \herschel data. This work was supported by the National Science Foundation under award CAREER-1454333 (PI: M. S. Povich). This work is based in part on observations made with the {\it Spitzer Space Telescope}, which is operated by the Jet Propulsion Laboratory, California Institute of Technology under a contract with NASA. This work is based in part on observations made with {\it Herschel}, an ESA space observatory with science instruments provided by European-led Principal Investigator consortia and with important participation from NASA. This research made use of data products from the {\it Midcourse Space Experiment}, with data processing funded by the Ballistic Missile Defense Organization with additional support from NASA Office of Space Science. This research used data products produced by the \planck Collaboration. This research has made extensive use of the NASA/IPAC Infrared Science Archive, which is operated by the Jet Propulsion Laboratory, California Institute of Technology, under contract with the National Aeronautics and Space Administration. This research has made use of the SVO Filter Profile Service supported from the Spanish MINECO through grant AyA2014-55216. \software{mpfitfun \citep{Markwardt09}}

\appendix
\section{The Massive Star-Forming Region Sample}\label{appendix:region_discussion}
Here we discuss details of the individual MSFRs, in order of increasing $L_{\rm TIR}$.

\begin{itemize}	
	\item {\bf The Flame Nebula}: The Flame Nebula (NGC~2024) includes both a dense cluster and the Horsehead Nebula, located in the Orion~B molecular cloud near the Orion Belt star $\eta$~Ori. Although it is often assumed to lie at the same distance as the Orion Nebula \citep{Menten+07}, our distance estimate from the Gaia DR2 database \citep{Gaia2018} is somewhat closer (0.33$\pm$0.01 kpc). A dark lane of cloud material obscures our view of the central cluster at optical wavelengths, with an average cluster extinction of $A_V\sim10$~mag \citep{Skinner+03}. The most luminous star in the cluster, NGC~2024~IRS~2b, is an extended, embedded infrared source, and its spectral type is uncertain. Our measured ionizing photon rate $N_C^{\prime}\sim3\times10^{47}$ s$^{-1}$ is broadly consistent with previous estimates of (4.6$\pm$0.7)$\times10^{47}$ s$^{-1}$ derived from 1667 MHz measurements by \citet[][corrected for the different distance assumed in that work]{Barnes+89}. The spectral type of the principal ionizing star, IRS~2b, is only constrained to be late-O or early-B type \citep{Bik+03}. We therefore calculate $N_C$ for the Flame Nebula assuming an O8~V, O9~V, or B0~V spectral type for the star, obtaining $3.1\times10^{48}$, $1.5\times10^{48}$, and $8.2\times10^{47}$ s$^{-1}$, respectively. The corresponding fraction of emitted stellar light that is reprocessed by dust in each case is 43\%, 68\%, and 108\%, respectively. Given these estimates, we favor the late-O-type interpretation of IRS~2b, and assume a spectral type of O9~V for this star. 

	\item {\bf W40}: The W40 \HII region is associated with the molecular cloud G28.74+3.52 in the Aquila Rift \citep{Westerhout58,Zeilik+Lada78}. The \HII region is powered by a a late-O star that is likely responsible for blowing the $\sim$4 pc bubble observed in wide-field infrared images \citep{Shuping+12}. The known stellar content of W40 was taken from \citet{Shuping+12}, and produces an ionizing photon rate $N_C=9.5\times10^{47}$ s$^{-1}$. These estimates are consistent with a previous upper limit of 1.5$\times10^{48}$ s$^{-1}$ predicted by \citet{Goss+Shaver70} and \citet{Smith+85} based on radio observations. Our distance estimate using Gaia parallaxes is 0.49$\pm$0.05 kpc, in excellent agreement with the distance estimated by \citet{Shuping+12}. More than half of the emitted stellar luminosity is reprocessed by dust.

	\item {\bf Westerlund 1}: Wd~1 has been intensively studied for its large, diverse population of evolved massive stars \citep{Crowther+06,Ritchie+09,Clark+11}, and {\em Hubble Space Telescope} observations have recently resolved its extremely rich, low-mass pre-main sequence population \citep{Andersen+17}. We adopt the 3.9~kpc distance derived by \citet{Koumpia+Bonanos12} from massive eclipsing binary systems. Although Wd~1 is indisputably one of the most massive clusters in the Galaxy, it has already lost its most massive initial members to supernovae, and much of the remaining massive stellar population consists of yellow hypergiants, WR stars, and red supergiants. The present-day bolometric luminosity and particularly Lyman continuum photon rates of Wd~1 are consequently much less than those of other comparably massive, but younger MSFRs included in our sample. We use the cataloged massive stellar population from \citet{Clark+05} to derive estimates of $L_{\star}$ and $N_C$. Virtually all of the original nebula has been dispersed by massive stellar feedback, likely including supernovae, and there is circumstantial evidence from \spitzer images that feedback from Wd~1 has disrupted a Galactic-scale massive molecular cloud filament that extends for several hundred pc along the Scutum-Centaurus spiral arm at the same heliocentric distance \citep{Goodman+14}. The current IR nebulosity is dominated by optically-thick wind emission from the evolved massive stars \citep{Dougherty+10}, making Wd~1 unique among our sample, and its SED models are commensurately unusual, peaking near 24~\micron\ with a much suppressed cool dust continuum at longer wavelengths. The radio continuum in \planck is unmeasurably low. We find that Wd~1 is an extreme outlier in our analysis of monochromatic IR SFRs, and we caution that individual MSFRs occupying the same transient phase of massive cluster evolution could be misinterpreted in spatially-resolved IR or radio studies of obscured star formation in external galaxies.

	\item {\bf RCW36}: RCW~36 is an hourglass-shaped \HII region. We measure a distance to RCW~36 from Gaia parallaxes of 1.09 kpc, marginally more distant than the ${\sim}700$~pc found by \citet{Baba+04}. It is powered primarily by two late-O type stars in an embedded cluster \citep{Ellerbroek+13,Minier+13}. Interstellar material appears to have been cleared away by already formed massive stars on either side of the cluster, and H$\alpha$ observations have revealed a dense, ionized shell ${\sim}1.8$~pc in extent \citep{Rodgers+60}. A complex, ring-like structure observed by \citet{Baba+04} and \citet{Minier+13} has been interpreted as the initial ionization of the surrounding molecular cloud by the central cluster and young stellar objects \citep{Ellerbroek+13}, leading to the expansion of the \HII region and the emergence of bright rims, pillars, and triggered star formation. We estimate an ionizing photon rate $N_C^{\prime}\sim1\times10^{48}$~s$^{-1}$ from the \planck observations. This is roughly consistent with the value derived from 4.8 GHz observations by \citet{Whiteoak+Gardner77}. Nearly all of the stellar luminosity is reprocessed by local dust, albeit with large uncertainties on $N_C^{\prime}/N_C$ and $L_{\rm TIR}/L_{\star}$.

	\item {\bf Berkeley~87}: Berkeley~87 is a sparse grouping of early-type stars. We estimate a distance to Berkeley~87 of 1.74 kpc using Gaia parallaxes, somewhat farther away than the 950$\pm$26 pc estimated by \citet{Turner+82} using stars the western edge of the Cygnus~X complex \citep{Cong77}. It appears to be associated with a group of compact \HII regions, and is notable for containing the a rare type of Wolf-Rayet star \citep[WR~142][]{Stephenson66,Sanduleak71,vanderHucht+81}. In addition, \citet{Polcaro+06} estimate Berkeley~87 hosts $\sim$8 late-O type stars. The current evolved massive stars in this region may not have been the most massive stars to form with this cluster. 

	\item {\bf The Orion Nebula}: The Orion Nebula is the best-studied MSFR in the Galaxy, due to its close proximity to Earth \citep[we estimate 0.41 kpc from Gaia parallaxes, consistent with the 414$\pm$17 pc estimate from][]{Menten+07}. The Orion Nebula Cluster (ONC), and especially the Trapezium group, is the dominant source of ionizing radiation that has produced the blister cavity on the surface of the parent molecular cloud \citep{ODell01a,ODell01b}. Due to their young age \citep[$\sim$1.1 Myr; ][]{Getman+14}, none of the stars in the ONC have evolved into the Wolf-Rayet phase or undergone supernovae. The \planck radio observations support this picture; we measure a radio spectral index consistent with free-free emission, and the measured ionizing photon rate $N_C^{\prime}\sim2\times10^{48}$ s$^{-1}$ roughly agrees with previous estimates of 5--8$\times10^{48}$ s$^{-1}$ \citep{Condon92}, derived from 1--25 GHz measurements of the most luminous portions of the OB1d region \citep{Felli+93,vanderWerf+Goss89}. Using the stellar populations cataloged by \citet{Voss+10}, we estimate an ionizing photon rate from the ONC of $N_C=7.7\times10^{48}$ s$^{-1}$. 

	\item {\bf The Lagoon Nebula}: The Lagoon Nebula (M8) is illuminated by the massive star cluster NGC~6530, which contains ${\ga}1100$ members \citep{Damiani+04,Damiani+06,Prisinzano+05,Prisinzano+07}. The nebula lies in the Sagittarius-Carina arm, close to the Trifid Nebula and the supernova remnant W28. The predicted Lyman continuum photon rate from the stellar content \citep{Skiff09} is $N_C=3.7\times10^{49}$ s$^{-1}$, which is roughly three times higher than our observed $N_C^{\prime}= (1.1\pm0.2)\times10^{49}$~s$^{-1}$. Meanwhile, Lagoon is moderately obscured, with ${\sim}30\%$ of the stellar luminosity reprocessed by dust, implying $f_{C, \rm abs}\sim0$. Three additional early-B-type candidates with were found by \citet{Povich+17}; we include them in our analysis for consistency with other MSFRs, but this increases the predicted ionizing photon rate and stellar luminosity by just ${\sim}2\%$ and ${\sim}10\%$, respectively. 

	\item {\bf The Trifid Nebula}: The Trifid Nebula (M20) is an active star-forming region near the Galactic center, and harbors one of the youngest known star clusters \citep{Cernicharo+98,Rho+08}. The ionization of the \HII region is dominated by HD~164492, a multiple star system composed of O7~V, B6~V, A2~Ia, and possibly a Be star \citep{Rho+04,Rho+06}. A blue reflection nebula can be seen at optical wavelengths $\sim$10$^{\prime}$ north of HD~164492. The distance to the Trifid Nebula is still debated, with estimates ranging from 1.67 kpc \citep{Lynds+85} to 2.8 kpc \citep{Kohoutek+99}, and even shorter distances (816 pc and 1093 pc) were estimated by \citet{Kharchenko+05}. Using Gaia parallaxes, we estimate a distance to the Trifid Nebula of 1.57$\pm$0.21 kpc, and find a bolometric luminosity of (3.7$\pm$1.1)$\times10^5$ \Lsun and an ionizing photon rate of (1.5$\pm$0.4)$\times10^{49}$ s$^{-1}$ from the \planck observations. These estimates exceed the predicted $N_C$ and $L_{\star}$ from the observed stellar population by more than a factor of two. Our estimated distance to the Trifid Nebula may be an overestimate, especially since it is derived from measurements of a single star. Although \citet{Povich+17} identify three new OB candidates in the vicinity of the Trifid Nebula, these stars are likely too far from the nebula to contribute significantly to the observed luminosity or ionizing photon rate from the region.
	
	\item {\bf W42}: W42 is an obscured giant \HII region (also known as G25.38--0.18) with a projected location toward the near end of the Galactic central bar. We adopt a distance to the W42 complex of 2.2 kpc, which was inferred by \citet{Blum+00} assuming that the principal ionizing star is a zero-age main sequence O5--6 V. This is considerably closer than the kinematic distance estimate of ${\sim}3.7$~kpc \citep{Anderson+Bania09}, but the larger distance would make our measured luminosity inconsistent with the cataloged massive stellar population. \spitzer imaging has revealed a bipolar nebula several parsecs wide. The evolved, supermassive cluster RSGC1 \citep{Figer+06} is located near the end of the bar, at a significantly larger distance than W42 but with a projected separation only 6.8\arcmin\ to the southeast (RSGC1 appears as the cluster of bright, blue stars to the lower-right of W42 in Figure~\ref{radial_fig:finding_charts}). We therefore only measure the radio flux towards W42 at 100 GHz and 70 GHz, and assume a spectral index of -0.1. Only the normalization is used to estimate $N_C^{\prime}$. Assuming the W42 \HII region is ionized principally by a O5~V star with a contribution from a B0~V star \citep{Blum+00}, we estimate a Lyman continuum flux $N_C=1.7\times10^{49}$ s$^{-1}$. This means ${\sim}100\%$ of the emitted stellar luminosity is reprocessed by dust.

	\item {\bf NGC~7538}: NGC~7538 in the Perseus spiral arm is an \HII region that has been well mapped across multiple wavelength regimes, from the optical to the submillimeter \citep{Momose+01,Ungerechts+00,Yao+00,Campbell+Persson88,Balog+04}. VLBA observations of 12 GHz methanol masers were used to determine a distance to NGC~7538 of 2.65$^{+0.12}_{-0.11}$ kpc \citep{Moscadelli+09}. The region can be divided into multiple prominent centers for star formation activity, aligned from northwest to southeast. It hosts massive stars and young stellar objects in a variety of early evolutionary stages, including at least eleven high-luminosity infrared sources \citep[NGC~7538 IRS~1--11][]{Kameya+90}. The northwest region contains IRS 4--6 and corresponds to the optical \HII region; IRS~6, an O6--7 star, is likely the principal source of ionizing photons for this region \citep{WynnWilliams+74,Moreno+Chavarria86,Ojha+04}. The most extensive, current star-forming activity is likely occurring in the central region, which surrounds IRS~1-3 \citep{Ojha+04}. The principal star in this region is IRS~2, which is thought to be either an O5~V \citep{Ojha+04,Luisi+16} or O3~V \citep{Puga+10}; we adopt the O5~V type. From the known stellar population of NGC~7538, we estimate $N_C=4.7\times 10^{49}$~s$^{-1}$.  
	
	\item {\bf W4}: The W4 \HII region is an example of a candidate Galactic chimney powered by the massive star cluster OCl~352 \citep{Normandeau+96}. The chimney is thought to have formed from a superbubble powered by the central star cluster \citep{Basu+99}. \citet{Terebey+03} found a swept-up, inhomogeneous and partially ionized shell of gas and dust surrounding OCl~352; an ionized halo provides direct evidence of significant Lyman continuum leakage.	Before extracting a surface brightness profile and performing aperture photometry on this region, we first masked out the circular aperture region for W3.

	\item {\bf The Eagle Nebula}: The Eagle Nebula (M16) is located in the Sagittarius-Carina. We estimate a distance to the Eagle Nebula of 1.71$\pm$0.18 kpc from Gaia parallaxes, consistent with the distance of 1.75 kpc found by \citet{Guarcello+07}. The \HII region is powered by the NGC~6611 cluster, which contains numerous OB stars and is largely concentrated within the central cavity of the molecular cloud \citep{Dufton+06}. We cannot rule out the possibility that supernovae have already occurred in the region \citep{Flagey+11}. We estimate an ionizing photon rate $N_C^{\prime}$ of (1.6$\pm$0.4)$\times10^{49}$ s$^{-1}$ from the \planck observations, in agreement with previous 15.4 GHz observations by \citet{Felli+Churchwell70}. Using only the previously cataloged OB stars associated with NGC~6611 yields $L_{\rm TIR}/L_{\star}\approx 0.56$, which seems reasonable given the morphology of the nebula, with its large, open cavity and famous, visibly-revealed pillars. \citet{Povich+17} identify seven X-ray emitting candidate OB stars associated with the Eagle Nebula. Some of these candidates are likely unassociated contaminants; specifically, Eagle-1, 3, and 7 from \citet{Povich+17} are located at large off-axis angles and corresponding large PSFs in the {\em Chandra} imaging, making mismatches with field giants likely. We therefore only include four of the seven \citep{Povich+17} OB candidates in our analysis. This increases the ionizing photon rate to $N_C= 4.3\times10^{49}$~s$^{-1}$. We therefore assume that these stellar candidates do indeed contribute to the ionization of the Eagle Nebula, and include them in our analysis. 
			
	\item {\bf W33}: W33 is located in the inner Galactic Plane at a distance of ${\sim}2.4$~kpc \citep{Immer+13}.  \citet{Haschick+Ho83} discovered an obscured proto-cluster containing a number of late-O to early-B stars through radio observations, and recently \citet{Messineo+15} presented a near infrared spectroscopic survey of bright stars in the region. We measure a radio spectral index $\alpha=-0.09\pm0.01$, consistent with thermal free-free emission. PSR~J1813--1749, one of the youngest and most energetic pulsars in the Milky Way \citep{Halpern+12}, is found to be in close proximity (in projection) to W42, as is the associated pulsar wind nebula HESS~J1813--178 \citep{Helfand+07,Messineo+08}. However, the lack of evidence for non-thermal contamination in the radio continuum suggests that this pulsar is unlikely to be significantly interacting with W42. From the known massive stellar population of W33, we estimate $N_C=1.1\times 10^{50}$~s$^{-1}$. The dust-processed luminosity measured from our SED modeling is $L_{\rm TIR}=(1.2\pm0.3)\times 10^6$~\Lsun, consistent with the ${\sim}8\times 10^5$~\Lsun estimated by \citet{Immer+14}. The ionizing photon rate from the radio observations is ${\sim}43\%$ the value expected from the known massive stars in the region, while we estimate ${\sim}60\%$ of the emitted stellar luminosity is reprocessed by the dust. 
	
	\item {\bf RCW~38}: The RCW~38 cluster is a relatively nearby \citep[1.7$\pm$0.9 kpc;][]{Schneider+10} site of active high-mass star formation containing more than a thousand members \citep{Wolk+06}. Using a combination of X-ray, infrared, and radio observations, \citet{Wolk+06} identified 31 candidate OB stars, four of which are likely earlier than B2. Two additional OB candidates were found in \citet{Povich+17}, but are consistent with $\sim$B2~V stars and therefore have minimal impact on the energy budget of the region. The energy output of the associated cluster is dominated by the bright source IRS~2, which is a likely O5-O5 binary \citep{DeRose+09}. Additional OB star candidates \citep[][their Table~13]{Wolk+06} are significantly fainter, with luminosities corresponding to $\sim$O9.5 or later. Based on these OB candidates, we include two O9.5 stars (their sources 112 and 251) and one B2 star (their source 396), although these are unlikely to have a significant impact on the derived $L_{\star}$ and $N_C$ values for the region. Given the young age of RCW~38 ($\leq$1 Myr), we assume all stars are on the main sequence. This yields $N_C$=3.5$\times10^{49}$ s$^{-1}$. We measure an ionizing photon rate $N_C^{\prime}$ of (2.6$\pm$0.1)$\times10^{49}$ s$^{-1}$ from the \planck observations. 

	\item {\bf W3}: W3 is a prominent MSFR that has undergone at least three episodes of recent star formation. The sequential star formation observed in W3 may have been triggered by the expansion of the neighboring W4 \HII region \citep{Lada+78,Thronson+85}. VLBI parallaxes to H$_2$O masers by \citet{Hachisuka+06} place the region at a distance of 2.04~kpc; this is consistent with our distance estimate from Gaia parallaxes of 2.18$\pm$0.12 kpc. We measure a thermal radio spectral index, in contrast to the 0.7--6 cm radio continuum measurements of the W3 hypercompact \HII region by \citet{Wilson+03}, who found significant departures from thermal free-free emission and suggested a synchrotron origin for the radio emission. An estimate of the ionizing photon rate of $1.4\times10^{49}$~s$^{-1}$ from previous radio continuum measurements can be inferred from \citet{Wilson+03}, which broadly agrees with the $N_C^{\prime}=(2.9\pm0.2)\times 10^{49}$~s$^{-1}$ we measure from \planck observations. The stellar content of this region is summarized in \citet{Navarete+11} and \citet{Povich+17}; using the known stellar content of W3 yields an expected Lyman continuum photon rate of $4.5\times 10^{49}$~s$^{-1}$. Five additional spectroscopically confirmed or candidate OB stars were identified by \citet{Kiminki+15} and \citet{Povich+17}; when these stars are included, the Lyman continuum rate increases to $N_C=5.9\times 10^{49}$~s$^{-1}$ and the stellar luminosity is $L_{\star}= 1.7\times 10^6$ \Lsun. These estimates suggest that $\sim$82\% of the emitted stellar luminosity is reprocessed by dust in the region, a number that likely reflects a mixture of the stellar contributions from the highly-embedded W3 Main cluster, which dominates the IR SED, and the adjacent, unobscured IC 1795 association. 

	\item {\bf NGC~3576}: NGC~3576 is an optically faint \HII region \citep{Goss+Radhakrishnan69}, powered by an apparently embedded massive stellar cluster \citep[5.4$\times10^3$ \Msun;][]{Figueredo+02,Maercker+06}. It sits ${\sim}30^{\arcmin}$ west of NGC~3603 but at significantly closer distance -- we estimate a distance of 2.77$\pm$0.31 kpc from Gaia parallaxes, consistent with the $\sim$2.8 kpc found by \citet{DePree+99} -- so the two regions are likely associated with different spiral arms. Despite its numerous cluster members, \citet{Figueredo+02} and \citet{Barbosa+03} have noted the currently known OB star content is insufficient to account for the strength of the radio emission from NGC~3576. A strong, north-south outflow has been observed in both radio and H$\alpha$ observations of the region \citep{DePree+99,Muller+98}. The angular separation between NGC~3576 and NGC~3603 is less than the beam size FWHM of \planck at low frequencies, leading to significant confusion when measuring the radio spectral index. We therefore fix the radio spectral index $\alpha$ to the thermal value of -0.1 for both regions. There is a nearby young pulsar PSR~J1112-6103 \citep{Manchester+01}; although its distance is not known, its apparent placement within an infrared bubble extending northward from NGC~3576 and the apparent filling of this cavity with hard X-rays \citep{Townsley+11b,Townsley+14} suggests a possible cavity supernova remnant. Previous radio measurements by \citet{Goss+Shaver70} suggest an ionizing photon rate 1.6$\times10^{50}$ s$^{-1}$, significantly higher than the $\sim$4$\times10^{49}$ s$^{-1}$ implied by our \planck measurements; it is possible the \citet{Goss+Shaver70} estimates at 5 GHz suffered from contamination from the nearby pulsar. The known stellar content of NGC~3576 \citep{Skiff09,Wenger+00,MaizApellaniz+16} can only account for roughly half the measured $L_{\rm TIR}$. Six additional candidate OB stars were found by \citet{Povich+17}. Including these stars brings our estimates of both $N_C^{\prime}/N_C$ and $L_{\rm TIR}/L_{\star}$ closer to values found for other regions, but is still insufficient to explain the radio and infrared observations of the \HII region. It is therefore likely that several important ionizing stars within the NGC~3576 cluster have yet to be identified, equivalent to ${\sim}4$ canonical O5~V stars. Due to the incompleteness of the known stellar content, we do not include this region in the portions of our analysis that require reliable estimates of the stellar content, and we cannot estimate $f_{C,{\rm abs}}$.

	\item {\bf NGC~6334}: NGC~6334 (the Cat's Paw Nebula) is a site of extensive and rapid star formation at a distance of 1.63$\pm$0.16 kpc (from Gaia parallaxes) in Scorpious. It is connected by a long, dusty filamentary structure to NGC~6357 \citep{Russeil+10,Russeil+12}, and the multiple-component giant \HII region complex is powered by several massive young stellar clusters, most of them obscured by or embedded within a long, dusty filament oriented parallel to the Galactic midplane \citep{Persi+00}. Hot, X-ray emitting gas pervades the giant \HII region, with a large bipolar outflow that shows a striking correlation with the bubble structures revealed by \spitzer \citep{Townsley+14}. Using only the previously cataloged OB stars associated with NGC~6334 yields a Lyman continuum comparable to the value measured from the radio continuum observations but fails to produce a stellar luminosity comparable to the observed bolometric luminosity of the \HII region by a factor of ${\sim}5$. When the ten OB candidates found in \citet{Povich+17} are included, the ionizing photon rate becomes $N_C=1.1\times10^{50}$~s$^{-1}$, which reduces the $N_C^{\prime}/N_C$ to ${\sim}0.7$ and gives $L_{\rm TIR}/L_{\star}\sim2.3$. The OB candidates therefore help move the energy budget in NGC~6334 in the direction of balance, but additional ionizing stars (or reclassification of the principal ionizing stars to earlier spectral types) are clearly required to explain the bolometric luminosity. We therefore exclude NGC~6334 from the parts of our analysis that require knowledge of the stellar content. The $N_C^{\prime}/N_C$ ratio would be driven lower by the discovery of additional ionizing stars, meaning that significant absorption of Lyman continuum photons by dust is likely. 
		
	\item {\bf G29.96--0.02}: G29.96--0.02 is an ultracompact \HII region complex \citep{Wood+Churchwell89,Cesaroni+94,DeBuizer+02} with active, ongoing star formation \citep{Beltran+13}. There is disagreement on the distance to the \HII region; following \citet{Beltran+11} we adopt a distance of 6.2 kpc, although \citet{Pratap+99} have estimated distances from 4.2 to 9 kpc. \citet{Townsley+14} found, surprisingly, large-scale diffuse X-ray emission surrounding the embedded massive young cluster, indicating that the young massive stars in the cluster are beginning to pierce their natal cloud and affect the surrounding hot ISM. Although only one massive star is bright enough in the infrared to have its spectral type estimated \citep[equivalent of an O5-6 dwarf,][]{Watson+Hanson97,MartinHernandez+03}, models of the G29.96--0.02 region indicate that a cluster of young stars must be present to account for its observed luminosity \citep{Lumsden+03}. The predicted Lyman continuum flux was estimated by \citet{Beltran+13} to be 1.08$\times10^{50}$ s$^{-1}$ from SED models of \herschel data alone. The lone catalogued star in the G29.96--0.02 cluster (which we assume has a spectral type O5) cannot provide the bolometric luminosity ($L_{\rm TIR}\sim4\times10^6$~\Lsun); we estimate $L_{\rm TIR}/L_{\star}\sim5.4$ (Table~\ref{table:energy_budget}). Even assuming a closer distance to the \HII region of 4.2~kpc \citep{Pratap+99} would not resolve the discrepancy. The missing stellar input luminosity is equivalent to ${\sim}20$ or ${\sim}7$ canonical O5~V stars assuming distances of 6.2 kpc or 4.2 kpc, respectively. Given the obvious incompleteness of the stellar content, we exclude this region from the portions of our analysis that require reliable estimates of the massive stellar content.
	
	\item {\bf NGC~6357}: NGC~6537 appears to be associated with the same giant molecular cloud complex as NGC~6334 \citep{Russeil+10}; our distance estimate of 1.78$\pm$0.18 kpc from Gaia parallaxes supports this picture. At least three massive, young stellar clusters have blown parsec-scale \HII region bubbles through the large NGC~6357 molecular cloud complex, and a prominent 60$^{\prime}$ diameter shell can be seen opening away from the Galactic plane in H$\alpha$ \citep{Lortet+84,Cappa+11}. This shell can be interpreted as a proto-superbubble, blown by a now-dissolved older stellar cluster and supernovae that resulted from its most massive stars \citep{Wang+07,Townsley+14}. The presence of diffuse X-rays and a Wolf-Rayet star within the H$\alpha$ shell supports this interpretation. Using only the previously cataloged OB stars associated with NGC~6357 yields an ionizing photon rate of 2.7$\times10^{50}$ s$^{-1}$; this is comparable to previous estimates of (1.4--3.3)$\times10^{49}$ s$^{-1}$ based on the stellar content from \citet[][and references therein]{Cappa+11}. \citet{Povich+17} identified twenty candidate OB stars belonging to NGC~6357, twelve of which have been spectroscopically confirmed as massive stars. When these new stars are included, $L_{\star}$ increases by ${\sim}40\%$. We find that $\sim61\%$ of the emitted stellar luminosity is reprocessed by local dust.

	\item {\bf M17}: M17 (also known as the Omega or Swan Nebula) is one of the brightest infrared and radio sources in the sky, comparable in both brightness and age to the Orion Nebula \citep{Povich+07,Getman+14}, but ${\sim}5$ times more distant \citep[][and this work]{Xu+11}. The \HII region is ionized by the open cluster NGC~6618, which contains over 100 OB stars \citep{Lada+91} including at least four O4~V stars \citep{Chini+80,Hoffmeister+08}. The \HII region forms a blister erupting from the side of the giant molecular cloud M17 SW. Bright, ``V''-shaped nebular bars of ionized gas and dust surround the central cluster \citep{Povich+07}, and the cavity between the bars is almost completely evacuated and filled with X-ray emitting plasma that exits the eastern opening in a ``champagne flow'' \citep{Townsley+03,Townsley+14}. An older stellar association containing O and B-type (super)giants has produced a large, diffuse \HII region, M17 EB, immediately to the north of M17 \citep{Povich+09,Povich+17}. Using only the previously cataloged OB stars associated with M17 would give $N_C=2.2\times10^{50}$ s$^{-1}$. Three additional OB candidates found in \citet{Povich+17} had spectral types earlier than B2 confirmed with follow-up spectroscopy; we include these candidates for completeness, but their impact on the derived $N_C$ and $L_{\star}$ are minimal.
	 
	\item {\bf G333}:  Despite being one of the brightest MSFRs regions in the southern sky at long wavelengths, the G333 complex is almost completely obscured at optical wavelengths \citep{Rank+78}. It hosts a compact, young massive star cluster within its core-halo structure \citep{Aitken+77, Retallack+Goss80, Fujiyoshi+05}. We performed aperture photometry on the three brightest embedded \HII regions in the G333 complex. More than 500 additional X-ray sources have been found in the regions surrounding the young cluster, and diffuse X-ray emission pervades the entire region \citep{Townsley+14}. \citet{Fujiyoshi+05} undertook a broadband, near-infrared imaging study of G333.6--0.2; however, spectroscopic observations are needed to securely identify the spectral types of the stars. The luminosity of of the G333 complex is equivalent to fifteen O5~V stars, which would yield $N_C^{\prime}/N_C\sim0.44$, consistent with the other MSFRs in our sample. Due to the ambiguity in the stellar population, we assume five O5~V equivalents in Table~\ref{table:energy_budget} \citep{Fujiyoshi+05}, and do not include the G333 complex in the portions of our analysis that require reliable estimates of the stellar content. \citet{Townsley+14} note the remarkable similarity between the embedded cluster in G333.6--0.2 to the one found in NGC~3576, including the strong extinction gradient ($A_V\sim 12$--36~mag), the ``champagne flow'' of diffuse X-ray emission spilling from the central cluster and eroding the natal cloud, and the ``distributed'' population of stars around the central embedded cluster that likely formed during a previous episode of star formation within the cloud. The evident erosion of the surrounding molecular cloud  is consistent with our finding that approximately half of the stellar luminosity escapes from this apparently embedded cluster. It is possible that the \HII region forms a blister that faces away from Earth, and we have a disadvantaged sightline from behind the remaining molecular cloud. 
	
	\item {\bf W43}: W43 is a highly-obscured \citep[$A_V\sim 30$~mag,][]{Blum+99}, giant \HII region at a heliocentric distance of 5.5~kpc \citep[based on VLBI parallaxes of masers toward the region][]{Reid+09,Zhang+14}. It is powered by a cluster of OB and Wolf-Rayet stars. High-resolution 1-8 GHz VLA observations of the W43 ionizing cluster revealed non-thermal radio sources \citep{LuqueEscamilla+11}; however at the higher \planck frequencies and larger angular scales have shown W43 to be dominated by synchrotron emission. The ionizing cluster is spatially coincident with the very high energy gamma ray source HESS~J1848--018 \citep{Chaves+09}; winds from very massive stars in the cluster may accelerate electrons to relativistic speeds \citep[e.g.,][]{vanLoo05}, and/or supernova remnants may contribute to the high-energy and radio synchrotron emission \citep{Romero+07}. Spectroscopic studies by \citet{Blum+99} and \citet{LuqueEscamilla+11} have identified the most luminous star in W43 is a WN7 + O4~III binary; the central cluster also contains an O3.5~V + O4~V binary and an O3.5~III star. Six later-type OB giant stars and three Wolf-Rayet stars are also listed in the SIMBAD database \citep{Wenger+00} and included in our analysis. The stellar content of W43 hence produces an estimated $N_C{\sim} 4.0\times10^{50}$~s$^{-1}$ and $L_{\star}= 7.1\times10^6$~\Lsun. Our radio continuum measurements yield an $N_C^{\prime}$ value $\sim$40\% higher than the expected stellar Lyman continuum rate.
	
	\item {\bf RCW49}: RCW~49 (NGC~3247) is a luminous Southern Galactic \HII region \citep{Paladini+03} ionized by the very massive, compact cluster Westerlund~2. Wd 2 contains dozens of O stars \citep{Wenger+00}. The distance to the region has been a subject of historical debate, with estimates ranging from 2.3 kpc \citep{Brand+Blitz93,Whiteoak+Uchida97} to 7.9 kpc \citep{Moffat+91}. Radio recombination line observation toward RCW~49 indicate a kinematic distance between 2.2--5.8 kpc \citep{Wilson+70,Caswell+Haynes87}. Recent studies of the stellar content using the {\em Hubble Space Telescope} \citep{VargasAlvarez+13,Zeidler+15} have converged near the ${\sim}4.0$~kpc distance adopted by \citet{Churchwell+04}. Using Gaia DR2 parallaxes, we estimate a distance to RCW~49 of 4.4 kpc (albeit with a large uncertainty, $\sim$1.0 kpc). \spitzer images of the region reveal a complex nebular structure filaments, knots, pillars, and bow shocks \citep{Churchwell+04}. The large bubble surrounding Wd~2 has been almost entirely evacuated of dust and gas by the winds and radiation of massive stars, while a second cavity to the southwest contains the massive eclipsing binary system WR 20a. We measure an ionizing photon rate of 2.4$\times10^{50}$ s$^{-1}$ from the \planck 30--100 GHz observations. We use the spectral classifications from \citet{VargasAlvarez+13} and \citet{Moffat+91} to estimate $N_C=5.6\times10^{50}$ s$^{-1}$ and $L_{\star}=10^7$ \Lsun. These estimates include the massive Wolf-Rayet binary system WR20 which, although technically not part of the Wd~2 cluster, is likely to contribute significantly to the energy budget of RCW~49.

	\item {\bf G305}: The G305 complex is located a distance of 3.59$\pm$0.85 kpc \citep[estimated from Gaia parallaxes, which is consistent with $\sim$4 kpc from][]{Clark+Porter04} in the Scutum-Crux spiral arm. It consists of several giant and compact \HII regions, and one ultracompact \HII region \citep{Hindson+13} powered by the open clusters Danks~1 and 2, and the Wolf-Rayet star WR~48a \citep{Danks+84}. The ionizing stars located in Danks~1 and 2 are driving the expansion of a large cavity into the surrounding molecular cloud, and ongoing star formation occurs along the rim of this cavity \citep{Clark+Porter04,Davies+12}. Numerous Wolf-Rayet stars are found in these very rich clusters. We measure $N_C^{\prime}=(2.0\pm0.3)\times10^{50}$~s$^{-1}$ from the \planck observations, in good agreement with the 5.5 GHz measurement of $2.4\times10^{50}$~s$^{-1}$ from \citet{Hindson+13}. The cataloged massive stellar population \citep{Davies+12} provides $N_C=2.9\times10^{50}$~s$^{-1}$ and $L_{\star}=6.7\times10^6$~\Lsun, which can account for the observed radio emission but not our the total infrared luminosity. Given the likely incompleteness in the stellar content of G305, it is not included in portions of our analysis requiring reliable $N_C$ and $L_{\star}$.

	\item {\bf W49A}: W49A is the most distant \citep[$\sim$11.4~kpc;][]{Gwinn+92} MSFR in our sample, and also one of the most luminous and massive \citep[$L>10^7$ \Lsun and $M\sim10^6$ \Msun][]{WardThompson+90,Sievers+91}. The region is undergoing vigorous star formation \citep{Roberts+11}. W49A is optically obscured \citep[$A_V>20$ mag,][]{Alves+Homeier03,Homeier+Alves05}, surrounded by both local molecular cloud material \citep{Mufson+Liszt77,Simon+01} and additional clouds associated with the Sagittarius spiral arm, which crosses the line-of-sight to the region twice \citep{Plume+04}. Roughly one hundred O star candidates have been identified in W49A \citep{Alves+Homeier03,Homeier+Alves05}, and a recent spectroscopic survey by \citet{Wu+16} confirmed spectral classifications for 22 massive stars and young stellar objects that are members of the central cluster. We measure $N_C^{\prime}\sim3.8\times10^{50}$~s$^{-1}$ from the \planck observations, marginally lower than previously reported values ($9.8\times10^{50}$ s$^{-1}$ and $\sim$10$^{51}$ s$^{-1}$ by \citealp{Kennicutt84} and \citealp{Gwinn+92}, respectively). We use the spectroscopically identified massive stars from \citet{Wu+16}, as well as additional massive stars listed in SIMBAD \citep{Wenger+00} for our analysis. The bolometric luminosity inferred from our SED modeling ($L_{\rm TIR}\sim 1.6\times10^7$~\Lsun) is ${\sim}48\%$ higher than the expected integrated stellar luminosity, a discrepancy equivalent to ${\sim}17$ missing canonical O5~V stars. Given the evident incompleteness of the massive stellar census our estimated $N_C^{\prime}/N_C=0.63$ is an {\it upper} limit. \citet{GalvanMadrid+13} inferred a large ionizing photon leakage rate, but we caution that this would require the addition of an even larger number of new ionizing stars, given the high dust-reprocessed luminosity of the region. Due the incomplete census of massive stars in W49A, we exclude this region from portions of our analysis requiring reliable estimates of the massive stellar content. 

	\item {\bf The Carina Nebula}: The Carina Nebula is one of the most famous and active regions of star formation in the Milky Way, containing more than 200 massive OB stars \citep{Gagne+11}, over 1,400 young stellar objects \citep{Povich+11a}, and ${>}10^4$ cataloged stellar members \citep{Broos+11a,Broos+11b}. \citet{Smith06} took a direct census of the energy input from the known OB stars in the Carina Nebula, while \citet{Smith+Brooks07} examined the global properties of the surrounding nebulosity, and recently \citet{Alexander+16} obtained spectroscopic observations of 141 known and 94 candidate OB stars in the Carina complex, confirming 23 new massive stars. Many  of these OB stars are contained within several dense clusters, including Trumpler (Tr)~14, Tr~16, Tr~15, Bochum (Bo)~10, and Bo~11. Energy input to the Carina complex is dominated by Tr~16, which hosts the famous luminous blue variable binary $\eta$~Carinae \citep{Duncan+95,Damineli+00,Corcoran05}. At present times, the dense, optically thick stellar wind \citep{Hillier+01,Smith+03b} and surrounding dusty Homunculus nebula absorbs nearly all ionizing radiation from $\eta$~Car itself and processes it into infrared radiation \citep{Cox+95,Smith+03c}. However, before entering its current evolutionary phase, $\eta$~Car likely contributed ${\sim}20\%$ of all UV radiation to the Carina complex \citep{Smith06,Smith+Brooks07}. Following \citet{Smith06}, we assume an O5~V spectral type for the $\eta$~Car companion. In addition to $\eta$~Car, three WNL stars reside in Tr~16; when on the main sequence, these three stars plus $\eta$~Car presumably had O2 spectral types, which contributed significantly to the ionization of the Carina complex over the 2--3 Myr lifetime of Tr~16. The prototypical O2~If* supergiant HD 93129A resides in Tr~14, a smaller and more compact stellar cluster than Tr~16. The open cluster Tr~15 contains several massive stars \citep[see][]{Smith06,Smith+Brooks07,Alexander+16}, including Tr15-18, which we assume has a spectral type O8~I. Bo~10 and 11 are relatively meager clusters, each containing a handful of stars. Most notable is the Wolf-Rayet star HD~92809 in Bo~10, and HD~93632 in Bo~11, which has a spectral type of O4.5~III.

Using Gaia parallaxes, we find the distance to the Carina Nebula to be 2.69$\pm$0.40 kpc. For the current state of the Carina Nebula, \citep{Smith+Brooks07} estimate a total Lyman continuum photon rate of $9\times10^{50}$~s$^{-1}$, in excellent agreement with our own estimate of $N_C=9.4\times10^{50}$~s$^{-1}$ using the updated spectroscopic and photometric results from \citet{Alexander+16}. Our measured bolometric luminosity of the Carina Nebula, $L_{\rm TIR}=(1.8\pm0.5)\times10^7$~\Lsun is in good agreement with the \citet{Smith+Brooks07} value of ${\sim}1.2\times 10^7$~\Lsun, and accounts for ${\sim}77\%$ of the known stellar luminosity. The radio continuum from the \planck 30--100 GHz observations predicts a marginally lower ionizing photon rate than previously reported, $N_C^{\prime}=(2.9\pm0.3)\times10^{50}$~s$^{-1}$. Multiple, independent lines of evidence indicate that supernovae have already occurred in the Carina Complex, including the discovery of a neutron star \citep{Hamaguchi+09}, a depleted upper IMF in Tr 15 \citep{Wang+11}, and the high luminosity of the diffuse X-ray emission filling the Carina \HII region \citep{Townsley+11b}. Note that we have assumed that $\eta$~Car contributes nothing to the present-day $N_C$, but in reality any Lyman continuum photons emitted by the LBV and its companion that are reprocessed by the Homunculus will be included in our measured $L_{\rm TIR}$.

	\item {\bf W51A}: The W51 giant molecular cloud complex located in the Carina-Sagittarius spiral arm consists of numerous bright, extended infrared and radio sources \citep{Kumar+04}. We focus on the W51A \HII region, which is the site of one of the most spectacular examples of recent massive star formation in the Galaxy. Just the principal massive young embedded stellar cluster, G49.5--0.4, contains ${>}30$ O stars \citep{Okumura+00}. We adopt a distance of $5.1^{+2.9}_{-1.4}$ kpc based on maser parallax measurements by \citet{Xu+09}, which is a high relative uncertainty for a parallax distance. Spectrophotometric distances of O stars by \citet{Figueredo+08} imply a considerably closer distance, $2.0\pm0.4$~kpc, while kinematic distance estimates \citep{Conti+Crowther04,Russeil03} give ${\sim}5.5$~kpc, consistent with the maser parallax. \citet[][]{Figueredo+08} estimate their cataloged massive stars produce an ionizing photon rate of $1.5\times10^{50}$~s$^{-1}$. We further include the spectroscopically cataloged massive stars from \citet{Okumura+00} in our analysis, which increases $N_C$ to 4.3$\times10^{50}$ s$^{-1}$ and $L_{\star}$ to 9.1$\times10^6$ \Lsun. The bolometric luminosity of the W51A \HII region is a factor of $\sim$2 higher than predicted for the cataloged stellar population, or $\sim$30 O5~V star equivalents. Given the distance uncertainty and likely incompletely cataloged stellar population in W51A, we do not include W51A in our analysis.

	\item {\bf NGC~3603}: NGC~3063 is a giant \HII region located in the Galactic plane. The dense cluster and surrounding nebula bear a striking morphological similarity to R136 in the 30~Doradus region of the Large Magellanic cloud, and, along with W49A, NGC~3603 is frequently noted as one of the most massive \HII regions in the Milky Way \citep{Eisenhauer+98}. Although near on the sky to NGC~3576, is it much more distant, at 7 kpc from the Sun \citep{Moffat83,Moffat+94,Drissen+95,Brandl+99}. We use the spectroscopic classifications from \citet{Melena+08} and \citet{Crowther+Dessart98} to estimate $N_C=1.4\times10^{51}$~s$^{-1}$ and $L_{\star}=2.3\times10^7$~\Lsun. Due to confusion with NGC~3576 at low \planck frequencies, we assume a (fixed) radio spectral index $\alpha=-0.1$ in our SED modeling and estimate $N_C^{\prime}$ from the 100 GHz and 70 GHz frequencies only. Simultaneously modeling the \herschel SPIRE and \planck observations allows us to decompose the relative contributions from dust and the free-free continuum at these frequencies. NGC 3603 is the most luminous MSFR in our sample, and we find  $L_{\rm TIR}/L_{\star}\sim100\%$. Given that the nebular morphology reveals at least one narrow, blowout channel toward the northwest (Figure~\ref{radial_fig:finding_charts}), it is likely that some fraction of the stellar luminosity does escape the region, so there may be additional massive stars or unresolved binary systems remaining to be classified in this very rich cluster.
\end{itemize}

\section{Supplemental Tables}\label{appendix:supplemental_figs}
We provide a summary of the second-best fit SED models for each region (Table~\ref{table:SED_global_2nd_best}) and the fluxes measured through our aperture photometry (Table~\ref{table:all_photometry}; available electronically from the journal).

\begin{table*}[ht]
\centering
\begin{scriptsize}
\caption{Global SED Second-Best Model Fits}
\begin{tabular}{ccccccccccccc}
\hline \hline
Name	& $U_1$			& $q_{\rm PAH,1}$	& $U_{\rm min,2}$	& $U_{\rm max,2}$	& $q_{\rm PAH,2}$	& $f_{\rm bol}$	& 1-$\gamma$		& $L_{\rm TIR}$		& $T_{\rm BB}$		& $\alpha$	& $f_{\rm Br\alpha}$		& $\Delta$\chisqr	\\ 
		&				& (\%)			& 				&				& (\%)			& (\%)		& (10$^{-5}$)		& 10$^6$ (L$_{\odot}$)	& (K)				&			& (\%)				&				\\
 (1)		& (2)				& (3)	 			& (4)	 			& (5)		 		& (6)				& (7)			& (8)				& (9)					& (10)			& (11)		& (12)				& (13)			\\
 \hline
Flame			& 10$^5$		& 2.50	& 0.50		& 10$^5$		& 3.90	& 37		& 5.8$\pm$0.4		& 0.05$\pm$0.01		& 35.6$\pm$5.3	& -0.10$\pm$0.01	& 2		& +0.007	\\
W40				& 10$^5$		& 2.50	& \nodata		& \nodata		& \nodata	& 25		& \nodata			& 0.05$\pm$0.01		& 26.2$\pm$1.1	& -0.10$\pm$0.01	& 2		& +0.092	\\ 
Wd~1			& 10$^5$		& 4.58	& 1.00		& 10$^5$		& 3.90	& 100	& 74.5$\pm$16.7	& 0.10$\pm$0.03		& 7.7$\pm$1.2		& \nodata			& \nodata	& +0.097	\\
RCW36			& 10$^5$		& 3.90	& 0.50		& 10$^5$		& 4.58	& 46		& 5.4$\pm$0.8		& 0.12$\pm$0.02		& 25.9$\pm$1.8	& -0.09$\pm$0.01	& 3		& +0.102	\\
Berkeley~87$^c$	& 10$^5$		& 0.47	& 0.50		& 10$^5$		& 4.58	& 44		& 5.4$\pm$0.8		& 0.17$\pm$0.03		& 27.7$\pm$1.3	& -0.10$\pm$0.02	& 10		& +0.040	\\
Orion$^c$			& 10$^5$		& 3.19	& 0.50		& 10$^5$		& 1.77	& 62		& 7.5$\pm$0.3		& 0.24$\pm$0.06		& 33.9$\pm$1.9	& -0.10$\pm$0.02	& 26		& +0.051	\\
Lagoon$^{a,c}$ 	& 10$^5$		& 2.50	& 0.50		& 10$^5$		& 0.47	& 78		& 5.4$\pm$0.8		& 0.34$\pm$0.10		& 30.0$\pm$1.4	& -0.08$\pm$0.01 	& 14		& +0.074	\\
Trifid				& 10$^3$		& 3.19	& 0.50		& 10$^3$		& 4.58	& 84		& 5.0$\pm$0.8		& 0.37$\pm$0.15		& 20.5$\pm$3.1	& -0.08$\pm$0.01	& 6		& +0.027	\\
W42				& 10$^5$		& 3.19	& 0.50		& 10$^5$		& 1.12	& 62		& 5.2$\pm$0.3		& 0.38$\pm$0.08		& 26.2$\pm$0.6	& -0.10 (fixed)		& 21		& +0.004	\\
NGC~7538$^b$	& 10$^5$		& 4.58	& 0.50		& 10$^5$		& 2.50	& 58		& 5.0$\pm$0.2		& 0.59$\pm$0.12		& 27.2$\pm$1.5	& -0.09$\pm$0.01	& 7		& +0.055	\\
W4$^c$			& 10$^5$		& 3.90	& 0.50		& 10$^5$		& 4.58	& 27		& 5.0$\pm$0.8		& 0.76$\pm$0.12		& 24.7$\pm$0.9	& \nodata			& \nodata	& +0.192	\\
Eagle$^c$			& 10$^5$		& 1.12	& \nodata		& \nodata		& \nodata	& 38		& \nodata			& 0.92$\pm$0.29 		& 22.4$\pm$2.0	& -0.08$\pm$0.02	& 84		& +1.733	\\
W33				& 10$^5$		& 0.47	& 0.50		& 10$^5$		& 1.77	& 29		& 5.4$\pm$0.3		& 1.18$\pm$0.29		& 25.7$\pm$1.6	& -0.08$\pm$0.01	& 32		& +0.116	\\
RCW~38			& 10$^5$		& 1.77	& 0.50		& 10$^5$		& 3.90	& 56		& 5.0$\pm$0.2		& 1.23$\pm$0.27		& 30.4$\pm$1.0	& -0.10$\pm$0.02	& 6		& +0.046	\\
W3$^c$			& 10$^5$		& 2.50	& 0.50		& 10$^5$		& 3.90	& 50		& 4.8$\pm$0.2		& 1.31$\pm$0.24		& 32.4$\pm$1.0	& -0.10$\pm$0.01	& 7		& +0.032	\\
NGC~3576		& 10$^5$		& 1.77	& 0.50		& 10$^5$		& 4.58	& 51		& 5.7$\pm$0.2		& 1.51$\pm$0.37		& 30.5$\pm$2.7	& -0.10 (fixed)		& 15		& +0.060	\\
NGC~6334$^c$ 	& 10$^5$		& 4.58	& 0.50		& 10$^5$		& 0.47	& 55		& 5.1$\pm$0.2		& 2.75$\pm$0.68		& 30.0$\pm$2.6	& -0.10$\pm$0.01	& 5		& +0.022	\\
G29.96--0.02		& 10$^5$		& 4.58	& 0.50		& 10$^5$		& 1.12	& 38		& 4.9$\pm$0.4		& 3.92$\pm$0.96		& 30.2$\pm$2.0	& -0.10 (fixed)		& 14		& +0.001	\\
NGC~6357 		& 10$^5$		& 0.47	& 0.50		& 10$^5$		& 3.90	& 58		& 5.0$\pm$0.8		& 4.48$\pm$1.31		& 27.7$\pm$2.6	& -0.10 (fixed)		& 7		& +0.047	\\
M17				& 10$^5$		& 2.50	& 0.50		& 10$^5$		& 3.90	& 66		& 5.8$\pm$0.3		& 4.56$\pm$1.22		& 32.5$\pm$4.9	& -0.10$\pm$0.02	& 5		& +0.030	\\
G333.6--0.02		& 10$^5$		& 3.90	& 0.50		& 10$^5$		& 1.77	& 52		& 5.9$\pm$0.9		& 4.91$\pm$1.48		& 37.2$\pm$5.6	& -0.08$\pm$0.01	& 18		& +0.009	\\
W43				& 10$^5$		& 2.50	& 0.50		& 10$^5$		& 3.90	& 46		& 5.5$\pm$0.8		& 5.05$\pm$1.17		& 28.3$\pm$1.7	& -0.10$\pm$0.01	& 50		& +0.004	\\
RCW49			& 10$^5$		& 1.77	& 0.50		& 10$^5$		& 3.19	& 61		& 5.0$\pm$0.2		& 9.05$\pm$1.56		& 34.2$\pm$1.0	& -0.11$\pm$0.01	& 10		& +0.019	\\
G305			& 10$^5$		& 1.12	& 0.50		& 10$^5$		& 3.90	& 45		& 4.9$\pm$0.7		& 13.31$\pm$3.62		& 26.0$\pm$1.6	& -0.11$\pm$0.02	& 8		& +0.075	\\
W49A			& 10$^5$		& 3.90	& 0.50		& 10$^5$		& 1.12	& 31		& 5.3$\pm$0.2		& 15.90$\pm$4.18		& 30.4$\pm$4.6	& -0.11$\pm$0.02	& 17		& +0.043	\\
Carina			& 10$^4$		& 2.50	& 0.50		& 10$^4$		& 1.77	& 93		& 30.7$\pm$3.6	& 16.36$\pm$4.98		& 29.0$\pm$4.4	& -0.07$\pm$0.01	& 5		& +0.565	\\
W51A			& 10$^5$		& 3.90	& 0.50		& 10$^5$		& 1.12	& 48		& 5.8$\pm$0.3		& 17.97$\pm$3.39		& 32.1$\pm$1.4	& -0.10$\pm$0.01	& 9		& +0.021	\\
NGC~3603		& 10$^5$		& 1.77	& 0.50		& 10$^5$		& 1.77	& 62		& 9.3$\pm$0.3		& 20.33$\pm$5.69		& 37.4$\pm$5.6	& -0.10 (fixed)		& 8		& +0.526	\\
\hline \hline
\end{tabular}
\end{scriptsize}
\label{table:SED_global_2nd_best}
\tablecomments{$^a$Missing \herschel PACS observations. $^b$Missing \spitzer [5.8] and [8.0] observations. $^c$\spitzer observations not used in fit due to incomplete coverage of the region.}
\end{table*}

\clearpage

\begin{sidewaystable}[ht]\setlength{\tabcolsep}{2pt}
\scriptsize
\centering
\caption{Photometry Data For All Massive Star-Forming Regions}
\begin{tabular}{ccccccccccccccccc}
\hline \hline
Region	& Center ($l$, $b$)	& \multicolumn{2}{c}{\spitzer}			&& \multicolumn{2}{c}{\msx}			&& \multicolumn{2}{c}{\iras}		&& \multicolumn{2}{c}{\herschel}	&& \multicolumn{2}{c}{\planck}			\\ \cline{3-4} \cline{6-7} \cline{9-10} \cline{12-13} \cline{15-17}
		& (J2000)			& Band	& $S_{\nu}$ (Jy)			&& Band	& $S_{\nu}$ (Jy)			&& Band	& $S_{\nu}$ (Jy)		&& Band	& $S_{\nu}$ (Jy)		&& Band	& $r_{\rm ap}$ ($^\circ$)	& $S_{\nu}$ (Jy)	\\
(1)		& (2)				& (3)		& (4)						&& (5)	& (6)						&& (7)	& (8)					&& (9)	& (10)				&& (11)	& (12)				& (13)			\\
\hline
		&			& IRAC1 (3.6 \micron)	& \nodata				&& $A$ (8.28 \micron)	& 12,830$\pm$100		&& 12 \micron	& 24,250$\pm$103		&& PACS 70 \micron		& 633,410$\pm$17,270	&& 100 GHz	& 0.38	& 110$\pm$20	\\
		& (209.013,	& IRAC2 (4.5 \micron)	& \nodata				&& $C$ (12.13 \micron)	& 25,480$\pm$150		&& 25 \micron	& 113,350$\pm$140		&& PACS 160 \micron	& 505,700$\pm$19,970	&& 70 GHz	& 0.38	& 110$\pm$2	\\
Orion	& -19.375)		& IRAC3 (5.8 \micron)	& \nodata				&& $D$ (14.65 \micron)	& 29,840$\pm$90		&& 60 \micron	& 381,710$\pm$880		&& SPIRE 250 \micron	& 93,860$\pm$1,130		&& 44 GHz	& 0.45	& 120$\pm$3	\\
		&			& IRAC4 (8.0 \micron)	& \nodata				&& $E$ (21.3 \micron)	& 157,460$\pm$130		&& 100 \micron	& 458,820$\pm$1,810	&& SPIRE 350 \micron	& 42,650$\pm$560		&& 30 GHz	& 0.54	& 130$\pm$6	\\
		&			&					&					&&					&					&&			&					&& SPIRE 500 \micron	& 14,730$\pm$220		&& 			&		&		 	\\
\hline
		&			& IRAC1 (3.6 \micron)	& 380$\pm$60			&& $A$ (8.28 \micron)	& 5,470$\pm$75		&& 12 \micron	& 6,660$\pm$50		&& PACS 70 \micron		& 201,770$\pm$3,050	&& 100 GHz	& 0.27	& 22$\pm$9	\\
		& (206.512,	& IRAC2 (4.5 \micron)	& 310$\pm$50			&& $C$ (12.13 \micron)	& 7,410$\pm$140		&& 25 \micron	& 26,890$\pm$90		&& PACS 160 \micron	& 180,360$\pm$13,360	&& 70 GHz	& 0.27	& 16$\pm$1	\\
Flame	& -16.349)		& IRAC3 (5.8 \micron)	& 610$\pm$90			&& $D$ (14.65 \micron)	& 4,400$\pm$90		&& 60 \micron	& 171,490$\pm$270		&& SPIRE 250 \micron	& 32,310$\pm$2,250		&& 44 GHz	& 0.45	& 23$\pm$2	\\
		&			& IRAC4 (8.0 \micron)	& 4,800$\pm$720		&& $E$ (21.3 \micron)	& 18,880$\pm$110		&& 100 \micron	& 209,380$\pm$1,050	&& SPIRE 350 \micron	& 15,080$\pm$1,320		&& 30 GHz	& 0.54	& 26$\pm$5	\\
		&			&					&					&& 					& 					&& 			& 					&& SPIRE 500 \micron	& 4,580$\pm$510		&& 			& 		&			\\
\hline
		&			& IRAC1 (3.6 \micron)	& 200$\pm$23			&& $A$ (8.28 \micron)	& 3,160$\pm$130		&& 12 \micron	& 3,810$\pm$90		&& PACS 70 \micron		& 80,280$\pm$6,670		&& 100 GHz	& 0.37	& 11$\pm$7	\\
		& (28.737,		& IRAC2 (4.5 \micron)	& 220$\pm$22			&& $C$ (12.13 \micron)	& 4,540$\pm$250		&& 25 \micron	& 8,930$\pm$130		&& PACS 160 \micron	& 114,720$\pm$25,650	&& 70 GHz	& 0.37	& 12$\pm$3	\\
W40		& 3.477)		& IRAC3 (5.8 \micron)	& 200$\pm$32			&& $D$ (14.65 \micron)	& 2,360$\pm$140		&& 60 \micron	& 75,550$\pm$720		&& SPIRE 250 \micron	& 39,000$\pm$3,210		&& 44 GHz	& 0.45	& 13$\pm$4	\\
		& 			& IRAC4 (8.0 \micron)	& 2,410$\pm$250		&& $E$ (21.3 \micron)	& 5,960$\pm$320		&& 100 \micron	& 105,090$\pm$2,700	&& SPIRE 350 \micron	& 18,590$\pm$1,750		&& 30 GHz	& 0.54	& 18$\pm$7	\\
		&			&					&					&& 					& 					&& 			& 					&& SPIRE 500 \micron	& 7,200$\pm$710		&& 			&		& 			\\
\hline
		&			& IRAC1 (3.6 \micron)	& 91$\pm$7			&& $A$ (8.28 \micron)	& 990$\pm$80			&& 12 \micron	& 1,340$\pm$70		&& PACS 70 \micron		& 38,840$\pm$2,450		&& 100 GHz	& 0.27	& 9$\pm$5	\\
		& (265.156	& IRAC2 (4.5 \micron)	& 68$\pm$4			&& $C$ (12.13 \micron)	& 1,580$\pm$140		&& 25 \micron	& 4,480$\pm$100		&& PACS 160 \micron	& 46,850$\pm$9,190		&& 70 GHz	& 0.27	& 6$\pm$1	\\
RCW36	& 1.449)	 	& IRAC3 (5.8 \micron)	& 520$\pm$20			&& $D$ (14.65 \micron)	& 1,030$\pm$60		&& 60 \micron	& 37,180$\pm$330		&& SPIRE 250 \micron	& 16,730$\pm$1,840		&& 44 GHz	& 0.45	& 9$\pm$3	\\
		& 			& IRAC4 (8.0 \micron)	& 1,170$\pm$70		&& $E$ (21.3 \micron)	& 3,750$\pm$130		&& 100 \micron	& 41,530$\pm$1,410		&& SPIRE 350 \micron	& 8,740$\pm$1,060		&& 30 GHz	& 0.54	& 8$\pm$5	\\
		&			&					&					&& 					& 					&& 			& 					&& SPIRE 500 \micron	& 3,670$\pm$430		&& 			& 		&			\\
\hline
		&				& IRAC1 (3.6 \micron)	& 220$\pm$68			&& $A$ (8.28 \micron)	& 2,320$\pm$370		&& 12 \micron	& 2,990$\pm$440		&& PACS 70 \micron		& 43,880$\pm$11,830	&& 100 GHz	& 0.28	& 27$\pm$1	\\
		& (7.001,			& IRAC2 (4.5 \micron)	& 210$\pm$60			&& $C$ (12.13 \micron)	& 3,440$\pm$650		&& 25 \micron	& 4,190$\pm$620		&& PACS 160 \micron	& 122,280$\pm$56,330	&& 70 GHz	& 0.28	& 34$\pm$2	\\
Trifid		& -0.271)			& IRAC3 (5.8 \micron)	& 1,080$\pm$330		&& $D$ (14.65 \micron)	& 1,730$\pm$340		&& 60 \micron	& 35,740$\pm$4,500		&& SPIRE 250 \micron	& 43,750$\pm$9,850		&& 44 GHz	& 0.45	& 60$\pm$10	\\
		&				& IRAC4 (8.0 \micron)	& 3,420$\pm$530		&& $E$ (21.3 \micron)	& 2,410$\pm$420		&& 100 \micron	& 77,610$\pm$13,360	&& SPIRE 350 \micron	& 19,520$\pm$4,510		&& 30 GHz	& 0.54	& 54$\pm$27	\\
		&				&					&					&& 					& 					&& 			& 					&& SPIRE 500 \micron	& 7,080$\pm$1,600		&& 			& 		&			\\
\hline
		&			& IRAC1 (3.6 \micron)	& 23$\pm$3			&& $A$ (8.28 \micron)	& 640$\pm$40			&& 12 \micron	& 1,080$\pm$40		&& PACS 70 \micron		& 20,490$\pm$1,270		&& 100 GHz	& 0.16	& 6$\pm$3	\\
		& (75.809,		& IRAC2 (4.5 \micron)	& 24$\pm$4			&& $C$ (12.13 \micron)	& 1,120$\pm$70		&& 25 \micron	& 3,570$\pm$60		&& PACS 160 \micron	& 28,000$\pm$5,150		&& 70 GHz	& 0.22	& 7$\pm$2	\\
Berkeley87 & 0.401)		& IRAC3 (5.8 \micron)	& 120$\pm$18			&& $D$ (14.65 \micron)	& 900$\pm$50			&& 60 \micron	& 20,610$\pm$230		&& SPIRE 250 \micron	& 9,060$\pm$710		&& 44 GHz	& 0.45	& 13$\pm$7	\\
		&			& IRAC4 (8.0 \micron)	& 350$\pm$50			&& $E$ (21.3 \micron)	& 2,630$\pm$70		&& 100 \micron	& 35,950$\pm$490		&& SPIRE 350 \micron	& 4,420$\pm$390		&& 30 GHz	& 0.54	& 7$\pm$6	\\
		&			&					&					&& 					& 					&& 			& 					&& SPIRE 500 \micron	& 1,590$\pm$150		&& 			& 		&			\\
\hline
		& 			& IRAC1 (3.6 \micron)	& 110$\pm$20			&& $A$ (8.28 \micron)	& 1,830$\pm$140		&& 12 \micron	& 3,960$\pm$150		&& PACS 70 \micron		& \nodata				&& 100 GHz	& 0.22	& 33$\pm$15	\\
		& (6.029,		& IRAC2 (4.5 \micron)	& 110$\pm$20			&& $C$ (12.13 \micron)	& 4,680$\pm$270		&& 25 \micron	& 23,570$\pm$220		&& PACS 160 \micron	& \nodata				&& 70 GHz	& 0.22	& 33$\pm$2	\\
Lagoon	& -1.211)		& IRAC3 (5.8 \micron)	& 610$\pm$50			&& $D$ (14.65 \micron)	& 5,640$\pm$150		&& 60 \micron	& 80,170$\pm$1,810		&& SPIRE 250 \micron	& 22,290$\pm$1,160		&& 44 GHz	& 0.45	& 73$\pm$11	\\
		& 			& IRAC4 (8.0 \micron)	& 2,190$\pm$130		&& $E$ (21.3 \micron)	& 17,600$\pm$160		&& 100 \micron	& 77,840$\pm$4,170		&& SPIRE 350 \micron	& 10,660$\pm$640		&& 30 GHz	& 0.54	& 79$\pm$24	\\
		&			&					&					&& 					& 					&& 			& 					&& SPIRE 500 \micron	& 4,150$\pm$190		&& 			& 		&			\\
\hline \hline
\end{tabular}
\label{table:all_photometry}
\tablecomments{The remainder of this table is available in the electronic journal.}
\end{sidewaystable}


\begin{thebibliography}{}
\expandafter\ifx\csname natexlab\endcsname\relax\def\natexlab#1{#1}\fi

\end{thebibliography}


\begin{thebibliography}{}
\expandafter\ifx\csname natexlab\endcsname\relax\def\natexlab#1{#1}\fi

\bibitem[{{Aitken} {et~al.}(1977){Aitken}, {Griffiths}, \& {Jones}}]{Aitken+77}
{Aitken}, D.~K., {Griffiths}, J., \& {Jones}, B. 1977, \mnras, 179, 179

\bibitem[{{Alexander} {et~al.}(2016){Alexander}, {Hanes}, {Povich}, \&
  {McSwain}}]{Alexander+16}
{Alexander}, M.~J., {Hanes}, R.~J., {Povich}, M.~S., \& {McSwain}, M.~V. 2016,
  \aj, 152, 190

\bibitem[{{Alonso-Herrero} {et~al.}(2006){Alonso-Herrero}, {Rieke}, {Rieke},
  {Colina}, {P{\'e}rez-Gonz{\'a}lez}, \& {Ryder}}]{AlonsoHerrero+06}
{Alonso-Herrero}, A., {Rieke}, G.~H., {Rieke}, M.~J., {et~al.} 2006, \apj, 650,
  835

\bibitem[{{Alves} \& {Homeier}(2003)}]{Alves+Homeier03}
{Alves}, J., \& {Homeier}, N. 2003, \apjl, 589, L45

\bibitem[{{Andersen} {et~al.}(2017){Andersen}, {Gennaro}, {Brandner}, {Stolte},
  {de Marchi}, {Meyer}, \& {Zinnecker}}]{Andersen+17}
{Andersen}, M., {Gennaro}, M., {Brandner}, W., {et~al.} 2017, \aap, 602, A22

\bibitem[{{Anderson} \& {Bania}(2009)}]{Anderson+Bania09}
{Anderson}, L.~D., \& {Bania}, T.~M. 2009, \apj, 690, 706

\bibitem[{{Andr{\'e}} {et~al.}(2010){Andr{\'e}}, {Men'shchikov}, {Bontemps},
  {K{\"o}nyves}, {Motte}, {Schneider}, {Didelon}, {Minier}, {Saraceno},
  {Ward-Thompson}, {di Francesco}, {White}, {Molinari}, {Testi}, {Abergel},
  {Griffin}, {Henning}, {Royer}, {Mer{\'{\i}}n}, {Vavrek}, {Attard},
  {Arzoumanian}, {Wilson}, {Ade}, {Aussel}, {Baluteau}, {Benedettini},
  {Bernard}, {Blommaert}, {Cambr{\'e}sy}, {Cox}, {di Giorgio}, {Hargrave},
  {Hennemann}, {Huang}, {Kirk}, {Krause}, {Launhardt}, {Leeks}, {Le Pennec},
  {Li}, {Martin}, {Maury}, {Olofsson}, {Omont}, {Peretto}, {Pezzuto}, {Prusti},
  {Roussel}, {Russeil}, {Sauvage}, {Sibthorpe}, {Sicilia-Aguilar}, {Spinoglio},
  {Waelkens}, {Woodcraft}, \& {Zavagno}}]{Andre+10}
{Andr{\'e}}, P., {Men'shchikov}, A., {Bontemps}, S., {et~al.} 2010, \aap, 518,
  L102

\bibitem[{{Baba} {et~al.}(2004){Baba}, {Nagata}, {Nagayama}, {Nagashima},
  {Kato}, {Kurita}, {Sato}, {Nakajima}, {Tamura}, {Nakaya}, \&
  {Sugitani}}]{Baba+04}
{Baba}, D., {Nagata}, T., {Nagayama}, T., {et~al.} 2004, \apj, 614, 818

\bibitem[{{Balog} {et~al.}(2004){Balog}, {Kenyon}, {Lada}, {Barsony},
  {Vink{\'o}}, \& {G{\'a}spa{\'r}}}]{Balog+04}
{Balog}, Z., {Kenyon}, S.~J., {Lada}, E.~A., {et~al.} 2004, \aj, 128, 2942

\bibitem[{{Barbosa} {et~al.}(2003){Barbosa}, {Damineli}, {Blum}, \&
  {Conti}}]{Barbosa+03}
{Barbosa}, C.~L., {Damineli}, A., {Blum}, R.~D., \& {Conti}, P.~S. 2003, \aj,
  126, 2411

\bibitem[{{Barnes} {et~al.}(1989){Barnes}, {Crutcher}, {Bieging}, {Storey}, \&
  {Willner}}]{Barnes+89}
{Barnes}, P.~J., {Crutcher}, R.~M., {Bieging}, J.~H., {Storey}, J.~W.~V., \&
  {Willner}, S.~P. 1989, \apj, 342, 883

\bibitem[{{Basu} {et~al.}(1999){Basu}, {Johnstone}, \& {Martin}}]{Basu+99}
{Basu}, S., {Johnstone}, D., \& {Martin}, P.~G. 1999, \apj, 516, 843

\bibitem[{{Beichman} {et~al.}(1998){Beichman}, {Chester}, {Skrutskie}, {Low},
  \& {Gillett}}]{Beichman+98}
{Beichman}, C.~A., {Chester}, T.~J., {Skrutskie}, M., {Low}, F.~J., \&
  {Gillett}, F. 1998, \pasp, 110, 480

\bibitem[{{Beltr{\'a}n} {et~al.}(2011){Beltr{\'a}n}, {Cesaroni}, {Neri}, \&
  {Codella}}]{Beltran+11}
{Beltr{\'a}n}, M.~T., {Cesaroni}, R., {Neri}, R., \& {Codella}, C. 2011, \aap,
  525, A151

\bibitem[{{Beltr{\'a}n} {et~al.}(2013){Beltr{\'a}n}, {Olmi}, {Cesaroni},
  {Schisano}, {Elia}, {Molinari}, {Di Giorgio}, {Kirk}, {Mottram},
  {Pestalozzi}, {Testi}, \& {Thompson}}]{Beltran+13}
{Beltr{\'a}n}, M.~T., {Olmi}, L., {Cesaroni}, R., {et~al.} 2013, \aap, 552,
  A123

\bibitem[{{Bendo} {et~al.}(2003){Bendo}, {Joseph}, {Wells}, {Gallais}, {Haas},
  {Heras}, {Klaas}, {Laureijs}, {Leech}, {Lemke}, {Metcalfe}, {Rowan-Robinson},
  {Schulz}, \& {Telesco}}]{Bendo+03}
{Bendo}, G.~J., {Joseph}, R.~D., {Wells}, M., {et~al.} 2003, \aj, 125, 2361

\bibitem[{{Benjamin} {et~al.}(2003){Benjamin}, {Churchwell}, {Babler}, {Bania},
  {Clemens}, {Cohen}, {Dickey}, {Indebetouw}, {Jackson}, {Kobulnicky},
  {Lazarian}, {Marston}, {Mathis}, {Meade}, {Seager}, {Stolovy}, {Watson},
  {Whitney}, {Wolff}, \& {Wolfire}}]{Benjamin+03}
{Benjamin}, R.~A., {Churchwell}, E., {Babler}, B.~L., {et~al.} 2003, \pasp,
  115, 953

\bibitem[{{Berriman} {et~al.}(2002){Berriman}, {Kong}, {Good}, {Lonsdale},
  {Voges}, {Henry}, {Bean}, \& {Blackwell}}]{Berriman+02}
{Berriman}, G.~B., {Kong}, M., {Good}, J.~C., {et~al.} 2002, in Astronomical
  Society of the Pacific Conference Series, Vol. 281, Astronomical Data
  Analysis Software and Systems XI, ed. D.~A. {Bohlender}, D.~{Durand}, \&
  T.~H. {Handley}, 63

\bibitem[{{Bik} {et~al.}(2003){Bik}, {Lenorzer}, {Kaper}, {Comer{\'o}n},
  {Waters}, {de Koter}, \& {Hanson}}]{Bik+03}
{Bik}, A., {Lenorzer}, A., {Kaper}, L., {et~al.} 2003, \aap, 404, 249

\bibitem[{{Blum} {et~al.}(2000){Blum}, {Conti}, \& {Damineli}}]{Blum+00}
{Blum}, R.~D., {Conti}, P.~S., \& {Damineli}, A. 2000, \aj, 119, 1860

\bibitem[{{Blum} {et~al.}(1999){Blum}, {Damineli}, \& {Conti}}]{Blum+99}
{Blum}, R.~D., {Damineli}, A., \& {Conti}, P.~S. 1999, \aj, 117, 1392

\bibitem[{{Boudet} {et~al.}(2005){Boudet}, {Mutschke}, {Nayral}, {J{\"a}ger},
  {Bernard}, {Henning}, \& {Meny}}]{Boudet+05}
{Boudet}, N., {Mutschke}, H., {Nayral}, C., {et~al.} 2005, \apj, 633, 272

\bibitem[{{Brand} \& {Blitz}(1993)}]{Brand+Blitz93}
{Brand}, J., \& {Blitz}, L. 1993, \aap, 275, 67

\bibitem[{{Brandl} {et~al.}(1999){Brandl}, {Brandner}, {Eisenhauer}, {Moffat},
  {Palla}, \& {Zinnecker}}]{Brandl+99}
{Brandl}, B., {Brandner}, W., {Eisenhauer}, F., {et~al.} 1999, \aap, 352, L69

\bibitem[{{Broos} {et~al.}(2011{\natexlab{a}}){Broos}, {Getman}, {Povich},
  {Townsley}, {Feigelson}, \& {Garmire}}]{Broos+11b}
{Broos}, P.~S., {Getman}, K.~V., {Povich}, M.~S., {et~al.} 2011{\natexlab{a}},
  \apjs, 194, 4

\bibitem[{{Broos} {et~al.}(2011{\natexlab{b}}){Broos}, {Townsley}, {Feigelson},
  {Getman}, {Garmire}, {Preibisch}, {Smith}, {Babler}, {Hodgkin}, {Indebetouw},
  {Irwin}, {King}, {Lewis}, {Majewski}, {McCaughrean}, {Meade}, \&
  {Zinnecker}}]{Broos+11a}
{Broos}, P.~S., {Townsley}, L.~K., {Feigelson}, E.~D., {et~al.}
  2011{\natexlab{b}}, \apjs, 194, 2

\bibitem[{{Broos} {et~al.}(2013){Broos}, {Getman}, {Povich}, {Feigelson},
  {Townsley}, {Naylor}, {Kuhn}, {King}, \& {Busk}}]{Broos+13}
{Broos}, P.~S., {Getman}, K.~V., {Povich}, M.~S., {et~al.} 2013, \apjs, 209, 32

\bibitem[{{Calzetti} {et~al.}(2005){Calzetti}, {Kennicutt}, {Bianchi},
  {Thilker}, {Dale}, {Engelbracht}, {Leitherer}, {Meyer}, {Sosey}, {Mutchler},
  {Regan}, {Thornley}, {Armus}, {Bendo}, {Boissier}, {Boselli}, {Draine},
  {Gordon}, {Helou}, {Hollenbach}, {Kewley}, {Madore}, {Martin}, {Murphy},
  {Rieke}, {Rieke}, {Roussel}, {Sheth}, {Smith}, {Walter}, {White}, {Yi},
  {Scoville}, {Polletta}, \& {Lindler}}]{Calzetti+05}
{Calzetti}, D., {Kennicutt}, Jr., R.~C., {Bianchi}, L., {et~al.} 2005, \apj,
  633, 871

\bibitem[{{Calzetti} {et~al.}(2007){Calzetti}, {Kennicutt}, {Engelbracht},
  {Leitherer}, {Draine}, {Kewley}, {Moustakas}, {Sosey}, {Dale}, {Gordon},
  {Helou}, {Hollenbach}, {Armus}, {Bendo}, {Bot}, {Buckalew}, {Jarrett}, {Li},
  {Meyer}, {Murphy}, {Prescott}, {Regan}, {Rieke}, {Roussel}, {Sheth}, {Smith},
  {Thornley}, \& {Walter}}]{Calzetti+07}
{Calzetti}, D., {Kennicutt}, R.~C., {Engelbracht}, C.~W., {et~al.} 2007, \apj,
  666, 870

\bibitem[{{Calzetti} {et~al.}(2010){Calzetti}, {Wu}, {Hong}, {Kennicutt},
  {Lee}, {Dale}, {Engelbracht}, {van Zee}, {Draine}, {Hao}, {Gordon},
  {Moustakas}, {Murphy}, {Regan}, {Begum}, {Block}, {Dalcanton}, {Funes}, {Gil
  de Paz}, {Johnson}, {Sakai}, {Skillman}, {Walter}, {Weisz}, {Williams}, \&
  {Wu}}]{Calzetti+10}
{Calzetti}, D., {Wu}, S.-Y., {Hong}, S., {et~al.} 2010, \apj, 714, 1256

\bibitem[{{Campbell} \& {Persson}(1988)}]{Campbell+Persson88}
{Campbell}, B., \& {Persson}, S.~E. 1988, \aj, 95, 1185

\bibitem[{{Cappa} {et~al.}(2011){Cappa}, {Barb{\'a}}, {Duronea}, {Vasquez},
  {Arnal}, {Goss}, \& {Fern{\'a}ndez Laj{\'u}s}}]{Cappa+11}
{Cappa}, C.~E., {Barb{\'a}}, R., {Duronea}, N.~U., {et~al.} 2011, \mnras, 415,
  2844

\bibitem[{{Carey} {et~al.}(2009){Carey}, {Noriega-Crespo}, {Mizuno}, {Shenoy},
  {Paladini}, {Kraemer}, {Price}, {Flagey}, {Ryan}, {Ingalls}, {Kuchar},
  {Pinheiro Gon{\c c}alves}, {Indebetouw}, {Billot}, {Marleau}, {Padgett},
  {Rebull}, {Bressert}, {Ali}, {Molinari}, {Martin}, {Berriman}, {Boulanger},
  {Latter}, {Miville-Deschenes}, {Shipman}, \& {Testi}}]{Carey+09}
{Carey}, S.~J., {Noriega-Crespo}, A., {Mizuno}, D.~R., {et~al.} 2009, \pasp,
  121, 76

\bibitem[{{Carlstrom} \& {Kronberg}(1991)}]{Carlstrom+91}
{Carlstrom}, J.~E., \& {Kronberg}, P.~P. 1991, \apj, 366, 422

\bibitem[{{Caswell} \& {Haynes}(1987)}]{Caswell+Haynes87}
{Caswell}, J.~L., \& {Haynes}, R.~F. 1987, Australian Journal of Physics, 40,
  215

\bibitem[{{Cernicharo} {et~al.}(1998){Cernicharo}, {Lefloch}, {Cox},
  {Cesarsky}, {Esteban}, {Yusef-Zadeh}, {Mendez}, {Acosta-Pulido}, {Garcia
  Lopez}, \& {Heras}}]{Cernicharo+98}
{Cernicharo}, J., {Lefloch}, B., {Cox}, P., {et~al.} 1998, Science, 282, 462

\bibitem[{{Cesaroni} {et~al.}(1994){Cesaroni}, {Olmi}, {Walmsley},
  {Churchwell}, \& {Hofner}}]{Cesaroni+94}
{Cesaroni}, R., {Olmi}, L., {Walmsley}, C.~M., {Churchwell}, E., \& {Hofner},
  P. 1994, \apjl, 435, L137

\bibitem[{{Chaves} \& {for the H.~E.~S.~S.~Collaboration}(2009)}]{Chaves+09}
{Chaves}, R.~C.~G., \& {for the H.~E.~S.~S.~Collaboration}. 2009, ArXiv
  e-prints, arXiv:0907.0768

\bibitem[{{Chini} {et~al.}(1980){Chini}, {Elsaesser}, \& {Neckel}}]{Chini+80}
{Chini}, R., {Elsaesser}, H., \& {Neckel}, T. 1980, \aap, 91, 186

\bibitem[{{Chomiuk} \& {Povich}(2011)}]{Chomiuk+Povich11}
{Chomiuk}, L., \& {Povich}, M.~S. 2011, \aj, 142, 197

\bibitem[{{Churchwell} {et~al.}(2004){Churchwell}, {Whitney}, {Babler},
  {Indebetouw}, {Meade}, {Watson}, {Wolff}, {Wolfire}, {Bania}, {Benjamin},
  {Clemens}, {Cohen}, {Devine}, {Dickey}, {Heitsch}, {Jackson}, {Kobulnicky},
  {Marston}, {Mathis}, {Mercer}, {Stauffer}, \& {Stolovy}}]{Churchwell+04}
{Churchwell}, E., {Whitney}, B.~A., {Babler}, B.~L., {et~al.} 2004, \apjs, 154,
  322

\bibitem[{{Churchwell} {et~al.}(2009){Churchwell}, {Babler}, {Meade},
  {Whitney}, {Benjamin}, {Indebetouw}, {Cyganowski}, {Robitaille}, {Povich},
  {Watson}, \& {Bracker}}]{Churchwell+09}
{Churchwell}, E., {Babler}, B.~L., {Meade}, M.~R., {et~al.} 2009, \pasp, 121,
  213

\bibitem[{{Clark} {et~al.}(2005){Clark}, {Negueruela}, {Crowther}, \&
  {Goodwin}}]{Clark+05}
{Clark}, J.~S., {Negueruela}, I., {Crowther}, P.~A., \& {Goodwin}, S.~P. 2005,
  \aap, 434, 949

\bibitem[{{Clark} \& {Porter}(2004)}]{Clark+Porter04}
{Clark}, J.~S., \& {Porter}, J.~M. 2004, \aap, 427, 839

\bibitem[{{Clark} {et~al.}(2011){Clark}, {Ritchie}, {Negueruela}, {Crowther},
  {Damineli}, {Jablonski}, \& {Langer}}]{Clark+11}
{Clark}, J.~S., {Ritchie}, B.~W., {Negueruela}, I., {et~al.} 2011, \aap, 531,
  A28

\bibitem[{{Cohen} {et~al.}(2007){Cohen}, {Parker}, {Green}, {Murphy},
  {Miszalski}, {Frew}, {Meade}, {Babler}, {Indebetouw}, {Whitney}, {Watson},
  {Churchwell}, \& {Watson}}]{Cohen+07}
{Cohen}, M., {Parker}, Q.~A., {Green}, A.~J., {et~al.} 2007, \apj, 669, 343

\bibitem[{{Condon}(1992)}]{Condon92}
{Condon}, J.~J. 1992, \araa, 30, 575

\bibitem[{{Cong}(1977)}]{Cong77}
{Cong}, H.~I.~L. 1977, PhD thesis, National Aeronautics and Space
  Administration.~Goddard Space Flight Center, Greenbelt, MD.

\bibitem[{{Conti} \& {Crowther}(2004)}]{Conti+Crowther04}
{Conti}, P.~S., \& {Crowther}, P.~A. 2004, \mnras, 355, 899

\bibitem[{{Corcoran}(2005)}]{Corcoran05}
{Corcoran}, M.~F. 2005, \aj, 129, 2018

\bibitem[{{Coupeaud} {et~al.}(2011){Coupeaud}, {Demyk}, {Meny}, {Nayral},
  {Delpech}, {Leroux}, {Depecker}, {Creff}, {Brubach}, \& {Roy}}]{Coupeaud+11}
{Coupeaud}, A., {Demyk}, K., {Meny}, C., {et~al.} 2011, \aap, 535, A124

\bibitem[{{Cox} {et~al.}(1995){Cox}, {Mezger}, {Sievers}, {Najarro},
  {Bronfman}, {Kreysa}, \& {Haslam}}]{Cox+95}
{Cox}, P., {Mezger}, P.~G., {Sievers}, A., {et~al.} 1995, \aap, 297, 168

\bibitem[{{Crocker} {et~al.}(2012){Crocker}, {Krips}, {Bureau}, {Young},
  {Davis}, {Bayet}, {Alatalo}, {Blitz}, {Bois}, {Bournaud}, {Cappellari},
  {Davies}, {de Zeeuw}, {Duc}, {Emsellem}, {Khochfar}, {Krajnovi{\'c}},
  {Kuntschner}, {Lablanche}, {McDermid}, {Morganti}, {Naab}, {Oosterloo},
  {Sarzi}, {Scott}, {Serra}, \& {Weijmans}}]{Crocker+12}
{Crocker}, A., {Krips}, M., {Bureau}, M., {et~al.} 2012, \mnras, 421, 1298

\bibitem[{{Crowther}(2007)}]{Crowther07}
{Crowther}, P.~A. 2007, \araa, 45, 177

\bibitem[{{Crowther} \& {Dessart}(1998)}]{Crowther+Dessart98}
{Crowther}, P.~A., \& {Dessart}, L. 1998, \mnras, 296, 622

\bibitem[{{Crowther} {et~al.}(2006){Crowther}, {Hadfield}, {Clark},
  {Negueruela}, \& {Vacca}}]{Crowther+06}
{Crowther}, P.~A., {Hadfield}, L.~J., {Clark}, J.~S., {Negueruela}, I., \&
  {Vacca}, W.~D. 2006, \mnras, 372, 1407

\bibitem[{{Dale} {et~al.}(2005){Dale}, {Bendo}, {Engelbracht}, {Gordon},
  {Regan}, {Armus}, {Cannon}, {Calzetti}, {Draine}, {Helou}, {Joseph},
  {Kennicutt}, {Li}, {Murphy}, {Roussel}, {Walter}, {Hanson}, {Hollenbach},
  {Jarrett}, {Kewley}, {Lamanna}, {Leitherer}, {Meyer}, {Rieke}, {Rieke},
  {Sheth}, {Smith}, \& {Thornley}}]{Dale+05}
{Dale}, D.~A., {Bendo}, G.~J., {Engelbracht}, C.~W., {et~al.} 2005, \apj, 633,
  857

\bibitem[{{Dale} \& {Bonnell}(2011)}]{Dale+Bonnell11}
{Dale}, J.~E., \& {Bonnell}, I. 2011, \mnras, 414, 321

\bibitem[{{Damiani} {et~al.}(2004){Damiani}, {Flaccomio}, {Micela},
  {Sciortino}, {Harnden}, \& {Murray}}]{Damiani+04}
{Damiani}, F., {Flaccomio}, E., {Micela}, G., {et~al.} 2004, \apj, 608, 781

\bibitem[{{Damiani} {et~al.}(2006){Damiani}, {Prisinzano}, {Micela}, \&
  {Sciortino}}]{Damiani+06}
{Damiani}, F., {Prisinzano}, L., {Micela}, G., \& {Sciortino}, S. 2006, \aap,
  459, 477

\bibitem[{{Damineli} {et~al.}(2000){Damineli}, {Kaufer}, {Wolf}, {Stahl},
  {Lopes}, \& {de Ara{\'u}jo}}]{Damineli+00}
{Damineli}, A., {Kaufer}, A., {Wolf}, B., {et~al.} 2000, \apjl, 528, L101

\bibitem[{{Danks} {et~al.}(1984){Danks}, {Wamsteker}, {Shaver}, \&
  {Retallack}}]{Danks+84}
{Danks}, A.~C., {Wamsteker}, W., {Shaver}, P.~A., \& {Retallack}, D.~S. 1984,
  \aap, 132, 301

\bibitem[{{Davies} {et~al.}(2012){Davies}, {Clark}, {Trombley}, {Figer},
  {Najarro}, {Crowther}, {Kudritzki}, {Thompson}, {Urquhart}, \&
  {Hindson}}]{Davies+12}
{Davies}, B., {Clark}, J.~S., {Trombley}, C., {et~al.} 2012, \mnras, 419, 1871

\bibitem[{{De Buizer} {et~al.}(2002){De Buizer}, {Watson}, {Radomski},
  {Pi{\~n}a}, \& {Telesco}}]{DeBuizer+02}
{De Buizer}, J.~M., {Watson}, A.~M., {Radomski}, J.~T., {Pi{\~n}a}, R.~K., \&
  {Telesco}, C.~M. 2002, \apjl, 564, L101

\bibitem[{{de Pree} {et~al.}(1999){de Pree}, {Nysewander}, \&
  {Goss}}]{DePree+99}
{de Pree}, C.~G., {Nysewander}, M.~C., \& {Goss}, W.~M. 1999, \aj, 117, 2902

\bibitem[{{Deeg} {et~al.}(1993){Deeg}, {Brinks}, {Duric}, {Klein}, \&
  {Skillman}}]{Deeg+93}
{Deeg}, H.-J., {Brinks}, E., {Duric}, N., {Klein}, U., \& {Skillman}, E. 1993,
  \apj, 410, 626

\bibitem[{{DeRose} {et~al.}(2009){DeRose}, {Bourke}, {Gutermuth}, {Wolk},
  {Megeath}, {Alves}, \& {N{\"u}rnberger}}]{DeRose+09}
{DeRose}, K.~L., {Bourke}, T.~L., {Gutermuth}, R.~A., {et~al.} 2009, \aj, 138,
  33

\bibitem[{{Dougherty} {et~al.}(2010){Dougherty}, {Clark}, {Negueruela},
  {Johnson}, \& {Chapman}}]{Dougherty+10}
{Dougherty}, S.~M., {Clark}, J.~S., {Negueruela}, I., {Johnson}, T., \&
  {Chapman}, J.~M. 2010, \aap, 511, A58

\bibitem[{{Draine}(1978)}]{Draine78}
{Draine}, B.~T. 1978, \apjs, 36, 595

\bibitem[{{Draine}(2003)}]{Draine03}
---. 2003, \araa, 41, 241

\bibitem[{{Draine} \& {Li}(2007)}]{Draine+Li07}
{Draine}, B.~T., \& {Li}, A. 2007, \apj, 657, 810

\bibitem[{{Drissen} {et~al.}(1995){Drissen}, {Moffat}, {Walborn}, \&
  {Shara}}]{Drissen+95}
{Drissen}, L., {Moffat}, A.~F.~J., {Walborn}, N.~R., \& {Shara}, M.~M. 1995,
  \aj, 110, 2235

\bibitem[{{Dufton} {et~al.}(2006){Dufton}, {Smartt}, {Lee}, {Ryans}, {Hunter},
  {Evans}, {Herrero}, {Trundle}, {Lennon}, {Irwin}, \& {Kaufer}}]{Dufton+06}
{Dufton}, P.~L., {Smartt}, S.~J., {Lee}, J.~K., {et~al.} 2006, \aap, 457, 265

\bibitem[{{Duncan} {et~al.}(1995){Duncan}, {White}, {Lim}, {Nelson}, {Drake},
  \& {Kundu}}]{Duncan+95}
{Duncan}, R.~A., {White}, S.~M., {Lim}, J., {et~al.} 1995, \apjl, 441, L73

\bibitem[{{Dunne} \& {Eales}(2001)}]{Dunne+Eales01}
{Dunne}, L., \& {Eales}, S.~A. 2001, \mnras, 327, 697

\bibitem[{{Dwek}(1986)}]{Dwek86}
{Dwek}, E. 1986, \apj, 302, 363

\bibitem[{{Eisenhauer} {et~al.}(1998){Eisenhauer}, {Quirrenbach}, {Zinnecker},
  \& {Genzel}}]{Eisenhauer+98}
{Eisenhauer}, F., {Quirrenbach}, A., {Zinnecker}, H., \& {Genzel}, R. 1998,
  \apj, 498, 278

\bibitem[{{Elbaz} {et~al.}(2011){Elbaz}, {Dickinson}, {Hwang},
  {D{\'{\i}}az-Santos}, {Magdis}, {Magnelli}, {Le Borgne}, {Galliano},
  {Pannella}, {Chanial}, {Armus}, {Charmandaris}, {Daddi}, {Aussel}, {Popesso},
  {Kartaltepe}, {Altieri}, {Valtchanov}, {Coia}, {Dannerbauer}, {Dasyra},
  {Leiton}, {Mazzarella}, {Alexander}, {Buat}, {Burgarella}, {Chary}, {Gilli},
  {Ivison}, {Juneau}, {Le Floc'h}, {Lutz}, {Morrison}, {Mullaney}, {Murphy},
  {Pope}, {Scott}, {Brodwin}, {Calzetti}, {Cesarsky}, {Charlot}, {Dole},
  {Eisenhardt}, {Ferguson}, {F{\"o}rster Schreiber}, {Frayer}, {Giavalisco},
  {Huynh}, {Koekemoer}, {Papovich}, {Reddy}, {Surace}, {Teplitz}, {Yun}, \&
  {Wilson}}]{Elbaz+11}
{Elbaz}, D., {Dickinson}, M., {Hwang}, H.~S., {et~al.} 2011, \aap, 533, A119

\bibitem[{{Ellerbroek} {et~al.}(2013){Ellerbroek}, {Podio}, {Kaper}, {Sana},
  {Huppenkothen}, {de Koter}, \& {Monaco}}]{Ellerbroek+13}
{Ellerbroek}, L.~E., {Podio}, L., {Kaper}, L., {et~al.} 2013, \aap, 551, A5

\bibitem[{{Engelbracht} {et~al.}(2005){Engelbracht}, {Gordon}, {Rieke},
  {Werner}, {Dale}, \& {Latter}}]{Engelbracht+05}
{Engelbracht}, C.~W., {Gordon}, K.~D., {Rieke}, G.~H., {et~al.} 2005, \apjl,
  628, L29

\bibitem[{{Engelbracht} {et~al.}(2008){Engelbracht}, {Rieke}, {Gordon},
  {Smith}, {Werner}, {Moustakas}, {Willmer}, \& {Vanzi}}]{Engelbracht+08}
{Engelbracht}, C.~W., {Rieke}, G.~H., {Gordon}, K.~D., {et~al.} 2008, \apj,
  678, 804

\bibitem[{{Everett} \& {Churchwell}(2010)}]{Everett+Churchwell10}
{Everett}, J.~E., \& {Churchwell}, E. 2010, \apj, 713, 592

\bibitem[{{Fazio} {et~al.}(2004){Fazio}, {Hora}, {Allen}, {Ashby}, {Barmby},
  {Deutsch}, {Huang}, {Kleiner}, {Marengo}, {Megeath}, {Melnick}, {Pahre},
  {Patten}, {Polizotti}, {Smith}, {Taylor}, {Wang}, {Willner}, {Hoffmann},
  {Pipher}, {Forrest}, {McMurty}, {McCreight}, {McKelvey}, {McMurray}, {Koch},
  {Moseley}, {Arendt}, {Mentzell}, {Marx}, {Losch}, {Mayman}, {Eichhorn},
  {Krebs}, {Jhabvala}, {Gezari}, {Fixsen}, {Flores}, {Shakoorzadeh}, {Jungo},
  {Hakun}, {Workman}, {Karpati}, {Kichak}, {Whitley}, {Mann}, {Tollestrup},
  {Eisenhardt}, {Stern}, {Gorjian}, {Bhattacharya}, {Carey}, {Nelson},
  {Glaccum}, {Lacy}, {Lowrance}, {Laine}, {Reach}, {Stauffer}, {Surace},
  {Wilson}, {Wright}, {Hoffman}, {Domingo}, \& {Cohen}}]{Fazio+04}
{Fazio}, G.~G., {Hora}, J.~L., {Allen}, L.~E., {et~al.} 2004, \apjs, 154, 10

\bibitem[{{Feigelson} {et~al.}(2013){Feigelson}, {Townsley}, {Broos}, {Busk},
  {Getman}, {King}, {Kuhn}, {Naylor}, {Povich}, {Baddeley}, {Bate},
  {Indebetouw}, {Luhman}, {McCaughrean}, {Pittard}, {Pudritz}, {Sills}, {Song},
  \& {Wadsley}}]{Feigelson+13}
{Feigelson}, E.~D., {Townsley}, L.~K., {Broos}, P.~S., {et~al.} 2013, \apjs,
  209, 26

\bibitem[{{Felli} \& {Churchwell}(1970)}]{Felli+Churchwell70}
{Felli}, M., \& {Churchwell}, E. 1970, \apj, 160, 43

\bibitem[{{Felli} {et~al.}(1993){Felli}, {Churchwell}, {Wilson}, \&
  {Taylor}}]{Felli+93}
{Felli}, M., {Churchwell}, E., {Wilson}, T.~L., \& {Taylor}, G.~B. 1993, \aaps,
  98, 137

\bibitem[{{Figer} {et~al.}(2006){Figer}, {MacKenty}, {Robberto}, {Smith},
  {Najarro}, {Kudritzki}, \& {Herrero}}]{Figer+06}
{Figer}, D.~F., {MacKenty}, J.~W., {Robberto}, M., {et~al.} 2006, \apj, 643,
  1166

\bibitem[{{Figuer{\^e}do} {et~al.}(2002){Figuer{\^e}do}, {Blum}, {Damineli}, \&
  {Conti}}]{Figueredo+02}
{Figuer{\^e}do}, E., {Blum}, R.~D., {Damineli}, A., \& {Conti}, P.~S. 2002,
  \aj, 124, 2739

\bibitem[{{Figuer{\^e}do} {et~al.}(2005){Figuer{\^e}do}, {Blum}, {Damineli}, \&
  {Conti}}]{Figueredo+05}
---. 2005, \aj, 129, 1523

\bibitem[{{Figuer{\^e}do} {et~al.}(2008){Figuer{\^e}do}, {Blum}, {Damineli},
  {Conti}, \& {Barbosa}}]{Figueredo+08}
{Figuer{\^e}do}, E., {Blum}, R.~D., {Damineli}, A., {Conti}, P.~S., \&
  {Barbosa}, C.~L. 2008, \aj, 136, 221

\bibitem[{{Flagey} {et~al.}(2011){Flagey}, {Boulanger}, {Noriega-Crespo},
  {Paladini}, {Montmerle}, {Carey}, {Gagn{\'e}}, \& {Shenoy}}]{Flagey+11}
{Flagey}, N., {Boulanger}, F., {Noriega-Crespo}, A., {et~al.} 2011, \aap, 531,
  A51

\bibitem[{{Fujiyoshi} {et~al.}(2005){Fujiyoshi}, {Smith}, {Moore}, {Lumsden},
  {Aitken}, \& {Roche}}]{Fujiyoshi+05}
{Fujiyoshi}, T., {Smith}, C.~H., {Moore}, T.~J.~T., {et~al.} 2005, \mnras, 356,
  801

\bibitem[{{Gagn{\'e}} {et~al.}(2011){Gagn{\'e}}, {Fehon}, {Savoy}, {Cohen},
  {Townsley}, {Broos}, {Povich}, {Corcoran}, {Walborn}, {Remage Evans},
  {Moffat}, {Naz{\'e}}, \& {Oskinova}}]{Gagne+11}
{Gagn{\'e}}, M., {Fehon}, G., {Savoy}, M.~R., {et~al.} 2011, \apjs, 194, 5

\bibitem[{{Gaia Collaboration} {et~al.}(2018){Gaia Collaboration}, {Brown},
  {Vallenari}, {Prusti}, {de Bruijne}, {Babusiaux}, \&
  {Bailer-Jones}}]{Gaia2018}
{Gaia Collaboration}, {Brown}, A.~G.~A., {Vallenari}, A., {et~al.} 2018, ArXiv
  e-prints, arXiv:1804.09365

\bibitem[{{Galliano} {et~al.}(2008){Galliano}, {Dwek}, \&
  {Chanial}}]{Galliano+08}
{Galliano}, F., {Dwek}, E., \& {Chanial}, P. 2008, \apj, 672, 214

\bibitem[{{Galv{\'a}n-Madrid} {et~al.}(2013){Galv{\'a}n-Madrid}, {Liu},
  {Zhang}, {Pineda}, {Peng}, {Zhang}, {Keto}, {Ho}, {Rodr{\'{\i}}guez},
  {Zapata}, {Peters}, \& {De Pree}}]{GalvanMadrid+13}
{Galv{\'a}n-Madrid}, R., {Liu}, H.~B., {Zhang}, Z.-Y., {et~al.} 2013, \apj,
  779, 121

\bibitem[{{Getman} {et~al.}(2014){Getman}, {Feigelson}, {Kuhn}, {Broos},
  {Townsley}, {Naylor}, {Povich}, {Luhman}, \& {Garmire}}]{Getman+14}
{Getman}, K.~V., {Feigelson}, E.~D., {Kuhn}, M.~A., {et~al.} 2014, \apj, 787,
  108

\bibitem[{{Giammanco} {et~al.}(2005){Giammanco}, {Beckman}, \&
  {Cedr{\'e}s}}]{Giammanco+05}
{Giammanco}, C., {Beckman}, J.~E., \& {Cedr{\'e}s}, B. 2005, \aap, 438, 599

\bibitem[{{Goodman} {et~al.}(2014){Goodman}, {Alves}, {Beaumont}, {Benjamin},
  {Borkin}, {Burkert}, {Dame}, {Jackson}, {Kauffmann}, {Robitaille}, \&
  {Smith}}]{Goodman+14}
{Goodman}, A.~A., {Alves}, J., {Beaumont}, C.~N., {et~al.} 2014, \apj, 797, 53

\bibitem[{{Gordon} {et~al.}(2008){Gordon}, {Engelbracht}, {Rieke}, {Misselt},
  {Smith}, \& {Kennicutt}}]{Gordon+08}
{Gordon}, K.~D., {Engelbracht}, C.~W., {Rieke}, G.~H., {et~al.} 2008, \apj,
  682, 336

\bibitem[{{Gordon} {et~al.}(2010){Gordon}, {Galliano}, {Hony}, {Bernard},
  {Bolatto}, {Bot}, {Engelbracht}, {Hughes}, {Israel}, {Kemper}, {Kim}, {Li},
  {Madden}, {Matsuura}, {Meixner}, {Misselt}, {Okumura}, {Panuzzo}, {Rubio},
  {Reach}, {Roman-Duval}, {Sauvage}, {Skibba}, \& {Tielens}}]{Gordon+10}
{Gordon}, K.~D., {Galliano}, F., {Hony}, S., {et~al.} 2010, \aap, 518, L89

\bibitem[{{Gordon} {et~al.}(2014){Gordon}, {Roman-Duval}, {Bot}, {Meixner},
  {Babler}, {Bernard}, {Bolatto}, {Boyer}, {Clayton}, {Engelbracht}, {Fukui},
  {Galametz}, {Galliano}, {Hony}, {Hughes}, {Indebetouw}, {Israel}, {Jameson},
  {Kawamura}, {Lebouteiller}, {Li}, {Madden}, {Matsuura}, {Misselt}, {Montiel},
  {Okumura}, {Onishi}, {Panuzzo}, {Paradis}, {Rubio}, {Sandstrom}, {Sauvage},
  {Seale}, {Sewi{\l}o}, {Tchernyshyov}, \& {Skibba}}]{Gordon+14}
{Gordon}, K.~D., {Roman-Duval}, J., {Bot}, C., {et~al.} 2014, \apj, 797, 85

\bibitem[{{Goss} \& {Radhakrishnan}(1969)}]{Goss+Radhakrishnan69}
{Goss}, W.~M., \& {Radhakrishnan}, V. 1969, \aplett, 4, 199

\bibitem[{{Goss} \& {Shaver}(1970)}]{Goss+Shaver70}
{Goss}, W.~M., \& {Shaver}, P.~A. 1970, Australian Journal of Physics
  Astrophysical Supplement, 14, 1

\bibitem[{{Griffin} {et~al.}(2010){Griffin}, {Abergel}, {Abreu}, {Ade},
  {Andr{\'e}}, {Augueres}, {Babbedge}, {Bae}, {Baillie}, {Baluteau}, {Barlow},
  {Bendo}, {Benielli}, {Bock}, {Bonhomme}, {Brisbin}, {Brockley-Blatt},
  {Caldwell}, {Cara}, {Castro-Rodriguez}, {Cerulli}, {Chanial}, {Chen},
  {Clark}, {Clements}, {Clerc}, {Coker}, {Communal}, {Conversi}, {Cox},
  {Crumb}, {Cunningham}, {Daly}, {Davis}, {de Antoni}, {Delderfield}, {Devin},
  {di Giorgio}, {Didschuns}, {Dohlen}, {Donati}, {Dowell}, {Dowell}, {Duband},
  {Dumaye}, {Emery}, {Ferlet}, {Ferrand}, {Fontignie}, {Fox}, {Franceschini},
  {Frerking}, {Fulton}, {Garcia}, {Gastaud}, {Gear}, {Glenn}, {Goizel},
  {Griffin}, {Grundy}, {Guest}, {Guillemet}, {Hargrave}, {Harwit}, {Hastings},
  {Hatziminaoglou}, {Herman}, {Hinde}, {Hristov}, {Huang}, {Imhof}, {Isaak},
  {Israelsson}, {Ivison}, {Jennings}, {Kiernan}, {King}, {Lange}, {Latter},
  {Laurent}, {Laurent}, {Leeks}, {Lellouch}, {Levenson}, {Li}, {Li},
  {Lilienthal}, {Lim}, {Liu}, {Lu}, {Madden}, {Mainetti}, {Marliani}, {McKay},
  {Mercier}, {Molinari}, {Morris}, {Moseley}, {Mulder}, {Mur}, {Naylor},
  {Nguyen}, {O'Halloran}, {Oliver}, {Olofsson}, {Olofsson}, {Orfei}, {Page},
  {Pain}, {Panuzzo}, {Papageorgiou}, {Parks}, {Parr-Burman}, {Pearce},
  {Pearson}, {P{\'e}rez-Fournon}, {Pinsard}, {Pisano}, {Podosek}, {Pohlen},
  {Polehampton}, {Pouliquen}, {Rigopoulou}, {Rizzo}, {Roseboom}, {Roussel},
  {Rowan-Robinson}, {Rownd}, {Saraceno}, {Sauvage}, {Savage}, {Savini},
  {Sawyer}, {Scharmberg}, {Schmitt}, {Schneider}, {Schulz}, {Schwartz},
  {Shafer}, {Shupe}, {Sibthorpe}, {Sidher}, {Smith}, {Smith}, {Smith},
  {Spencer}, {Stobie}, {Sudiwala}, {Sukhatme}, {Surace}, {Stevens}, {Swinyard},
  {Trichas}, {Tourette}, {Triou}, {Tseng}, {Tucker}, {Turner}, {Vaccari},
  {Valtchanov}, {Vigroux}, {Virique}, {Voellmer}, {Walker}, {Ward}, {Waskett},
  {Weilert}, {Wesson}, {White}, {Whitehouse}, {Wilson}, {Winter}, {Woodcraft},
  {Wright}, {Xu}, {Zavagno}, {Zemcov}, {Zhang}, \& {Zonca}}]{Griffin+10}
{Griffin}, M.~J., {Abergel}, A., {Abreu}, A., {et~al.} 2010, \aap, 518, L3

\bibitem[{{Guarcello} {et~al.}(2007){Guarcello}, {Prisinzano}, {Micela},
  {Damiani}, {Peres}, \& {Sciortino}}]{Guarcello+07}
{Guarcello}, M.~G., {Prisinzano}, L., {Micela}, G., {et~al.} 2007, \aap, 462,
  245

\bibitem[{{Gwinn} {et~al.}(1992){Gwinn}, {Moran}, \& {Reid}}]{Gwinn+92}
{Gwinn}, C.~R., {Moran}, J.~M., \& {Reid}, M.~J. 1992, \apj, 393, 149

\bibitem[{{Hachisuka} {et~al.}(2006){Hachisuka}, {Brunthaler}, {Menten},
  {Reid}, {Imai}, {Hagiwara}, {Miyoshi}, {Horiuchi}, \& {Sasao}}]{Hachisuka+06}
{Hachisuka}, K., {Brunthaler}, A., {Menten}, K.~M., {et~al.} 2006, \apj, 645,
  337

\bibitem[{{Haffner} {et~al.}(2003){Haffner}, {Reynolds}, {Tufte}, {Madsen},
  {Jaehnig}, \& {Percival}}]{Haffner+03}
{Haffner}, L.~M., {Reynolds}, R.~J., {Tufte}, S.~L., {et~al.} 2003, \apjs, 149,
  405

\bibitem[{{Halpern} {et~al.}(2012){Halpern}, {Gotthelf}, \&
  {Camilo}}]{Halpern+12}
{Halpern}, J.~P., {Gotthelf}, E.~V., \& {Camilo}, F. 2012, \apjl, 753, L14

\bibitem[{{Hamaguchi} {et~al.}(2009){Hamaguchi}, {Corcoran}, {Ezoe},
  {Townsley}, {Broos}, {Gruendl}, {Vaidya}, {White}, {Strohmayer}, {Petre}, \&
  {Chu}}]{Hamaguchi+09}
{Hamaguchi}, K., {Corcoran}, M.~F., {Ezoe}, Y., {et~al.} 2009, \apjl, 695, L4

\bibitem[{{Harayama} {et~al.}(2008){Harayama}, {Eisenhauer}, \&
  {Martins}}]{Harayama+08}
{Harayama}, Y., {Eisenhauer}, F., \& {Martins}, F. 2008, \apj, 675, 1319

\bibitem[{{Haschick} \& {Ho}(1983)}]{Haschick+Ho83}
{Haschick}, A.~D., \& {Ho}, P.~T.~P. 1983, \apj, 267, 638

\bibitem[{{Helfand} {et~al.}(2007){Helfand}, {Gotthelf}, {Halpern}, {Camilo},
  {Semler}, {Becker}, \& {White}}]{Helfand+07}
{Helfand}, D.~J., {Gotthelf}, E.~V., {Halpern}, J.~P., {et~al.} 2007, \apj,
  665, 1297

\bibitem[{{Hillier} {et~al.}(2001){Hillier}, {Davidson}, {Ishibashi}, \&
  {Gull}}]{Hillier+01}
{Hillier}, D.~J., {Davidson}, K., {Ishibashi}, K., \& {Gull}, T. 2001, \apj,
  553, 837

\bibitem[{{Hindson} {et~al.}(2013){Hindson}, {Thompson}, {Urquhart}, {Faimali},
  {Johnston-Hollitt}, {Clark}, \& {Davies}}]{Hindson+13}
{Hindson}, L., {Thompson}, M.~A., {Urquhart}, J.~S., {et~al.} 2013, \mnras,
  435, 2003

\bibitem[{{Hoffmeister} {et~al.}(2008){Hoffmeister}, {Chini}, {Scheyda},
  {Schulze}, {Watermann}, {N{\"u}rnberger}, \& {Vogt}}]{Hoffmeister+08}
{Hoffmeister}, V.~H., {Chini}, R., {Scheyda}, C.~M., {et~al.} 2008, \apj, 686,
  310

\bibitem[{{Hollenbach} \& {Tielens}(1997)}]{Hollenbach+97}
{Hollenbach}, D.~J., \& {Tielens}, A.~G.~G.~M. 1997, \araa, 35, 179

\bibitem[{{Homeier} \& {Alves}(2005)}]{Homeier+Alves05}
{Homeier}, N.~L., \& {Alves}, J. 2005, \aap, 430, 481

\bibitem[{{Hunt} {et~al.}(2015){Hunt}, {Draine}, {Bianchi}, {Gordon}, {Aniano},
  {Calzetti}, {Dale}, {Helou}, {Hinz}, {Kennicutt}, {Roussel}, {Wilson},
  {Bolatto}, {Boquien}, {Croxall}, {Galametz}, {Gil de Paz}, {Koda},
  {Mu{\~n}oz-Mateos}, {Sandstrom}, {Sauvage}, {Vigroux}, \&
  {Zibetti}}]{Hunt+15}
{Hunt}, L.~K., {Draine}, B.~T., {Bianchi}, S., {et~al.} 2015, \aap, 576, A33

\bibitem[{{Immer} {et~al.}(2014){Immer}, {Galv{\'a}n-Madrid}, {K{\"o}nig},
  {Liu}, \& {Menten}}]{Immer+14}
{Immer}, K., {Galv{\'a}n-Madrid}, R., {K{\"o}nig}, C., {Liu}, H.~B., \&
  {Menten}, K.~M. 2014, \aap, 572, A63

\bibitem[{{Immer} {et~al.}(2013){Immer}, {Reid}, {Menten}, {Brunthaler}, \&
  {Dame}}]{Immer+13}
{Immer}, K., {Reid}, M.~J., {Menten}, K.~M., {Brunthaler}, A., \& {Dame}, T.~M.
  2013, \aap, 553, A117

\bibitem[{{Inoue}(2001)}]{Inoue01}
{Inoue}, A.~K. 2001, \aj, 122, 1788

\bibitem[{{Inoue} {et~al.}(2001){Inoue}, {Hirashita}, \& {Kamaya}}]{Inoue+01}
{Inoue}, A.~K., {Hirashita}, H., \& {Kamaya}, H. 2001, \apj, 555, 613

\bibitem[{{Jager} {et~al.}(1998){Jager}, {Mutschke}, \& {Henning}}]{Jager+98}
{Jager}, C., {Mutschke}, H., \& {Henning}, T. 1998, \aap, 332, 291

\bibitem[{{Kameya} {et~al.}(1990){Kameya}, {Morita}, {Kawabe}, \&
  {Ishiguro}}]{Kameya+90}
{Kameya}, O., {Morita}, K.-I., {Kawabe}, R., \& {Ishiguro}, M. 1990, \apj, 355,
  562

\bibitem[{{Kennicutt} \& {Evans}(2012)}]{Kennicutt+12}
{Kennicutt}, R.~C., \& {Evans}, N.~J. 2012, \araa, 50, 531

\bibitem[{{Kennicutt}(1984)}]{Kennicutt84}
{Kennicutt}, Jr., R.~C. 1984, \apj, 287, 116

\bibitem[{{Kennicutt}(1998)}]{Kennicutt98}
---. 1998, \araa, 36, 189

\bibitem[{{Kennicutt} {et~al.}(2009){Kennicutt}, {Hao}, {Calzetti},
  {Moustakas}, {Dale}, {Bendo}, {Engelbracht}, {Johnson}, \&
  {Lee}}]{Kennicutt+09}
{Kennicutt}, Jr., R.~C., {Hao}, C.-N., {Calzetti}, D., {et~al.} 2009, \apj,
  703, 1672

\bibitem[{{Kharchenko} {et~al.}(2005){Kharchenko}, {Piskunov}, {R{\"o}ser},
  {Schilbach}, \& {Scholz}}]{Kharchenko+05}
{Kharchenko}, N.~V., {Piskunov}, A.~E., {R{\"o}ser}, S., {Schilbach}, E., \&
  {Scholz}, R.-D. 2005, \aap, 438, 1163

\bibitem[{{Kiminki} {et~al.}(2015){Kiminki}, {Kim}, {Bagley}, {Sherry}, \&
  {Rieke}}]{Kiminki+15}
{Kiminki}, M.~M., {Kim}, J.~S., {Bagley}, M.~B., {Sherry}, W.~H., \& {Rieke},
  G.~H. 2015, \apj, 813, 42

\bibitem[{{Klein} {et~al.}(1988){Klein}, {Wielebinski}, \& {Morsi}}]{Klein+88}
{Klein}, U., {Wielebinski}, R., \& {Morsi}, H.~W. 1988, \aap, 190, 41

\bibitem[{{Kohoutek} {et~al.}(1999){Kohoutek}, {Mayer}, \&
  {Lorenz}}]{Kohoutek+99}
{Kohoutek}, L., {Mayer}, P., \& {Lorenz}, R. 1999, \aaps, 134, 129

\bibitem[{{Koumpia} \& {Bonanos}(2012)}]{Koumpia+Bonanos12}
{Koumpia}, E., \& {Bonanos}, A.~Z. 2012, \aap, 547, A30

\bibitem[{{Kroupa} \& {Weidner}(2003)}]{K+W03}
{Kroupa}, P., \& {Weidner}, C. 2003, \apj, 598, 1076

\bibitem[{{Kuhn} {et~al.}(2013){Kuhn}, {Povich}, {Luhman}, {Getman}, {Busk}, \&
  {Feigelson}}]{Kuhn+13}
{Kuhn}, M.~A., {Povich}, M.~S., {Luhman}, K.~L., {et~al.} 2013, \apjs, 209, 29

\bibitem[{{Kumar} {et~al.}(2004){Kumar}, {Kamath}, \& {Davis}}]{Kumar+04}
{Kumar}, M.~S.~N., {Kamath}, U.~S., \& {Davis}, C.~J. 2004, \mnras, 353, 1025

\bibitem[{{Lada} {et~al.}(1991){Lada}, {Depoy}, {Merrill}, \&
  {Gatley}}]{Lada+91}
{Lada}, C.~J., {Depoy}, D.~L., {Merrill}, K.~M., \& {Gatley}, I. 1991, \apj,
  374, 533

\bibitem[{{Lada} {et~al.}(1978){Lada}, {Elmegreen}, {Cong}, \&
  {Thaddeus}}]{Lada+78}
{Lada}, C.~J., {Elmegreen}, B.~G., {Cong}, H.-I., \& {Thaddeus}, P. 1978,
  \apjl, 226, L39

\bibitem[{{Lawton} {et~al.}(2010){Lawton}, {Gordon}, {Babler}, {Block},
  {Bolatto}, {Bracker}, {Carlson}, {Engelbracht}, {Hora}, {Indebetouw},
  {Madden}, {Meade}, {Meixner}, {Misselt}, {Oey}, {Oliveira}, {Robitaille},
  {Sewilo}, {Shiao}, {Vijh}, \& {Whitney}}]{Lawton+10}
{Lawton}, B., {Gordon}, K.~D., {Babler}, B., {et~al.} 2010, \apj, 716, 453

\bibitem[{{Lee} {et~al.}(2012){Lee}, {Murray}, \& {Rahman}}]{Lee+12}
{Lee}, E.~J., {Murray}, N., \& {Rahman}, M. 2012, \apj, 752, 146

\bibitem[{{Leitherer} {et~al.}(1999){Leitherer}, {Schaerer}, {Goldader},
  {Delgado}, {Robert}, {Kune}, {de Mello}, {Devost}, \&
  {Heckman}}]{Leitherer+99}
{Leitherer}, C., {Schaerer}, D., {Goldader}, J.~D., {et~al.} 1999, \apjs, 123,
  3

\bibitem[{{Li} {et~al.}(2010){Li}, {Calzetti}, {Kennicutt}, {Hong},
  {Engelbracht}, {Dale}, \& {Moustakas}}]{Li+10}
{Li}, Y., {Calzetti}, D., {Kennicutt}, R.~C., {et~al.} 2010, \apj, 725, 677

\bibitem[{{Li} {et~al.}(2013){Li}, {Crocker}, {Calzetti}, {Wilson},
  {Kennicutt}, {Murphy}, {Brandl}, {Draine}, {Galametz}, {Johnson}, {Armus},
  {Gordon}, {Croxall}, {Dale}, {Engelbracht}, {Groves}, {Hao}, {Helou}, {Hinz},
  {Hunt}, {Krause}, {Roussel}, {Sauvage}, \& {Smith}}]{Li+13}
{Li}, Y., {Crocker}, A.~F., {Calzetti}, D., {et~al.} 2013, \apj, 768, 180

\bibitem[{{Lortet} {et~al.}(1984){Lortet}, {Testor}, \& {Niemela}}]{Lortet+84}
{Lortet}, M.~C., {Testor}, G., \& {Niemela}, V. 1984, \aap, 140, 24

\bibitem[{{Luisi} {et~al.}(2016){Luisi}, {Anderson}, {Balser}, {Bania}, \&
  {Wenger}}]{Luisi+16}
{Luisi}, M., {Anderson}, L.~D., {Balser}, D.~S., {Bania}, T.~M., \& {Wenger},
  T.~V. 2016, \apj, 824, 125

\bibitem[{{Lumsden} {et~al.}(2003){Lumsden}, {Puxley}, {Hoare}, {Moore}, \&
  {Ridge}}]{Lumsden+03}
{Lumsden}, S.~L., {Puxley}, P.~J., {Hoare}, M.~G., {Moore}, T.~J.~T., \&
  {Ridge}, N.~A. 2003, \mnras, 340, 799

\bibitem[{{Luque-Escamilla} {et~al.}(2011){Luque-Escamilla},
  {Mu{\~n}oz-Arjonilla}, {S{\'a}nchez-Sutil}, {Mart{\'{\i}}}, {Combi}, \&
  {S{\'a}nchez-Ayaso}}]{LuqueEscamilla+11}
{Luque-Escamilla}, P.~L., {Mu{\~n}oz-Arjonilla}, A.~J., {S{\'a}nchez-Sutil},
  J.~R., {et~al.} 2011, \aap, 532, A92

\bibitem[{{Lynds} {et~al.}(1985){Lynds}, {Canzian}, \& {Oneil}}]{Lynds+85}
{Lynds}, B.~T., {Canzian}, B.~J., \& {Oneil}, Jr., E.~J. 1985, \apj, 288, 164

\bibitem[{{Madden}(2000)}]{Madden00}
{Madden}, S.~C. 2000, \nar, 44, 249

\bibitem[{{Madden} {et~al.}(2006){Madden}, {Galliano}, {Jones}, \&
  {Sauvage}}]{Madden+06}
{Madden}, S.~C., {Galliano}, F., {Jones}, A.~P., \& {Sauvage}, M. 2006, \aap,
  446, 877

\bibitem[{{Maercker} {et~al.}(2006){Maercker}, {Burton}, \&
  {Wright}}]{Maercker+06}
{Maercker}, M., {Burton}, M.~G., \& {Wright}, C.~M. 2006, \aap, 450, 253

\bibitem[{{Ma{\'{\i}}z Apell{\'a}niz} {et~al.}(2016){Ma{\'{\i}}z
  Apell{\'a}niz}, {Sota}, {Arias}, {Barb{\'a}}, {Walborn},
  {Sim{\'o}n-D{\'{\i}}az}, {Negueruela}, {Marco}, {Le{\~a}o}, {Herrero},
  {Gamen}, \& {Alfaro}}]{MaizApellaniz+16}
{Ma{\'{\i}}z Apell{\'a}niz}, J., {Sota}, A., {Arias}, J.~I., {et~al.} 2016,
  \apjs, 224, 4

\bibitem[{{Makovoz} {et~al.}(2006){Makovoz}, {Roby}, {Khan}, \&
  {Booth}}]{Makovoz+06}
{Makovoz}, D., {Roby}, T., {Khan}, I., \& {Booth}, H. 2006, in \procspie, Vol.
  6274, Society of Photo-Optical Instrumentation Engineers (SPIE) Conference
  Series, 62740C

\bibitem[{{Manchester} {et~al.}(2001){Manchester}, {Lyne}, {Camilo}, {Bell},
  {Kaspi}, {D'Amico}, {McKay}, {Crawford}, {Stairs}, {Possenti}, {Kramer}, \&
  {Sheppard}}]{Manchester+01}
{Manchester}, R.~N., {Lyne}, A.~G., {Camilo}, F., {et~al.} 2001, \mnras, 328,
  17

\bibitem[{{Marble} {et~al.}(2010){Marble}, {Engelbracht}, {van Zee}, {Dale},
  {Smith}, {Gordon}, {Wu}, {Lee}, {Kennicutt}, {Skillman}, {Johnson}, {Block},
  {Calzetti}, {Cohen}, {Lee}, \& {Schuster}}]{Marble+10}
{Marble}, A.~R., {Engelbracht}, C.~W., {van Zee}, L., {et~al.} 2010, \apj, 715,
  506

\bibitem[{{Markwardt}(2009)}]{Markwardt09}
{Markwardt}, C.~B. 2009, in Astronomical Society of the Pacific Conference
  Series, Vol. 411, Astronomical Data Analysis Software and Systems XVIII, ed.
  D.~A. {Bohlender}, D.~{Durand}, \& P.~{Dowler}, 251

\bibitem[{{Mart{\'{\i}}n-Hern{\'a}ndez}
  {et~al.}(2003){Mart{\'{\i}}n-Hern{\'a}ndez}, {Bik}, {Kaper}, {Tielens}, \&
  {Hanson}}]{MartinHernandez+03}
{Mart{\'{\i}}n-Hern{\'a}ndez}, N.~L., {Bik}, A., {Kaper}, L., {Tielens},
  A.~G.~G.~M., \& {Hanson}, M.~M. 2003, \aap, 405, 175

\bibitem[{{Martins} {et~al.}(2005){Martins}, {Schaerer}, \&
  {Hillier}}]{Martins+05b}
{Martins}, F., {Schaerer}, D., \& {Hillier}, D.~J. 2005, \aap, 436, 1049

\bibitem[{{Mathis} {et~al.}(1983){Mathis}, {Mezger}, \& {Panagia}}]{Mathis+83}
{Mathis}, J.~S., {Mezger}, P.~G., \& {Panagia}, N. 1983, \aap, 128, 212

\bibitem[{{McKee} \& {Williams}(1997)}]{McKee+Williams97}
{McKee}, C.~F., \& {Williams}, J.~P. 1997, \apj, 476, 144

\bibitem[{{Melena} {et~al.}(2008){Melena}, {Massey}, {Morrell}, \&
  {Zangari}}]{Melena+08}
{Melena}, N.~W., {Massey}, P., {Morrell}, N.~I., \& {Zangari}, A.~M. 2008, \aj,
  135, 878

\bibitem[{{Mennella} {et~al.}(1998){Mennella}, {Brucato}, {Colangeli},
  {Palumbo}, {Rotundi}, \& {Bussoletti}}]{Mennella+98}
{Mennella}, V., {Brucato}, J.~R., {Colangeli}, L., {et~al.} 1998, \apj, 496,
  1058

\bibitem[{{Mennella} {et~al.}(1995){Mennella}, {Colangeli}, {Bussoletti},
  {Merluzzi}, {Monaco}, {Palumbo}, \& {Rotundi}}]{Mennella+95}
{Mennella}, V., {Colangeli}, L., {Bussoletti}, E., {et~al.} 1995, \planss, 43,
  1217

\bibitem[{{Menten} {et~al.}(2007){Menten}, {Reid}, {Forbrich}, \&
  {Brunthaler}}]{Menten+07}
{Menten}, K.~M., {Reid}, M.~J., {Forbrich}, J., \& {Brunthaler}, A. 2007, \aap,
  474, 515

\bibitem[{{Messineo} {et~al.}(2008){Messineo}, {Figer}, {Davies}, {Rich},
  {Valenti}, \& {Kudritzki}}]{Messineo+08}
{Messineo}, M., {Figer}, D.~F., {Davies}, B., {et~al.} 2008, \apjl, 683, L155

\bibitem[{{Messineo} {et~al.}(2015){Messineo}, {Clark}, {Figer}, {Kudritzki},
  {Najarro}, {Rich}, {Menten}, {Ivanov}, {Valenti}, {Trombley}, {Chen}, \&
  {Davies}}]{Messineo+15}
{Messineo}, M., {Clark}, J.~S., {Figer}, D.~F., {et~al.} 2015, \apj, 805, 110

\bibitem[{{Minier} {et~al.}(2013){Minier}, {Tremblin}, {Hill}, {Motte},
  {Andr{\'e}}, {Lo}, {Schneider}, {Audit}, {White}, {Hennemann}, {Cunningham},
  {Deharveng}, {Didelon}, {Di Francesco}, {Elia}, {Giannini}, {Nguyen Luong},
  {Pezzuto}, {Rygl}, {Spinoglio}, {Ward-Thompson}, \& {Zavagno}}]{Minier+13}
{Minier}, V., {Tremblin}, P., {Hill}, T., {et~al.} 2013, \aap, 550, A50

\bibitem[{{Miville-Desch{\^e}nes} \& {Lagache}(2005)}]{Miville+Lagache05}
{Miville-Desch{\^e}nes}, M.-A., \& {Lagache}, G. 2005, \apjs, 157, 302

\bibitem[{{Moffat}(1983)}]{Moffat83}
{Moffat}, A.~F.~J. 1983, \aap, 124, 273

\bibitem[{{Moffat} {et~al.}(1994){Moffat}, {Drissen}, \& {Shara}}]{Moffat+94}
{Moffat}, A.~F.~J., {Drissen}, L., \& {Shara}, M.~M. 1994, \apj, 436, 183

\bibitem[{{Moffat} {et~al.}(1991){Moffat}, {Shara}, \& {Potter}}]{Moffat+91}
{Moffat}, A.~F.~J., {Shara}, M.~M., \& {Potter}, M. 1991, \aj, 102, 642

\bibitem[{{Molinari} {et~al.}(2010){Molinari}, {Swinyard}, {Bally}, {Barlow},
  {Bernard}, {Martin}, {Moore}, {Noriega-Crespo}, {Plume}, {Testi}, {Zavagno},
  {Abergel}, {Ali}, {Andr{\'e}}, {Baluteau}, {Benedettini}, {Bern{\'e}},
  {Billot}, {Blommaert}, {Bontemps}, {Boulanger}, {Brand}, {Brunt}, {Burton},
  {Campeggio}, {Carey}, {Caselli}, {Cesaroni}, {Cernicharo}, {Chakrabarti},
  {Chrysostomou}, {Codella}, {Cohen}, {Compiegne}, {Davis}, {de Bernardis}, {de
  Gasperis}, {Di Francesco}, {di Giorgio}, {Elia}, {Faustini}, {Fischera},
  {Fukui}, {Fuller}, {Ganga}, {Garcia-Lario}, {Giard}, {Giardino}, {Glenn},
  {Goldsmith}, {Griffin}, {Hoare}, {Huang}, {Jiang}, {Joblin}, {Joncas},
  {Juvela}, {Kirk}, {Lagache}, {Li}, {Lim}, {Lord}, {Lucas}, {Maiolo},
  {Marengo}, {Marshall}, {Masi}, {Massi}, {Matsuura}, {Meny}, {Minier},
  {Miville-Desch{\^e}nes}, {Montier}, {Motte}, {M{\"u}ller}, {Natoli}, {Neves},
  {Olmi}, {Paladini}, {Paradis}, {Pestalozzi}, {Pezzuto}, {Piacentini},
  {Pomar{\`e}s}, {Popescu}, {Reach}, {Richer}, {Ristorcelli}, {Roy}, {Royer},
  {Russeil}, {Saraceno}, {Sauvage}, {Schilke}, {Schneider-Bontemps},
  {Schuller}, {Schultz}, {Shepherd}, {Sibthorpe}, {Smith}, {Smith},
  {Spinoglio}, {Stamatellos}, {Strafella}, {Stringfellow}, {Sturm}, {Taylor},
  {Thompson}, {Tuffs}, {Umana}, {Valenziano}, {Vavrek}, {Viti}, {Waelkens},
  {Ward-Thompson}, {White}, {Wyrowski}, {Yorke}, \& {Zhang}}]{Molinari+10}
{Molinari}, S., {Swinyard}, B., {Bally}, J., {et~al.} 2010, \pasp, 122, 314

\bibitem[{{Momose} {et~al.}(2001){Momose}, {Tamura}, {Kameya}, {Greaves},
  {Chrysostomou}, {Hough}, \& {Morino}}]{Momose+01}
{Momose}, M., {Tamura}, M., {Kameya}, O., {et~al.} 2001, \apj, 555, 855

\bibitem[{{Moreno} \& {Chavarria-K.}(1986)}]{Moreno+Chavarria86}
{Moreno}, M.~A., \& {Chavarria-K.}, C. 1986, \aap, 161, 130

\bibitem[{{Moscadelli} {et~al.}(2009){Moscadelli}, {Reid}, {Menten},
  {Brunthaler}, {Zheng}, \& {Xu}}]{Moscadelli+09}
{Moscadelli}, L., {Reid}, M.~J., {Menten}, K.~M., {et~al.} 2009, \apj, 693, 406

\bibitem[{{Motte} {et~al.}(2010){Motte}, {Zavagno}, {Bontemps}, {Schneider},
  {Hennemann}, {di Francesco}, {Andr{\'e}}, {Saraceno}, {Griffin}, {Marston},
  {Ward-Thompson}, {White}, {Minier}, {Men'shchikov}, {Hill}, {Abergel},
  {Anderson}, {Aussel}, {Balog}, {Baluteau}, {Bernard}, {Cox}, {Csengeri},
  {Deharveng}, {Didelon}, {di Giorgio}, {Hargrave}, {Huang}, {Kirk}, {Leeks},
  {Li}, {Martin}, {Molinari}, {Nguyen-Luong}, {Olofsson}, {Persi}, {Peretto},
  {Pezzuto}, {Roussel}, {Russeil}, {Sadavoy}, {Sauvage}, {Sibthorpe},
  {Spinoglio}, {Testi}, {Teyssier}, {Vavrek}, {Wilson}, \&
  {Woodcraft}}]{Motte+10}
{Motte}, F., {Zavagno}, A., {Bontemps}, S., {et~al.} 2010, \aap, 518, L77

\bibitem[{{Mu{\~n}oz-Mateos} {et~al.}(2009){Mu{\~n}oz-Mateos}, {Gil de Paz},
  {Boissier}, {Zamorano}, {Dale}, {P{\'e}rez-Gonz{\'a}lez}, {Gallego},
  {Madore}, {Bendo}, {Thornley}, {Draine}, {Boselli}, {Buat}, {Calzetti},
  {Moustakas}, \& {Kennicutt}}]{MunozMateos+09}
{Mu{\~n}oz-Mateos}, J.~C., {Gil de Paz}, A., {Boissier}, S., {et~al.} 2009,
  \apj, 701, 1965

\bibitem[{{Mufson} \& {Liszt}(1977)}]{Mufson+Liszt77}
{Mufson}, S.~L., \& {Liszt}, H.~S. 1977, \apj, 212, 664

\bibitem[{{Muller} {et~al.}(1998){Muller}, {Reed}, {Armandroff}, {Boroson}, \&
  {Jacoby}}]{Muller+98}
{Muller}, G.~P., {Reed}, R., {Armandroff}, T., {Boroson}, T.~A., \& {Jacoby},
  G.~H. 1998, in \procspie, Vol. 3355, Optical Astronomical Instrumentation,
  ed. S.~{D'Odorico}, 577--585

\bibitem[{{Murphy} {et~al.}(2011){Murphy}, {Condon}, {Schinnerer}, {Kennicutt},
  {Calzetti}, {Armus}, {Helou}, {Turner}, {Aniano}, {Beir{\~a}o}, {Bolatto},
  {Brandl}, {Croxall}, {Dale}, {Donovan Meyer}, {Draine}, {Engelbracht},
  {Hunt}, {Hao}, {Koda}, {Roussel}, {Skibba}, \& {Smith}}]{Murphy+11}
{Murphy}, E.~J., {Condon}, J.~J., {Schinnerer}, E., {et~al.} 2011, \apj, 737,
  67

\bibitem[{{Murray} \& {Rahman}(2010)}]{Murray+Rahman10}
{Murray}, N., \& {Rahman}, M. 2010, \apj, 709, 424

\bibitem[{{Navarete} {et~al.}(2011){Navarete}, {Figueredo}, {Damineli},
  {Mois{\'e}s}, {Blum}, \& {Conti}}]{Navarete+11}
{Navarete}, F., {Figueredo}, E., {Damineli}, A., {et~al.} 2011, \aj, 142, 67

\bibitem[{{Normandeau} {et~al.}(1996){Normandeau}, {Taylor}, \&
  {Dewdney}}]{Normandeau+96}
{Normandeau}, M., {Taylor}, A.~R., \& {Dewdney}, P.~E. 1996, \nat, 380, 687

\bibitem[{{O'Dell}(2001)}]{ODell01a}
{O'Dell}, C.~R. 2001, \pasp, 113, 29

\bibitem[{{O'dell}(2001)}]{ODell01b}
{O'dell}, C.~R. 2001, \araa, 39, 99

\bibitem[{{Oey} \& {Kennicutt}(1997)}]{Oey+Kennicutt97}
{Oey}, M.~S., \& {Kennicutt}, Jr., R.~C. 1997, \mnras, 291, 827

\bibitem[{{Ojha} {et~al.}(2004){Ojha}, {Tamura}, {Nakajima}, {Fukagawa},
  {Sugitani}, {Nagashima}, {Nagayama}, {Nagata}, {Sato}, {Vig}, {Ghosh},
  {Pickles}, {Momose}, \& {Ogura}}]{Ojha+04}
{Ojha}, D.~K., {Tamura}, M., {Nakajima}, Y., {et~al.} 2004, \apj, 616, 1042

\bibitem[{{Okumura} {et~al.}(2000){Okumura}, {Mori}, {Nishihara}, {Watanabe},
  \& {Yamashita}}]{Okumura+00}
{Okumura}, S.-i., {Mori}, A., {Nishihara}, E., {Watanabe}, E., \& {Yamashita},
  T. 2000, \apj, 543, 799

\bibitem[{{Ott}(2010)}]{Ott10}
{Ott}, S. 2010, in Astronomical Society of the Pacific Conference Series, Vol.
  434, Astronomical Data Analysis Software and Systems XIX, ed. Y.~{Mizumoto},
  K.-I. {Morita}, \& M.~{Ohishi}, 139

\bibitem[{{Paladini} {et~al.}(2003){Paladini}, {Burigana}, {Davies}, {Maino},
  {Bersanelli}, {Cappellini}, {Platania}, \& {Smoot}}]{Paladini+03}
{Paladini}, R., {Burigana}, C., {Davies}, R.~D., {et~al.} 2003, \aap, 397, 213

\bibitem[{{Paradis} {et~al.}(2009){Paradis}, {Bernard}, \&
  {M{\'e}ny}}]{Paradis+09}
{Paradis}, D., {Bernard}, J.-P., \& {M{\'e}ny}, C. 2009, \aap, 506, 745

\bibitem[{{Peeters} {et~al.}(2004){Peeters}, {Spoon}, \&
  {Tielens}}]{Peeters+04}
{Peeters}, E., {Spoon}, H.~W.~W., \& {Tielens}, A.~G.~G.~M. 2004, \apj, 613,
  986

\bibitem[{{Pellegrini} {et~al.}(2012){Pellegrini}, {Oey}, {Winkler}, {Points},
  {Smith}, {Jaskot}, \& {Zastrow}}]{Pellegrini+12}
{Pellegrini}, E.~W., {Oey}, M.~S., {Winkler}, P.~F., {et~al.} 2012, \apj, 755,
  40

\bibitem[{{P{\'e}rez-Gonz{\'a}lez} {et~al.}(2006){P{\'e}rez-Gonz{\'a}lez},
  {Kennicutt}, {Gordon}, {Misselt}, {Gil de Paz}, {Engelbracht}, {Rieke},
  {Bendo}, {Bianchi}, {Boissier}, {Calzetti}, {Dale}, {Draine}, {Jarrett},
  {Hollenbach}, \& {Prescott}}]{PerezGonzalez+06}
{P{\'e}rez-Gonz{\'a}lez}, P.~G., {Kennicutt}, Jr., R.~C., {Gordon}, K.~D.,
  {et~al.} 2006, \apj, 648, 987

\bibitem[{{Persi} {et~al.}(2000){Persi}, {Tapia}, \& {Roth}}]{Persi+00}
{Persi}, P., {Tapia}, M., \& {Roth}, M. 2000, \aap, 357, 1020

\bibitem[{{Plume} {et~al.}(2004){Plume}, {Kaufman}, {Neufeld}, {Snell},
  {Hollenbach}, {Goldsmith}, {Howe}, {Bergin}, {Melnick}, \&
  {Bensch}}]{Plume+04}
{Plume}, R., {Kaufman}, M.~J., {Neufeld}, D.~A., {et~al.} 2004, \apj, 605, 247

\bibitem[{{Poglitsch} {et~al.}(2010){Poglitsch}, {Waelkens}, {Geis},
  {Feuchtgruber}, {Vandenbussche}, {Rodriguez}, {Krause}, {Renotte}, {van
  Hoof}, {Saraceno}, {Cepa}, {Kerschbaum}, {Agn{\`e}se}, {Ali}, {Altieri},
  {Andreani}, {Augueres}, {Balog}, {Barl}, {Bauer}, {Belbachir}, {Benedettini},
  {Billot}, {Boulade}, {Bischof}, {Blommaert}, {Callut}, {Cara}, {Cerulli},
  {Cesarsky}, {Contursi}, {Creten}, {De Meester}, {Doublier}, {Doumayrou},
  {Duband}, {Exter}, {Genzel}, {Gillis}, {Gr{\"o}zinger}, {Henning},
  {Herreros}, {Huygen}, {Inguscio}, {Jakob}, {Jamar}, {Jean}, {de Jong},
  {Katterloher}, {Kiss}, {Klaas}, {Lemke}, {Lutz}, {Madden}, {Marquet},
  {Martignac}, {Mazy}, {Merken}, {Montfort}, {Morbidelli}, {M{\"u}ller},
  {Nielbock}, {Okumura}, {Orfei}, {Ottensamer}, {Pezzuto}, {Popesso},
  {Putzeys}, {Regibo}, {Reveret}, {Royer}, {Sauvage}, {Schreiber}, {Stegmaier},
  {Schmitt}, {Schubert}, {Sturm}, {Thiel}, {Tofani}, {Vavrek}, {Wetzstein},
  {Wieprecht}, \& {Wiezorrek}}]{Poglitsch+10}
{Poglitsch}, A., {Waelkens}, C., {Geis}, N., {et~al.} 2010, \aap, 518, L2

\bibitem[{{Polcaro} {et~al.}(2006){Polcaro}, {Norci}, \&
  {Miroshnichenko}}]{Polcaro+06}
{Polcaro}, V.~F., {Norci}, L., \& {Miroshnichenko}, A.~S. 2006, in Astronomical
  Society of the Pacific Conference Series, Vol. 355, Stars with the B[e]
  Phenomenon, ed. M.~{Kraus} \& A.~S. {Miroshnichenko}, 197

\bibitem[{{Povich} {et~al.}(2017){Povich}, {Busk}, {Feigelson}, {Townsley}, \&
  {Kuhn}}]{Povich+17}
{Povich}, M.~S., {Busk}, H.~A., {Feigelson}, E.~D., {Townsley}, L.~K., \&
  {Kuhn}, M.~A. 2017, \apj, 838, 61

\bibitem[{{Povich} {et~al.}(2016){Povich}, {Townsley}, {Robitaille}, {Broos},
  {Orbin}, {King}, {Naylor}, \& {Whitney}}]{Povich+16}
{Povich}, M.~S., {Townsley}, L.~K., {Robitaille}, T.~P., {et~al.} 2016, \apj,
  825, 125

\bibitem[{{Povich} {et~al.}(2007){Povich}, {Stone}, {Churchwell}, {Zweibel},
  {Wolfire}, {Babler}, {Indebetouw}, {Meade}, \& {Whitney}}]{Povich+07}
{Povich}, M.~S., {Stone}, J.~M., {Churchwell}, E., {et~al.} 2007, \apj, 660,
  346

\bibitem[{{Povich} {et~al.}(2009){Povich}, {Churchwell}, {Bieging}, {Kang},
  {Whitney}, {Brogan}, {Kulesa}, {Cohen}, {Babler}, {Indebetouw}, {Meade}, \&
  {Robitaille}}]{Povich+09}
{Povich}, M.~S., {Churchwell}, E., {Bieging}, J.~H., {et~al.} 2009, \apj, 696,
  1278

\bibitem[{{Povich} {et~al.}(2011){Povich}, {Smith}, {Majewski}, {Getman},
  {Townsley}, {Babler}, {Broos}, {Indebetouw}, {Meade}, {Robitaille},
  {Stassun}, {Whitney}, {Yonekura}, \& {Fukui}}]{Povich+11a}
{Povich}, M.~S., {Smith}, N., {Majewski}, S.~R., {et~al.} 2011, \apjs, 194, 14

\bibitem[{{Pratap} {et~al.}(1999){Pratap}, {Megeath}, \& {Bergin}}]{Pratap+99}
{Pratap}, P., {Megeath}, S.~T., \& {Bergin}, E.~A. 1999, \apj, 517, 799

\bibitem[{{Price} {et~al.}(2001){Price}, {Egan}, {Carey}, {Mizuno}, \&
  {Kuchar}}]{Price+01}
{Price}, S.~D., {Egan}, M.~P., {Carey}, S.~J., {Mizuno}, D.~R., \& {Kuchar},
  T.~A. 2001, \aj, 121, 2819

\bibitem[{{Prisinzano} {et~al.}(2007){Prisinzano}, {Damiani}, {Micela}, \&
  {Pillitteri}}]{Prisinzano+07}
{Prisinzano}, L., {Damiani}, F., {Micela}, G., \& {Pillitteri}, I. 2007, \aap,
  462, 123

\bibitem[{{Prisinzano} {et~al.}(2005){Prisinzano}, {Damiani}, {Micela}, \&
  {Sciortino}}]{Prisinzano+05}
{Prisinzano}, L., {Damiani}, F., {Micela}, G., \& {Sciortino}, S. 2005, \aap,
  430, 941

\bibitem[{{Puga} {et~al.}(2010){Puga}, {Mar{\'{\i}}n-Franch}, {Najarro},
  {Lenorzer}, {Herrero}, {Acosta Pulido}, {Chavarr{\'{\i}}a}, {Bik}, {Figer},
  \& {Ram{\'{\i}}rez Alegr{\'{\i}}a}}]{Puga+10}
{Puga}, E., {Mar{\'{\i}}n-Franch}, A., {Najarro}, F., {et~al.} 2010, \aap, 517,
  A2

\bibitem[{{Rank} {et~al.}(1978){Rank}, {Dinerstein}, {Lester}, {Bregman},
  {Aitken}, \& {Jones}}]{Rank+78}
{Rank}, D.~M., {Dinerstein}, H.~L., {Lester}, D.~F., {et~al.} 1978, \mnras,
  185, 179

\bibitem[{{Reid} {et~al.}(2009){Reid}, {Menten}, {Zheng}, {Brunthaler},
  {Moscadelli}, {Xu}, {Zhang}, {Sato}, {Honma}, {Hirota}, {Hachisuka}, {Choi},
  {Moellenbrock}, \& {Bartkiewicz}}]{Reid+09}
{Reid}, M.~J., {Menten}, K.~M., {Zheng}, X.~W., {et~al.} 2009, \apj, 700, 137

\bibitem[{{Rela{\~n}o} {et~al.}(2007){Rela{\~n}o}, {Lisenfeld},
  {P{\'e}rez-Gonz{\'a}lez}, {V{\'{\i}}lchez}, \& {Battaner}}]{Relano+07}
{Rela{\~n}o}, M., {Lisenfeld}, U., {P{\'e}rez-Gonz{\'a}lez}, P.~G.,
  {V{\'{\i}}lchez}, J.~M., \& {Battaner}, E. 2007, \apjl, 667, L141

\bibitem[{{Retallack} \& {Goss}(1980)}]{Retallack+Goss80}
{Retallack}, D.~S., \& {Goss}, W.~M. 1980, \mnras, 193, 261

\bibitem[{{Rho} {et~al.}(2004){Rho}, {Ram{\'{\i}}rez}, {Corcoran}, {Hamaguchi},
  \& {Lefloch}}]{Rho+04}
{Rho}, J., {Ram{\'{\i}}rez}, S.~V., {Corcoran}, M.~F., {Hamaguchi}, K., \&
  {Lefloch}, B. 2004, \apj, 607, 904

\bibitem[{{Rho} {et~al.}(2006){Rho}, {Reach}, {Lefloch}, \& {Fazio}}]{Rho+06}
{Rho}, J., {Reach}, W.~T., {Lefloch}, B., \& {Fazio}, G.~G. 2006, \apj, 643,
  965

\bibitem[{{Rho} {et~al.}(2008){Rho}, {Kozasa}, {Reach}, {Smith}, {Rudnick},
  {DeLaney}, {Ennis}, {Gomez}, \& {Tappe}}]{Rho+08}
{Rho}, J., {Kozasa}, T., {Reach}, W.~T., {et~al.} 2008, \apj, 673, 271

\bibitem[{{Ritchie} {et~al.}(2009){Ritchie}, {Clark}, {Negueruela}, \&
  {Crowther}}]{Ritchie+09}
{Ritchie}, B.~W., {Clark}, J.~S., {Negueruela}, I., \& {Crowther}, P.~A. 2009,
  \aap, 507, 1585

\bibitem[{{Roberts} {et~al.}(2011){Roberts}, {van der Tak}, {Fuller}, {Plume},
  \& {Bayet}}]{Roberts+11}
{Roberts}, H., {van der Tak}, F.~F.~S., {Fuller}, G.~A., {Plume}, R., \&
  {Bayet}, E. 2011, \aap, 525, A107

\bibitem[{{Rodgers} {et~al.}(1960){Rodgers}, {Campbell}, \&
  {Whiteoak}}]{Rodgers+60}
{Rodgers}, A.~W., {Campbell}, C.~T., \& {Whiteoak}, J.~B. 1960, \mnras, 121,
  103

\bibitem[{{Romero} {et~al.}(2007){Romero}, {Okazaki}, {Orellana}, \&
  {Owocki}}]{Romero+07}
{Romero}, G.~E., {Okazaki}, A.~T., {Orellana}, M., \& {Owocki}, S.~P. 2007,
  \aap, 474, 15

\bibitem[{{Russeil}(2003)}]{Russeil03}
{Russeil}, D. 2003, \aap, 397, 133

\bibitem[{{Russeil} {et~al.}(2010){Russeil}, {Zavagno}, {Motte}, {Schneider},
  {Bontemps}, \& {Walsh}}]{Russeil+10}
{Russeil}, D., {Zavagno}, A., {Motte}, F., {et~al.} 2010, \aap, 515, A55

\bibitem[{{Russeil} {et~al.}(2011){Russeil}, {Pestalozzi}, {Mottram},
  {Bontemps}, {Anderson}, {Zavagno}, {Beltr{\'a}n}, {Bally}, {Brand}, {Brunt},
  {Cesaroni}, {Joncas}, {Marshall}, {Martin}, {Massi}, {Molinari}, {Moore},
  {Noriega-Crespo}, {Olmi}, {Thompson}, {Wienen}, \& {Wyrowski}}]{Russeil+11}
{Russeil}, D., {Pestalozzi}, M., {Mottram}, J.~C., {et~al.} 2011, \aap, 526,
  A151

\bibitem[{{Russeil} {et~al.}(2012){Russeil}, {Zavagno}, {Adami}, {Anderson},
  {Bontemps}, {Motte}, {Rodon}, {Schneider}, {Ilmane}, \&
  {Murphy}}]{Russeil+12}
{Russeil}, D., {Zavagno}, A., {Adami}, C., {et~al.} 2012, \aap, 538, A142

\bibitem[{{Sanduleak}(1971)}]{Sanduleak71}
{Sanduleak}, N. 1971, \apjl, 164, L71

\bibitem[{{Schneider} {et~al.}(2010){Schneider}, {Csengeri}, {Bontemps},
  {Motte}, {Simon}, {Hennebelle}, {Federrath}, \& {Klessen}}]{Schneider+10}
{Schneider}, N., {Csengeri}, T., {Bontemps}, S., {et~al.} 2010, \aap, 520, A49

\bibitem[{{Shuping} {et~al.}(2012){Shuping}, {Vacca}, {Kassis}, \&
  {Yu}}]{Shuping+12}
{Shuping}, R.~Y., {Vacca}, W.~D., {Kassis}, M., \& {Yu}, K.~C. 2012, \aj, 144,
  116

\bibitem[{{Sievers} {et~al.}(1991){Sievers}, {Mezger}, {Bordeon}, {Kreysa},
  {Haslam}, \& {Lemke}}]{Sievers+91}
{Sievers}, A.~W., {Mezger}, P.~G., {Bordeon}, M.~A., {et~al.} 1991, \aap, 251,
  231

\bibitem[{{Simon} {et~al.}(2001){Simon}, {Jackson}, {Clemens}, {Bania}, \&
  {Heyer}}]{Simon+01}
{Simon}, R., {Jackson}, J.~M., {Clemens}, D.~P., {Bania}, T.~M., \& {Heyer},
  M.~H. 2001, \apj, 551, 747

\bibitem[{{Skibba} {et~al.}(2011){Skibba}, {Engelbracht}, {Dale}, {Hinz},
  {Zibetti}, {Crocker}, {Groves}, {Hunt}, {Johnson}, {Meidt}, {Murphy},
  {Appleton}, {Armus}, {Bolatto}, {Brandl}, {Calzetti}, {Croxall}, {Galametz},
  {Gordon}, {Kennicutt}, {Koda}, {Krause}, {Montiel}, {Rix}, {Roussel},
  {Sandstrom}, {Sauvage}, {Schinnerer}, {Smith}, {Walter}, {Wilson}, \&
  {Wolfire}}]{Skibba+11}
{Skibba}, R.~A., {Engelbracht}, C.~W., {Dale}, D., {et~al.} 2011, \apj, 738, 89

\bibitem[{{Skiff}(2009)}]{Skiff09}
{Skiff}, B.~A. 2009, VizieR Online Data Catalog, 1

\bibitem[{{Skinner} {et~al.}(2003){Skinner}, {Gagn{\'e}}, \&
  {Belzer}}]{Skinner+03}
{Skinner}, S., {Gagn{\'e}}, M., \& {Belzer}, E. 2003, \apj, 598, 375

\bibitem[{{Smith} {et~al.}(1985){Smith}, {Bentley}, {Castelaz}, {Gehrz},
  {Grasdalen}, \& {Hackwell}}]{Smith+85}
{Smith}, J., {Bentley}, A., {Castelaz}, M., {et~al.} 1985, \apj, 291, 571

\bibitem[{{Smith} {et~al.}(1978){Smith}, {Biermann}, \& {Mezger}}]{SBM78}
{Smith}, L.~F., {Biermann}, P., \& {Mezger}, P.~G. 1978, \aap, 66, 65

\bibitem[{{Smith}(2006)}]{Smith06}
{Smith}, N. 2006, \mnras, 367, 763

\bibitem[{{Smith} \& {Brooks}(2007)}]{Smith+Brooks07}
{Smith}, N., \& {Brooks}, K.~J. 2007, \mnras, 379, 1279

\bibitem[{{Smith} {et~al.}(2003{\natexlab{a}}){Smith}, {Davidson}, {Gull},
  {Ishibashi}, \& {Hillier}}]{Smith+03b}
{Smith}, N., {Davidson}, K., {Gull}, T.~R., {Ishibashi}, K., \& {Hillier},
  D.~J. 2003{\natexlab{a}}, \apj, 586, 432

\bibitem[{{Smith} {et~al.}(2003{\natexlab{b}}){Smith}, {Gehrz}, {Hinz},
  {Hoffmann}, {Hora}, {Mamajek}, \& {Meyer}}]{Smith+03c}
{Smith}, N., {Gehrz}, R.~D., {Hinz}, P.~M., {et~al.} 2003{\natexlab{b}}, \aj,
  125, 1458

\bibitem[{{Stephens} {et~al.}(2014){Stephens}, {Evans}, {Xue}, {Chu},
  {Gruendl}, \& {Segura-Cox}}]{Stephens+14}
{Stephens}, I.~W., {Evans}, J.~M., {Xue}, R., {et~al.} 2014, \apj, 784, 147

\bibitem[{{Stephenson}(1966)}]{Stephenson66}
{Stephenson}, C.~B. 1966, \aj, 71, 477

\bibitem[{{Str{\"o}mgren}(1939)}]{Stromgren39}
{Str{\"o}mgren}, B. 1939, \apj, 89, 526

\bibitem[{{Subrahmanyan} \& {Goss}(1996)}]{Subrahmanyan+Goss96}
{Subrahmanyan}, R., \& {Goss}, W.~M. 1996, \mnras, 281, 239

\bibitem[{{Terebey} {et~al.}(2003){Terebey}, {Fich}, {Taylor}, {Cao}, \&
  {Hancock}}]{Terebey+03}
{Terebey}, S., {Fich}, M., {Taylor}, R., {Cao}, Y., \& {Hancock}, T. 2003,
  \apj, 590, 906

\bibitem[{{Thronson} {et~al.}(1985){Thronson}, {Lada}, \&
  {Hewagama}}]{Thronson+85}
{Thronson}, Jr., H.~A., {Lada}, C.~J., \& {Hewagama}, T. 1985, \apj, 297, 662

\bibitem[{{Tielens} {et~al.}(2005){Tielens}, {Waters}, \&
  {Bernatowicz}}]{Tielens+05}
{Tielens}, A.~G.~G.~M., {Waters}, L.~B.~F.~M., \& {Bernatowicz}, T.~J. 2005, in
  Astronomical Society of the Pacific Conference Series, Vol. 341, Chondrites
  and the Protoplanetary Disk, ed. A.~N. {Krot}, E.~R.~D. {Scott}, \&
  B.~{Reipurth}, 605

\bibitem[{{Tothill} {et~al.}(2008){Tothill}, {Gagn{\'e}}, {Stecklum}, \&
  {Kenworthy}}]{Tothill+08}
{Tothill}, N.~F.~H., {Gagn{\'e}}, M., {Stecklum}, B., \& {Kenworthy}, M.~A.
  2008, {The Lagoon Nebula and its Vicinity}, ed. B.~{Reipurth}, 533

\bibitem[{{Townsley} {et~al.}(2011){Townsley}, {Broos}, {Chu}, {Gruendl},
  {Oey}, \& {Pittard}}]{Townsley+11b}
{Townsley}, L.~K., {Broos}, P.~S., {Chu}, Y.-H., {et~al.} 2011, \apjs, 194, 16

\bibitem[{{Townsley} {et~al.}(2014){Townsley}, {Broos}, {Garmire}, {Bouwman},
  {Povich}, {Feigelson}, {Getman}, \& {Kuhn}}]{Townsley+14}
{Townsley}, L.~K., {Broos}, P.~S., {Garmire}, G.~P., {et~al.} 2014, \apjs, 213,
  1

\bibitem[{{Townsley} {et~al.}(2003){Townsley}, {Feigelson}, {Montmerle},
  {Broos}, {Chu}, \& {Garmire}}]{Townsley+03}
{Townsley}, L.~K., {Feigelson}, E.~D., {Montmerle}, T., {et~al.} 2003, \apj,
  593, 874

\bibitem[{{Treyer} {et~al.}(2010){Treyer}, {Schiminovich}, {Johnson}, {O'Dowd},
  {Martin}, {Wyder}, {Charlot}, {Heckman}, {Martins}, {Seibert}, \& {van der
  Hulst}}]{Treyer+10}
{Treyer}, M., {Schiminovich}, D., {Johnson}, B.~D., {et~al.} 2010, \apj, 719,
  1191

\bibitem[{{Turner} \& {Forbes}(1982)}]{Turner+82}
{Turner}, D.~G., \& {Forbes}, D. 1982, \pasp, 94, 789

\bibitem[{{Ungerechts} {et~al.}(2000){Ungerechts}, {Umbanhowar}, \&
  {Thaddeus}}]{Ungerechts+00}
{Ungerechts}, H., {Umbanhowar}, P., \& {Thaddeus}, P. 2000, \apj, 537, 221

\bibitem[{{van der Hucht} {et~al.}(1981){van der Hucht}, {Conti}, {Lundstrom},
  \& {Stenholm}}]{vanderHucht+81}
{van der Hucht}, K.~A., {Conti}, P.~S., {Lundstrom}, I., \& {Stenholm}, B.
  1981, \ssr, 28, 227

\bibitem[{{van der Werf} \& {Goss}(1989)}]{vanderWerf+Goss89}
{van der Werf}, P.~P., \& {Goss}, W.~M. 1989, \aap, 224, 209

\bibitem[{{van Loo}(2005)}]{vanLoo05}
{van Loo}, S. 2005, in Massive Stars and High-Energy Emission in OB
  Associations, ed. G.~{Rauw}, Y.~{Naz{\'e}}, R.~{Blomme}, \& E.~{Gosset},
  61--64

\bibitem[{{Vargas {\'A}lvarez} {et~al.}(2013){Vargas {\'A}lvarez},
  {Kobulnicky}, {Bradley}, {Kannappan}, {Norris}, {Cool}, \&
  {Miller}}]{VargasAlvarez+13}
{Vargas {\'A}lvarez}, C.~A., {Kobulnicky}, H.~A., {Bradley}, D.~R., {et~al.}
  2013, \aj, 145, 125

\bibitem[{{Voss} {et~al.}(2010){Voss}, {Diehl}, {Vink}, \&
  {Hartmann}}]{Voss+10}
{Voss}, R., {Diehl}, R., {Vink}, J.~S., \& {Hartmann}, D.~H. 2010, \aap, 520,
  A51

\bibitem[{{Vutisalchavakul} {et~al.}(2016){Vutisalchavakul}, {Evans}, \&
  {Heyer}}]{VEH16}
{Vutisalchavakul}, N., {Evans}, Neal~J., I., \& {Heyer}, M. 2016, \apj, 831, 73

\bibitem[{{Vutisalchavakul} \& {Evans}(2013)}]{V+E13}
{Vutisalchavakul}, N., \& {Evans}, II, N.~J. 2013, \apj, 765, 129

\bibitem[{{Wang} {et~al.}(2007){Wang}, {Townsley}, {Feigelson}, {Getman},
  {Broos}, {Garmire}, \& {Tsujimoto}}]{Wang+07}
{Wang}, J., {Townsley}, L.~K., {Feigelson}, E.~D., {et~al.} 2007, \apjs, 168,
  100

\bibitem[{{Wang} {et~al.}(2011){Wang}, {Feigelson}, {Townsley}, {Broos},
  {Getman}, {Wolk}, {Preibisch}, {Stassun}, {Moffat}, {Garmire}, {King},
  {McCaughrean}, \& {Zinnecker}}]{Wang+11}
{Wang}, J., {Feigelson}, E.~D., {Townsley}, L.~K., {et~al.} 2011, \apjs, 194,
  11

\bibitem[{{Ward-Thompson} \& {Robson}(1990)}]{WardThompson+90}
{Ward-Thompson}, D., \& {Robson}, E.~I. 1990, \mnras, 244, 458

\bibitem[{{Watson} \& {Hanson}(1997)}]{Watson+Hanson97}
{Watson}, A.~M., \& {Hanson}, M.~M. 1997, \apjl, 490, L165

\bibitem[{{Weingartner} \& {Draine}(2001)}]{Weingartner+01}
{Weingartner}, J.~C., \& {Draine}, B.~T. 2001, \apj, 548, 296

\bibitem[{{Wenger} {et~al.}(2000){Wenger}, {Ochsenbein}, {Egret}, {Dubois},
  {Bonnarel}, {Borde}, {Genova}, {Jasniewicz}, {Lalo{\"e}}, {Lesteven}, \&
  {Monier}}]{Wenger+00}
{Wenger}, M., {Ochsenbein}, F., {Egret}, D., {et~al.} 2000, \aaps, 143, 9

\bibitem[{{Westerhout}(1958)}]{Westerhout58}
{Westerhout}, G. 1958, \bain, 14, 215

\bibitem[{{Whiteoak} \& {Gardner}(1977)}]{Whiteoak+Gardner77}
{Whiteoak}, J.~B., \& {Gardner}, F.~F. 1977, Proceedings of the Astronomical
  Society of Australia, 3, 147

\bibitem[{{Whiteoak} \& {Uchida}(1997)}]{Whiteoak+Uchida97}
{Whiteoak}, J.~B.~Z., \& {Uchida}, K.~I. 1997, \aap, 317, 563

\bibitem[{{Wilson} {et~al.}(2003){Wilson}, {Boboltz}, {Gaume}, \&
  {Megeath}}]{Wilson+03}
{Wilson}, T.~L., {Boboltz}, D.~A., {Gaume}, R.~A., \& {Megeath}, S.~T. 2003,
  \apj, 597, 434

\bibitem[{{Wilson} {et~al.}(1970){Wilson}, {Mezger}, {Gardner}, \&
  {Milne}}]{Wilson+70}
{Wilson}, T.~L., {Mezger}, P.~G., {Gardner}, F.~F., \& {Milne}, D.~K. 1970,
  \aap, 6, 364

\bibitem[{{Wolk} {et~al.}(2006){Wolk}, {Spitzbart}, {Bourke}, \&
  {Alves}}]{Wolk+06}
{Wolk}, S.~J., {Spitzbart}, B.~D., {Bourke}, T.~L., \& {Alves}, J. 2006, \aj,
  132, 1100

\bibitem[{{Wood} \& {Churchwell}(1989)}]{Wood+Churchwell89}
{Wood}, D.~O.~S., \& {Churchwell}, E. 1989, \apjs, 69, 831

\bibitem[{{Wu} {et~al.}(2016){Wu}, {Bik}, {Bestenlehner}, {Henning},
  {Pasquali}, {Brandner}, \& {Stolte}}]{Wu+16}
{Wu}, S.-W., {Bik}, A., {Bestenlehner}, J.~M., {et~al.} 2016, \aap, 589, A16

\bibitem[{{Wynn-Williams} {et~al.}(1974){Wynn-Williams}, {Becklin}, \&
  {Neugebauer}}]{WynnWilliams+74}
{Wynn-Williams}, C.~G., {Becklin}, E.~E., \& {Neugebauer}, G. 1974, \apj, 187,
  473

\bibitem[{{Xu} {et~al.}(2011){Xu}, {Moscadelli}, {Reid}, {Menten}, {Zhang},
  {Zheng}, \& {Brunthaler}}]{Xu+11}
{Xu}, Y., {Moscadelli}, L., {Reid}, M.~J., {et~al.} 2011, \apj, 733, 25

\bibitem[{{Xu} {et~al.}(2009){Xu}, {Reid}, {Menten}, {Brunthaler}, {Zheng}, \&
  {Moscadelli}}]{Xu+09}
{Xu}, Y., {Reid}, M.~J., {Menten}, K.~M., {et~al.} 2009, \apj, 693, 413

\bibitem[{{Yao} {et~al.}(2000){Yao}, {Ishii}, {Nagata}, {Nakaya}, \&
  {Sato}}]{Yao+00}
{Yao}, Y., {Ishii}, M., {Nagata}, T., {Nakaya}, H., \& {Sato}, S. 2000, \apj,
  542, 392

\bibitem[{{Zasowski} {et~al.}(2009){Zasowski}, {Majewski}, {Indebetouw},
  {Meade}, {Nidever}, {Patterson}, {Babler}, {Skrutskie}, {Watson}, {Whitney},
  \& {Churchwell}}]{Zasowski+09}
{Zasowski}, G., {Majewski}, S.~R., {Indebetouw}, R., {et~al.} 2009, \apj, 707,
  510

\bibitem[{{Zeidler} {et~al.}(2015){Zeidler}, {Sabbi}, {Nota}, {Grebel}, {Tosi},
  {Bonanos}, {Pasquali}, {Christian}, {de Mink}, \& {Ubeda}}]{Zeidler+15}
{Zeidler}, P., {Sabbi}, E., {Nota}, A., {et~al.} 2015, \aj, 150, 78

\bibitem[{{Zeilik} \& {Lada}(1978)}]{Zeilik+Lada78}
{Zeilik}, II, M., \& {Lada}, C.~J. 1978, \apj, 222, 896

\bibitem[{{Zhang} {et~al.}(2014){Zhang}, {Moscadelli}, {Sato}, {Reid},
  {Menten}, {Zheng}, {Brunthaler}, {Dame}, {Xu}, \& {Immer}}]{Zhang+14}
{Zhang}, B., {Moscadelli}, L., {Sato}, M., {et~al.} 2014, \apj, 781, 89

\bibitem[{{Zubko} {et~al.}(1996){Zubko}, {Kre{\l}owski}, \&
  {Wegner}}]{Zubko+96}
{Zubko}, V.~G., {Kre{\l}owski}, J., \& {Wegner}, W. 1996, \mnras, 283, 577

\bibitem[{{Zurita} {et~al.}(2002){Zurita}, {Beckman}, {Rozas}, \&
  {Ryder}}]{Zurita+02}
{Zurita}, A., {Beckman}, J.~E., {Rozas}, M., \& {Ryder}, S. 2002, \aap, 386,
  801

\end{thebibliography}
\end{document}